
\magnification = \magstep0

\input eplain.tex

 \hsize=18.6truecm   \hoffset=-1.0truecm  \vsize=25.1truecm \voffset=-0.7truecm

\font\sc=cmr8   
\font\vsc=cmr7

\font\ninerm=cmr9
\font\eightrm=cmr8
\font\sixrm=cmr6

\font\ninei=cmmi9
\font\eighti=cmmi8
\font\sixi=cmmi6

\font\ninesy=cmsy9
\font\eightsy=cmsy8
\font\sixsy=cmsy6

\font\ninebf=cmbx9
\font\eightbf=cmbx8
\font\sixbf=cmbx6

\font\ninett=cmtt9
\font\eighttt=cmtt8

\hyphenchar\tentt=-1 
\hyphenchar\ninett=-1
\hyphenchar\eighttt=-1

\font\ninesl=cmsl9
\font\eightsl=cmsl8

\font\nineit=cmti9
\font\eightit=cmti8

\def\ninepoint{\def\rm{\fam0\ninerm}%
  \textfont0=\ninerm \scriptfont0=\sixrm \scriptscriptfont0=\fiverm
  \textfont1=\ninei \scriptfont1=\sixi \scriptscriptfont1=\fivei
  \textfont2=\ninesy \scriptfont2=\sixsy \scriptscriptfont2=\fivesy
  \textfont3=\tenex \scriptfont3=\tenex \scriptscriptfont3=\tenex
  \def\it{\fam\itfam\nineit}%
  \textfont\itfam=\nineit
  \def\sl{\fam\slfam\ninesl}%
  \textfont\slfam=\ninesl
  \def\bf{\fam\bffam\ninebf}%
  \textfont\bffam=\ninebf \scriptfont\bffam=\sixbf
   \scriptscriptfont\bffam=\fivebf
  \def\tt{\fam\ttfam\ninett}%
  \textfont\ttfam=\ninett
  \tt 
  \normalbaselineskip=11pt
  \let\sc=\sevenrm
  \let\big=\ninebig
  \setbox\strutbox=\hbox{\vrule height8pt depth3pt width1pt}%
  \normalbaselines\rm}

\def\eightpoint{\def\rm{\fam0\eightrm}%
  \textfont0=\eightrm \scriptfont0=\sixrm \scriptscriptfont0=\fiverm
  \textfont1=\eighti \scriptfont1=\sixi \scriptscriptfont1=\fivei
  \textfont2=\eightsy \scriptfont2=\sixsy \scriptscriptfont2=\fivesy
  \textfont3=\tenex \scriptfont3=\tenex \scriptscriptfont3=\tenex
  \def\it{\fam\itfam\eightit}%
  \textfont\itfam=\eightit
  \def\sl{\fam\slfam\eightsl}%
  \textfont\slfam=\eightsl
  \def\bf{\fam\bffam\eightbf}%
  \textfont\bffam=\eightbf \scriptfont\bffam=\sixbf
   \scriptscriptfont\bffam=\fivebf
  \def\tt{\fam\ttfam\eighttt}%
  \textfont\ttfam=\eighttt
  \tt 
  \normalbaselineskip=9pt
  \let\sc=\sixrm
  \let\big=\eightbig
  \setbox\strutbox=\hbox{\vrule height7pt depth2pt width1pt}%
  \normalbaselines\rm}

\def\sevenpoint{\def\rm{\fam0\sevenrm}%
  \textfont0=\sevenrm \scriptfont0=\sixrm \scriptscriptfont0=\fiverm
  \textfont1=\seveni \scriptfont1=\sixi \scriptscriptfont1=\fivei
}

\def\mathhexbox#1#2#3{\leavevmode\hbox{$\mathsurround=0pt 
                                       \mathchar"#1#2#3$}}
\def\copyright{{\ooalign{\hfil\raise.07ex\hbox{c}\hfil\crcr\mathhexbox20D}}}
\def\xleft{$\phantom{{\rm Vol.}~0}$} 

\def\absbaselines{\baselineskip=11pt \lineskip=0pt \lineskiplimit=0pt}
\def\sglbaselines{\baselineskip=10.4pt \lineskip=0pt \lineskiplimit=0pt}
\def\medbaselines{\baselineskip=10pt \lineskip=0pt \lineskiplimit=0pt}
\def\smlbaselines{\baselineskip=8pt \lineskip=0pt \lineskiplimit=0pt}
 \def\vss{\vskip 6pt} \def\vsss{\vskip 2pt}
\parskip = 0pt 
\def\bb{\kern -2pt}  \nopagenumbers
\def\makeheadline{\vbox to 0pt{\vskip-30pt\line{\vbox to8.5pt{}\the
                               \headline}\vss}\nointerlineskip}

\def\footnoterule{\kern-3pt \hrule width \hsize \kern 2.6pt \vskip 3pt}

\pretolerance=15000  \tolerance=15000
\def\ts{\thinspace}  \def\cl{\centerline}
\def\ni{\noindent}   \def\nhi{\noindent \hangindent=10pt}
                     \def\nnhi{\noindent \hangindent=17pt}
                     \def\ihi{\indent \hangindent=22pt}
                     \def\iihi{\parindent=22pt \indent \hangindent=22pt}
       \def\bk{\kern -0.3em}  \def\b{\kern -0.1em}
\def\r0{$\rho_0$}    
\def\0{\phantom{0}}  \def\1{\phantom{1}}  
\def\etal{{et~al.\ }}  

\def\gapprox{$_>\atop{^\sim}$}  \def\lapprox{$_<\atop{^\sim}$}
\def\ltapprox{\hbox{$<\mkern-19mu\lower4pt\hbox{$\sim$}$}}
\def\gtapprox{\hbox{$>\mkern-19mu\lower4pt\hbox{$\sim$}$}}

\def\mltapprox{\raise2pt\hbox{$<\mkern-19mu\lower5pt\hbox{$\sim$}$}}

\newdimen\sa  \def\sd{\sa=.1em  \ifmmode $\rlap{.}$''$\kern -\sa$
                                \else \rlap{.}$''$\kern -\sa\fi}
              \def\dgd{\sa=.1em \ifmmode $\rlap{.}$^\circ$\kern -\sa$
                                \else \rlap{.}$^\circ$\kern -\sa\fi}
\newdimen\sb  \def\md{\sa=.06em \ifmmode $\rlap{.}$'$\kern -\sa$
                                \else \rlap{.}$'$\kern -\sa\fi}
\def\s{\ifmmode ^{\prime\prime} \else $^{\prime\prime}$ \fi}
\def\min{\ifmmode ^{\prime} \else $^{\prime}$ \fi}

\def\etal{et~al.\ }

\def\omit#1{\empty}
\def\o#1{\empty}

\input colordvi
\def\B{\Blue}
\def\G{\Green}
\def\R{\Red}

\parindent=0pt

\headline={\leftskip = -0.15in
           \medbaselines\vbox to 0pt{\sc THE ASTROPHYSICAL JOURNAL, 
           000:000--000, Received 2011 September 12; Accepted 2011, October 18 \hfill\null

           \copyright\/ \vsc 2012. The American Astronomical Society. All Rights
           Reserved. Printed in U.S.A. \hfill\null}}

\cl {\null}

\sglbaselines

\vskip -7pt

\omit{Revised Parallel Sequence Hubble Classification}

\omit{Kormendy \& Bender}

\cl{A REVISED PARALLEL-SEQUENCE MORPHOLOGICAL CLASSIFICATION OF GALAXIES:}  
\cl{STRUCTURE AND FORMATION OF S0 AND SPHEROIDAL GALAXIES}

\vss

\cl{J{\sc OHN} K{\sc ORMENDY}$^{1,2,3}$ {\sc AND} R{\sc ALF} B{\sc ENDER}$^{2,3}$}\vsss

\cl{\vsc $^1$Department of Astronomy, University of Texas, Austin, TX 78712, USA;
         kormendy@astro.as.utexas.edu} 

\cl{\vsc $^2$Universit\"ats-Sternwarte, Scheinerstrasse 1, D-81679 M\"unchen, Germany}

\cl{\vsc $^3$Max-Planck-Institut f\"ur Extraterrestrische Physik,
                 Giessenbachstrasse, D-85748 Garching-bei-M\"unchen, Germany; 
                 bender@mpe.mpg.de}

\vss

\pretolerance=15000  \tolerance=15000

\parindent = 22pt
\cl {ABSTRACT}
\vsss
{\narrower\absbaselines

\ni\quad We update van den Bergh's parallel sequence galaxy classification in which S0 galaxies
form a sequence \hbox{S0a--S0b--S0c} that parallels the sequence Sa--Sb--Sc of spiral galaxies.  
The ratio $B/T$ of bulge to total light defines the position of a galaxy in this tuning fork diagram.  
Our classification makes one major improvement.  We extend 
the \hbox{S0a--S0b--S0c} sequence to spheroidal (``Sph'') galaxies that are positioned in parallel to
irregular galaxies in a similarly extended Sa--Sb--Sc--Im sequence.  This provides a natural ``home'' 
for spheroidals, which previously were omitted from galaxy classification schemes or inappropriately
combined with ellipticals.

      To motivate our juxtaposition of Sph and Im galaxies, we present photometry and bulge-disk 
decompositions of four rare, late-type S0s that bridge the gap between the more common S0b and Sph galaxies.  
NGC~4762 is an edge-on SB0bc galaxy with a very small classical-bulge-to-total ratio of $B/T = 0.13 \pm 0.02$.   
NGC~4452 is an edge-on SB0c galaxy with an even tinier pseudobulge-to-total ratio of $PB/T = 0.017 \pm 0.004$.  
It is therefore an S0c.  VCC 2048, whose published classification is S0, contains an edge-on disk, but its
``bulge'' plots in the structural parameter sequence of spheroidals.  It is therefore a disky~Sph.  
And NGC 4638 is similarly~a~``missing~link'' between S0s and Sphs -- it has a tiny bulge and an edge-on disk 
embedded in a Sph halo.  In an Appendix, we present photometry and bulge-disk decompositions
of all HST ACS Virgo Cluster Survey S0s that do not have published decompositions.  We use these data to
update the structural parameter correlations of Sph, S$+$Im, and E galaxies.  We show that Sph galaxies of 
increasing luminosity form a continuous sequence with the disks (but not bulges) of S0c--S0b--S0a galaxies.
Remarkably, the Sph{\ts}--{\ts}S0-disk sequence is almost identical to that of Im galaxies $+$ spiral galaxy
disks.  We review published observations for galaxy transformation processes, particularly ram-pressure 
stripping of cold gas.  We suggest that Sph galaxies are transformed, ``red and dead'' Scd--Im galaxies in the 
same way that many S0 galaxies are transformed, red and dead Sa--Sc spiral galaxies.  

\vsss


\cl{~~~~~{\it Subject{\ts\ts}headings:}{\kern 4pt}galaxies:{\ts}{\ts}elliptical{\ts}and{\ts}lenticular,{\ts}cD{\ts}--{\ts}galaxies:{\ts}evolution{\ts}--{\ts}galaxies:{\ts}photometry{\ts}--{\ts}galaxies:{\ts}structure}}

\parindent = 10pt

\doublecolumns\sglbaselines

\cl{\null}
\vskip -16pt
\cl {1.~\sc INTRODUCTION}
\vss

\pretolerance=15000  \tolerance=15000

\centerline{\null} \vskip -15pt

      Sidney van den Bergh's (1976) alternative to Hubble types puts
S0s in a sequence S0a--S0b--S0c that parallels the sequence \hbox{Sa--Sb--Sc} of spiral galaxies.
Only the ratio $B/T$ of bulge
to total light and not (e.{\ts}g.)~spiral arm pitch angle defines the position of a galaxy 
in the classification.  The motivation was the observation that many S0s 
have small $B/T$ values that are not consistent with the traditional interpretation that 
they~are a transition class between E and Sa galaxies. Rather,~they~are structurally 
similar to \hbox{Sa--Sc} galaxies.  Their lack of spiral structure, 
of substantial H{\ts}I~gas, and of obvious star formation was attributed to S $\rightarrow$ S0
conversion processes such as ram pressure stripping of cold gas by hot gas in galaxy clusters. 

\vss

       Van{\ts}den{\ts}Bergh's classification encodes important aspects of galaxy evolution and therefore
is a valuable complement to Hubble types.  It has had substantial impact.  Still, we believe 
that it has been underappreciated, because the importance of galaxy transformation has been unclear.  
Now, new observations and theoretical developments on a variety of 
evolution processes make galaxy transformation ``an idea whose time has come.''
With small revisions, re-introduction of parallel sequence classification is timely.  

      Figure 1 shows our proposed classification scheme.  The most important revision
is the addition of spheroidal (``Sph'') galaxies in parallel to irregulars.  The rest of
this paper explains and justifies this change.

\singlecolumn

\vskip 1truein

\vfill

\includegraphics{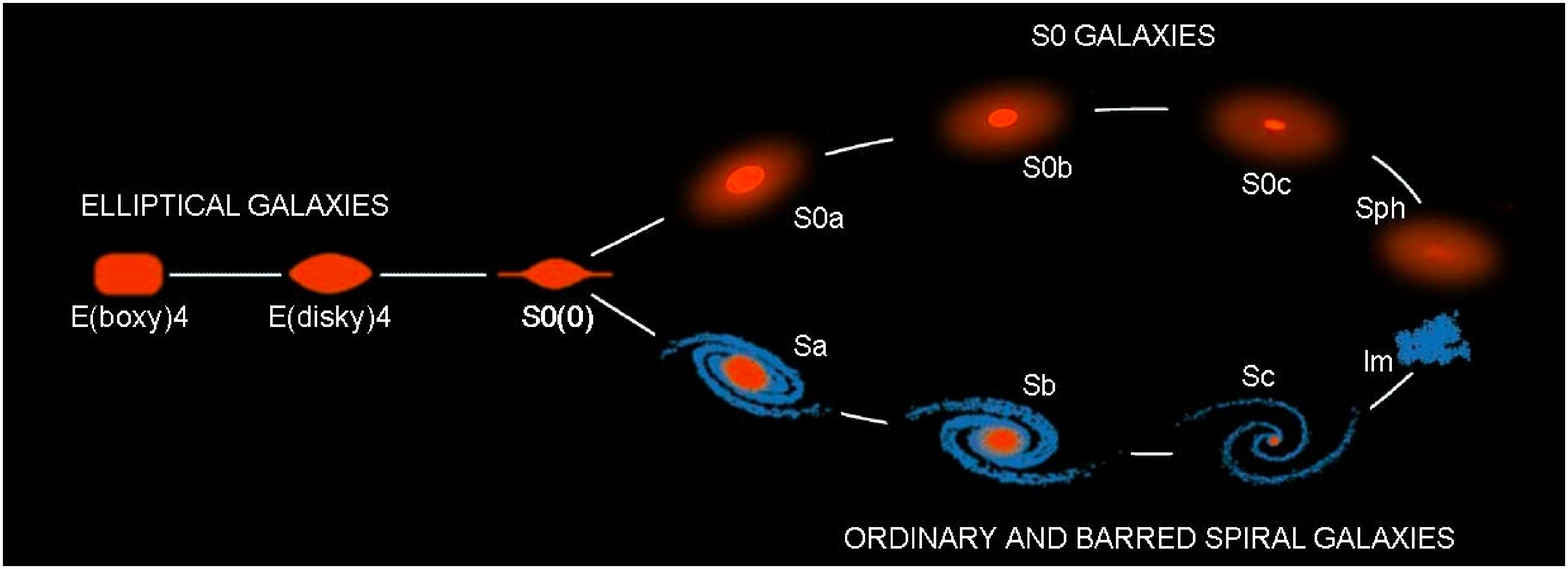}

Fig.~1~-- Revised parallel-sequence morphological classification of galaxies.
E types are from Kormendy \& Bender (1996).  Transition objects between
spirals and S0s (van den Bergh's anemic galaxies) exist but are not illustrated.
Bulge-to-total ratios decrease toward the right; Sc and S0c galaxies have tiny or no
pseudobulges. Sph and Im galaxies are bulgeless.

\eject

\doublecolumns

\headline={\vbox to 0pt{\leftskip = -0.15in
                        \folio\xleft \hfill KORMENDY \& BENDER \hfill Vol.~00}}

\headline={\vbox to 0pt{\leftskip = -0.15in
                        \folio\xleft \hfill PARALLEL-SEQUENCE GALAXY CLASSIFICATION \hfill Vol.~00}}

\def\sglbaselines{\baselineskip=11pt \lineskip=0pt \lineskiplimit=0pt}

\sglbaselines

      To motivate our juxtaposition of Sph and Im galaxies at the late end of the Hubble sequence, 
we update in \S\S\ts3\ts--\ts4 the observational evidence that Sphs are not dwarf examples
of elliptical galaxies.  With much larger samples of published parameters, we confirm the
results of Kormendy (1985, 1987, 2009) and Kormendy \etal (2009) that the correlations between 
half-light radius $r_e$ and surface brightness $\mu_e$ with each other and with absolute 
magnitude $M_V$ show that spheroidal and elliptical galaxies form disjoint sequences.  We show
for the first time that the sequence of Sphs is continuous with that of S0 disks (\S\ts5). 
And (\S\ts6) we confirm the results of the above papers that the Sph+S0 disk sequence is essentially 
indistinguishable from the sequence of spiral galaxy disks.  As in the above papers, we
conclude that spheroidal galaxies are ``red and dead'' dwarf S+Im galaxies that have been
transformed by a variety of internal and external evolution processes.  Thus they are closely 
similar to S0a--S0c galaxies; we summarize the evidence that these are transformed spiral galaxies.

       Adding Sphs resolves a puzzle with van den Bergh's classification that
has become more acute since 1976.  If~S0s are defunct spirals, where are the defunct Scs?
Van den Bergh (1976) classifies only the Hubble Atlas~galaxies (Sandage 1961).  They are few.    
But he lists 13 ellipticals (omitting two spheroidals), 11 E/S0a
and S0a galaxies, 5 S0a/b and S0b galaxies, and no S0c galaxies.  Among anemics, there are 4 Aa 
galaxies, 22 Aab and Ab galaxies, 5 Ab/c galaxies, and two Ac galaxies,~both~uncertain.
These statistics invite the interpretation that partial transformation from spiral to anemic galaxies
is easier than complete transformation from spiral to S0 galaxies and that it is easier to transform 
earlier-type galaxies.  Nevertheless, the complete lack of S0cs is remarkable considering that 
bulgeless progenitors are common and (in many cases) low in total mass.  Specifically, in 
terms of the finer-binned de Vaucouleurs (1959) classification, one could resonably expect to find
defunct Sd, Sm, and Im galaxies.  These are progressively fainter objects with progressively shallower 
gravitational potential wells.  Moreover, Sc galaxies do not contain classical bulges,
and later-type galaxies scarcely even contain pseudobulges (Kormendy \& Kennicutt 2004; 
Kormendy \etal 2010).  If spirals can be turned into S0s, then should it not be easier to 
transform the latest-type galaxies into ``bulgeless S0s''?  Where are they?

      We suggest that they are the spheroidal galaxies.

      In \S\ts3, we present new \hbox{photometry and bulge-disk} decompositions of rare, 
late-type S0 galaxies that further bridge the gap between the more common S0b galaxies and the
exceedingly common Sph galaxies.  NGC~4762 is an edge-on SB(lens)0bc galaxy with a
very small classical-bulge-to-total ratio of $B/T = 0.13 \pm 0.02$.   
NGC 4452 is an edge-on SB(lens)0 galaxy with an even tinier pseudobulge-to-total
ratio of $PB/T = 0.017 \pm 0.004$.  It is the first known true SB0c.  VCC 2048, whose 
published classification also is S0, proves to contain an edge-on disk, but its ``bulge'' 
parameters lie in the structural parameter sequence of Sph galaxies.  It is therefore an edge-on
Sph that still contains a disk.  Finally, NGC 4638 is an edge-on S0 with a spectacularly boxy Sph
halo.  In all respects, galaxy structural parameters are continuous 
between Sph galaxies and the disks (but not the bulges)~of S0c -- S0b -- S0a galaxies.  
This is consistent with S $\rightarrow$ Sph transformation.

\cl{\null}

\cl{\null}

      After this research was finished but before~this~paper was written, we became
aware that the ATLAS3D group independently propose a \hbox{parallel-sequence} classification
motivated by their kinematic results (Cappellari \etal 2011b; Krajnovi\'c 2011).  Also, after
this paper was refereed and resubmitted, Laurikainen \etal (2011) was posted; it includes
additional examples of S0c galaxies and discusses the connection with van den Bergh's (1976)
classification.  We kept our paper separate from the above to emphasize how three groups 
independently reach similar conclusions.  This is a sign that the ideas that we all 
discuss are robust.

\vss\vsss
\cl {2.~\sc E -- S0c GALAXIES }
\vss

      Before we focus on spheroidal galaxies, we update the motivation that underlies the E -- S0(0)
part of Figure 1.

\vss\vsss
\cl {2.1.~\it Elliptical Galaxies}
\vss

      The classification of elliptical galaxies is from Kormendy \& Bender (1996).
The physically important distinction is not among galaxies with different apparent flattenings,
which mostly reflect our viewing geometry.  Rather, it is between the two varieties of ellipticals,
as reviewed in Kormendy \& Bender (1996), Kormendy \etal (2009), and Kormendy (2009).  From this
last paper,

\nnhi \underbar{``Giant ellipticals} ($M_V$ \lapprox \ts$-21.5$; $H_0 = 72$ km s$^{-1}$ Mpc$^{-1}$)

\nnhi (1) have cores,  i.{\thinspace}e., central missing light with respect to and inward
          extrapolation of the outer S\'ersic profile;
  
\nnhi (2) rotate slowly, so rotation is unimportant dynamically; 

\nnhi (3) therefore are moderately anisotropic and triaxial; 

\nnhi (4) are less flattened (ellipticity $\sim${\thinspace}0.15) than smaller Es; 

\nnhi (5) have boxy-distorted isophotes; 

\nnhi (6) have S\'ersic (1968) function outer profiles with $n > 4$;

\nnhi (7) mostly are made of very old stars that are enhanced in $\alpha$ elements;

\nnhi (8) often contain strong radio sources, and 

\nnhi (9) contain X-ray-emitting gas, more of it in bigger Es.

\nnhi \underbar{Normal and dwarf true ellipticals} ($M_V$ \gapprox \ts{$-21.5$) generally 

\nnhi (1) are coreless and have central extra light with respect to an inward extrapolation
          of the outer S\'ersic profile;

\nnhi (2) rotate rapidly, so rotation is dynamically important to their structure; 

\nnhi (3) are nearly isotropic and oblate spheroidal, with axial dispersions $\sigma_z$
          that are somewhat smaller than the $\sigma_{r,\phi}$ in the equatorial plane; 

\nnhi (4) are flatter than giant ellipticals (ellipticity $\sim${\thinspace}0.3); 

\nnhi (5) have disky-distorted isophotes; 

\nnhi (6) have S\'ersic function outer profiles with $n$ \lapprox \ts4;

\nnhi (7) are made of (still old but) younger stars with only modest or no $\alpha$-element enhancement; 

\nnhi (8) rarely contain strong radio sources, and

\nnhi (9) rarely contain X-ray-emitting gas.''

      Here, ``dwarf true elliptical'' means compact ellipticals like M{\ts}32 that extend the
fundamental plane correlations of bigger elliptical galaxies to the lowest luminosities (\S\ts3).  

      We do not repeat here the many references to the papers that derived the above results. 
They are listed in the above reviews.  However, it is important to note the generally good agreement
between the above dichotomy and the results of the SAURON and ATLAS$^{3\rm D}$ surveys (e.{\ts}g.,
de Zeeuw \etal 2002;
Emsellem \etal 2004, 2007, 2011;
Cappellari \etal 2007, 2011a, b;
Krajnovi\'c \etal 2008, 2011;
McDermid \etal 2006): 

\cl{\null}

\cl{\null}

A -- The important difference between our analysis and that of
the SAURON group is that we decompose S0s into (pseudo)bulge and disk parts and then
treat each component separately, whereas the SAURON team treats S0s as single-component 
systems and derives one set of parameters (e.{\ts}g., anisotropy measure $\beta$) for each
galaxy.  Point C, below, results from this difference in analysis.

B -- The SAURON division of ellipticals into slow and fast rotators is almost 
identical to ours.  Instead of just choosing a value of their rotation parameter $\lambda$ 
at which to divide slow and fast rotators, we find the value that
is most consistent with the core--no-core division.  This value is $\lambda = 0.13$ instead
of $\lambda = 0.10$.  If we divide the Emsellem \etal (2007) sample of ellipticals at $\lambda = 0.13$, 
then the only exception to the core-rotation correlation (points 1~and~2 above) is that NGC 4458 
is an extra light galaxy that rotates slowly.  However, it is almost exactly circular, so it can 
be a rapid rotator that is viewed face-on.

C -- The SAURON group conclude that extra light ellipticals are very anisotropic, with
$\sigma_z \ll \sigma_r$ and $\sigma_\phi$.  We agree that $\sigma_z$ is in general 
smaller than the other two dispersion components, but we suspect that the large difference
found by the SAURON team results from the inclusion of disk light in S0s, which are all
coreless.  

D -- The SAURON group conclude that core galaxies are nearly isotropic.  They find moderate
triaxiality in some galaxies, but they omit the most anisotropic galaxies from their
statistics, because they cannot be fitted with three-integral models.  In fact,
they have analyzed their most anisotropic galaxies with triaxial models (e.{\ts}g., NGC 4365: 
van den Bosch \etal 2008), and these reveal the triaxiality.

We believe that the agreements in our pictures of the two varieties of ellipticals far outweigh
the differences. We also emphasize that many kinds of physical properties other than kinematics
combine to create the dichotomy listed above.  We therefore believe that the difference between 
the two kinds of ellipticals is robust.

The formation physics that underlies the E--E dichotomy is suggested in Kormendy{\ts}et{\ts}al.{\ts}(2009).
The ``smoking~gun'' differences are 1 (cores versus extra light) and 9 \hbox{(x-ray gas} is
or is not present).   In coreless galaxies, the distinct, extra light component above the inward 
extrapolation of the outer S\'ersic profile first seen by Kormendy (1999) strongly resembles
the distinct central components predicted in numerical simulations of mergers of galaxies that 
contain gas. In the simulations, the gas dissipates, falls toward the center, undergoes a starburst, 
and builds a compact stellar component that is distinct from the S\'ersic-function main body of the 
elliptical (e.{\ts}g.,
Mihos \& Hernquist 1994;
Hopkins \etal 2009a).
This led Kormendy (1999, see also C\^ot\'e \etal 2007; Kormendy \etal 2009) to suggest \underbar{how} 
the \hbox{E--E} dichotomy arose: The most recent major merger that made extra light ellipticals 
involved cold gas dissipation and a central starburst (it was ``wet''), whereas the most recent major
merger that made core ellipticals was dissipationless (``dry'').  Central to this picture is our
understanding that cores (i.{\ts}e., missing light with respect to the inward extrapolation of the
outer S\'ersic function) were scoured by binary BHs that were made in mergers and that flung stars
away from the center as they sunk toward their own eventual merger 
(Begelman \etal 1980; 
Ebisuzaki \etal 1991; 
Makino \& Ebisuzaki 1996; 
Quinlan \& Hernquist 1997; 
Faber \etal 1997; 
Milosavljevi\'c \& Merritt 2001; 
Milosavljevi\'c et al.~2002; 
Merritt 2006).  
This is important because it underscores the need for {\it major} mergers: Only they and not
minor mergers with progenitor mass ratios of (say) \lapprox \ts1/10 -- and therefore BH mass
ratios of \lapprox \ts1/10 -- can scour the large amounts of light (not) observed in cores
(Kormendy \& Bender 2009).
Observation 9 led Kormendy \etal (2009) to suggest \underbar{why} the E--E dichotomy arose. 
If energy feedback from (for example) active galactic nuclei (AGNs) requires a working surface of hot gas
(Kauffmann \etal 2008), then this is present in core galaxies but absent in extra light galaxies.  
This suggests that effects of energy feedback are a strong function of galaxy mass: they are weak 
enough in small Es not to prevent merger starbursts but strong enough in giant Es and their progenitors 
to make dry mergers dry.

      An additional aspect of the formation puzzle has become clearer since Kormendy \etal 
(2009):~It~may~explain~the difference between the small S\'ersic indices $n \sim 3 \pm 1$ of extra light
ellipticals and the much larger \hbox{$n \sim 5$ to 12} of core galaxies.~The S\'ersic indices of extra
light Es are consistent with those found in simulations of single~major merger events
(Hopkins \etal 2009a).  The large S\'ersic indices of core galaxies are not explained in
Kormendy \etal (2009).  The extra light in these galaxies that converts
$n \simeq 3 \pm 1$ into $n \gg 4$ may be the debris accumulated in many minor mergers (e.{\ts}g.,
Naab \etal 2009;
Hopkins \etal 2010;
van Dokkum \etal 2010;
Oser \etal 2010, 2011).

\vss\vsss
\cl {2.2.~\it S0(0) Galaxies}
\vss

      We retain class S0(0) for the largest-$B/T$ S0s that really are transition galaxies between 
ellipticals and spirals.  It has long been clear that there is a complete continuum
between ellipticals and \hbox{large-$B/T$ S0s}.  However, we do not suggest a $B/T$ value at
which to divide S0(0) and S0a galaxies.  To determine a physically meaningful value requires
bulge-disk decompositions of large numbers of Sa galaxies.  These are not available.  On the
other hand, we know of no physics that depends on the exact definition of S0(0) galaxies
or on their distinction from disky-distorted ellipticals 
(Bender 1987;
Bender \etal 1987, 1988, 1989).

      Note that S0(0) galaxies are -- to our knowledge -- never barred.  Their disk-to-total
luminosity ratios $D/T$ are so small that their disks are not self-gravitating.  Under these
circumstances, a bar instability is impossible.  It is similarly impossible to have a bar in
an elliptical galaxy.

\vss\vsss
\cl {2.3.~\it Comments About the Parallel S and S0 Sequences}
\vss

      Unbarred and barred galaxies are not distinguished in Figure 1.  This difference is important 
in many contexts; it is embodied in Hubble\ts--{\ts}Sandage\ts--{\ts}de{\ts}Vaucouleurs classes.  We 
do not suggest that~Fig.~1 should replace Hubble classes.  Each classification
is designed for the astrophysical context in which it is useful.  We focus on the difference 
between gas-rich, star forming galaxies and gas-poor, mostly non-star-forming galaxies.  The 
tines of the tuning-fork diagram correspond roughly to the red sequence and blue cloud 
in the SDSS color-magnitude diagram 
(Strateva et al.~2001;
Bernardi \etal 2003;
Hogg \etal 2002, 2004;
Blanton \etal 2003, 2005;
Baldry \etal 2004).

      Minor comments:

In Fig.~1, E -- S0(0) galaxies are illustrated edge-on but S0a -- Sph and Sa -- Im
galaxies are illustrated at viewing angles intermediate between edge-on and face-on.
This is consistent with past versions of the tuning fork diagram.

Figure 1 shows all galaxies similar in size.  In reality, Sph and Sd{\ts}--{Im} 
galaxies are smaller than earlier types.

The stage-dependent separation of the tuning-fork tines in Fig.~1 is deliberate.
Sph galaxies and irregulars are~more similar than (say) Sc and S0c galaxies.  {\it Unlike\/} 
spirals, some Im galaxies look like Sphs at 3.6\ts$\mu$m; i.{\ts}e., the 
underlying old galaxy is Sph-like (Buta \etal 2010).  The increasing similarity of all 
morphologies at the lowest luminosities has been emphasized by van den Bergh
(1977, 2007).

The relative numbers of galaxies depend importantly on stage along each sequence.  This is
discussed in \S\ts8.

\vss\vsss\vsss
\cl {3.~\sc FAMILIES OF STELLAR SYSTEMS: THE E -- Sph DICHOTOMY}

\vss\vsss
\cl {3.1.~\it From Classical Morphology to Physical Morphology:}
\cl {\it Reasons for the Name ``Spheroidal Galaxy''}
\vss

      At the start of research on a new kind of object -- fish, rocks, planets, stars, or
galaxies -- it is useful to classify the objects under study into ``natural groups''~(Morgan~1951)
that isolate common features.  Sandage \& Bedke (1994) and Sandage (2004) emphasize that,
at this stage, no attempt must be made to attach physical interpretation to the classification.  
Nevertheless, the morphologist must make choices about which features to use in constructing the classification. 
{\it A classification scheme remains useful as the subject matures only if the natural groups succeed in 
ordering objects in a physically interpretable way.}   Hubble knew this.  He chose to use parameters that 
later became central to our understanding when he set up his galaxy classification (Hubble 1936; Sandage 1961, 
1975; Sandage \& Bedke 1994; see de Vaucouleurs 1959 for refinements).    
Sandage recognizes Hubble's genius in the Carnegie Atlas of 
Galaxies: ``Hubble correctly guessed that the presence or absence of a disk, the openness of the spiral-arm 
pattern, and the degree of resolution of the arms into stars, would be highly relevant.  It was an indefinable 
genius of Hubble that enabled him to understand in an unknown way \dots\ that this start to galaxy 
classification had relevance to nature itself.''  That is, in setting up his classification, Hubble made 
choices with future interpretation in mind. The result is successful precisely because it orders galaxies by 
properties that reflect essential physics.

      Nevertheless, it is no surprise that a purely descriptive classification can miss essential physics.
One reason is that convergent evolution happens.  In the animal kingdom on Earth, dolphins look like fish
but instead are mammals.  We understand how convergent evolution has engineered this similarity.  In the
same way, we argue that convergent evolution has engineered two kinds of galaxies -- ellipticals and
spheroidals -- that look similar but that have different origins.  That difference is the main subject of
this paper.

      The similarity between E and Sph galaxies that gets classical morphology into trouble 
is that both lack cold gas and (by and large) stellar disks. Gas dissipation cannot make lumpy 
structure or young stars.  Phase mixing and relaxation quickly \hbox{smooth isophotes into near-ellipses}.
Then both kinds of galaxies satisfy the definition (no~disk; smooth, nearly elliptical isophotes;
and no young~stars: Sandage 1961) of an elliptical galaxy.  In most papers, spheroidals are called
``dwarf elliptical'' or ``dE'' galaxies.  However, beginning in \S\ts3.2, we argue that ellipticals
are made via major mergers whereas ``dE galaxies'' are defunct late-type 
galaxies that lost their gas by transformation processes that do not involve mergers.  And 
we show that ``dE galaxies'' are recognizably different from ellipticals when we look beyond descriptive
morphology and measure structural parameters.  Because ``dE galaxies'' are not dwarf examples of
ellipticals -- as, e.{\ts}g., dwarf elephants should be dwarf examples of elephants -- we do not call
them ``dwarf ellipticals''.  However, it is impractical and unnecessary to drastically change 
established names.  The term ``dwarf spheroidal'' has long been used for the smallest ``dEs''
that are companions of our Galaxy (e.{\ts}g., Draco).  We therefore call all ``dE galaxies''
spheroidals.  Tiny objects like Draco are dwarf spheroidals.  Large ones like NGC 205 in the Local Group
and the many similar objects in the Virgo cluster are just called spheroidals.  This terminology
follows Kormendy (1985, 1987, 2009), Kormendy \& Bender (1994), and Kormendy \etal (2009).  Recognizing
that E and Sph galaxies are different is the first step in understanding why
we place Sphs at the late end of the parallel-sequence classification.

      We emphasize:~the name ``spheroidal'' is not meant to imply anything about the intrinsic shapes of these
galaxies.

      Readers will note -- perhaps with concern -- that we appear to be injecting interpretation into~our~classification.  
However:~{\it We use interpretation mainly to guide~our~choice of which observations to use in constructing our 
classification. The main difference from classical morphology is that we use quantitative measurements of structural 
parameters and not just visual impressions to define the classification.}  Hubble, Sandage, and de Vaucouleurs 
also made (slightly different) choices about which observations to embody in their classifications.

      At the same time, we do not try to fix~what~isn't~broken.  Our aim is not to replace Hubble
classification.  Rather, we try to improve a conceptual tool that is useful~in~parallel (no pun intended) with
other tools.  We propose a step in the development of a ``physical morphology'' 
(Kormendy 1979a, b, 1982; Kormendy \& Bender 1996) whose aim is to construct a classification that is based on 
our physical understanding of galaxies.  Quoting Kormendy (2004), ``I believe that we are now approaching the end 
of what we can accomplish by separating morphology and physics.  I would like~to~argue that we must break down the 
wall between morphology and interpretation.  Doing this successfully has always been a sign of the maturity of a 
subject.~For~example, it has happened in stellar astronomy.  I cannot imagine that people who observe and classify 
phenomena without interpretation would ever discover solar oscillations.  Without guidance from a theory, how would 
one ever conceive of the complicated measurements required to see solar oscillations or to use them to study the 
interior structure of the Sun?  In the same way, we need the guidance of a theory to make sense of the bewildering 
variety of phenomena associated with galaxies and to recognize what is fundamental and what is not.  Sandage (2004)
\dots~opened the door to such a phase when he wrote \dots~that morphology and interpretation must be kept separate 
`at least until the tension between induction and deduction' gets mature enough.  [Our aim here is to develop] 
physical morphology, \dots~not as a replacement for classical morphology -- which remains vitally important~-- 
but as a step beyond it.  Physical morphology is an iteration in detail that is analogous to de Vaucouleurs's 
iteration beyond the Hubble tuning fork diagram.''

     If we are successful, then parallel sequence classification will turn out to embody essential physics in
ways that Hubble types do not.

\eject

\hsize=18.6truecm   \hoffset=-0.8truecm  \vsize=24.42truecm \voffset=-0.3truecm

\singlecolumn

\cl{\null} \vskip 6.5truein

\vfill

\includegraphics{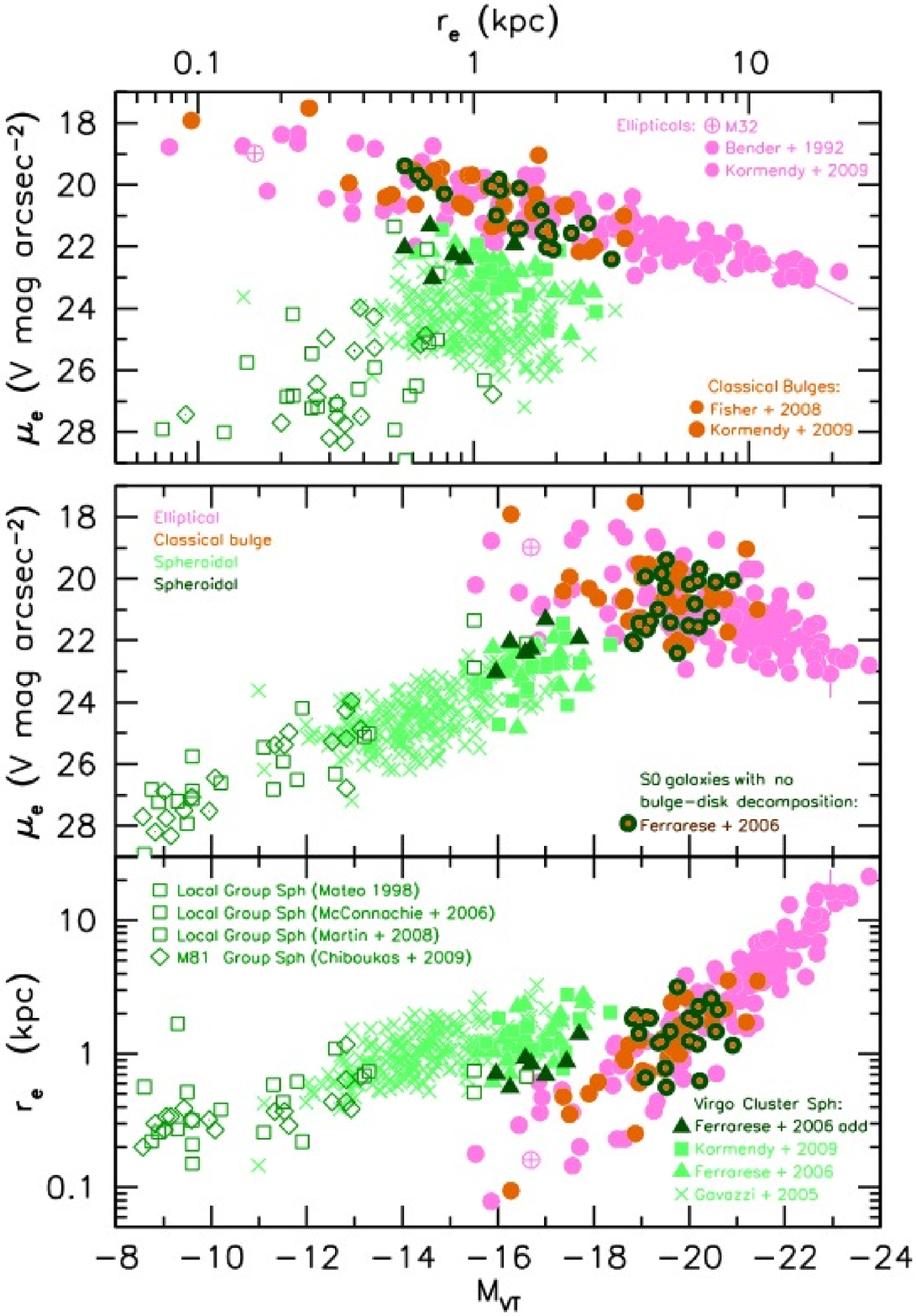}

Fig.~2~-- Global parameter correlations for ellipticals (pink), classical bulges (light brown),
and spheroidals (light green) from Kormendy \etal (2009: KFCB) and from Kormendy (2009).  Local
Group Sphs have been updated and M{\ts}81~Group Sphs have been added. 
The bottom panels show effective radius $r_e$ and surface brightness $\mu_e$ at the effective 
radius versus galaxy absolute magnitude.  The top panel is the $\mu_e$\ts--\ts$r_e$ or Kormendy (1977b)
relation, which shows the fundamental plane almost edge-on.  Data sources are given in the keys.  
Light green filled squares show the Sph galaxies whose brightness profiles were measured in KFCB.  
Green triangles show all galaxies from Ferrarese et al.~(2006, hereafter F2006) that were classified 
as dE there and that were not remeasured in KFCB.  Crosses show all spheroidals from Gavazzi et al.~(2005:
``dE'' there) that are not in KFCB or in F2006.  Dark green triangles show the 7 galaxies from F2006 that
were were omitted from KFCB because they were typed E, S0, or dS0 but that turn out to be spheroidals.  
Dark green filled circles with brown centers show the 23 remaining galaxies from F2006 that were omitted
in KFCB because they were typed S0 by F2006 but did not have bulge-disk decompositions published
there.  They are plotted here using the F2006 parameters measured for the whole galaxy, i.{\ts}e., 
for the bulge and disk together.  To properly study bulge and disk scaling relations, bulge-disk 
decompositions are required; these are carried out in \S\ts4 and in the Appendix.  However, even the 
whole-galaxy parameters do not conflict with the KFCB conclusion that spheroidals define a parameter
sequence that is very different from the fundamental plane of classical bulges and 
ellipticals.  This point was emphasized in Appendix~B~of~KFCB.  Section\ts4 uses these correlations to 
illustrate results of bulge-disk decompositions, to demonstrate that Sphs are continuous in 
parameter space with the disks (but not the bulges) of S0 galaxies, and to show that the disks of 
S0 and S$+$Im galaxies have similar correlations.  This provides justification for juxtaposing Sph 
and Im galaxies in the parallel-sequence classification.
\pretolerance=15000  \tolerance=15000 

\eject

\doublecolumns

\cl {3.2.~\it The Elliptical -- Spheroidal Dichotomy}
\vss

\cl {3.2.1.~\it Discovery}
\vss

       Wirth \& Gallagher (1984) were the first to suggest that compact dwarfs like M{\ts}32 and not diffuse
dwarfs~like NGC 205 are the low-luminosity versions of giant ellipticals.~The existence of free-flying analogs 
of M{\ts}32 implied to them that the compactness of better known dwarfs such as M{\ts}32, NGC 4486B, and NGC 5846A 
(Faber 1973) is not due only to tidal truncation by giant galaxy neighbors.  Wirth and Gallagher suggested that 
E and Sph galaxies form disjoint luminosity sequences that overlap for $-15$ \gapprox \ts$M_B$ \gapprox \ts$-18$ but 
that differ in mean surface brightness at $M_B = -15$ ``by nearly two orders of magnitude''.  An implication 
is that the luminosity function of true ellipticals is bounded and that M{\ts}32 is one of the faintest examples. 
Sandage et al.~(1985a, b) and Binggeli et al.~(1988) confirmed this result for Virgo cluster galaxies (\S\ts3.2.3 here).

      Kormendy (1985, 1987) used high-spatial-resolution photometry from the 
Canada-France-Hawaii Telescope to demonstrate that elliptical and spheroidal galaxies show a clearcut dichotomy in 
parameter space.    Ellipticals form a well defined sequence from cD galaxies to dwarfs like M{\ts}32.  
Lower-luminosity ellipticals have higher central surface brightnesses, whereas lower-luminosity spheroidals have
lower central surface brightnesses.  Far from extending the E parameter correlations to low luminosities, 
spheroidals show almost the same correlations as spiral-galaxy disks and Magellanic irregulars.  The above results
are based on near-central galaxy properties, but they are also seen in global properties (Kormendy 1987;
Binggeli \& Cameron 1991; Bender \etal 1992, 1993).  They are also confirmed with larger galaxy samples (Kormendy \&
Bender 1994).

      Kormendy (1985, 1987) concluded that E and Sph galaxies are distinct types of stellar systems that formed
differently.  Spheroidals are not dwarf ellipticals; they are physically related to S+Im galaxies.~They may be 
late-type galaxies that lost their gas or processed it all into stars.  Relevant evolution processes already known 
at that time included supernova-driven energy feedback (Saito 1979; Dekel \& Silk 1986); ram-pressure gas 
stripping 
(Gunn \& Gott 1972;
Lin \& Faber 1983; 
Kormendy 1985, 1987), 
and stochastic starbursts (Gerola et al.~1980, 1983). 
Section 8 discusses these processes.

      Kormendy \etal (2009:~KFCB) update and confirm the above results with a sample that includes all known 
elliptical galaxy members of the Virgo cluster and a large number of spheroidal galaxies.  They carry out photometry 
on a variety of images with different spatial scales, fields of view, and PSF resolutions; this provides more robust 
composite profiles over larger radius ranges than were available before.  The resulting correlations between effective 
radius $r_e$, surface brightness $\mu_e$ at $r_e$, and absolute magnitude $M_{VT}$ confirm the above conclusions.  
KFCB refute criticisms of the dichotomy as summarized here in \S\ts3.2.2.  And they provide a detailed review of 
formation processes -- major mergers for ellipticals and gas-removal transformation processes for spheroidals.  
This paper enlarges on that work.

      Figure 2 shows the $r_e$ -- $\mu_e$ -- $M_{VT}$ correlations from Kormendy (2009).  Ellipticals from Bender 
\etal (1992) and classical bulges from Fisher \& Drory (2008) are added to increase the sample size further.  
Figure 2 is the starting point for the present study.

\vss
\cl {3.2.2.~\it Published Criticisms and Our Responses}
\vsss\vsss

      The E -- Sph dichotomy has been challenged by many papers in the past decade.
The main arguments and our responses to them are as follows: \vss

\ihi  1. Sph and E galaxies have surface brightness profiles that are well described by S\'ersic functions;
         both together show a continuous correlation of S\'ersic index $n$ with galaxy luminosity (Jerjen \& 
         Binggeli 1997).
\vss

\iihi We agree with this observation (Figure 33 of KFCB).  However, conclusions about which objects are or are
      not physically related should not be based on just one measured parameter.  
\vss\parindent=10pt

\ihi  2. ``The striking dichotomy observed by Kormendy (1985) could be due to the lack, in Kormendy's sample,
         of galaxies in the $-20$ mag $<$ $M_B$ $<$ $-18$ mag range, corresponding precisely to the transition
         region between the two families.''  This quote is from Ferrarese \etal (2006), but the same criticism
         was also made by Binggeli (1994) and by Graham \& Guzm\'an (2003, 2004).
\vss

\iihi There was a 2-mag range in $M_B$ in which Kormendy (1985, 1987) had no bulges and only one spheroidal,
      but the two sequences were clearly diverging outside this magnitude range.  In any case, sample size and
      $M_{VT}$ gaps are not an issue in Kormendy \etal (2009) or in Figure 2 here.
\vss\parindent=10pt

\ihi  3. Spheroidal galaxies and coreless (``power-law'')~Es form a single sequence in fundamental
         plane parameter space from which core ellipticals deviate because of the light missing in cores
         (Graham \& Guzm\'an 2003, 2004; Gavazzi \etal 2006).
\vss

\iihi This is wrong.  The fraction of the galaxy light that is missing in cores ranges from almost 0\ts\%
      to just over 2\ts\% (Table 1 and Figure 41 in KFCB).  The effects of the missing light on global parameters
      is negligible.  In any case, bulges and coreless ellipticals together define a sequence that extends toward
      more compact objects (to the left, in Figure 2) than any spheroidals.  That sequence also is extended
      seamlessly toward more fluffy objects by other coreless ellipticals and by ellipticals with cores.  The E
      and Sph sequences do not quite join up in Figure 2, and they are more disjoint when more central parameters
      are measured.
\vss\parindent=10pt

\ihi  4. The ellipticals that are more compact than (in Fig.~2, to the left of) the continuous Sph-coreless-E
      sequence are pathological: They all have close, giant-galaxy companions, and they are all tidally stripped
      remnants of much bigger ellipticals (Ferrarese \etal 2006; Chen \etal 2010).
\vss

\iihi There are many reasons why we believe that tidal truncation cannot be the reason why the E sequence extends
      to more compact objects than Sphs: (i) Not all compact ellipticals are companions of bright galaxies.  Some
      are fairly isolated (e.{\ts}g., VCC 1871, which is $\sim 12 r_e({\rm NGC\ts4621)}$ from the giant elliptical
      NGC 4621; IC 76 = VCC 32).  (ii) Compact ellipticals do not  systematically have small S\'ersic indices
      indicative of outer truncation; instead, they have the same range of S\'ersic indices $n \sim 2$ to
      3.5 as isolated coreless ellipticals.  In particular, M{\ts}32 has $n \simeq 2.9$, larger than the median
      value for isolated coreless ellipticals.  These S\'ersic indices are exactly as found by $n$-body simulations
      of major galaxy mergers (Hopkins \etal 2009a; see also van Albada 1982).  (iii) Many Sph galaxies also are 
      companions of bright galaxies, but we do not argue that they have been truncated into compact 
      objects.  An example is NGC~205.  (iv) The compact end of the E sequence is also defined by tiny bulges 
      (Figure~2).  Classical bulges and ellipticals have closely similar FP correlations.  The classical bulges
      that appear in our correlation diagrams do not have bright companion galaxies.  (v) Our intuition about
      tidal truncation comes from globular clusters.  They have no ``protective'' dark matter halos, and the 
      truncator is overwhelmingly more massive than the victim.  Elliptical galaxies certainly can be tidally
      distorted, but the dark matter tends to be more distorted, so the galaxies merge relatively quickly 
      once tidal effects start.  We conclude that some compact Es may have been pruned slightly but that tidal
      truncation is not the reason why the E sequence extends to the left of where it is approached by the
      Sph sequence in Figure 2.
\vss\parindent=10pt

\cl{\null}

\ihi  5.\ts~C\^ot\'e \etal (2007), Chen \etal 2010, and Glass \etal (2011) argue that KFCB included only a 
      biased subsample of the ACS Virgo Cluster Survey galaxies.  Quoting Glass \etal (2011): ``K09 excluded 
      60\ts\% of the ACSVCS sample -- in particular, the vast majority of the galaxies in the 
      $-21.5$ \lapprox \ts$M_B$ \lapprox \ts$-18.5$ range.'' 
\vss

\iihi This statement is wrong.  KFCB did surface photometry of 40 of the 100 ACS VCS galaxies,
      i.{\ts}e., all ellipticals known to be cluster members when their survey was begun plus 
      five S0s and 10 Sphs.  The latter were included because, absent detailed photometry, it was 
      not known whether they are Es.  {\it In addition, correlation diagrams such as Figure~2 here 
      included 26 ACS VCS Sph galaxies~with parameters from Ferrarese \etal (2006:~{\it green triangles
      in KFCB and in this paper\/}).}  Two more galaxies did not have parameters in Ferrarese \etal (2006) 
      because of dust.
      Therefore KFCB omitted 32 -- not 60 -- viable ACS VCS galaxies.  Most are S0s.  KFCB studied 
      ellipticals and objects that get confused with ellipticals.  The study of S0s -- including the 
      necessary bulge-disk decomposition -- was postponed until future papers.  This is the first of
      those papers.  Adding ACS VCS S0s is the subject of \S\ts4 and the Appendix of this~paper. 
\vss\parindent=10pt

\singlecolumn

\vskip -0pt
\cl {3.2.3.~\it The Luminosity Functions of Elliptical and Spheroidal Galaxies Are Different}
\vsss\vsss

      Figure 3 shows the luminosity functions of Virgo cluster elliptical and spheroidal galaxies
as determined by Sandage \etal (1985b).  As these authors emphasize, the luminosity functions
are remarkably different.  And as these authors recognize, this is additional evidence that 
ellipticals and spheroidals are different kinds of galaxies.  Particularly remarkable is the fact
that the E and Sph luminosity functions overlap.  This means that {\it Sandage and collaborators 
can distinguish between E and Sph galaxies even when they have the same brightness.}
Consider how remarkable this statement is:

\cl{\null} \vskip 3.6truein

\vfill

\includegraphics{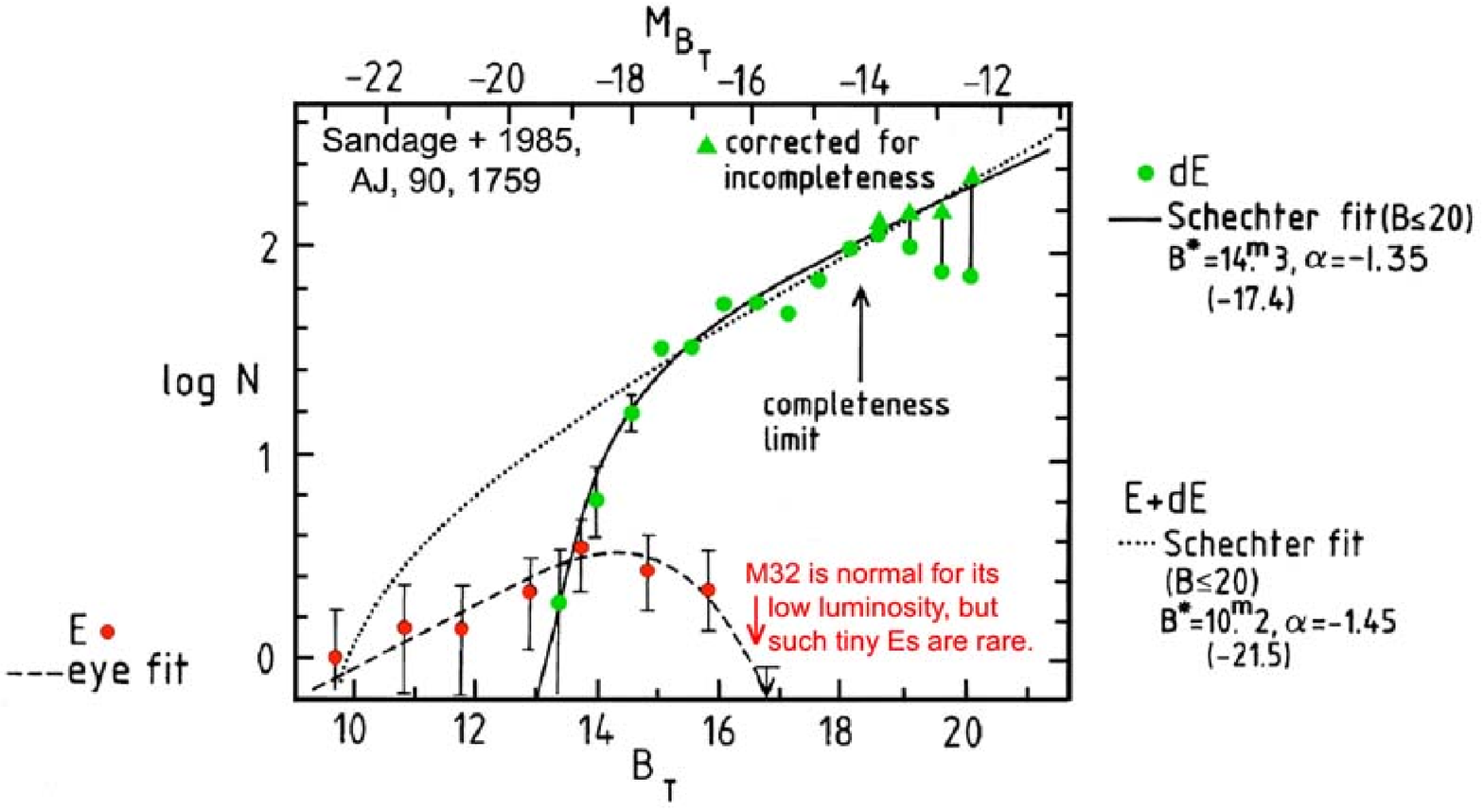}

Fig.~3~-- Luminosity functions of elliptical and spheroidal galaxies in the Virgo cluster.
This figure is adapted from Sandage \etal (1985b), who used the traditional name ``dE'' for
spheroidals.  We have updated the Hubble constant from $H_0 = 50$ to 70 km s$^{-1}$ Mpc$^{-1}$. 
Note that magnitudes are in $B$ band here but in $V$ band in the rest of this paper.
\pretolerance=15000  \tolerance=15000 
\eject

\hsize=18.6truecm   \hoffset=-0.8truecm  \vsize=24.42truecm \voffset=-0.3truecm

\doublecolumns

        A dwarf version of a creature is one that, when mature, is smaller than the normal sizes of
non-dwarf versions of that creature.  ``Shetland ponies are the {\it dwarfs} of the horse world.''~is 
a quote from the online Webster dictionary, {\tt http://www.merriam-webster.com/dictionary/}.
Their definition of ``dwarf'' recognizes that ``bodily proportions [may be] abnormal'', too, but the 
main defining feature is small stature.  And yet, Figure 3 invites us to imagine that the smallest
non-dwarf ellipticals are 20 times less luminous than the brightest ``dwarf ellipticals.''  

      In fact, Sandage and collaborators distinguish between elliptical and dwarf elliptical 
galaxies of the same luminosity with remarkable accuracy.  Quoting Sandage \& Binggeli (1984):
``The distinction between E and dE is made on morphological grounds alone, using surface brightness
as the criterion.  Normal E galaxies have a steep radial profle (generally following an $r^{1/4}$ law)
with high central brightness.  The typical dE has a {\it nearly flat} radial profile, following
either a King [1966] model with a small concentration index or equally well an exponential law.
\dots~The morphological transition from E to dE is roughly at $M_B \simeq -18$, but there is overlap.''
KFCB classify E and Sph $\simeq$ dE galaxies using a more quantitative version of the above criteria, i.{\ts}e.,
the effective-radius fundamental~plane correlations updated here in Figures 2, 7, 9, 12, 16\ts--\ts18, and 20,
and, in difficult cases, they use parameters measured at the radius that contains 10\ts\% of the total
light of the galaxy (Figure 34 in KFCB).

      At first, the Sandage group was ambivalent about whether or not the E -- dE distinction has a physical basis.
This is evident in Figure 7 of (Sandage \etal 1985a), where they argue for both
opposing points of view in the same sentence.  Similarly, Sandage \etal (1985b) admit that ``We are not
certain if this [dichotomy] is totally a tautology due merely to the arbitrary classification criteria
that separate E from dE types \dots~or if the faint cutoff in the [E luminosity function] has physical
meaning related to the properties of E and dE types.  In the first case, the problem would be only one
of definition.~In the~second, the fundamental difference in the forms of the luminosity functions of E and
dE types\ts\dots{\ts}would suggest that two separate physical families may, in fact, exist with {\it no\/} 
continuity between them (cf.~Kormendy 1985 for a similar conclusion).''  Revising a long-held picture can
be uncomfortable.

      Later, Binggeli, Sandage, \& Tammann (1988) came to recognize that ``The distinction [between] Es and dEs 
must almost certainly mean that the two classes are of different origin [Kormendy 1985, Dekel \& Silk 1986].  
This is also supported by the fact that the luminosity functions of Virgo Es and dEs [are different].''

\vss
\cl {3.2.4.~\it The E -- Sph Dichotomy: Perspective}
\vsss\vsss


      Finally, we emphasize that the criticisms of the \hbox{E{\ts}--{\ts}Sph} dichotomy in
\S\ts3.2.2 treat the issue as nothing more than an exercise in the analysis of surface photometry.  It is 
much more than this, as the difference in luminosity functions illustrates.  In addition, the E and Sph 
parameter sequences in Figure 2 are consistent with what we know about galaxy formation.  The E sequence 
constitutes the classical ``fundamental plane'' parameter correlations
(Djorgovski \& Davis 1987;
Faber \etal 1987; 
Djorgovski \etal 1988;
Bender \etal 1992, 1993).
Its interpretation is well known: galaxy structure is controlled by the \hbox{Virial theorem,
$r_e \propto \sigma^2 ~ I_e^{-1}$}, modified by small nonhomologies.
The scatter in the E fundamental plane is small (Saglia \etal 1993; Jorgensen~et~al.~1996).
And simulations of major galaxy mergers reproduce the elliptical-galaxy fundamental plane,
not a parameter correlation that is almost perpendicular to it (e.{\ts}g.,
Robertson \etal 2006;
Hopkins \etal 2008, 2009b).  Equating spheroidals with low-luminosity ellipticals would imply 
that they formed similarly, but we are confident that ellipticals 
formed via major galaxy mergers, and we believe that dwarf spheroidals cannot have formed 
by mergers (Tremaine 1981).  The E sequence is well understood.  Dekel \& Silk (1986) 
suggest that spheroidal and S$+$Im galaxies together form a sequence of decreasing baryon retention 
at lower galaxy luminositites.  We agree (\S\ts8).

\looseness=0

\vss\vsss
\cl {4.~\sc THE LARGE RANGE IN S0 BULGE-TO-TOTAL RATIOS}
\vss

      A study of S0s that is as thorough as the KFCB~study of ellipticals requires photometry
of a large sample of galaxies, each observed with multiple telescopes to provide redundancy and
large dynamic range.  That study is in progress.  Here, explanation and justification of the
proposed parallel sequence classification requires only a proof-of-concept study that
explores the observed range of bulge-to-total ratios and bulge and disk parameters.  We do this 
with photometry and bulge-disk decomposition of Virgo cluster S0s.  This also serves the need 
of \S\ts3 by adding all ACS VCS galaxies to our parameter correlations.

      Figure 2 includes all omitted ACS VCS galaxies with parameters taken from F2006.  KFCB 
omitted 3 peculiar Es, 26 S0s, and 2 dS0s.  All 3 peculiar Es, both dS0s, and 3 of 
the S0s turn out to be one-component systems that plot as Sphs.  They are shown as 
dark green triangles in Fig.~2.  They have slightly bright $\mu_e$ and small $r_e$;
this is not surprising, given that they were classified E, S0, and dS0.  But they lie within the 
scatter of Sph points and affect no conclusions.  They appear in all further parameter plots.

      The green circles with brown centers show the remaining ACS VCS S0s that KFCB omitted.
Adding these points does not invalidate the E -- Sph dichotomy.  However, they also do not give 
a realistic view of bulge and disk scaling relations for S0 galaxies.  In particular, since 
whole-galaxy parameters lump tiny bulges with large disks, they make it harder to see that the 
E correlation extends leftward of the Sph sequence in Figure 2.  This section provides the 
results of bulge-disk decompositions.

\vss\vsss
\cl {4.1. \it The ``Missing'' Latest-Type S0 Galaxies}
\vss

      Our aim in this section is to solve the puzzle of the ``missing'' S0c
galaxies.  Recall from \S\ts1 that van den Bergh (1976) lists 13 Hubble Atlas ellipticals,
11 E/S0a and S0a galaxies, 5 S0a/b and S0b galaxies, and no S0c galaxies.  In contrast, along
the spiral sequence, Sc galaxies are more common than Sas.  If spiral and S0 galaxies are proposed
to form parallel sequences with stage along the sequence defined by $B/T$ ratios,
then this is already enough to make us wonder how small the $B/T$ ratios of S0s can be.  If
in addition we inject interpretation and suspect that S0s are transformed spirals, then where
are the defunct Scs?  Especially when we have already identified Sph galaxies as defunct versions
of later-type, bulgeless Sd -- Im galaxies.

\eject

\singlecolumn

\cl{\null} \vskip 2.5truein

\includegraphics{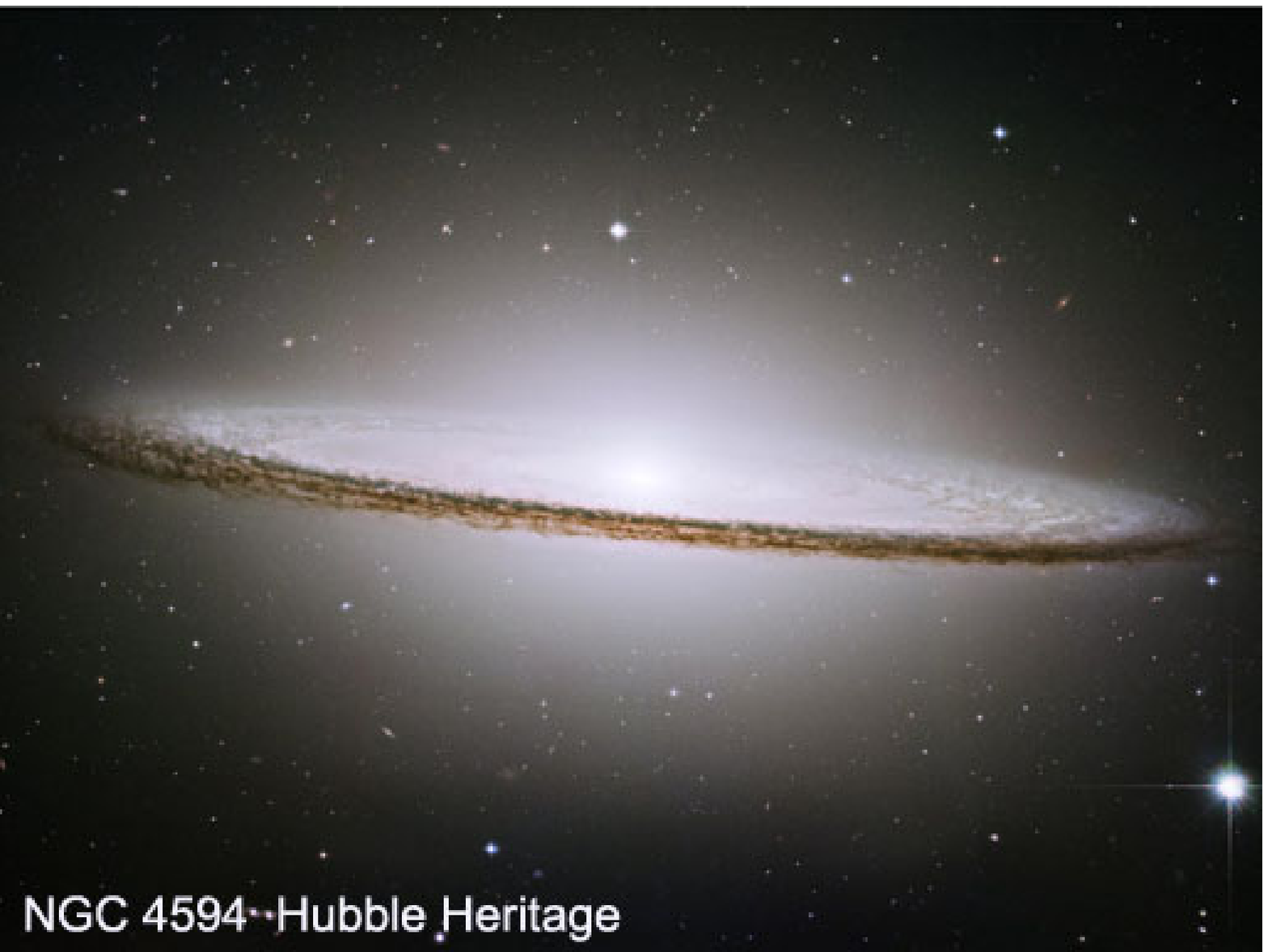}

\includegraphics{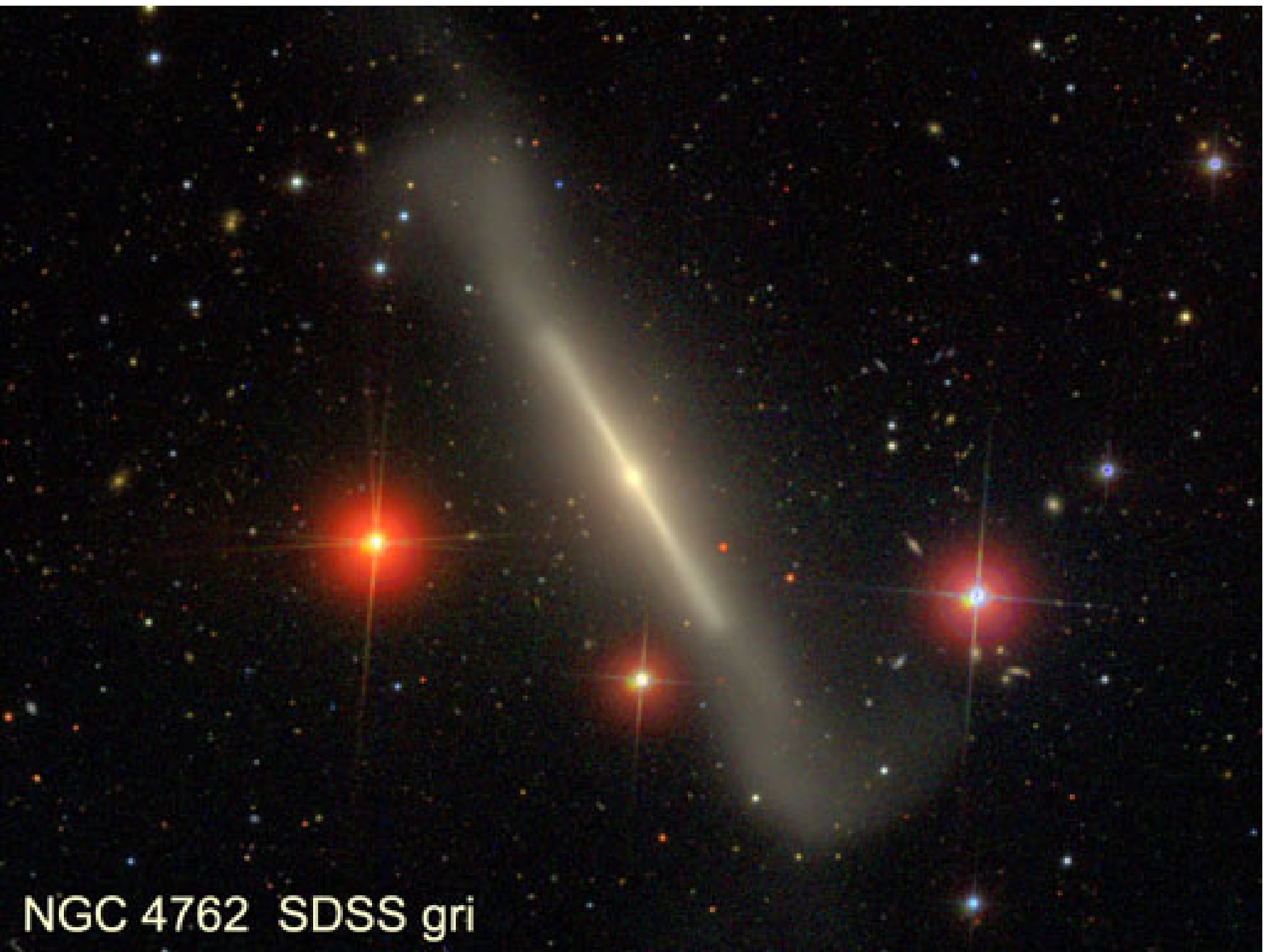}

Fig.~4~-- ({\it left\/}) Hubble Heritage image ({\tt 
http://hubblesite.org/newscenter/archive/releases/2003/28/image/a/}) of the Sombrero Galaxy,
NGC 4594.  ({\it right\/}) SDSS gri color image of NGC 4762, the second-brightest S0 galaxy
in the Virgo cluster.  These galaxies illustrate why bulge-disk decomposition is necessary.
NGC 4594 is an Sa galaxy with $B/T = 0.93 \pm 0.02$ (Kormendy 2011b).  Without
photometric decomposition, we measure essentially only the bulge.  We learn nothing about the
disk.  If an S0 version of this galaxy -- e.{\ts}g., NGC 3115 -- were viewed face-on, it would
be difficult even to see the disk (Hamabe 1982), and whole-galaxy parameters would not measure
it at all.  In contrast, NGC 4762 is one of the ``missing'' late-type S0s: we find in this 
section that $B/T = 0.13 \pm 0.02$.  Without photometric decomposition, we measure essentially 
only the disk.  We learn nothing about the bulge.  Moreover, the azimuthally averaged parameters 
used in F2006 are particularly difficult to interpret. 
\pretolerance=15000  \tolerance=15000 

\doublecolumns

\vss\vsss
\cl {4.1.1.~\it NGC 4762 (SB0bc):}
\cl {\it The Need for Bulge-Disk Decomposition}
\vss

      We begin with the well known, edge-on SB0 NGC 4762 
(Sandage 1961; 
Kormendy \& Kennicutt 2004).
It was recognized as a small-bulge S0 by van den Bergh (1976) and classified as S0b.  We will 
tweak the classification slightly to (R)SB0(lens)bc.

      Figure 4 contrasts the tiny bulge in NGC 4762 with the dominant bulge in NGC 4594.
It illustrates why bulge-disk decomposition is necessary.  Whole-galaxy parameters 
measure only the dominant component.  At all $B/T$, they measure 
each separate component incorrectly.  S0 galaxies prove to exist with all
$B/T$ values from~0~to~$\sim$\ts1, so {\it using
whole-disk parameters~for S0s is guaranteed to imply continuity between 
pure-bulge and pure-disk galaxies.}  We regret the need to belabor this point, because
bulge-disk decomposition was developed long ago (Kormendy 1977a) and has been standard analysis
machinery ever since (e.{\ts}g., 
Burstein 1979;
Kent 1985;
Byun \& Freeman 1995;
Scorza \& Bender 1995;
Baggett \etal 1998;
GIM2D: Simard \etal 2002;
GALFIT: Peng \etal 2002;
BUDA: de Souza \etal 2004;
Laurikainen \etal 2005, 2007;
Courteau \etal 2007;
GASP2D: M\'endez-Abreu \etal 2008;
Fisher \& Drory 2008;
Weinzirl~et~al.~2009). 
Several groups now analyze composite (bulge\ts$+${\ts}disk) galaxies
as single-component systems.  This mixes up the different formation physics and properties
of these very different components.

      Before we analyze surface photometry, we need to understand what kind of galaxy NGC 4762 is.  
This~is both easy and well known (Wakamatsu \& Hamabe 1984, Kormendy \& Kennicutt 2004).  
Figure 4 shows the essential features that are discussed in the Hubble Atlas and in the above papers:
the galaxy has three ``shelves'' in its major-axis brightness profile outside its central bulge. 

\cl{\null} \vskip 2.48truein
\vfill
\includegraphics{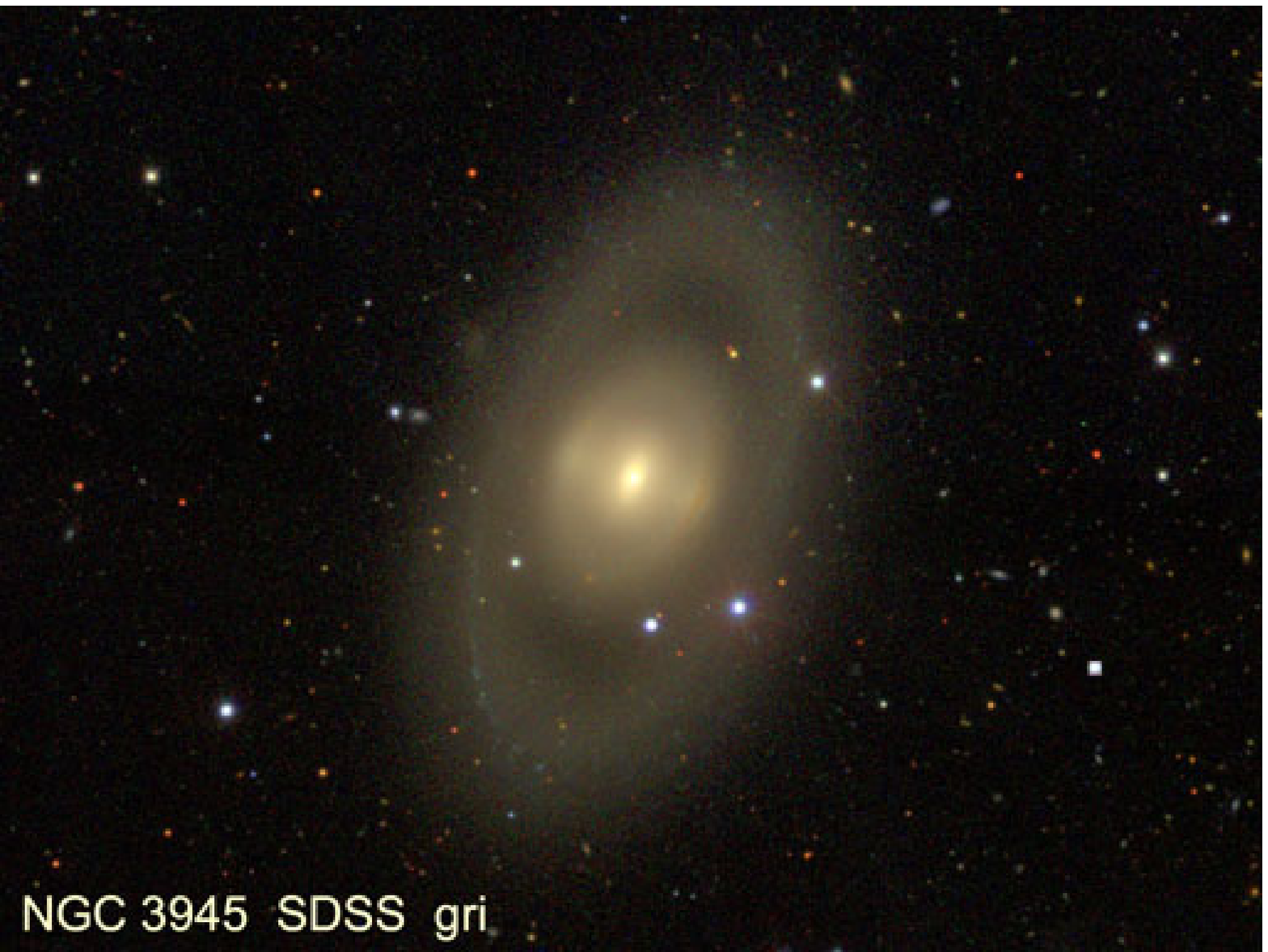}

Fig.~5~-- (R)SB(lens)0 galaxy NGC\ts3945, a more face-on analog of NGC 4762.  The bar fills 
the lens component in one dimension.  The lens is the elliptical ``shelf'' in the brightness 
distribution interior to the outer ring.  Note that it has a sharp outer edge.  Such profiles 
are well described by S\'ersic functions with $n < 0.5$; that is, ones that cut off at large 
radii faster than a Gaussian.  If NGC 3945 were rotated clockwise slightly and then rotated 
about a horizontal axis until it is seen edge-on, then the bar would look shorter than the
major-axis radius of the lens.  Then the major-axis brightness profile would show three
shelves as in NGC 4762 -- the bar, the lens, and the outer ring.

\vskip 7.5pt

\ni This means that it is an edge-on (R)SB(lens)0 galaxy, as illustrated by the more face-on,
prototypical (R)SB(lens)0 galaxy NGC 3945 (Kormendy 1979b) in Figure~5. 


\eject

\cl{\null} \vskip 3.7truein

\includegraphics{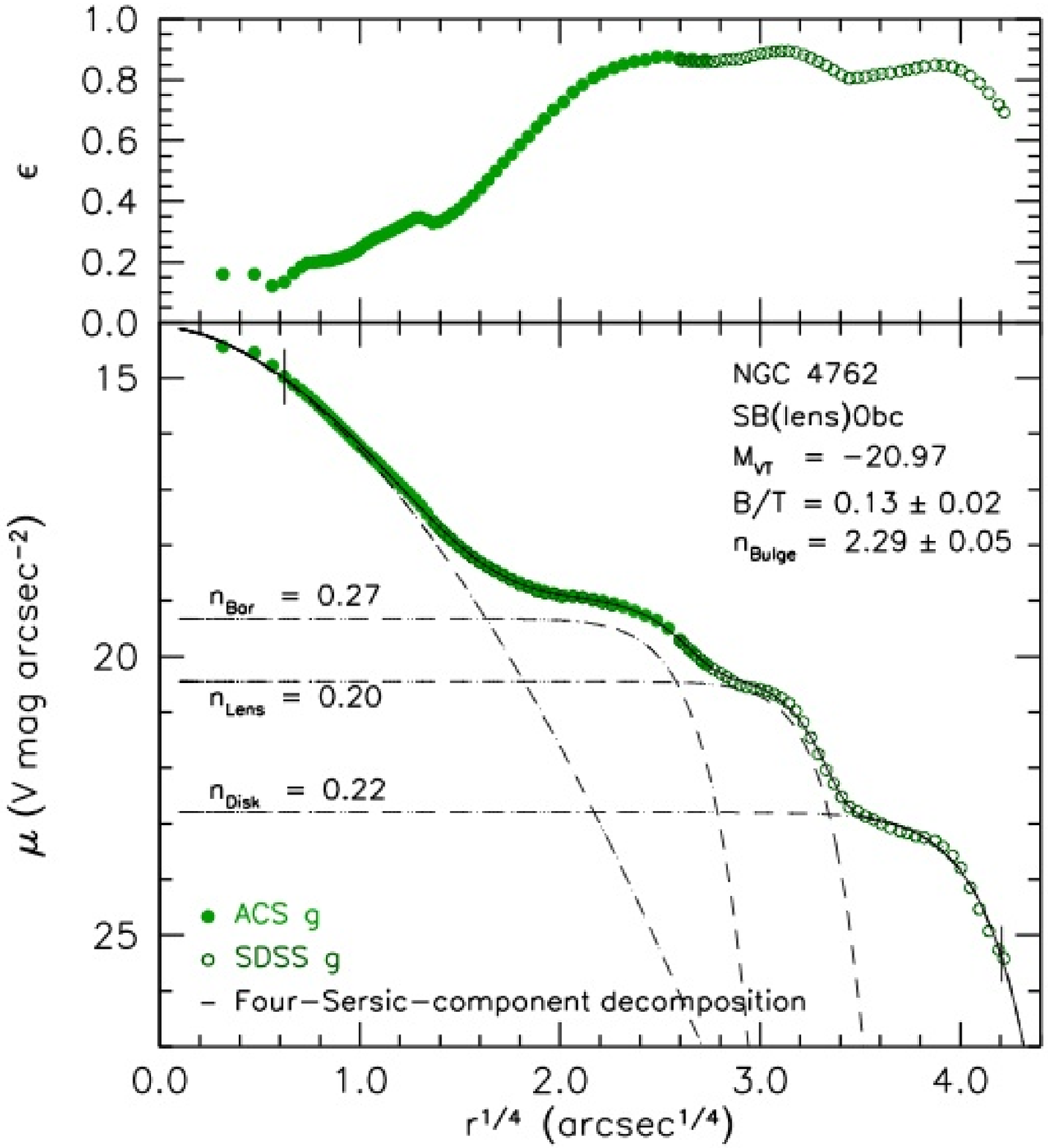}

Fig.~6~-- Ellipticity $\epsilon$ and surface brightness $\mu_V$ along the major axis of NGC 4762
measured by fitting elliptses to the isophotes in the ACS and SDSS $g$-band images.  We used the
transformation $V = g + 0.320 - 0.399(g-z)$ from KFCB and the galaxy color $g - z = 2.076$ from
F2006.  All magnitudes and colors are VEGAmag.  The dashed curves show a decomposition of the profile 
inside the fit range ({\it vertical dashes across the profile\/}).  The bulge, bar, lens, and disk 
are represented by S\'ersic functions with indices $n$ given in the figure.  Their intensity sum 
({\it solid curve}) fits the data with an RMS of 0.033 V mag arcsec$^{-2}$.

\vskip 8pt

      We measure the composite, major-axis brightness profile of NGC 4762 shown in Figure 6.
It clearly shows the three shelves in surface brightness discussed above.  

      Normally, a four-component photometric decomposition would involve ferocious parameter coupling;
Appendix~A of KFCB shows how serious this problem can be even for one-component fits of the three-parameter
S\'ersic function.  However, parameter coupling is serious when $n$ is large.  The more nearly each shelf
has constant brightness interior to an infinitely sharp outer cutoff, the less coupling there is between
components.  Bars, lenses, and outer rings have sharp outer cutoffs (Kormendy 1979b).  Four-component
decomposition shows that the bar, lens, and outer ring of
NGC 4762 are best fitted with S\'ersic functions that have extraordinarily small $n \sim 0.2$ to 0.3 
(a Gaussian profile has $n = 0.5$).  That is, each component looks to all components interior to it as
being almost constant in surface brightness.  The decomposition is therefore robust.  In
particular, the bulge parameters are well determined.  Table 1 in \S\ts4.2 lists the bulge and 
disk parameters for the 13 ACS VCS S0 galaxies for which we do photometry in this paper.  Gavazzi \etal (2000)
provide decompositions of the remaining S0s. 

      We measure a bulge S\'ersic index of $n = 2.29 \pm 0.05$.  Also, the bulge ellipticity 
$\epsilon \simeq 0.3$ is like that of a typical elliptical.  All this suggests that the 
bulge is classical (Kormendy \& Kennicutt 2004; Fisher \& Drory 2008). 

\cl{\null} 

\vskip 4.75truein

\includegraphics{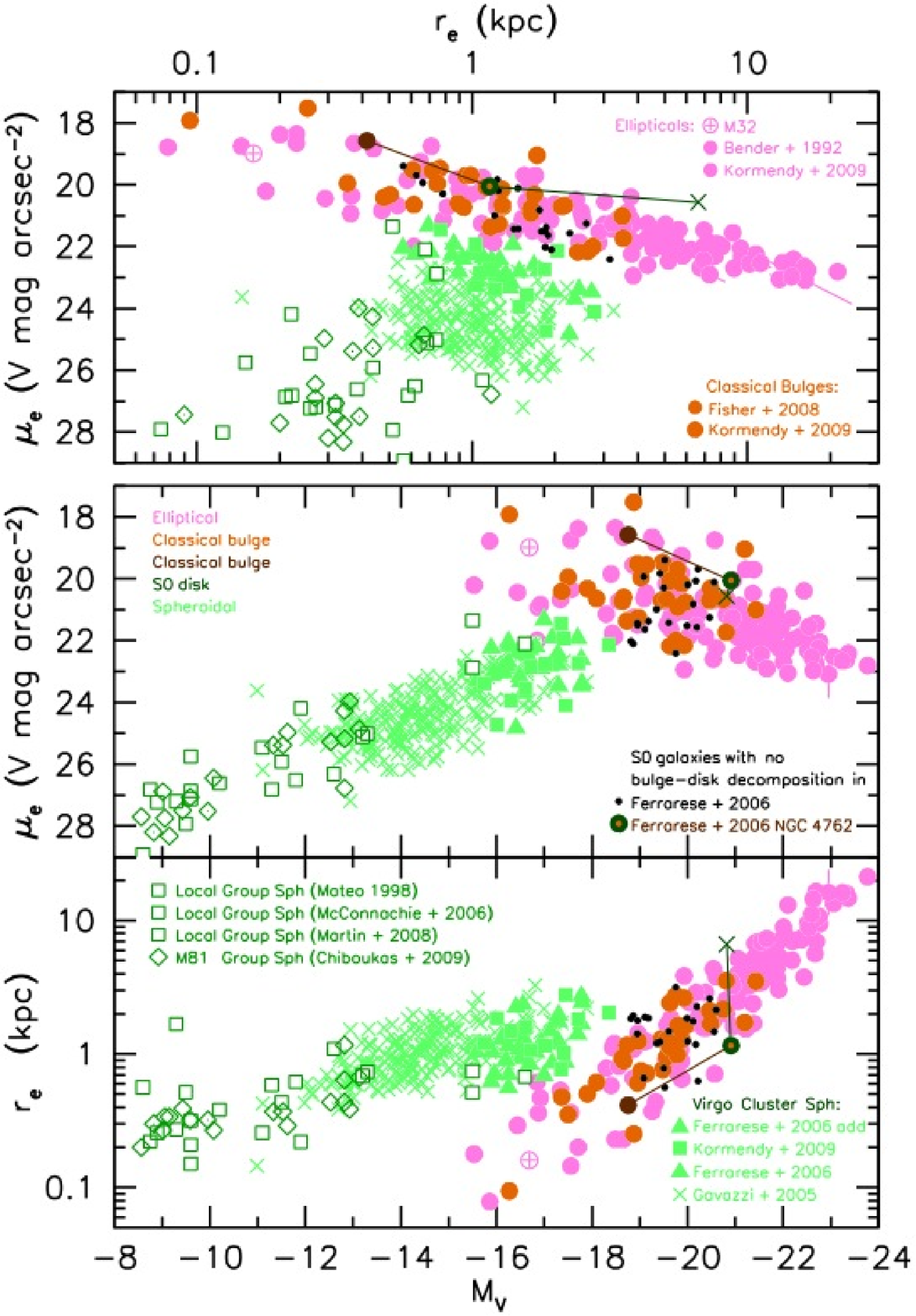}

Fig.~7~-- Parameter correlations from Figure 2 showing the results of the bulge-disk 
decomposition of NGC 4762.  The green filled circles with the brown centers show the 
total parameters measured by F2006 for the bulge and disk together.  These points are connected
by straight lines to the bulge parameters ({\it dark brown filled circles\/}) and the disk parameters 
({\it dark green crosses\/}).

\vskip 8pt

      Figure 7 shows the bulge and disk of NGC 4762 in the parameter correlations of Figure 2.  
The disk parameters are determined by integrating the total disk model out to half of the disk luminosity
taking component flattening into account.  They refer to all disk components together.  They look 
counterrintuitive: $r_e$ is 7 times larger than for the galaxy as a whole.  This is explained in the Appendix.

      The bulge parameters illustrate how the decomposition results strengthen our understanding of 
the parameter correlations.  In the top panel, the bulge of NGC 4762 contributes to the compact
extension ({\it leftward\/}) of the E$+$bulge correlations.  Taking the different flattenings of the
bulge and disk into account, the bulge absolute magnitude is $M_{V,\rm bulge} = -18.76$.  NGC 4762
helps to define the compact end of the correlations in the bottom panels.  

      The bulge-to-total ratio is $B/T = 0.13 \pm 0.02$, similar to $B/T = 0.12 \pm 0.02$ in the
SABbc galaxy NGC\ts4258 (Kormendy \etal 2010).~Given the observation that classical bulges do not 
occur in Sc galaxies (Kormendy \& Kennicutt 2004), the morphological type of NGC 4762 is well
constrained to be (R)SB0(lens)bc.  It begins to extend the parallel sequence classification beyond S0b.

\eject

\vss\vsss
\cl {4.1.2.~\it NGC 4452 (S0c)}
\vss

      NGC 4452 is closely similar to NGC 4762 but even later in type.  Figure 8 shows that it
has an edge-on thin disk with a sharp outer cutoff again indicative of a bar or lens.

\vfill

\cl{\null} \vskip 6.4truein

\includegraphics{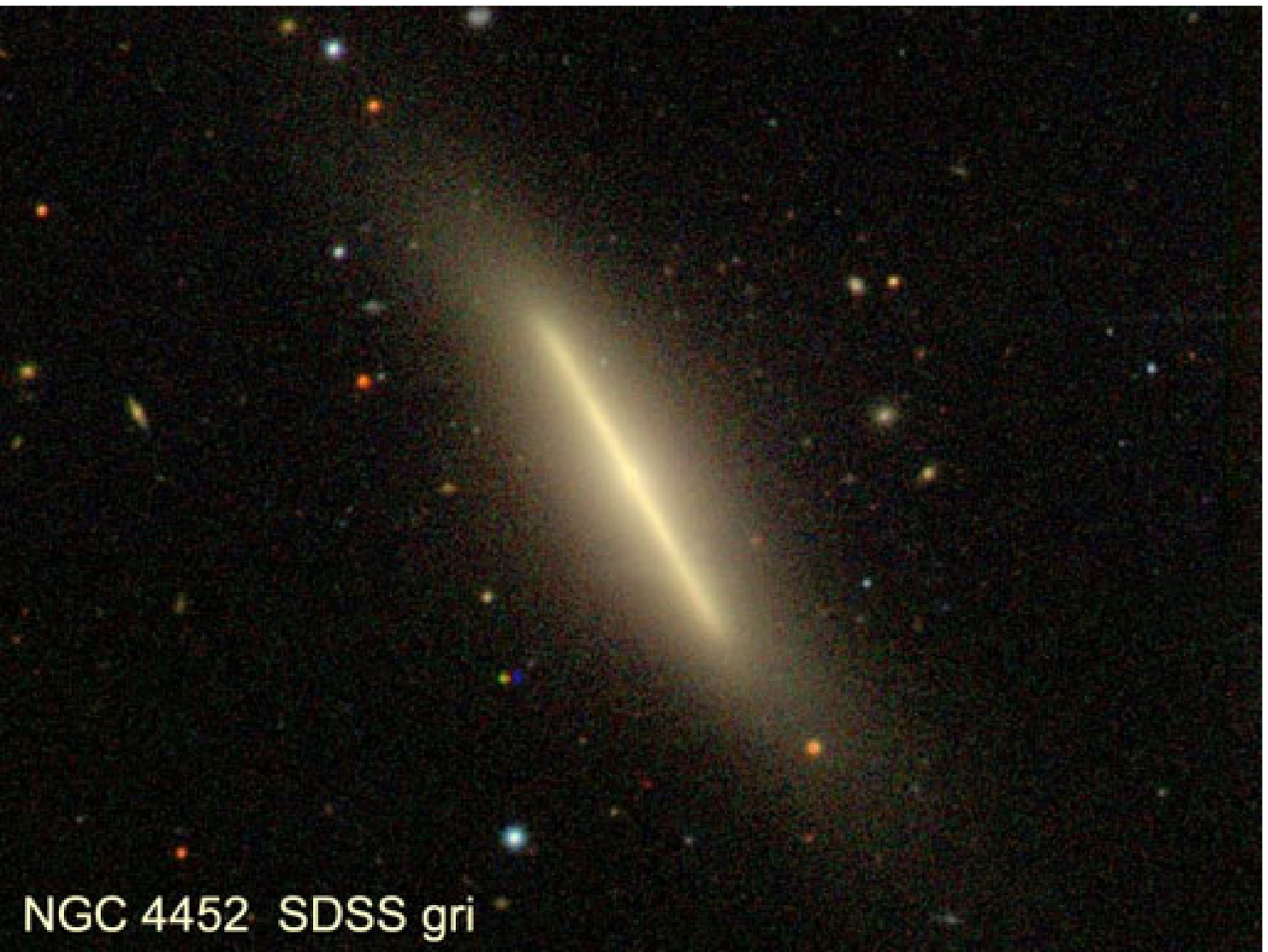}

\includegraphics{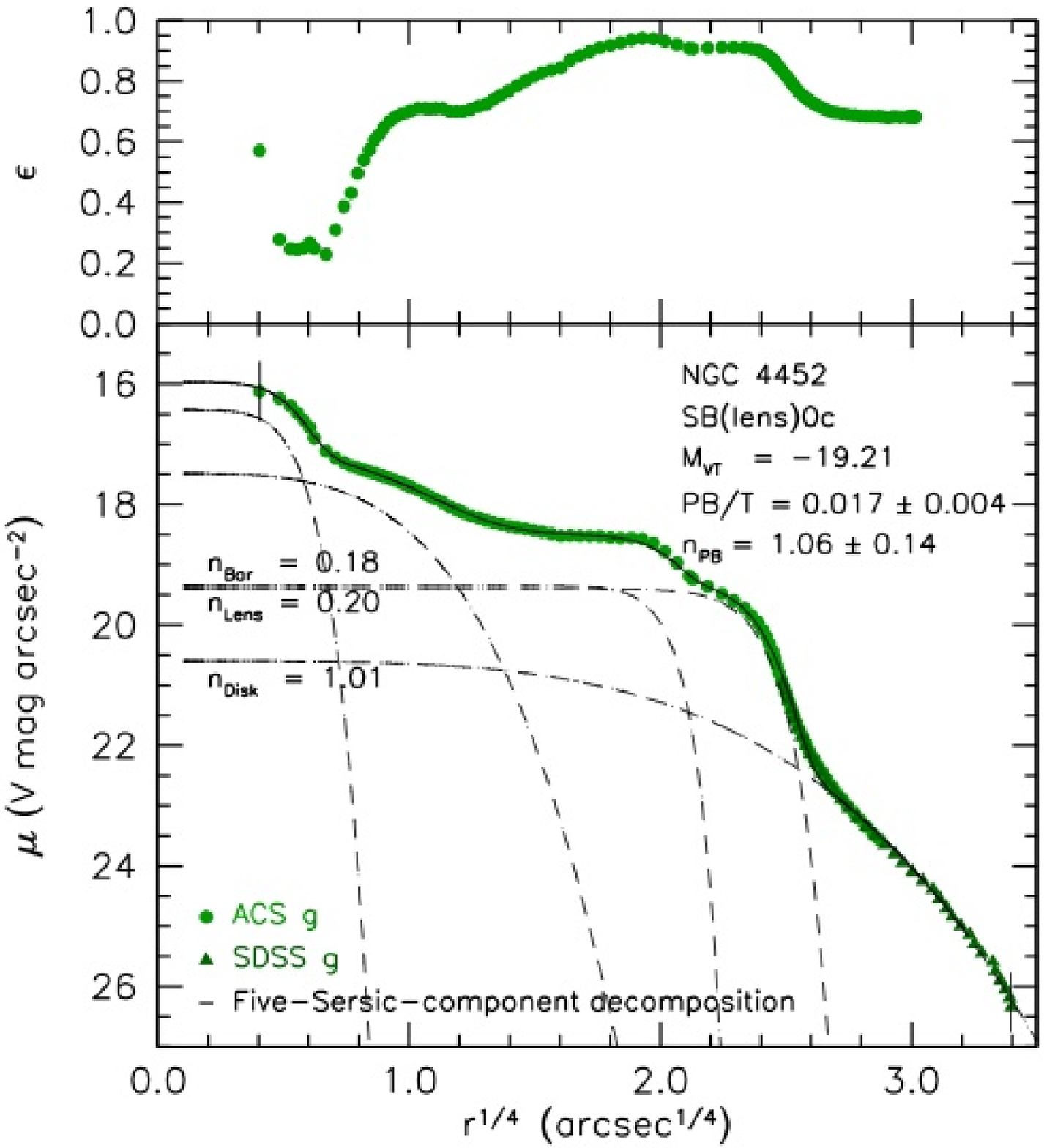}

Fig.~8~-- ({\it top\/}) Color image of SB(lens)0 galaxy NGC\ts4452 from the SDSS {\it g,r,i}
images via {\tt http://www.wikisky.org}.  The bulge -- which proves to be pseudo -- is so tiny
that it is almost invisible.  The inner disk is edge-on and very flat; it again consists of two 
shelves in surface~brightness.  Including the outer, thicker disk, these three shelves are 
signatures of a bar, lens, and disk.  ({\it bottom\/}) Ellipticity $\epsilon$ and 
surface brightness $\mu_V$ along the major axis of NGC 4452 measured by fitting ellipses to the 
isophotes in the ACS and SDSS $g$-band images.  The five dashed curves show a decomposition of 
the profile inside the fit range ({\it vertical dashes\/}).  The nucleus, 
bulge, bar, lens, and disk are represented by S\'ersic functions with indices $n$ as given in 
the figure.  The sum of the components ({\it solid curve}) fits the data with an RMS of 
0.044 V mag arcsec$^{-2}$.

\noindent Our photometry -- also illustrated in Figure 8 -- shows that the thin disk consists of
two shelves as in NGC 4762.  Together with the (thick and warped) outer disk, this implies that
NGC 4452 is an edge-on SB(lens)0 galaxy.  

      The brightness profile is more complicated than that of NGC 4762 in that NGC 4452 also
contains a nuclear star cluster.  We derive an AB $g$ magnitude of this nucleus of 20.56, in
good agreement with the F2006 value of 20.49.  Fortunately, the nucleus has a steep outer 
profile (formally, $n = 0.68 \pm 0.09$, but this is consistent with a Gaussian, and it applies 
to the PSF-convolved, observed profile).  The point is that the nuclear profile does not much 
influence the photometric decomposition.

      The bar and lens also have very cut~off~profiles ($n$\ts=\ts0.18 and $n = 0.20$,
respectively); both have essentially constant surface brightness underlying the
(pseudo)bulge.

      Thus the five-component decomposition in Figure 8 is surprisingly robust.  NGC 4452 contains
a pseudobulge with $n = 1.06 \pm 0.14$.  Pseudobulges have fundamental plane correlations similar to 
those of classical bulges but with larger scatter (Fisher \& Drory 2008; Kormendy \& Fisher 2008). 
The bulge and disk parameters of NGC 4452 are shown in Figure 9.  The pseudobulge provides further
support for the compact extension of the E correlations.

\vfill

\includegraphics{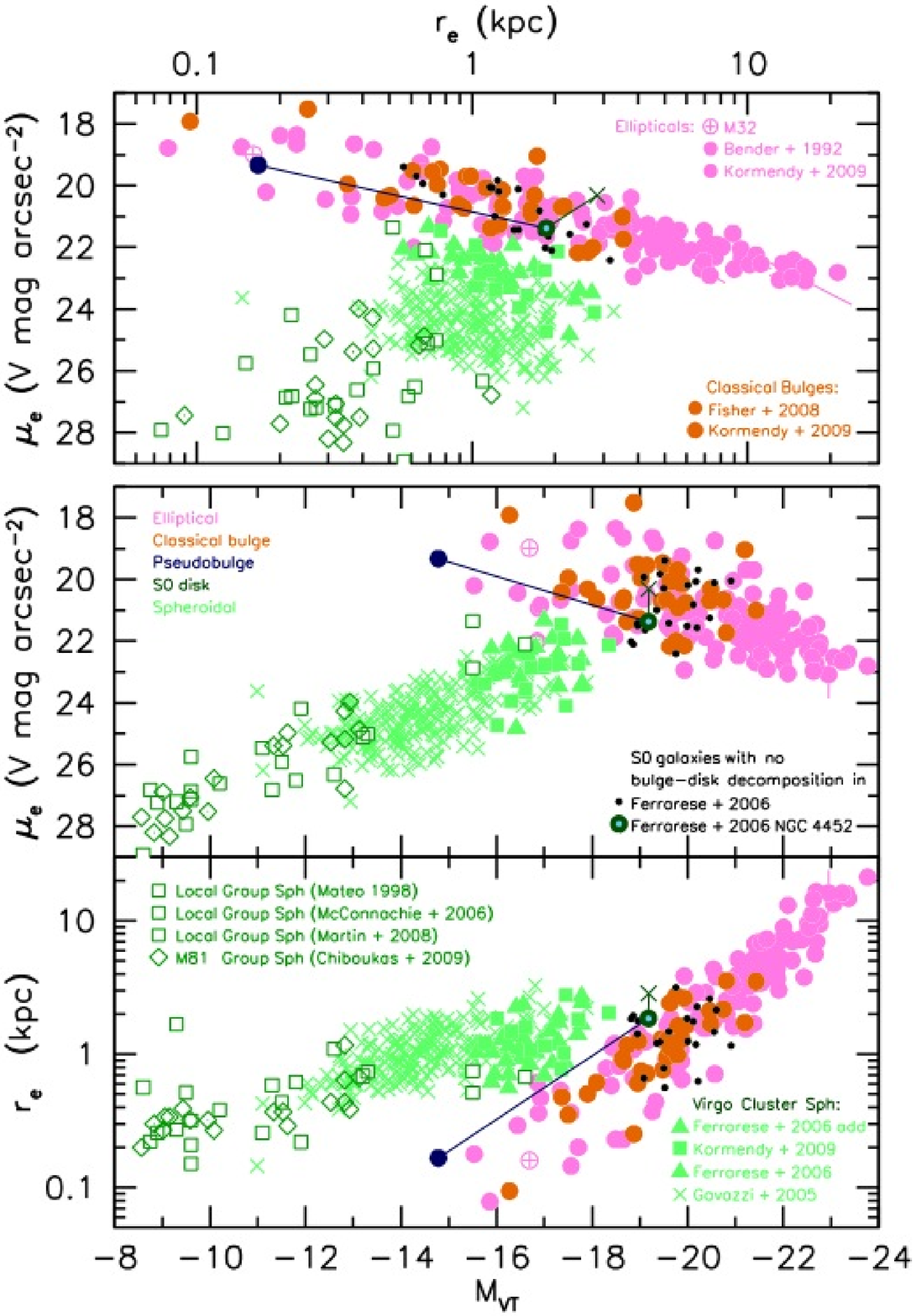}

Fig.~9~-- Parameter correlations from Figure 2 showing the results of the bulge-disk 
decomposition of NGC 4452.  The green filled circles with the blue centers show the 
total parameters measured by F2006 for the bulge and disk together.  These points are connected
by straight lines to the pseudobulge parameters ({\it dark blue filled circles\/}) and the 
disk parameters ({\it dark green crosses\/}).

\eject

      NGC 4452 is important for two additional reasons.  First, the pseudobulge-to-total
luminosity ratio is only $PB/T = 0.017 \pm 0.004$.  This is closely similar to $PB/T$ in the
Scd galaxies M{\ts}101, NGC 6946, and IC 342 (Kormendy \etal 2010).  Thus NGC 4452 is an S0c 
galaxy.  (If we had chosen the finer bins of the de Vaucouleurs
classes, then NGC 4452 would be S0cd.)
Similarly, with $B/T \simeq 0.08$,  NGC 1411 is a more face-on S0c (Laurikainen \etal 2006). 

      Second, the outer disk of NGC 4452 is warped and thicker than the ``superthin'' 
edge-on, late-type galaxies seen in isolated environments (van der Kruit \& Freeman 2011).  
Similarly, NGC 4762's outer disk is thick, warped, and tidally distorted.  Gravitational encounters 
may be at fault (NGC\ts4762 with NGC\ts4754; NGC\ts4452 with IC\ts3381).  We suggest that these are signs of 
environmental heating that helps  to convert flat disks into less flat spheroidals. 

\vss\vsss
\cl {4.1.3.~\it The ``Rosetta Stone'' Galaxy VCC 2048:}
\cl {\it An Edge-On Sph Galaxy that ``Still'' Contains a Disk}
\vss\vsss

      VCC 2048 is -- we suggest -- just such a galaxy.  It is classified E in RC3, but it contains an edge-on 
disk, suggesting that it is a dS0 (F2006).  The disk is illustrated in Figure 10.  However, we will show  
that the rounder component of VCC 2048 is not a bulge.  Rather, it is a more extreme version of the fat outer 
disks of NGC 4452 and NGC 4762.  In fact, the outer component of VCC 2048 plots within the Sph sequence of 
the fundamental plane parameter correlations.  This is therefore a Sph galaxy that ``still'' contains a small 
edge-on disk.  It provides a compelling connection between Sphs and S0 disks, showing properties of both.
VCC 2048 is a ``Rosetta Stone galaxy'' that supports the results of this paper especially clearly.

\singlecolumn

\cl{\null} 

\vfill

\includegraphics{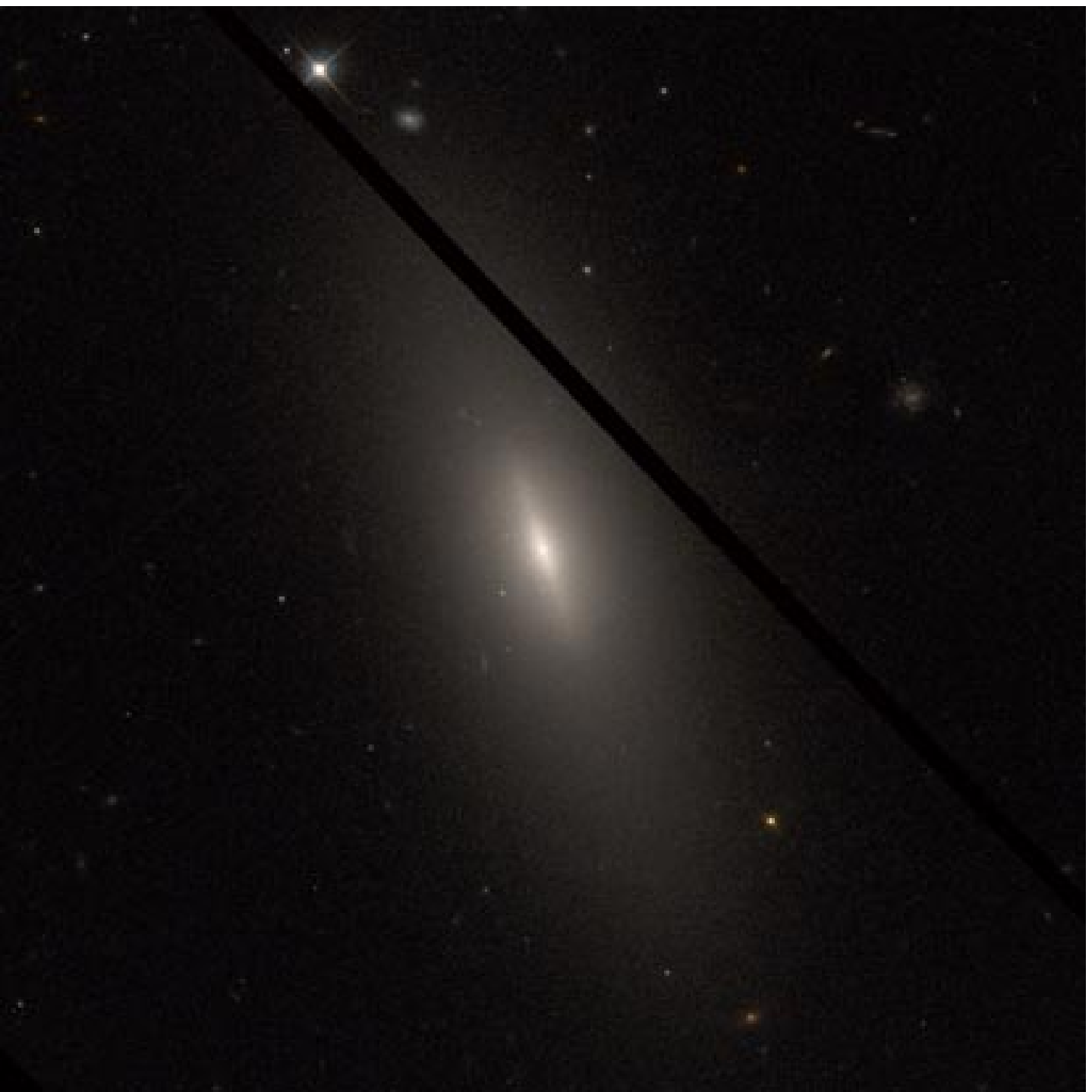}
\includegraphics{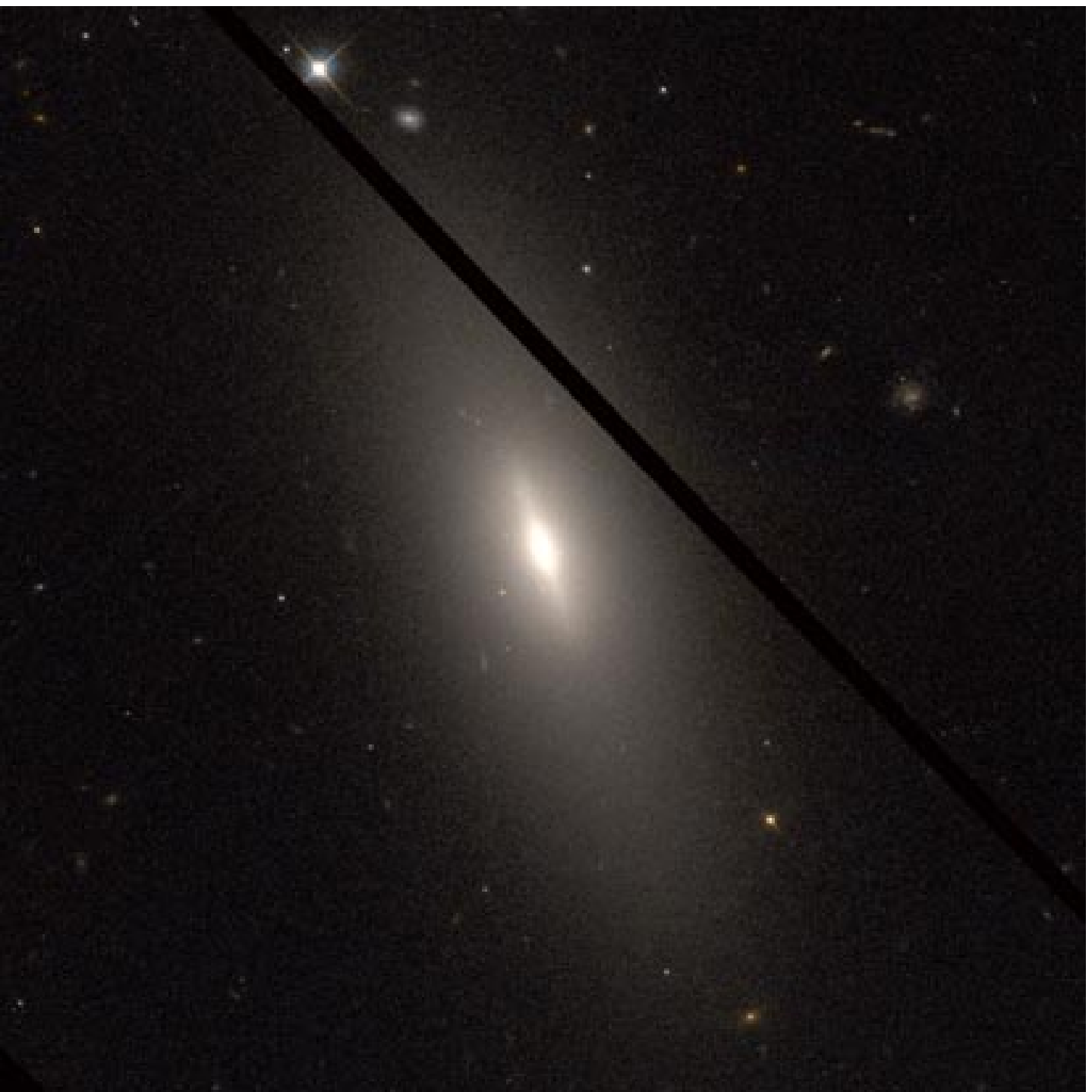}
\includegraphics{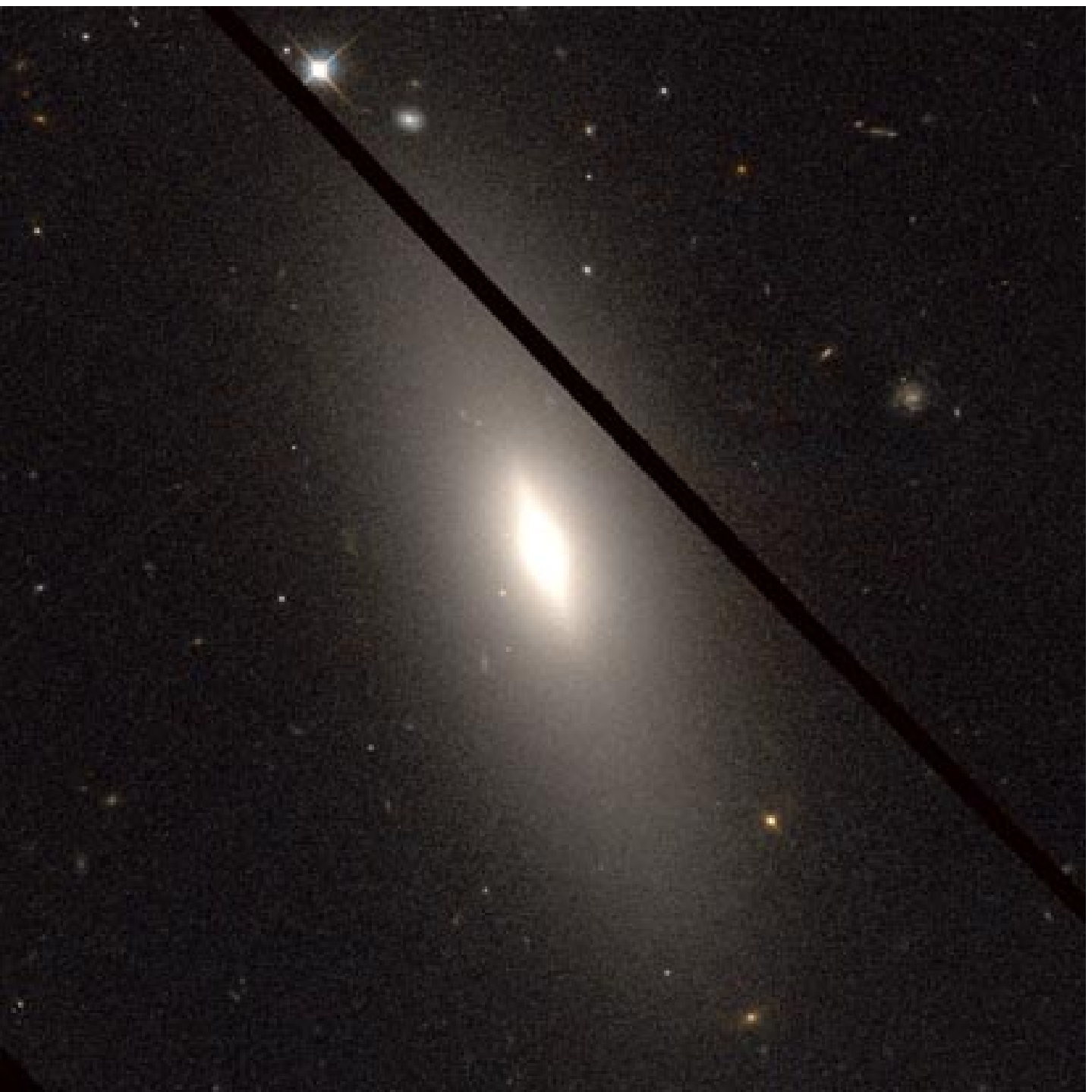}

\includegraphics{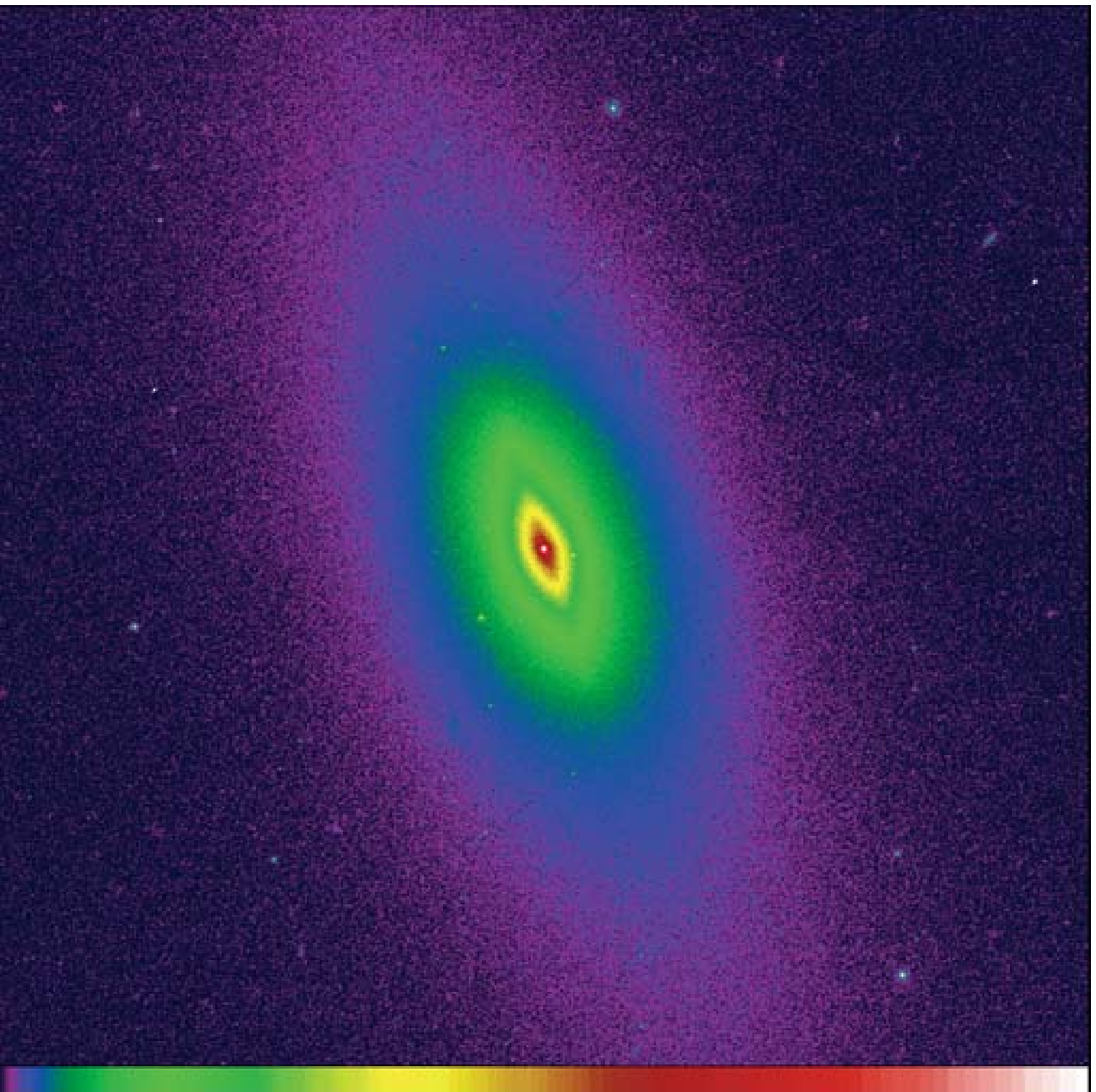}
\includegraphics{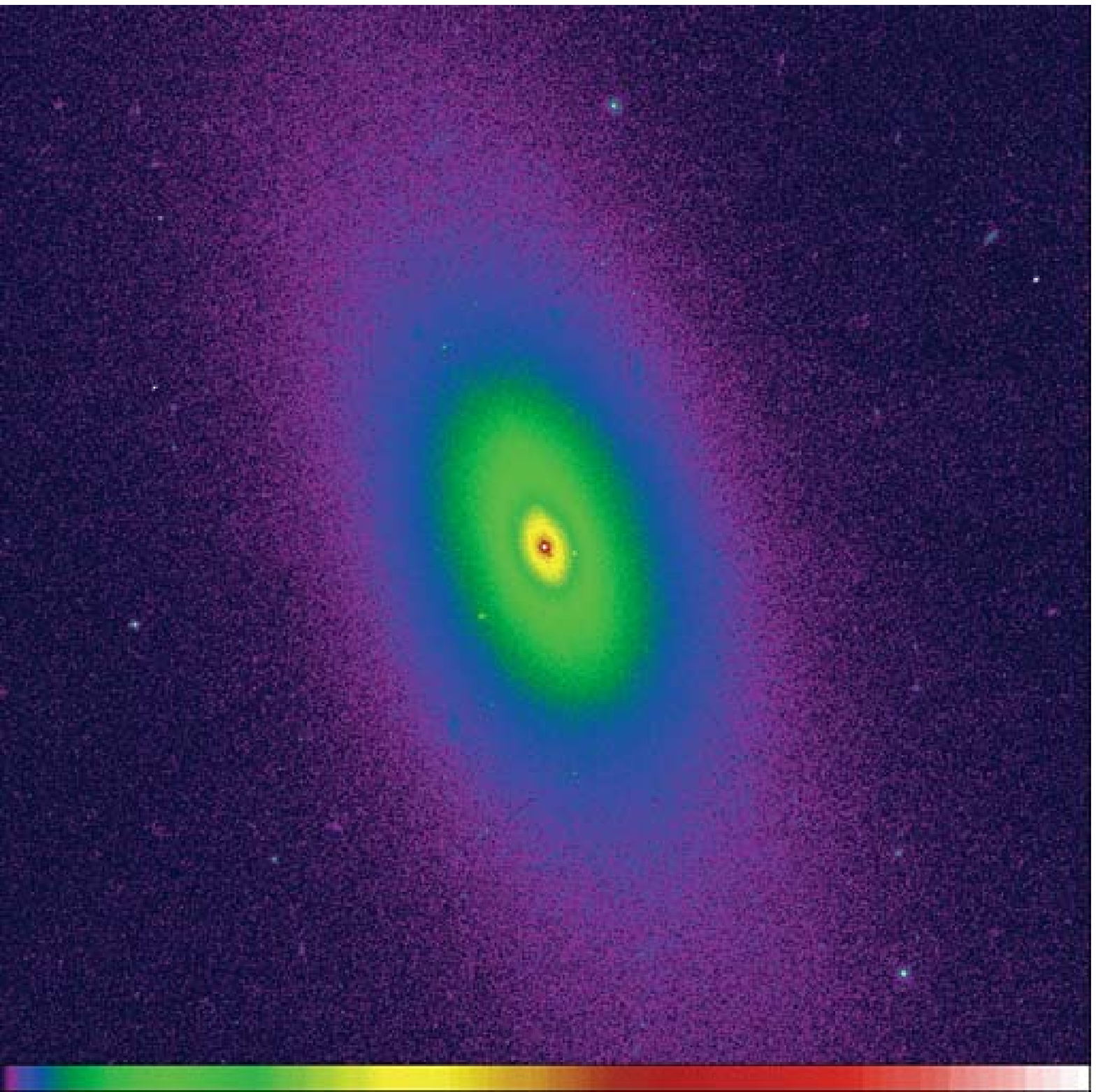}

Fig.~10~-- ({\it top\/}) Color images of VCC\ts2048 constructed from the HST ACS $g$ image ({\it blue}), 
the mean of the $g$ and $z$ images ({\it green\/}), and the $z$ image ({\it red\/}).  Different
brightness ``stretches'' emphasize the embedded disk and nucleus ({\it left\/}), the disk
({\it center\/}) and the outer envelope ({\it right\/}).  VCC 2048 is classified as an edge-on S0
by Binggeli \etal (1985) and Ferrarese \etal (2006).  It clearly has an edge-on disk
embedded in an ellipsoidal halo.  ({\it bottom\/}) False-color images of VCC 2048 as observed 
({\it left\/}) and after subtraction of the model, almost-edge-on ($i \simeq 81^\circ$) disk 
that renders the residual isophotes as nearly elliptical as possible.
This remnant envelope proves to have structural parameters characteristic of
a Sph galaxy, not a bulge (Figure 12).  VCC 2048 is therefore not an S0 galaxy but rather is
a Sph,N galaxy with an embedded disk.
\pretolerance=15000  \tolerance=15000 

\eject

\doublecolumns

      Figures 10 and 11 show our photometry and photometric decomposition of VCC 2048.
The photometry uses the algorithm of Bender (1987), Bender \& M\"ollenhoff (1987),
and Bender \etal (1987, 1988) to fit the 2-dimensional \phantom{00000000}

\vskip 6.25truein

\includegraphics{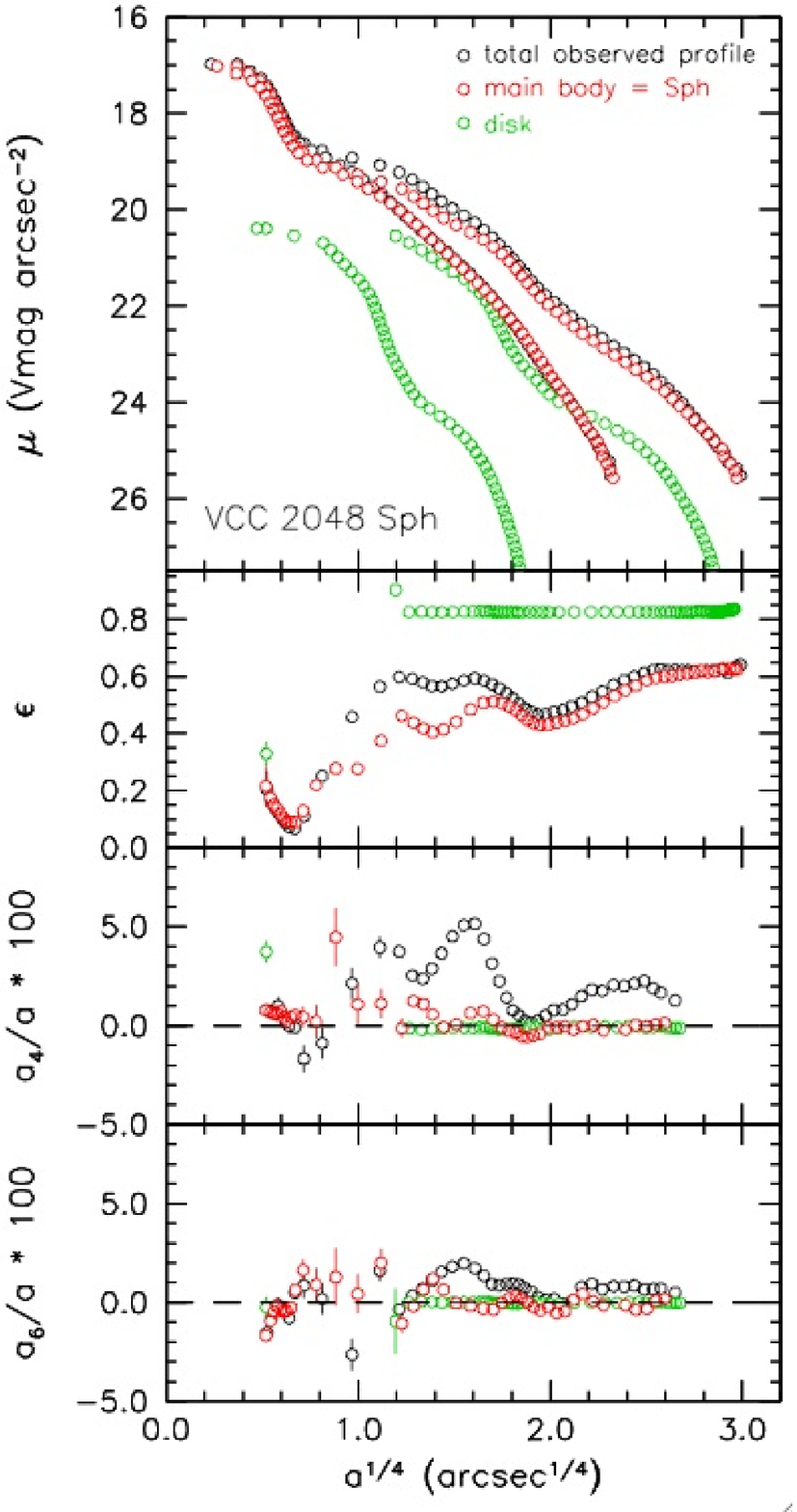}

Fig.~11~-- Profiles in VCC 2048 of ({\it top -- bottom\/})~\hbox{$V$-band} surface brightness,
ellipticity, and the isophote shape parameters $a_4/a$ and $a_6/a$ as functions of the
\hbox{major-axis} radius $a$.~Both major- and minor-axis profiles are shown.  The observed
profiles are shown by the black circles.  The green and blue circles show the results of a
photometric decomposition following the procedure of Scorza \& Bender (1995) and illustrated 
here in Figure 10.  A thin disk profile ({\it blue points\/}) is constructed non-parametrically
such that, when inclined at the optimum inclination $i = 81^\circ$ and subtracted from the
two-dimensional light distribution, it renders the residual isophotes as nearly elliptical as
possible.  This is shown by the bottom two panels here: disk subtraction removes the large 
$a_4$ and $a_6$ disky signature in the original isophotes ({\it black points\/}) and leaves
residual $a_4$ and $a_6$ profiles ({\it green circles\/}) that are consistent with zero, 
i.{\ts}e., elliptical isophotes.

\noindent isophotes with ellipses plus deviations from ellipses that are expanded in a Fourier 
series in $a_k \cos{k\theta}$ and $b_k \sin{k\theta}$ where $\theta$ is the eccentric anomaly 
of the ellipse (Appendix, \S{\ts}A1).  The $a_4$ and $a_6$ terms measure disky ($a_4 > 0$)
and boxy ($a_4 < 0$) distortions from exactly elliptical isophotes.  The strong and weaker 
signatures of an inner and an outer part of the embedded disk are clear in the raw $a_4$ and $a_6$
profiles ({\it black points\/}) in Figure 11.

      Photometric decomposition was carried out using the procedure of Scorza \& Bender (1995);
Scorza \etal (1998).  A nonparametric profile is calculated for an infinitely thin disk with
inclination $i = 81^\circ$ such that subtraction of the disk from the two-dimensional image leaves behind 
residual isophotes that are as nearly elliptical as possible (Fig.~10).  Figure 11 ({\it green points\/})
shows that we succeed in removing the $a_4 > 0$ and $a_6 > 0$ disk signatures. 
  The disk profile is shown blue in Fig.\ts11.
The effective parameters of the disk and main body of the galaxy are calculated by integrating 
each profile.   Results are shown in Fig.~12.  
{\it We conclude that the main body of VCC 2048 is not 
a bulge.  It is a spheroidal, and we classify the galaxy as Sph,N.
 Taking component flattening into account, $Sph/T = 0.92 \pm 0.02$.}

\vskip 1.0truein

\cl{\null} \vskip 1.5truein

\vfill

\includegraphics{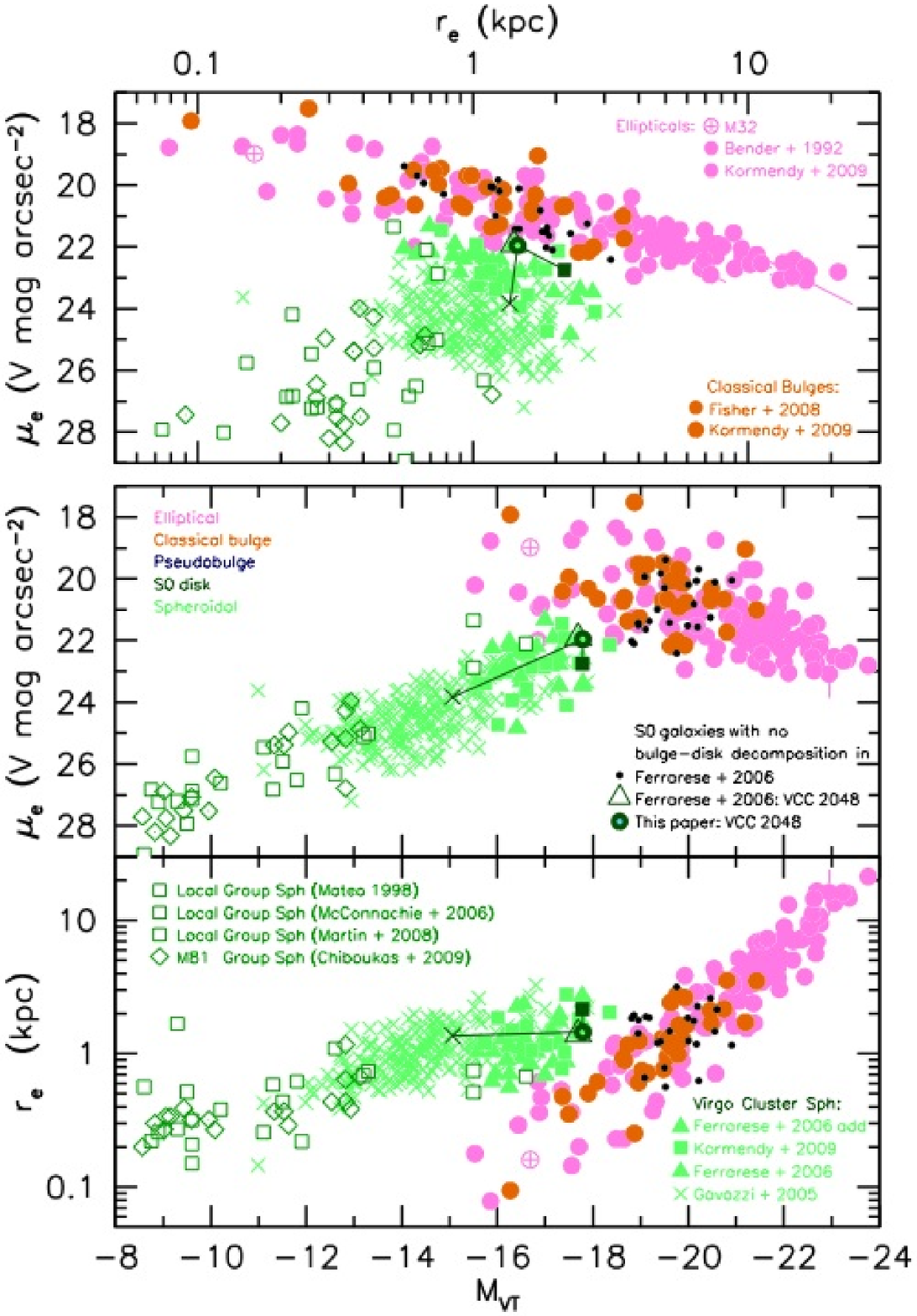}

Fig.~12~-- Parameter correlations from Fig.~2 showing~the results of the photometric 
decomposition of VCC\ts2048.  Large green triangles show the total parameters measured 
by F2006 for the disk and envelope together.  Our corresponding measurement for the whole
galaxy agrees with F2006; it is shown by the green filled circle with the light green center.
Our points are connected by straight lines to the envelope parameters ({\it dark green filled 
squares\/}) and the disk parameters ({\it dark green crosses\/}).

\eject

      We emphasize that the difference between VCC 2048 and NGC 4452 is mainly quantitative
and fairly subtle.  NGC 4452 has a disk-to-total ratio of $D/T \simeq 0.43$,
typical of S0 galaxies (e.{\ts}g., Simien \& de Vaucouleurs 1986).  The inner parts of 
NGC 4452 contain a bar and a lens, these are common and characterisic components in SB0 galaxies 
(Kormendy 1979b, 1981, 1982;
Buta \& Crocker 1991;
Sellwood \& Wilkinson 1993;
Buta 1995, 2011;
Buta \& Combes 1996;
Buta \etal 2007, 2010).
VCC 2048 has $Sph/T = 0.90$, $D/T \simeq 0.10$, and $B/T = 0$.  Most of the light is in a component 
that is indistinguishable from spheroidal galaxies.   Absent the embedded disk and given its effective 
parameters, the galaxy would certainly be classified as Sph,N.  On the other hand, the similarities
to NGC 4452 and NGC 4762 are compelling, too.  The thick, main component of VCC 2048 does not look
very different from the thick outer disks of the above S0 galaxies.  We see signs that those outer S0
disks are even now being heated and thickened by tidal encounters with neighbors.  We will find in the
next section that S0 disks are continuous with spheroidals in their parameter correlations.  This is 
an early sign of that result (see \S{\ts}A.12 for further discussion).

      We conclude that VCC 2048 is a ``missing link'' between S0 galaxies and spheroidals.  We
interpret the main body of the galaxy as a dynamically heated and therefore thick version of
an S0 disk.  We interpret the embedded disk as the inner, most robust remnant of the former S0
disk (although it is also possible that a small disk grew after the formation of the Sph by late
infall of cold gas).

\cl{\null}

      We note that similar edge-on disks have been detected in Fornax cluster Sph galaxies by
De Rijcke \etal (2003).

      The galaxies NGC 4762 (S0bc), NGC 4452 (S0c), and VCC 2048 (disky Sph,N) provide a continuous
link between earlier-type S0 galaxies and Sphs.  They include examples of the formerly-missing, 
latest-type S0 galaxies.  They are part of our motivation in placing Sph galaxies at the \hbox{late-type}
end of the S0a -- S0b -- S0c sequence in Figure 1.

\vss\vsss
\cl {4.1.4~\it Another Missing Link Between S0 and Sph Galaxies:}
\cl {\it NGC 4638 = VCC 1938}
\vss

      With the experience of the previous section, we are better prepared to interpret NGC 4638 = VCC 1938.
It is illustrated in Figures 13 and 14.  At first sight, it appears to be yet another ACS VCS S0 galaxy,
that is, an edge-on S0 with a bulge that dominates at both small and large radii.  However, the photometry
reveals something much more remarkable.  At the center, there is a normal, small, classical bulge (it is
overexposed in Figure 13 but evident in Figure 14).  The edge-on disk is essentially normal, too;
it has a higher-than-normal apparent surface brightness only because of the long path length resulting
from the fact that we see it edge-on.  The remarkable thing about NGC 4638 is the diffuse, boxy halo.
It proves to be a separate component from the disk and bulge.  And its shallow surface
brightness gradient is characteristic of a large spheroidal galaxy, not a bulge.  NGC 4638 therefore has 
both S0 and Sph characteristics.

\singlecolumn

\vfill

\includegraphics{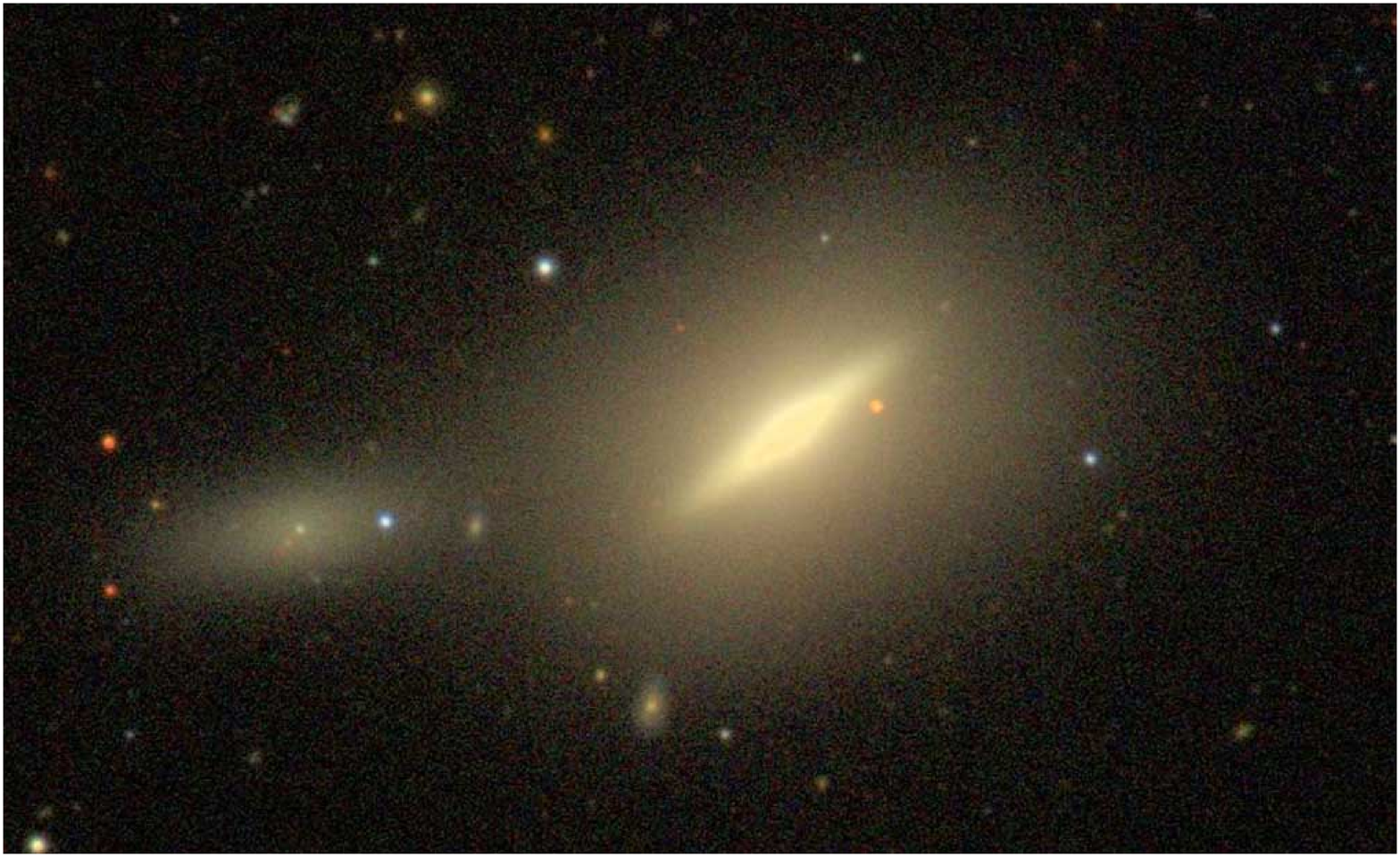}

Fig.~13~-- Color image of NGC 4638 = VCC 1938 from {\tt WIKISKY}.  The brightness ``stretch'' here emphasizes
faint features, i.{\ts}e., the extremely boxy, low-surface-brightness halo in which the S0 disk and bulge are
embedded.  The latter are illustrated using the HST ACS images in Figure 14.  The elongated dwarf to the west
of NGC 4638 is the Sph,N galaxy NGC 4637.

\eject

\doublecolumns

\cl{\null} \vskip 2.7truein

\includegraphics{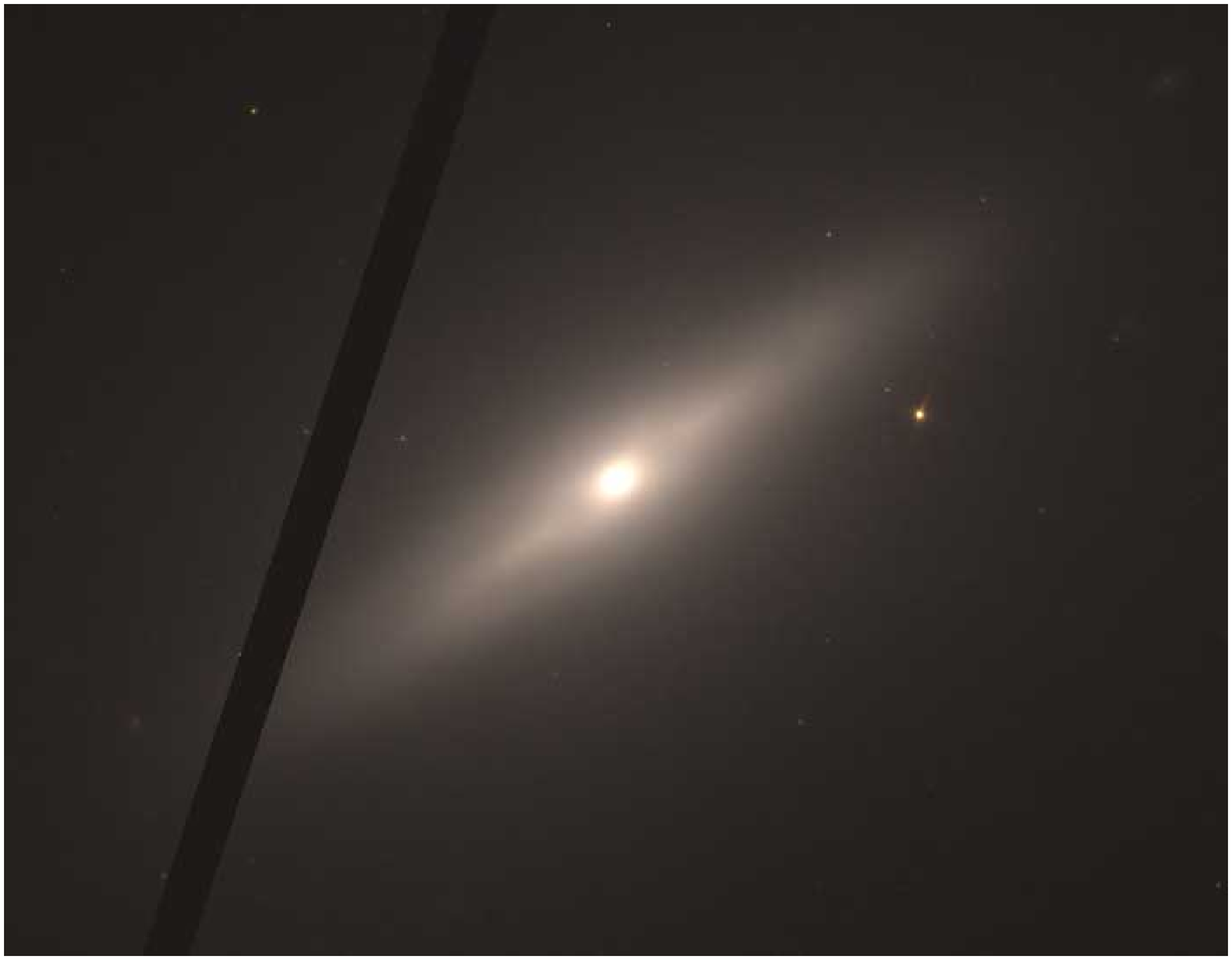}

Fig.~14~-- Color image of NGC 4638 = VCC 1938 made from the HST ACS $g$, mean of $g$ and $z$, and $z$ images.  
This image shows the edge-on disk and central bulge.  Brightness is proportional to the square root of intensity, 
so the brightness gradient in the bulge is much steeper than that in the boxy halo.  The very red foreground star 
near the NE side of the disk is also evident in Figure 13.

\vskip 4truein


 \includegraphics{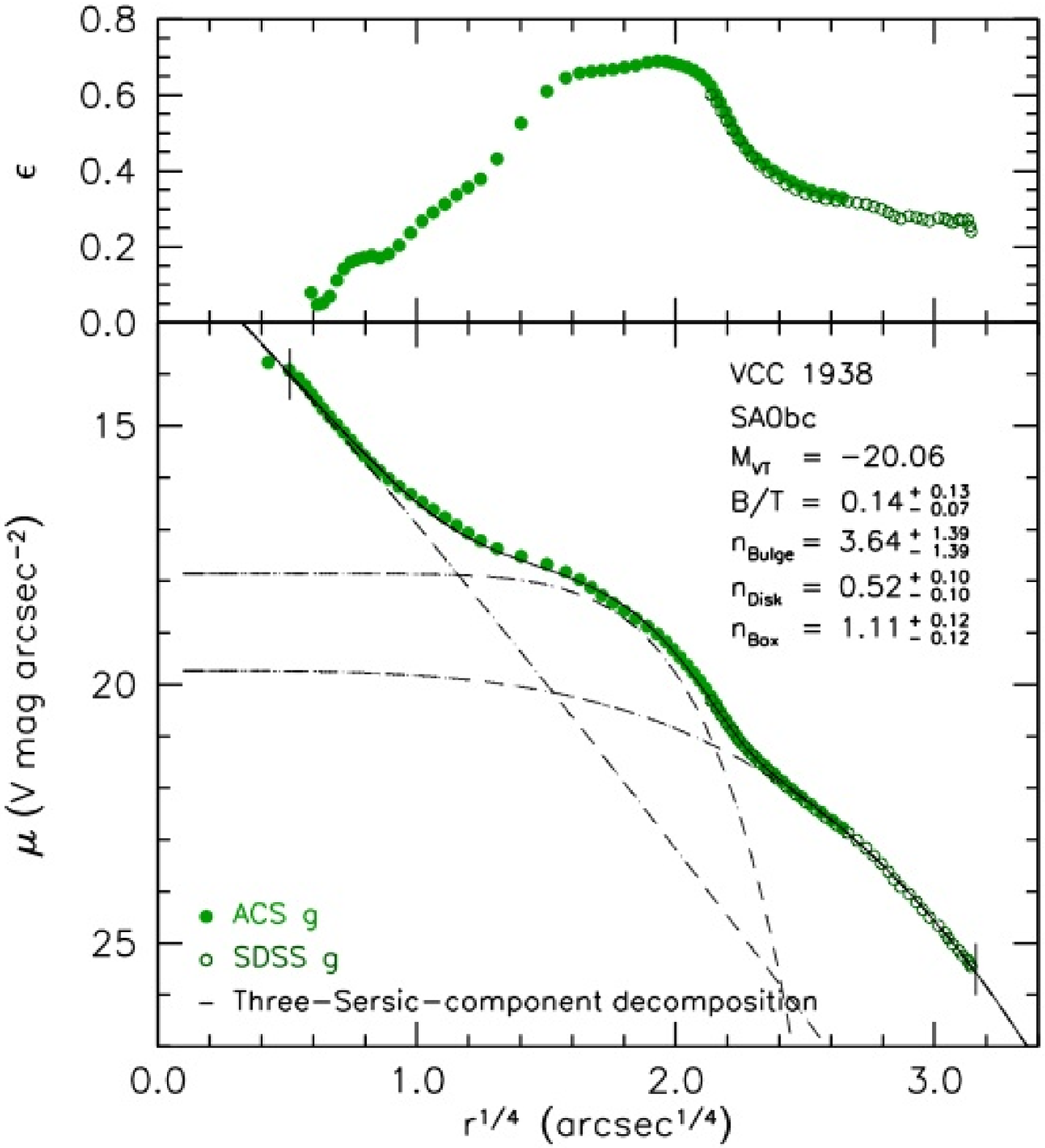}

Fig.~15~-- Ellipticity $\epsilon$ and surface brightness $\mu_V$ along the major axis of NGC 4638 as measured 
on the HST ACS and SDSS $g$ images.  Dashed curves show a three-S\'ersic-function decomposition of 
the profile inside the fit range ({\it vertical dashes\/}).  The bulge is small, but it is (at least:) mainly classical.
The disk has a Gaussian profile, as do many other S0s discussed in the Appendix.  Remarkably, the outer, boxy halo 
is clearly distinct from the bulge and disk and has a S\'ersic index $n = 1.11 \pm 0.12$ characteristic of a Sph galaxy.
The sum of the components ({\it solid curve}) fits the data with an RMS of 0.054 V mag arcsec$^{-2}$.

\vskip 12pt

      Figure 15 shows our photometry of NGC 4638.  The bulge has the flattening and the large S\'ersic index
(albeit with large uncertainties) of a classical bulge.  The disk is Gaussian, as are all lenses and
many outer S0 disks discussed in the previous sections and in the Appendix.  

      The boxy halo is clearly distinct from the disk and bulge.  Its profile is robustly concave-downward in 
Figure~15, indicating a small S\'ersic index $n \simeq 1$.  In contrast, if the bulge and halo formed a single
component with a large S\'ersic index, the halo profile would look concave-up in Figure 15.  The small S\'ersic 
index is characteristic of a Sph galaxy or disk.  Figure 16 shows how the F2006 whole-galaxy parameters are
decomposed into bulge, disk, and halo parts.  The halo plots within
the bulge sequence, but it will prove to be consistent with S0 disks (Figure 18).

      So NGC 4638 has properties of both S0 and Sph galaxies.  We suggest that it was produced from a relatively
normal SB(lens)0 or similar spiral galaxy by dynamical heating (i.{\ts}e., harassment:~\S\ts8.3) in a dense part
of the Virgo cluster (in projection, near NGC 4649 and surrounded on three sides by other galaxies: Fig.~40).  Its remaining
disk may be the (robust) remnant of the lens and the boxy halo may be the heavily heated remnant
of a (much less robust) disk.  Variants of this interpretation are possible (e.{\ts}g., the disk may have formed
after the Sph by late infall of cold gas).  

\vfill

\includegraphics{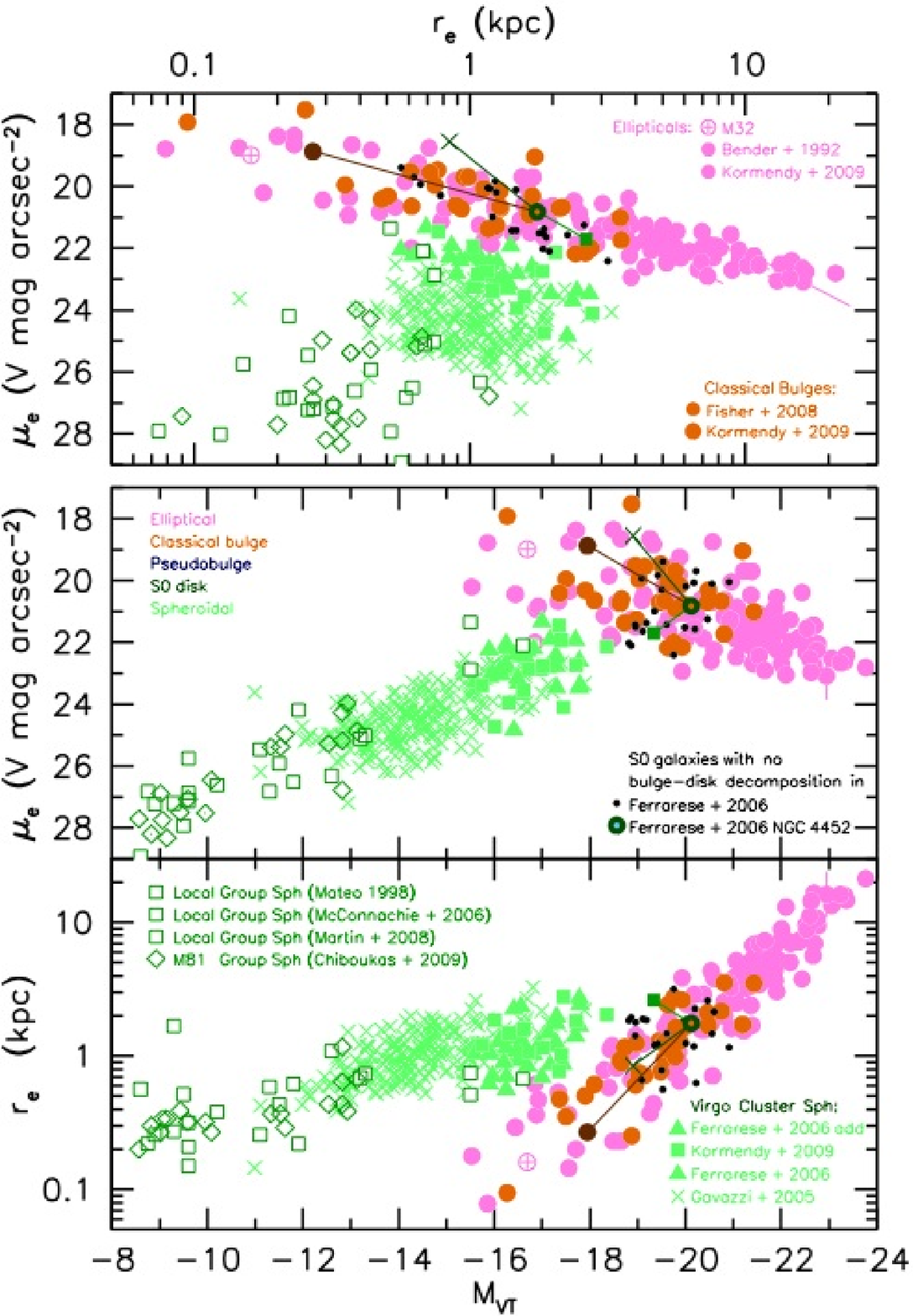}

Fig.~16~-- Parameter correlations showing the results~of the photometric decomposition of NGC 4638.  
The green filled circles with the brown centers show the whole-galaxy parameters measured by F2006.  Lines connect them to 
our measurements of the bulge ({\it brown filled circles\/}), disk ({\it green crosses\/}), and boxy halo 
({\it dark green filled squares\/}).

\eject

\singlecolumn
\cl{\null}

\vskip -24pt

\cl {4.2.~\it Photometric Parameters of Virgo Cluster S0 Galaxies}
\vss

      Table 1 lists the photometric parameters of Virgo cluster S0 galaxies measured in this paper.
All ACS VCS S0s that are not measured in this paper have bulge-disk decompositions in 
Gavazzi et al.~(2000),
Laurikainen \etal (2010: VCC 1030 = NGC 4435; this $K$-band measurement is insensitive to dust), or
Baggett \etal (1998: VCC 1535~= NGC 4526).
Thus all 100 galaxies from Ferrarese \etal (2006) appear in all of our parameter correlation diagrams.

      Galaxies in Table 1 are listed in order of (pseudo)bulge-to-total luminosity ratio $(P)B/T$ (Column 5).
Recall that all of these objects except the E galaxy NGC 4551 were classified S0 in F2006 and in standard
catalogs.  We base our conclusions on Table 1 and on the Virgo cluster S0s discussed in KFCB.
Additional $B/T$ values for S0 galaxies analyzed in KFCB are
$0.77 \pm 0.02$ in NGC 4660 (S0a);
$0.67 \pm 0.04$ in NGC 4564 (S0a);
$0.52$          in NGC 4570 (S0ab);
$0.33$          in NGC 4489 (S0b);
$0.13$          in NGC 4318 (S0bc).
{\it We conclude that S0 galaxies span the complete range of $(P)B/T$ ratios from 1 to 0.  This provides
quantitative justification for the parallel sequence classification and for the addition of Sph galaxies,
which have $(P)/B = 0$, at the end of the sequence.}

      A provisional correspondence between $(P)B/T$ and stage along the Hubble sequence is proposed in 
Column (3).  It was constructed by comparing these S0 galaxies to prototypical spiral galaxies with known
$(P)B/T$ ratios.  In addition, we choose to convert an {\it observational conclusion} about spiral galaxies
into a {\it classification criterion} for S0s.  Kormendy \& Kennicutt (2004) conclude that no Sc galaxy
known to them contains a classical bulge.  Consequently, we choose here to classify any galaxy that contains
even a small classical bulge as Sbc or earlier.  In fact, the smallest $B/T = 0.13 \pm 0.02$ in NGC 4762 that
we associate with type S0bc here corresponds well to $B/T = 0.12 \pm 0.02$ in the Sbc galaxy NGC 4258 (Kormendy 2011a).
A detailed study of bulge classifications and $(P)B/T$ ratios in larger samples of spiral and S0 galaxies will be
required to suggest more accurate Hubble stage classifications for S0s.  Such studies are in progress.
\vskip -24pt
\cl{\null}

%
%
\def\endtable{\endgroup}
\def\tableheight{\vrule width 0pt height 8.5pt depth 3.5pt}
{\catcode`|=\active \catcode`&=\active 
    \gdef\tabledelim{\catcode`|=\active \let|=\vbar
                     \catcode`&=\active \let&=\nobar} }
\def\table{\begingroup
    \def\twidth{\hsize}
    \def\tablewidth##1{\def\twidth{##1}}
    \def\defaultheight{\vrule width 0pt height 8.5pt depth 3.5pt}
    \def\heightdepth##1{\dimen0=##1
        \ifdim\dimen0>5pt 
            \divide\dimen0 by 2 \advance\dimen0 by 2.5pt
            \dimen1=\dimen0 \advance\dimen1 by -5pt
            \vrule width 0pt height \the\dimen0  depth \the\dimen1
        \else  \divide\dimen0 by 2
            \vrule width 0pt height \the\dimen0  depth \the\dimen0 \fi}
    \def\spacing##1{\def\defaultheight{\heightdepth{##1}}}
    \def\nextheight##1{\noalign{\gdef\tableheight{\heightdepth{##1}}}}
    \def\end{\cr\noalign{\gdef\tableheight{\defaultheight}}}
    \def\zerowidth##1{\omit\hidewidth ##1 \hidewidth}    
    \def\hline{\noalign{\hrule}}
    \def\skip##1{\noalign{\vskip##1}}
    \def\bskip##1{\noalign{\hbox to \twidth{\vrule height##1 depth 0pt \hfil
        \vrule height##1 depth 0pt}}}
    \def\header##1{\noalign{\hbox to \twidth{\hfil ##1 \unskip\hfil}}}
    \def\bheader##1{\noalign{\hbox to \twidth{\vrule\hfil ##1 
        \unskip\hfil\vrule}}}
    \def\spanloop{\span\omit \advance\mscount by -1}
    \def\extend##1##2{\omit
        \mscount=##1 \multiply\mscount by 2 \advance\mscount by -1
        \loop\ifnum\mscount>1 \spanloop\repeat \ \hfil ##2 \unskip\hfil}
    \def\vbar{&\vrule&}
    \def\nobar{&&}
    \def\hdash##1{ \noalign{ \relax \gdef\tableheight{\heightdepth{0pt}}
        \toks0={} \count0=1 \count1=0 \putout##1\end 
        \toks0=\expandafter{\the\toks0 &\end} \xdef\piggy{\the\toks0} }
        \piggy}
    \let\e=\expandafter
    \def\putspace{\ifnum\count0>1 \advance\count0 by -1
        \toks0=\e\e\e{\the\e\toks0\e&\e\multispan\e{\the\count0}\hfill} 
        \fi \count0=0 }
    \def\putrule{\ifnum\count1>0 \advance\count1 by 1
        \toks0=\e\e\e{\the\e\toks0\e&\e\multispan\e{\the\count1}\leaders\hrule\hfill}
        \fi \count1=0 }
    \def\putout##1{\ifx##1\end \putspace \putrule \let\next=\relax 
        \else \let\next=\putout
            \ifx##1- \advance\count1 by 2 \putspace
            \else    \advance\count0 by 2 \putrule \fi \fi \next}   }
\def\tablespec#1{
    \def\vdimens{\noexpand\tableheight}
    \def\tabby{\tabskip=0pt plus100pt minus100pt}
    \def\r{&################\tabby&\hfil################\unskip}
    \def\c{&################\tabby&\hfil################\unskip\hfil}
    \def\l{&################\tabby&################\unskip\hfil}
    \edef\templ{\noexpand\vdimens ########\unskip  #1 
         \unskip&########\tabskip=0pt&########\cr}
    \tabledelim
    \edef\body##1{ \vbox{
        \tabskip=0pt \offinterlineskip
        \halign to \twidth {\templ ##1}}} }

\input colordvi
\def\B{\Blue}
\def\G{\Green}
\def\R{\Red}
\def\Br{\Brown}
\def\Bl{\Black}

\def\0{$\phantom{0}$}
\def\m{$\phantom{\ts\ts.\ts}$}

\def\sss{\skip{5pt}}



{\sevenpoint
\sevenrm

$$
\table
\tablewidth{18.6truecm}
\tablespec{\l\l\l\c\c\c\c\c\c\c\c\c\c}
\body{
\header{TABLE 1}
\skip{5pt}
\header{STRUCTURAL PARAMETERS OF VIRGO CLUSTER S0 AND Sph GALAXIES}
\skip{10pt}
\hline \skip{0.001truein} \hline 
\skip{.2truecm}\hline \skip{0.001truein} \hline 
\skip{5pt}
& Galaxy    & Galaxy& Type  & $D$   & $\B{(P)}\Br{B/T$          &\Br{$M_{V,\rm bulge}$}& $n_{\rm bulge}$ & $\mu_{eV,\rm bulge}$      &\0$\log{r_{e,\rm bulge}}$}    &\B{$M_{V,\rm disk}$ & $n_{\rm disk}$          &$\mu_{eV,\rm disk}$         &\0$\log{r_{e,\rm disk}}$ }      & \end
& NGC       & VCC   &       & [Mpc] &\Br{                 }     &         \Br{         &                 & [mag arcsec$^{-2}$]       & [kpc]                   }    &\B{                 &                         &[mag arcsec$^{-2}$]         & [kpc]                   }      & \end
& (1)       & (2)   & (3)   & (4)   &\Br{(5)              }     &         \Br{(6)      & (7)             & (8)                       & (9)                     }    &\B{(10)             & (11)                    & (12)                       & (13)      }                    & \end
\skip{5pt}
\hline \skip{0.001truein} \hline
\skip{5pt}
\Bl& N4551 & V1630  & E     & 16.14 &\R {$1.00^{+0.00}_{-0.00}$& $-19.11$   & $1.968^{+0.056}_{-0.056}$ &$20.715^{+0.032}_{-0.032}$ &$ 0.080^{+0.005}_{-0.005}   $} &\B{ \dots        &\0\dots                     & \dots                      & \dots                    }     & \end \sss 
\Bl& N4417 & V944   & SA0a  & 16.00 &\Br{$0.88^{+0.06}_{-0.11}$& $-19.89$   & $3.80^{+0.11}_{-0.11}$    &$20.51^{+0.07}_{-0.07}$    &$ 0.186^{+0.012}_{-0.012}   $} &\B{ $-17.73$     &\0$0.52^{+0.06}_{-0.06}$    & $21.70^{+0.12}_{-0.12}$    & $0.495^{+0.014}_{-0.015}$}     & \end \sss 
\Bl& N4442 & V1062  & SB0a  & 15.28 &\Br{$0.78^{+0.09}_{-0.11}$& $-20.44$   & $3.20^{+0.31}_{-0.31}$    &$20.06^{+0.38}_{-0.38}$    &$ 0.193^{+0.095}_{-0.121}   $} &\B{ $-19.07$     &\0$0.48^{+0.15}_{-0.15}$    & $22.47^{+0.32}_{-0.32}$    & $0.718^{+0.021}_{-0.022}$}     & \end \sss 
\Bl& N4352 & V698   & SA0a  & 18.7  &\Br{$0.71^{+0.13}_{-0.05}$& $-18.49$   & $3.70^{+0.5}_{-0.9}$      & $21.89^{+0.05}_{-0.50}   $&$ 0.191^{+0.053}_{-0.107}   $} &\B{ $-17.52$     &\0$0.50^{+0.05}_{-0.12}$    & $22.23^{-0.02}_{+0.15}$    & $\m0.388^{+0.027}_{-0.010}$\0} & \end \sss
\Bl& N4578 & V1720  & SA0ab & 16.29 &\Br{$0.56^{+0.12}_{-0.12}$& $-19.09$   & $3.13^{+0.44}_{-0.44}$    &$20.70^{+0.47}_{-0.47}$    &$ 0.035^{+0.110}_{-0.147}   $} &\B{ $-18.83$     &\0$0.56^{+0.11}_{-0.11}$    & $22.99^{+0.25}_{-0.25}$    & $0.628^{+0.016}_{-0.017}$}     & \end \sss 
\Bl& N4483 & V1303  & SB0ab & 16.75 &\Br{$0.47^{+0.09}_{-0.09}$& $-18.20$   & $4.5^{+0.5}_{-0.5}$       &$21.34^{+0.35}_{-0.35}$    &$-0.066^{+0.120}_{-0.167}\0\0$}&\B{ $-18.33$     &\0$1.1^{+0.2}_{-0.2}$\0\0   & $21.89^{+0.29}_{-0.29}$    & $0.266^{+0.032}_{-0.035}$}     & \end \sss 
\Bl& N4528 & V1537  & SB0ab & 15.8  &\Br{$0.38^{+0.13}_{-0.13}$& $-18.04$   & $2.55^{+0.61}_{-0.20}$    & $18.83^{+0.66}_{-0.26}   $&$-0.533^{+0.169}_{-0.076}\0\0$}&\B{ $-18.43$     &\0$1.00^{+0.05}_{-0.05}$    & $20.69^{+0.09}_{-0.14}$    & $0.089^{+0.012}_{-0.012}$   }  & \end \sss
\Bl& N4623 & V1913  & SA0b  & 17.38 &\B {$0.26^{+0.12}_{-0.09}$& $-17.47$   & $3.34^{+0.76}_{-0.76}$    &$21.18^{+0.85}_{-0.85}$    &$-0.166^{+0.188}_{-0.339}\0\0$}&\B{ $-18.61$     &\0$1.00$\0\0\0\0\0          & $21.41^{+0.08}_{-0.08}$    & $0.387^{+0.008}_{-0.008}$}     & \end \sss 
\Bl& N4638 & V1938  & SA0bc & 17.46 &\Br{$0.14^{+0.13}_{-0.07}$& $-17.94$   & $3.64^{+1.39}_{-1.39}$    &$18.87^{+1.54}_{-1.54}$    &$-0.570^{+0.224}_{-0.490}\0\0$}&\B{ $-18.90$     &\0$0.52^{+0.10}_{-0.10}$    & $18.55^{+0.14}_{-0.14}$    &$-0.074^{+0.014}_{-0.014}$}\0\0 & \end \sss 
\Bl& N4638 & V1938  & SA0bc & 17.46 &\Br{$0.14^{+0.13}_{-0.07}$}& \dots     & \dots                     & \dots                     &  \dots                        &\G{ $-19.34$     &\0$1.11^{+0.12}_{-0.12}$    & $21.70^{+0.16}_{-0.16}$    & $0.422^{+0.025}_{-0.027}$}     & \end \sss 
\Bl& N4762 & V2095  & SB0bc & 16.53 &\Br{$0.13^{+0.02}_{-0.02}$& $-18.76$   & $2.29^{+0.05}_{-0.05}$    & $18.58^{+0.05}_{-0.05}   $&$-0.382^{+0.014}_{-0.015}\0\0$}&\B{ $-20.82$     &  \dots                     & $20.57^{+0.09}_{-0.09}$    & $\m0.822^{+0.043}_{-0.048}$ }  & \end \sss
\Bl& N4550 & V1619  & SA0c  & 15.49 &\B{$0.018^{+0.003}_{-0.005}$&$-15.04$  & $1.54^{+0.20}_{-0.38}$    &$17.16^{+0.25}_{-0.45}$    &$-1.382^{+0.065}_{-0.160}\0\0$}&\B{ $-19.38$     &\0$1.69^{+0.13}_{-0.08}$    & $20.21^{+0.33}_{-0.33}$    & $0.240^{+0.006}_{-0.005}$}     & \end \sss 
\Bl& N4452 & V1125  & SB0c  & 16.53 &\B{$0.017^{+0.004}_{-0.004}$& $-14.78$ & $1.06^{+0.14}_{-0.14}$    & $19.33^{+0.17}_{-0.17}   $&$-0.778^{+0.046}_{-0.051}\0\0$}&\B{ $-19.19$     &  \dots                     & $20.31^{+0.05}_{-0.05}$    & $\m0.454^{+0.005}_{-0.005}$ }  & \end \sss
\G{& \dots & V2048  & Sph,N & 16.53 &    $0.000^{+0.00}_{-0.00}$&  \dots    & \dots                     & \dots                     &  \dots                        &\B{ $-15.06$     &  \dots                     & $23.83^{+0.20}_{-0.15}$    & $\m0.134^{+0.057}_{-0.013}$}}  & \end \sss
\G{& \dots & V2048  & Sph,N & 16.53 &\G {$0.000^{+0.00}_{-0.00}$&  \dots    & \dots                     & \dots                     &  \dots                        &    $-17.78$     &  \dots                     & $22.75^{+0.20}_{-0.09}$    & $\m0.332^{+0.052}_{-0.025}$ }} & \end 
\skip{5pt}
\hline \skip{0.001truein} \hline
\skip{5pt}
}
\endtable
$$

\pretolerance=15000  \tolerance=15000
\lineskip=0pt \lineskiplimit=0pt

\vskip -10pt

      NOTES -- We adopt individual distances $D$ (Column 4) from Mei \etal (2007) when available. Otherwise, we use the mean distance $D = 16.53$ Mpc for ``all [79] galaxies (no W$^\prime$
cloud)'' given in Table 3 of Mei's paper.
Colors encode structural component types to match colors used for symbols in correlation plots: 
\Br{brown for classical bulges,}
\B{blue for pseudobulges and disks,} \Bl{and} 
\G{green for spheroidals}\Bl{.   }
Column (5) gives the \Br{classical-bulge-to-total luminosity ratio $B/T$} \Bl{or the} \B{pseudobulge-to-total luminosity ration $PB/T$}. \Bl
Columns (6) -- (9) list the (pseudo)bulge parameters bulge absolute magnitude $M_{V,\rm bulge}$, S\'ersic index $n$, effective brightness $\mu_{eV, \rm bulge}$ at the effective
radius and the base-10 logarithm of the major-axis effective radius $r_{e,{\rm bulge}}$ that contains half of the light of the bulge.
Columns (10) -- (13) similarly list the disk parameters.  Two galaxies contain a component that looks indistinguishable from spheroidal galaxies, it is listed in the disk
columns in {\G{green}}.  Both of these galaxies also have disks and so appear in two lines.  
Parameter errors are the internal errors given by the S\'ersic-S\'ersic decomposition program combined with errors given by comparing decompositions made with 
different assumptions (e.{\ts}g., disk fixed as exponential, or different fitting ranges).  
Parameter errors for photometric decompositions are uncertain because of strong (e.{\ts}g., bulge-disk) parameter coupling and because they depend on the
fitting functions that are assumed to describe the components.
}

\vfill\eject

\singlecolumn

\vss\vsss
\cl {4.3.~\it Parameter Correlations Including Virgo Cluster S0 Galaxies}
\vss

      Figure 17 shows the $\mu_e$ -- $r_e$ -- $M_V$ correlations with the bulges of Virgo cluster S0s added.  
All ACS VCS galaxies except two that are clobbered by dust are included here and in all further figures.  
Also added are S0 bulges from the bulge-disk decompositions of Baggett \etal (1998).  Bulge results are discussed here; 
S0$+$S galaxy disks are added in the next section.

      Figure 17 shows that classical bulges and elliptical galaxies have indistinguishable parameter correlations.  
The few pseudobulges that happen to be in our S0 sample deviate only a little from the above correlations, but larger 
samples of pseudobulges in later-type galaxies show larger scatter than classical bulges toward both high and low surface 
brightnesses.  These results confirm the conclusions of many previous studies (e.{\ts}g., 
Kormendy \& Fisher 2008;
Fisher \& Drory 2008).
The important new result is that adding the ACS VCS and Baggett S0s strengthens our derivation of the E$+$bulge correlations,
especially at the compact end of the luminosity sequence.  The conclusion that Sph galaxies are not \hbox{dwarf $\equiv$
low-luminosity} ellipticals is correspondingly strengthened also.

\vfill

\cl{\null} \vskip 6.5truein

\includegraphics{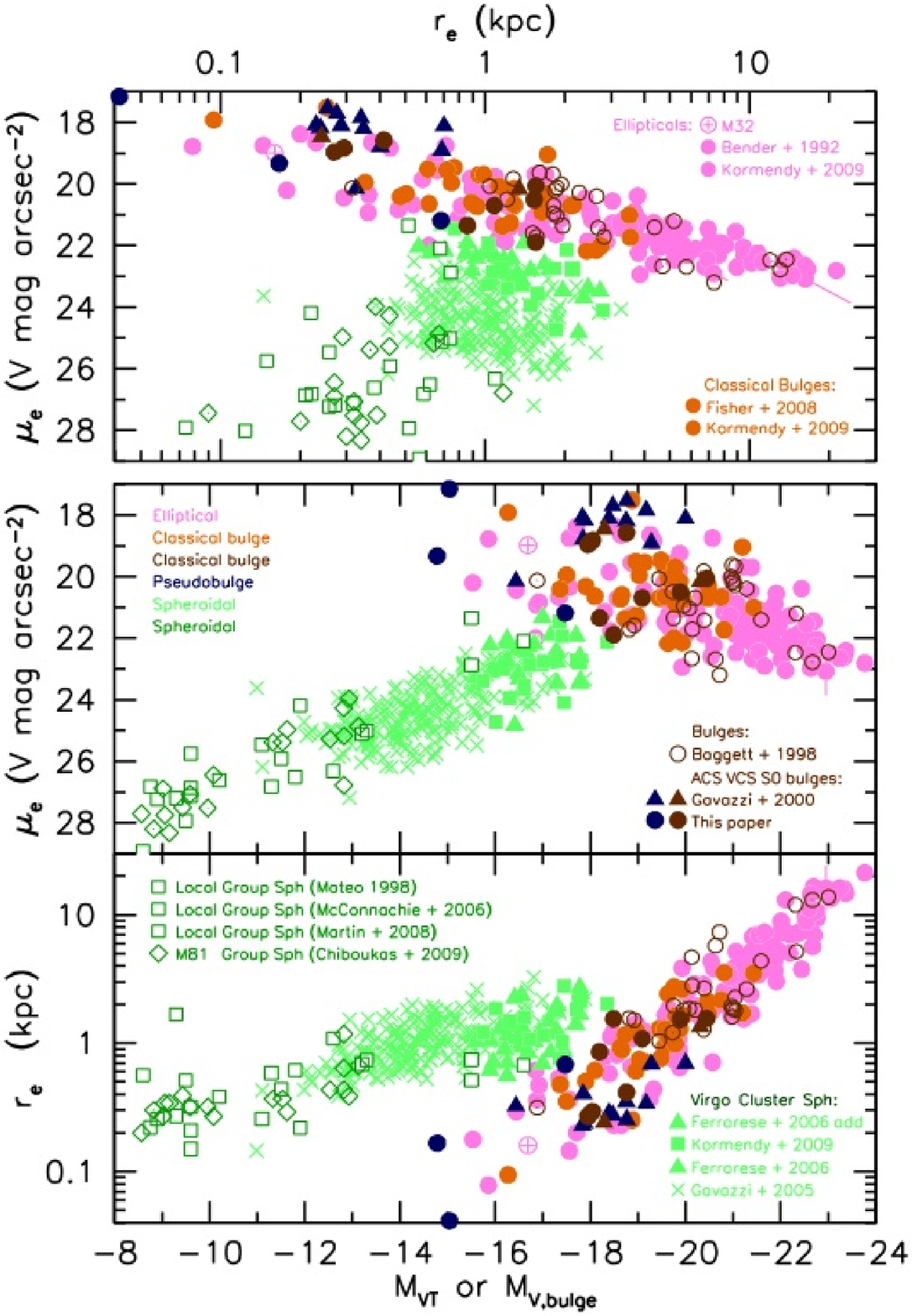}

Fig.~17~-- Global parameter correlations from Figure 2 including the augmented sample of bulges from 
this study.  All 26 ACS VCS S0s that were omitted in KFCB are included here, three as Sphs and 
23 as bulges.  Galaxies with profile measurements and photometric decompositions in this paper
are shown as dark brown or dark blue filled circles.  The remaining ACS VCS S0s have photometric
decompositions in Gavazzi \etal (2000) and are similarly shown as filled triangles.  In addition,
S0 bulges from the photometric decompositions of Baggett \etal (1998) are added as open
circles.  We do not have bulge-pseudobulge classifications for them, but most are likely 
to be classical based on their bright $M_{V,\rm bulge}$.  This is our final E$+$bulge
sample; for simplicity, points in further figures encode bulge type but not the source of the data.
\pretolerance=15000  \tolerance=15000 

\eject

\vss\vsss
\cl {5.~THE PARAMETER CORRELATIONS OF Sph GALAXIES ARE CONTINUOUS WITH THOSE OF S0 DISKS}
\vss

      Section 4 establishes complete continuity between early-type S0 galaxies with large bulges and Sph galaxies 
with no bulges.  It includes examples of the formerly missing, late-type (S0bc -- S0c) galaxies with tiny bulges.  
Also, the Sph VCC 2048 and the S0 NGC 4638 show aspects of both kind of galaxy, i.{\ts}e., edge-on disks embedded in 
Sph-like halos.  In this section, we make this link more quantitative with a larger sample of galaxies.  Figure 18 
shows the \hbox{$\mu_e$ -- $r_e$ -- $M_V$} correlations for Es$+$bulges and Sph galaxies with S0 disks added.  
The data come from the bulge-disk decomposition papers listed in the key to the middle panel.  Extending results from 
the previous section, Figure 18 shows that the Sph galaxy sequence is continuous with the sequence of S0 galaxy
disks.  There is a kink where bulges disappear and where (we suggest) the correlations turn into a sequence of decreasing 
baryon retention at lower galaxy luminosity (\S\S\ts3.2.4,\ts8,\ts9).

\singlecolumn

\cl{\null} \vskip 6.5truein

\vfill

\includegraphics{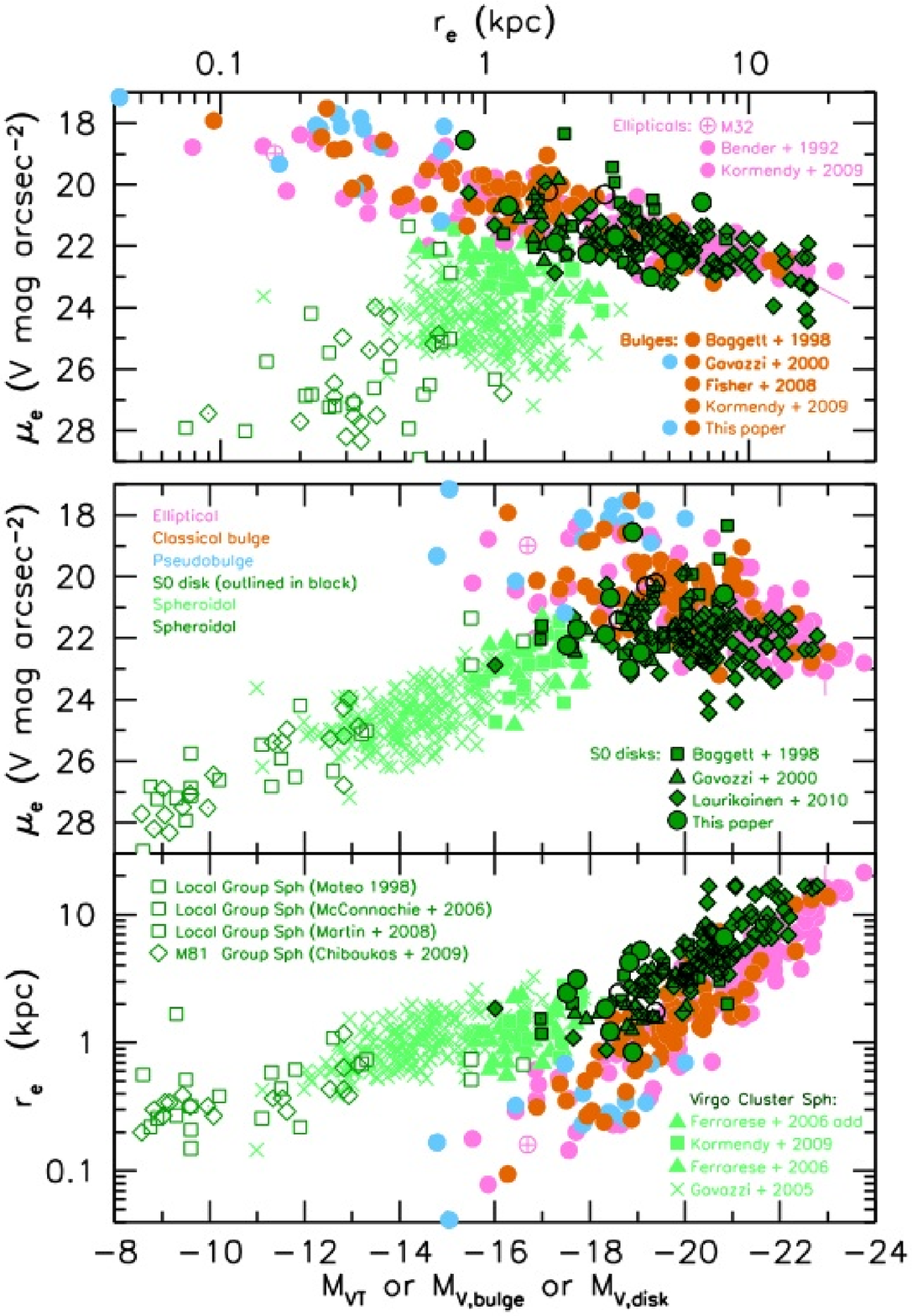}

Fig.~18~-- Parameter correlations for ellipticals, bulges, and Sphs with S0 disks added 
({\it green points outlined in black\/}).  Bulges and disks of S0 galaxies are plotted 
separately.  All S0 galaxies from the ACS VCS survey (Ferrarese \etal 2006) are plotted.  
The middle panel shows the Freeman (1970) result that disks
of high-luminosity galaxies tend to have the same central surface brightness $\mu_0$.
Here, $\mu_e \simeq 22.0$ $V$ mag arcsec$^{-2}$ corresponds to $\mu_0 = \mu_e - 1.82 \simeq
20.2$ $V$ mag arcsec$^{-2}$ or about 21.1 $B$ mag arcsec$^{-2}$.  This is slightly brighter
than the Freeman value of 21.65 $B$ mag arcsec$^{-2}$ because many of the disk parameters are
not corrected to face-on orientation.  Most disks that have higher-than-normal surface brightnesses 
are edge-on (e.{\ts}g., our measurements of NGC 4762 and NGC 4452).  Most disks that 
have lower-than-normal surface brightnesses are in galaxies that have outer rings; $\mu_e$ is
faint because the outer ring is included in the measurements.  We conclude from this figure
that Sph galaxies are continuous with the disks but not the bulges of S0 galaxies.
\pretolerance=15000  \tolerance=15000 

\eject

\doublecolumns


      Figure 18 includes 127 S0 galaxy disks from five sources:  

      Five disks are from decompositions in KFCB.  This paper provides 4 from the main text and 8
from the Appendix.

      Gavazzi \etal (2000) provide $H$-band photometry and photometric decompositions for 19 Virgo cluster 
S0s, 13 of which are also in the ACS VCS.  The 19 galaxies that we use here do not include
NGC 4489 = VCC 1321 (we use KFCB results), or 3 more galaxies that we remeasure here.  One additional
galaxy was discarded because it is severely tidally distorted, and one is clearly an Sa.
 
      Baggett \etal (1998) provide 16 S0 disk parameters.  These decompositions are somewhat less accurate 
than the more recent ones, because they use $r^{1/4}$ laws to describe the bulges.  We therefore apply 
somewhat stricter quality cuts: we keep the galaxy only if $r_e/{\rm seeing} > 4.5$ and if $T < 0$, where 
``seeing'' is tabulated in Baggett's paper.

      Laurikainen \etal (2010) provide 85 S0 galaxies; i.{\ts}e., de Vaucouleurs type $T \leq 0$ and 
yet not elliptical and not a merger in progress (i.{\ts}e., a disturbed E with shells and
tidal tails).  We were conservative in correcting Hubble types.  The most common correction is that 
many of the brightest catalogued ``S0s'' are really ellipticals.  They get misclassified as S0s for
two reasons: (1) A prominent dust disk is enough to earn an S0 classification in
many papers.  An example in the Virgo cluster is  NGC 4459. KFCB show that its main body is well 
described by a single S\'ersic function with, at small radii, extra light over the inward 
extrapolation of the outer S\'ersic~fit.  (2) The brightest ellipticals have S\'ersic profiles 
with \hbox{$n \gg 4$}.  Morphologists sometimes see these as core-halo objects and so classify the 
galaxies as S0.  NGC 4406 is an example.  KFCB discuss these classification problems. 
In the present sample, some of the brightest ``S0s'' probably are ellipticals.  Figure 18 shows that
such objects are consistent with the E correlations even when they are treated as S0s.

      Only a limited number of S0s are near enough for detailed component studies.  It is
inevitable that various authors' samples overlap.  We made sure that each galaxy is plotted only
once in Fig.~18.  In cases of duplication, we usually kept the results from our work, Gavazzi, 
Baggett, and Laurikainen in this order.  This is why numbers of galaxies from some sources
look surprisingly small.

      We conclude from Fig.~18 that the main bodies of Sph galaxies (not including 
nuclei) form a continuous sequence in parameter space with the disks (but not the bulges) of S0 galaxies. 
This is consistent with and one of the motivations for our suggestion that Sph galaxies
belong at the late-type end of the Figure 1 tuning-fork diagram next to S0cs.  

      The Sph and S0 disk sequences overlap a little~but~not very much.  This is partly a selection effect.
The~faintest true S0s do not make it into most galaxy samples.  Objects that are traditionally 
classified as dS0 almost always turn out to be bulgeless and therefore (by definition) are Sph.

      Nevertheless, S0 galaxies with disk absolute magnitudes $M_{V,\rm disk}$ \gapprox \ts$-18$ are
rare.  Figure 19 (from Kormendy \& Freeman 2011) shows why.  It quantifies the ``rotation~curve~conspiracy'' 
that visible and dark matter are arranged in galaxies so as to produce approximately featureless, flat 
flat rotation curves (Bahcall \& Casertano 1985; van Albada \& Sancisi 1986; Sancisi \& van Albada 1987).  
The rotation velocities produced by the bulge, disk, and halo are nearly equal 
($V_{\rm circ,bulge} \simeq V_{\rm circ,disk} \simeq V_{\rm circ}$) 
in galaxies with $V_{\rm circ} \sim 200$~km~s$^{-1}$.  In smaller galaxies, $V_{\rm circ,disk} \rightarrow 0$
at finite $V_{\rm circ} \simeq 42 \pm 4$ km s$^{-1}$.~~This~tells~us the mass scale below which dark halos
generally cannot capture or retain baryons (Kormendy \& Freeman 2011).  It is in good agreement with the theoretical
prediction that the formation of visible dwarfs is supressed below $V_{\rm circ} \sim 30$ to 40 km s$^{-1}$ 
because few such galaxies accrete enough gas before cosmological reionization to become discoverable 
(Bullock \etal 2000;
Cattaneo \etal 2011).  

      The important point here is that the correlation for bulges is steeper than the one for disks and 
reaches zero at $V_{\rm circ} \sim 104 \pm 16$ km s$^{-1}$. Progenitor late-type galaxies like M{\ts}33
and fainter do not -- by and large -- contain bulges.  

      The Tully-Fisher (1977) relation for S0 galaxies tells us that $V_{\rm circ} = 104$ km s$^{-1}$
at $M_V \simeq -17$ to $-18$ 
(Neistein \etal 1999;
Hinz, \etal 2003;
Bedregal \etal 2006;
Williams \etal 2010).
This is in excellent agreement with Figure~18: it is the disk absolute magnitude where S0s stop and
spheroidals take over.  When no bulge~is~visible, morphologists call the galaxy a spheroidal.
S0s and Sphs overlap a little in luminosity, because
the disappearance of bulges does not happen exactly at some magic disk $M_{V,\rm disk}$.

      In summary, the structural parameter correlations of spheroidal galaxies are continuous with the
disks of S0 galaxies.  The changeover in nomenclature does not reflect some 
fundamental change in galaxy properties but rather happens at the disk absolute magnitude where 
rotation curve decompositions tell us that bulges disappear.

\vfill

\includegraphics{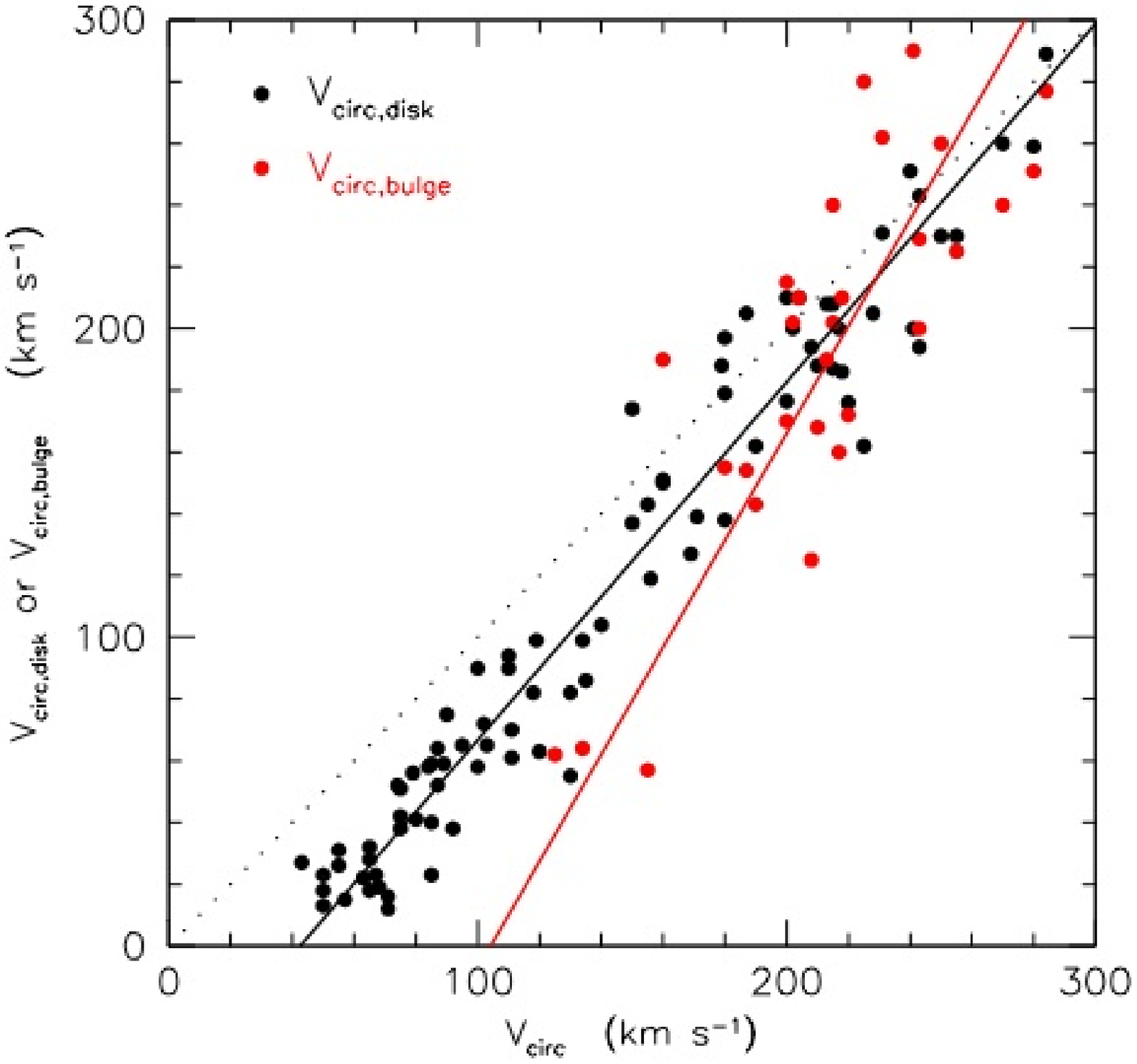}

      Fig.~19 -- Maximum rotation velocity of the bulge $V_{\rm circ,bulge}$ ({\it red points})
and disk $V_{\rm circ,disk}$ ({\it black points}) given in bulge-disk-halo decompositions of 
observed rotation curves $V(r)$ whose outer, DM rotation velocities~are~$V_{\rm circ}$.  This figure 
is from Kormendy \& Freeman (2011, updapted from Fig.~S2 in Kormendy \& Bender 2011); references to 
the $V(r)$ decomposition papers are given there.  The dotted line indicates that the rotation velocities 
of the visible and dark matter are equal.  Every~red~point has a corresponding black point, but many 
late-type galaxies are bulgeless, and then the plot shows only a black point.  The lines are symmetric
least-squares fits; the disk fit is
$V_{\rm circ,disk}  = (1.16 \pm 0.03)(V_{\rm circ} - 200) + (183 \pm 3)$ km s$^{-1}$;
$V_{\rm circ,bulge} = (1.73 \pm 0.29)(V_{\rm circ} - 200) + (166 \pm 9)$ km s$^{-1}$
is the bulge fit.  The correlation for bulges is steeper than that for disks; bulges disappear at 
$V_{\rm circ} \sim 104 \pm 16$ km s$^{-1}$.

\eject

\singlecolumn

\vskip -20pt

\cl {6.~THE PARAMETER CORRELATIONS OF Sph GALAXIES AND S0 GALAXY DISKS}
\cl {ARE INDISTINGUISHABLE FROM THOSE OF SPIRAL-GALAXY DISKS AND MAGELLANIC IRREGULARS}
\vss

      The result which suggested that Sph galaxies are defunct S$+$Im galaxies was the
observation by Kormendy (1985,~1987) that their parameter correlations are indistinguishable over
the absolute magnitude range in which they overlap.  However, galaxy samples were small in the
1980s.  Also, we now have to understand a sequence of Sphs $+$ S0 disks that is continuous from
S0s with $M_{V,\rm disk} \simeq -22$ to Sphs with $M_{VT} \simeq -8$ and probably much fainter.  Figure 20
brings the results of Kormendy (1985, 1987) up to date with a much larger sample of spiral-galaxy disks 
and Im galaxies.  It confirms that early- and late-type galaxies have similar disk parameter correlations
from the brightest disks to the faintest spheroidals and irregulars.

\cl{\null} \vskip 6.5truein

\vfill

\includegraphics{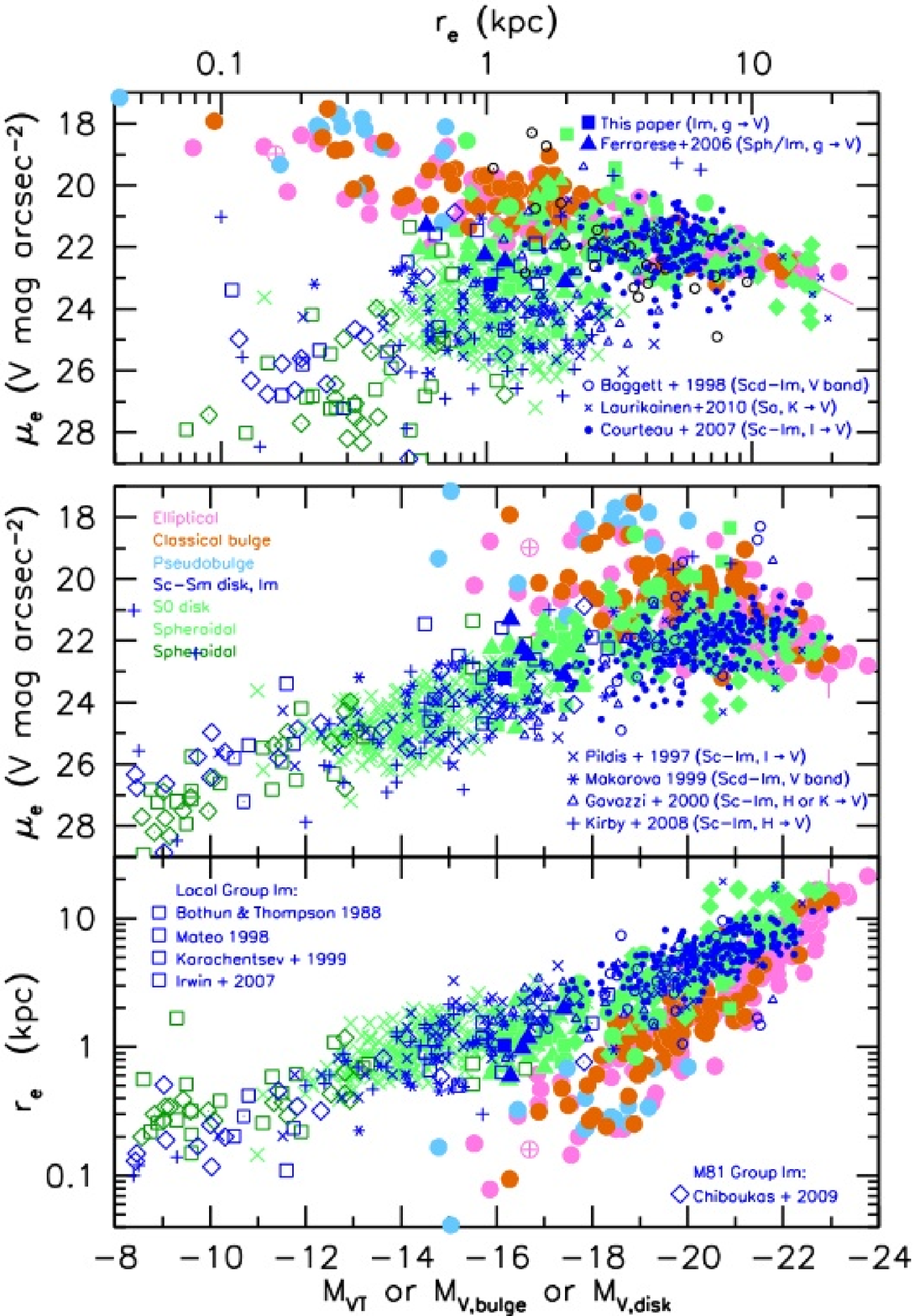}

Fig.~20~-- Parameter correlations for ellipticals, bulges, Sphs, and S0 disks, plus disks of 
Sa{\ts}--{\ts}Im galaxies ({\it blue points\/}).  When (pseudo)bulge-disk decomposition is
necessary, the two components are plotted separately.  All S0 galaxies from the ACS VCS survey 
(Ferrarese \etal 2006) are plotted ({\it light brown and light green points\/}).  Also, blue points 
show four galaxies from Ferrarese \etal (2006) that they note are Sph/Im transition galaxies and one 
galaxy (VCC 1512) from the above paper that we classify as Im and that we remeasured.  Most disks 
that have lower-than-normal surface brightnesses are in galaxies that have outer rings, and most disks 
that have higher-than-normal surface brightnesses are either edge-on or starbursting.  However, 
starbursting blue compact dwarf galaxies (BCGs) are omitted; they would add a few galaxies that 
scatter to higher surface brightnesses.   The blue points represent 407 galaxes from 14 sources listed in the keys.
\pretolerance=15000  \tolerance=15000 

\eject

\doublecolumns


      Figure 20 includes 407 Sa -- Im galaxy disks from~14 sources.  We concentrate on
Sc{\ts}--{\ts}Im galaxies, because our purpose is to compare S0 disks with late-type galaxies that may be
progenitors.  Also, disk parameters of \hbox{Sc{\ts}--{\ts}Im} galaxies are relatively well determined, 
because decompositions do not have to deal with large bulges.  However, we do include a sample of Sa galaxy 
disks from Laurikainen \etal (2010) to demonstrate that early-type disks are not very different from 
late-type disks in their structural parameters.  As in \S\ts5, the number of nearby galaxies is limited, 
so there is overlap between papers.  We ensure that each galaxy is plotted only once and take galaxies from 
various samples in the following order of preference (highest first):

      Parameters for the irregular galaxies in the Local Group were taken from 
Bothun \& Thompson (1988: LMC and SMC), 
Mateo (1998: 10 objects), 
Karachentsev \etal (1999: Sag DIG), 
Irwin \etal (2007: Leo T), and
Kirby \etal (2008: DDO 210 and IC 5152).

      Chiboukas \etal (2009) measured 19 dwarf Sph and 16 dS$+$Im galaxies in the M{\ts}81 group.  The 
spheroidals were added earlier to all parameter correlation figures; the late-type galaxies are added here.  
These objects greatly strengthen the derivation of the Sph and disk sequences at the lowest luminosities 
and show that the sequences remain remarkably similar all the way down to $M_V \simeq -9$.

      Turning next to the Virgo cluster, this paper contributes one object, VCC 1512.  It was recognized 
by F2006 as a Sph/Im transition object.  They ``deemed \dots~all [their measurements of] integrated quantities
for this galaxy [as] unreliable''.  We classify it as Im, although the difference is small and not important 
for our interpretation.  We remeasured the ACS~$g$ image and derive the parameters shown in Figure~20 by the
blue filled square.

      F2006 recognized four additional galaxies as Sph/Im transition objects, VCC 21, VCC 571, VCC 1499, and
VCC 1779.  All have some star formation, and several show prominent blue star clusters.  These objects were
plotted as Sphs in KFCB, but it is more appropriate to plot them as blue filled triangles here.

      Gavazzi \etal (2000) published $K^{\prime}$-band photometry and photometric decompositions of late-type
VCC galaxies.  This contributes 29 Sc -- Im galaxies to our sample.  About 1/3 of the galaxies are giants
($M_{V,\rm disk} < -18$), so this sample provides good overlap between giants and dwarfs.

      Courteau \etal (2007) is our primary source of disk parameters.  Figure 20 includes 181 Courteau
Sc -- Im galaxies measured in $I$ band and transformed to $V$ band using $V - I$ colors tabulated in his paper.
We kept galaxies with distances $D < 70$ Mpc.  Courteau studied mostly giant galaxies but a few dwarfs
overlap the Sph sequence.

      Baggett \etal (1998) is the source for 27 Scd -- Im disk parameters.  As we did for S0 disks, we keep a
galaxy only if the {\it bulge\/} $r_e/{\rm seeing} > 4.5$, where ``seeing'' is tabulated in Baggett's paper.  
This eliminates galaxies that show larger-than-normal scatter, presumably because they are not well enough
resolved for reliable bulge-disk decomposition.  Our Hubble type selection is $T > 5$.  Recall that Baggett's 
measurements are in $V$ band,

      Laurikainen \etal (2010) provide 28 mostly Sa galaxies; i.{\ts}e., de Vaucouleurs type $T = 1$.  Sa 
galaxies are invariably giants (e.{\ts}g., Sandage 1975; van den Bergh 2009a), so they help to constrain the 
S0 -- Sc comparison, but they do not overlap the spheroidals. 

      Pildis \etal (1997) measured 46 Sc -- (mostly) Im galaxies in $I$ band.  Most galaxies are faint and
help to define the dwarf part of the disk sequence.  We kept galaxies with heliocentric velocities
$< 3000$ km s$^{-1}$.

      Makarova (1999) measured 26 Scd -- (mostly) Im galaxies; all except one are at distances $D \leq 8.6$ Mpc.
This paper helps greatly to define the dwarf part of the disk sequence.  The measurements were made in $V$ band.

      Kirby \etal (2008) measured 33 Sc -- (mostly) Im and mostly dwarf galaxies in $H$ band.    
 
      Figure 20 updates parameter correlations of late-type galaxy disks studied previously by many
authors beginning with Freeman (1970).  As noted also in Figure 18, we confirm yet again the Freeman (1970)
result that giant galaxy disks generally have nearly uniform central surface brightnesses that vary little with
disk luminosity.  In contrast, $r_e$ or equivalently the exponential scale length $h = r_e/1.678$ varies 
with $L_{V,\rm disk}$ over the whole luminosity range, accounting in giant galaxies for most of the
variation in $L_{V,\rm disk} \simeq 2 \pi I_0 h^2$, where $\mu_0 = -2.5 \log{I_0} = \mu_e - 1.822$ is the
central surface brightess in magnitude units.

      Figure 20 confirms the results of Kormendy (1985,~1987) that the parameter correlations for
\hbox{late-type} and spheroidal galaxies are indistinguishable.  It extends this result to the brightest S0 and
Sc -- Scd disks.  This is consistent with our suggestion that Sph and late-type galaxies are closely related.  
The nature of the relationship is complicated: ongoing star formation increases the disk surface brightness 
whereas dust extinction reduces it, and the quenching of star formation by ram-pressure
stripping results in fading of the young stellar population and hence of the disk as a whole.  We return
to this issue in \S\ts8.  Here, we emphasize that the similarity of the S and S0 disk correlations forms
part of our motivation for suggesting parallel S0 -- Sph and spiral galaxy sequences in Figure 1.

      Deriving these results accurately over large ranges in $M_V$ requires detailed photometry and 
photometric decomposition of nearby, well-resolved galaxies.  However, we note that the difference between
the E$+$bulge and Sph$+$disk sequences is also seen in an SDSS study of 140,000 galaxies by Shen \etal (2003).

      Figure 20 has implications for the formation of elliptical galaxies by major mergers.  In terms
of effective parameters (i.{\ts}e., those relevant to Virial theorem arguments),~making present-day 
normal-luminosity giant ellipticals out of present-day giant disks requires relatively little dissipation.  
We observe that such mergers happen, and their remnants are consistent with extra-light Es 
(see~KFCB~for~a review).  This is plausible:~present-day disk galaxies contain little enough gas so that 
rearranging it during a merger cannot have a large effect on effective parameters.  However, making the
high-density centers of ellipticals requires substantial dissipation and star formation, as reviewed in KFCB.  
And making present-day small ellipticals out of present-day progenitors requires much more dissipation as 
galaxy luminosity decreases (Kormendy 1989).

      We emphasize that these reassuring consistency checks apply only to the relatively few ellipticals 
that are made recently enough so that the properties of present-day progenitors are relevant.  Most ellipticals
formed much longer ago out of progenitors that we are only beginning to observe.

\vss\vsss
\cl {7.~ADDITIONAL CONNECTIONS BETWEEN Sph}
\cl {GALAXIES AND S0 AND SPIRAL GALAXY DISKS}
\vss\vsss

      In Section 7.1, we review additional observations which show that Sph galaxies are related to
S0 galaxy disks, especially at high luminosities.  Then (Section 7.2), we review observations which
support the idea that Sph and S0 galaxies are transformed, ``red and dead'' spiral and irregular galaxies.

\vss\vsss
\cl {7.1.~\it Further Evidence That Higher-Luminosity Sphs}
\cl {\it Are More Disk-Like}
\vss

      The fact that spheroidal galaxies can contain disks was emphasized in \S\ts4.1.3 using VCC 2048,
an edge-on Sph that shows the disk directly.  A close relation to S0 disks is also implied by the
observation that the main body of VCC 2048 is flatter than any elliptical (E6.2 plus an additional
2\ts\% disky distortion -- see Figure 11).  Two additional observations show more indirectly but for
a larger sample of objects that Sphs are related to disks.

      First, some of the brighter Sph galaxies in Virgo show low-amplitude spiral structure in their
otherwise-smooth light distributions.  This was first -- and still best -- seen by Jerjen
\etal (2000) in IC 3328.  Their result is reproduced here in
Figure 21.  No dynamically hot stellar system such as an elliptical galaxy can produce 
fine-scale spiral structure.  Barazza \etal (2002) confirmed Jerjen's result and concluded that 
``This is unambiguous evidence for the presence of a disk.'' 

      Similar but less obvious spiral structure has been seen in other spheroidals
(Jerjen \etal 2001;
Barazza \etal 2002;
De Rijcke \etal 2003;
Graham \etal 2003;
Ferrarese \etal 2006;
Lisker \etal 2006, 2007, 2009).
Not all cases are significant.  Barazza \etal (2002) caution~us~that, when the spirals are
weak, features in the ellipticity and position angle profiles that are not spiral structure ``can
indeed produce amazingly spiral-like twisting isophtoes and thus mimic genuine spiral structure''.
The weakest observed spiral signals should be interpreted with caution, but the large number of
detections implies that some spirals are real.  Again, this is compelling evidence
that bright Sph galaxies are related to S0 disks.

      The second kind of evidence comes from measurements of
rotation velocities and velocity dispersions.  At a time when it was already understood that 
low-luminosity ellipticals rotate roughly like isotropic spheroids (Davies \etal 1983; see \S\ts2.1,
here), it was a surprise when Bender \& Nieto (1990) found that the Sphs Fornax, NGC 205, IC 794,
and VCC 351 rotate so little that they must be anisotropic.  VCC 351 is particularly interesting: 
it is flatter than any elliptical (E7), but it has the smallest ratio of rotation velocity to
velocity dispersion of any galaxy in their sample.  Some spheroidals must be anisotropic.  
Similar studies followed, with mixed results.  The Local Group spheroidals NGC 147, NGC 185, and 
NGC 205 are moderately (Bender \etal 1991) but not extremely (De Rijcke \etal 2006; Geha \etal 2006)
anisotropic.  More generally, some Sphs rotate rapidly 
(de Rijcke \etal 2001;
Simien \& Prugniel 2002;
Pedraz \etal 2002;
Tolstoy \etal 2009);
others do not
(Geha \etal 2003;
Thomas~et~al.~2006).
Most tellingly, van Zee \etal (2004) find that ``the rotation amplitudes of the rotating [Sphs] are 
comparable to those of similar-brightness dwarf irregular galaxies (dIs).  Evidence of a relationship 
between the rotation amplitude and galaxy luminosity is found and, in fact, agrees well with the 
Tully-Fisher relation. \dots~These observations reaffirm the possibility that some cluster [Sphs] may 
be formed when the neutral gaseous medium is stripped from dIs in the cluster environment. We hypothesize 
that several different mechanisms are involved in the creation of the overall population of [Sphs] and 
that stripping of infalling dIs may be the dominant process in the creation of [Sphs] in clusters like
Virgo.''  We agree with all of these statements.

\singlecolumn

\vfill


\includegraphics{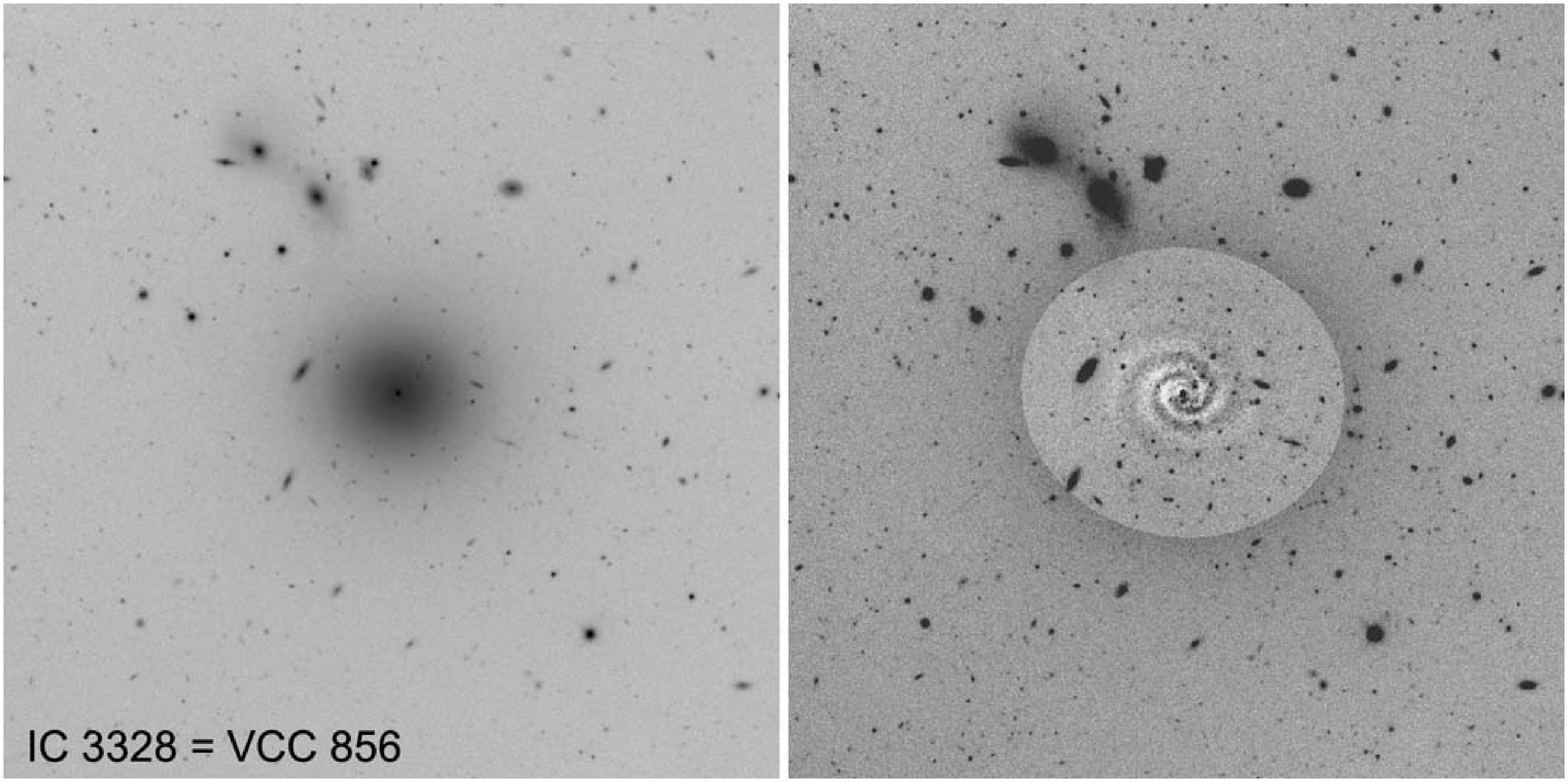}

Fig.~21~-- Spiral structure in the Sph,N galaxy IC 3328 = VCC 856.  At $M_V \simeq -17.8$, this 
is one of the brightest spheroidals in the Virgo cluster.  An $R$-band image ({\it left\/}) 
shows the low-surface-brightness, shallow-brightness-gradient, but nucleated light distribution 
that is characteristic of Sph galaxies.  After subtraction of the overall, elliptically symmetric 
light distribution ({\it right\/}), the residual image shows tightly-wound, two-armed spiral 
structure.  This means that IC 3328 is -- or at least: contains -- a disk.  This figure is taken 
from and kindly provided by Jerjen, Kalnajs, \& Binggeli (2000).
\pretolerance=15000  \tolerance=15000 

\eject

\doublecolumns

\vss\vsss
\cl {7.2.~\it Stellar Population Evidence that}
\cl {\it Sphs are Related to Irregulars}
\vss

      We have known for many years that Local Group dwarf spheroidals have episodically been converted 
into irregulars (Kormendy \& Bender 1994).  Their intermediate-age 
stellar populations (Da Costa 1994) tell us that they had a variety of different
bursty star formation histories ending, in some cases, only a short time ago.  For example, 
Hurley-Keller \etal (1998) concluded that the Carina dSph is made up of three stellar populations:  
\hbox{10\ts--\ts20\ts\%} of its stars are $\sim 12$ Gy old, but at least 50\ts\% of the stars are 6\ts--\ts8 
Gy old, and $\sim$\ts30\ts\% formed only 3 Gyr ago.  Hernandez \etal (2000) used an HST
color-magnitude diagram to get qualitatively similar results: bursts happened $\sim 8$, 5, and 3 Gyr ago
with some star formation extending to 1~Gyr~ago (see also Dolphin 2002).    There must have been gas 
at all of these times in order to make these stars.
Gas-rich, star-forming dwarfs are Magellanic irregulars.

      Mateo (1998) and Tolstoy \etal (2009) provide thorough reviews of star formation histories.  
A few Sphs consist almost exclusively of old stars (e.{\ts}g., Draco and UMi).  But in general, the 
star formation histories of Sph and Im galaxies look similarly bursty over most of cosmic time.  Again, the 
main difference is that the Sphs have essentially no star formation now.  That's why we call them Sphs.

      Metal abundance distributions also imply heterogeneous star formation and abundance enrichment
histories in Sph and (albeit with sparser data) dIm galaxies (see, e.{\ts}g., 
Venn \& Hill 2008;
Tolstoy \etal 2009, and
Frebel 2010
for reviews).  The $\alpha$ element abundances in dSph galaxies are not much enhanced with respect 
to solar, also consistent with prolonged star formation histories 
(Shetrone \etal 1998, 2001, 2003, 2009;
Tolstoy \etal 2003;
Venn \etal 2004;
Geisler \etal 2005).
However, {\it ``there are no examples of any dwarf systems that do not contain an `ancient' population
of stars.''\/} (Mateo 2008).   Tolstoy \etal (2009) agree: ``No genuinely young galaxy (of any type) has 
ever been found; stars are always found at the oldest look-back times observed.''  So these relics from
the earliest days of galaxy formation sputtered along, forming stars for billions of years before
changing into Sphs.

      Measurements of star formation histories have now been extended to larger samples of
galaxies outside the Local Group.  Figures 22 and 23 show~the~individual cumulative star formation 
histories and the mean specific star formation histories of galaxies in the HST ACS Nearby 
Galaxy Survey Treasury (ANGST:~Weisz~et~al.~2011a).  The survey covers 60 nearby ($D$ \lapprox \ts4 Mpc)
dwarfs~of~both early and late types.  Weisz \etal (2011a) conclude~that ``the average dwarf 
formed \gapprox \ts50\ts\% of its stars by $z \sim 2$ and 60\ts\% of its stars by $z \sim 1$, regardless
of its current morphological type'' and that ``the mean [star formation histories] of dIs,
[dwarf Sph/Im transition galaxies], and dSphs are similar over most of cosmic time, and only begin
to diverge a few Gyr ago, with the clearest differences between the three appearing during the most
recent 1 Gyr.''  These results echo the results obtained in the Local Group (Weisz \etal 2011b).
The conversion of irregular to spheroidals happened at different times for different spheroidals; this 
conversion does not seem to correlate with galaxy parameters, but it correlates strongly with environment (\S\ts8).

\vskip 1.7truein

\includegraphics{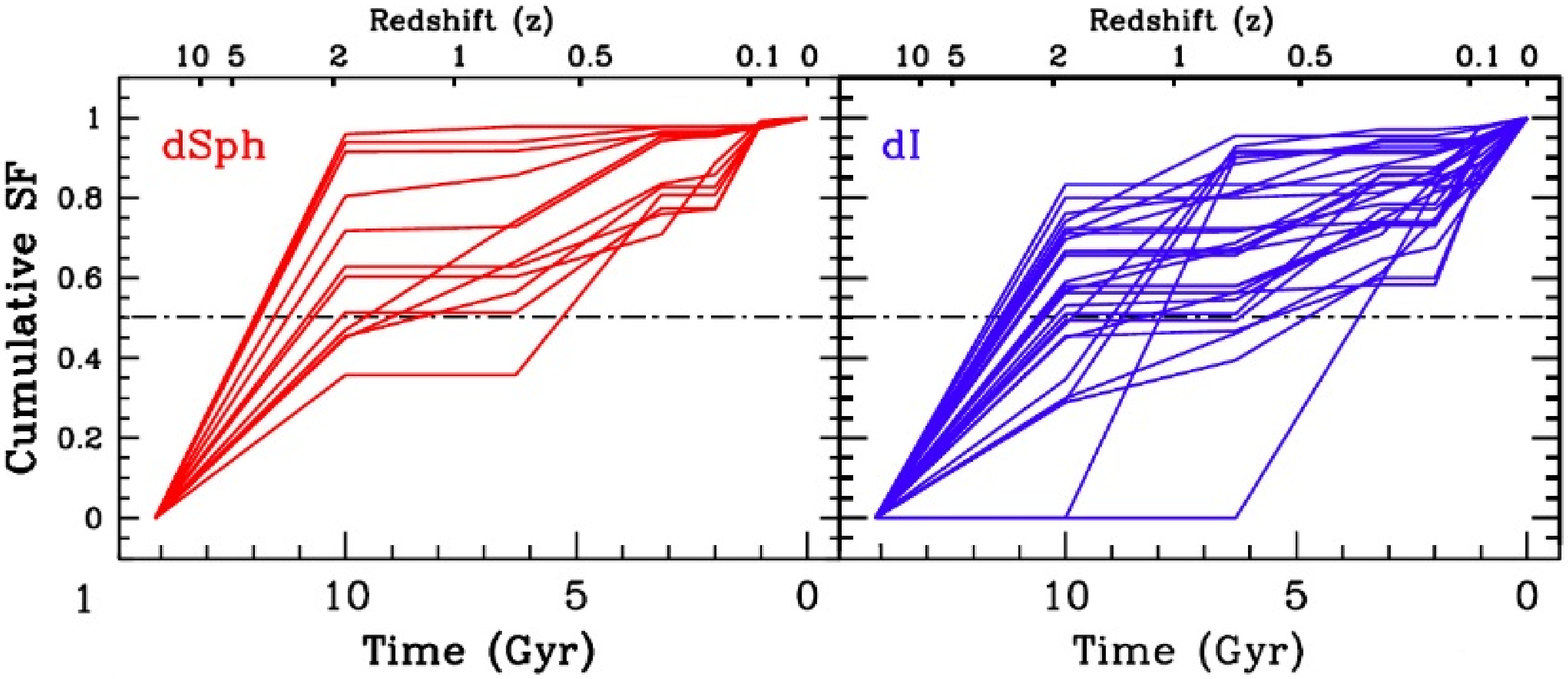}

Fig.~22~-- Individual star formation histories, i.{\ts}e., the cumulative fraction of all current
stars that were already formed as functions of lookback time and redshift $z$ for ({\it left\/}) 
Sph galaxies and ({\it right\/}) Im galaxies.  A horizontal line is drawn at 50\ts\% of the current 
stellar~mass.  The means of these star formation histories of Sph, Im, spiral, Sph/Im transition galaxies 
are virtually indistinguishable.  This figure is adapted from Figure 5 of Weisz \etal (2011a).
\pretolerance=15000  \tolerance=15000 

\singlecolumn

\vskip 2.7truein

\vfill

\includegraphics{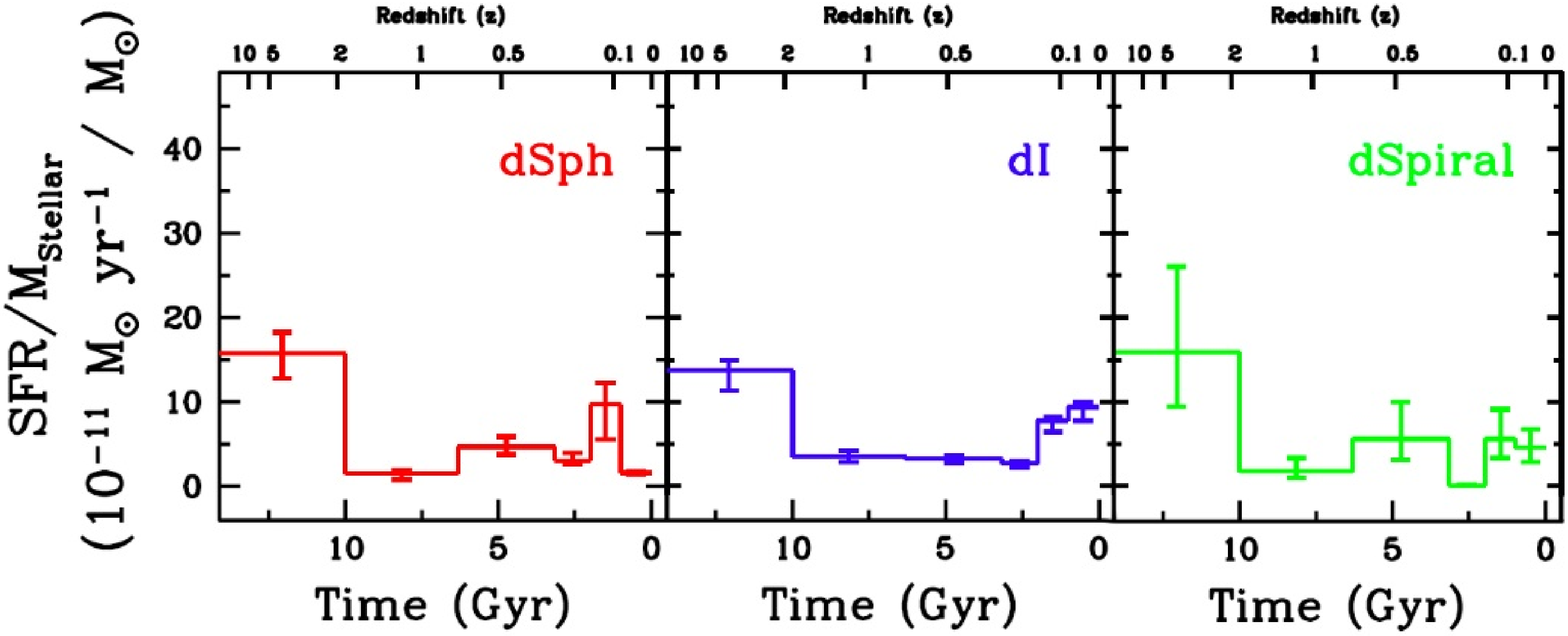}

Fig.~23~-- Mean specific star formation histories, i.{\ts}e., star formation rate divided by
the integrated stellar mass, for ANGST dwarf galaxies divided up by morphological type.  Different
types have essentially indistinguishable star formation histories, except that Sphs do not form
stars now.  This figure is adapted from Figure 4 in Weisz \etal (2011a).
\pretolerance=15000  \tolerance=15000 

\eject

\doublecolumns

\vss\vsss
\cl {8.~\sc OBSERVATIONAL EVIDENCE FOR}
\cl {\sc ENVIRONMENTAL GALAXY TRANSFORMATION PROCESSES}
\vss

     The observations presented so far suggest that S0$+$Sph galaxies are red and dead S\ts$+${\ts}Im 
galaxies, but they mostly do not point to any specific transformation process.  This section reviews
possible transformation processes and the observations that support them.  A comprehensive review is 
beyond the scope of this paper.  However, we discuss the most important results that help to justify 
and explain the parallel-sequence galaxy classification scheme.

      The central observational results presented in this paper are (1) the continuous structural parameter
correlations of S0 disks and Sph galaxies and (2) the close similarity between the S0 disk $+$ Sph parameter
sequences and those of spiral galaxy disks and Magellanic irregulars.  This section focuses on environmental 
galaxy transformation processes that may explain this similarity.  However, at least one internal process 
is more fundamental than all external processes, because it affects both star-forming and non-star-forming 
galaxies similarly and independently of environment:

      Dekel \& Silk (1986) ``suggest that {\it both the dI's and the [dSphs] have lost most of their mass\/}
in winds after the first burst of star formation, and that this process determined their final structural
relations.  The dI's somehow managed to retain a small fraction of their original gas, while the [dSphs]
either have lost all of their gas at the first burst of star formation or passed through a dI stage before
they lost the rest of the gas and turned  [dSph].''  That is: {\it The~Sph $+$ Im sequence of decreasing surface 
brightness with decreasing galaxy luminosity is a sequence of decreasing baryon retention.}
The idea of baryonic mass loss via winds had been suggested earlier by Larson~(1974) and Saito (1979).  
It has become more plausible as evidence has accumulated that smaller dwarfs are more dominated by 
dark matter and hence that potential wells exist that can retain the miniscule amounts of visible matter that 
remain in dSphs such as Draco (Kormendy \& Freeman 2004, 2011).  Otherwise, if more than 
half of the total mass were expelled, the galaxy would have been unbound.

\vss
\cl {8.1~\it Ram-Pressure Stripping.{\ts}I.{\ts}Evidence for Ongoing Stripping}
\vss

      Gunn \& Gott (1972) suggested that, given the density of hot gas in the Coma cluster implied by the
then-recent detection of X-ray emission (Meekins \etal 1971; Gursky \etal 1971), {\it ``a typical galaxy 
moving in it will be stripped of its interstellar material.}  We expect {\it no~normal spirals}
in the central regions of clusters like Coma.  The lack of such systems is, of course, observed.''

      The idea of ram-pressure stripping has fluctuated in popularity, never retreating very far into the 
background but never enjoying universal acceptance, either.  This situation is changing rapidly.  Observations 
of the ongoing stripping of H{\ts}I and H$\alpha$-emitting gas are turning ram-pressure stripping into An 
Idea Whose Time Has Come.  Van Gorkom and Kenney (2011) provide a comprehensive review of these developments.
Here, we concentrate on some of the most direct evidence for ram-pressure stripping in action.  We tie 
these results together with observations of the morphology-density relation that contribute to a more 
compelling picture of the importance of stripping.  These ideas underlay the parallel-sequence 
classification from the beginning (van den Bergh 1976); they still~do~so here.  In later subsections, 
we argue that the situation is only a little more complicated. i.{\ts}e., that ram-pressure stripping is the 
principal S$+$Im $\rightarrow$ S0$+$Sph transformation process but that other, mostly heating processes also 
help to engineer that galaxy structure that we observe.

      Figure 24 shows some of the best evidence for ongoing ram-pressure stripping in the Virgo cluster (adapted 
from Chung et al.\ts2007; Kenney et al.\ts2004,\ts2008).  Many spiral galaxies near the center of the cluster show 
H{\ts}I tails; the above authors interpret them as gas that is being stripped by the hot, X-ray-emitting gas that 
pervades the cluster (see Chung \etal 2009 for an update and van Gorkom \& Kenney 2011 for a review).  
If tails trail behind their galaxies, then they imply that most of these spirals are falling into the cluster.  
A spectacular example 
(Kenney et al.~2008; see 
Kotanyi \etal 1983;
Combes \etal 1988;
Veilleux et al.~1999;
Vollmer \etal 2005
for progressive improvements in the data) is the tidally disturbed spiral NGC 4438, which shows H$\alpha$ filaments 
extending all the way (we assume:)~back to the giant elliptical NGC 4406.  These galaxies have recession velocities 
of $-1000$ and $-1300$ km s$^{-1}$ with respect to the Virgo cluster core; it is usually assumed that they form part 
of a subgroup that is falling into the Virgo cluster from behind.  In addition, NGC 4438 has likely just had an 
encounter with NGC 4406 and is still both tidally distorted and shedding cold gas into the combined hot gas of NGC~4406 
and the Virgo cluster.

      A related result is the observation that spiral galaxies near the center of Virgo are smaller and more depleted 
in H{\ts}I gas than galaxies in the cluster outskirts (e.{\ts}g., Cayette \etal 1990, 
1994; Chung \etal 2009).  Chung and collaborators add that ``most of these galaxies in the [cluster] core also show gas 
displaced from the disk which is either currently being stripped or falling back after a stripping event''.
The three most depleted galaxies illustrated in Figure 8 of Chung \etal (2009) are NGC 4402, NGC~4405,~and~NGC~4064.
They have a mean absolute magnitude $M_V = -19.4 \pm 0.2$.  Virtually all Sphs are fainter than this (Figure 17).
If even the deep gravitational potential wells of still-spiral galaxies suffer H{\ts}I stripping, then the shallow
potential wells of dS$+$ Im galaxies are more likely to be stripped.  Moreover, while NGC 4402 is close to NGC 4406
and NGC 4405 is only 20$^\prime$ north of NGC 4396 (see Fig.~24), NGC 4064 is almost $9^\circ$ from M\ts87.  Most
galaxies in the cluster outskirts have relatively normal H{\ts}I content, but a few are H{\ts}I depleted even
there.  

      Thus, finding observational evidence for ram-pressure stripping in action has become a substantial industry.
An incomplete list of additional papers includes
Kenney \& Koopmann (1999:~NGC 4522);
Yoshida \etal (2004:~NGC 4388);
Oosterloo \& van Gorkom (2005:~NGC 4388);
Vollmer \etal (2008:~NGC 4501), and
Abramson \etal (2011:~NGC 4330).
Such observations and theoretical developments both suggest that ram-pressure stripping is more effective 
than we have thought 
(Mori \& Burkert 2000;
Quilis \etal 2000;
Grebel \etal 2003;
Roediger \& Hensler 2005;
Tonnesen \etal 2007, 2008, 2009, 2010;
Boselli \etal 2008;
van Gorkom \& Kenney 2011).

      This supports early suggestions that Sph galaxies are ram-pressure-stripped dS$+$Im galaxies~(Faber~\&~Lin~1983;
Lin \& Faber 1983; Kormendy 1987; van den Bergh 1994c).

\vfill\eject

\singlecolumn

\cl{\null}

\vfill

\includegraphics{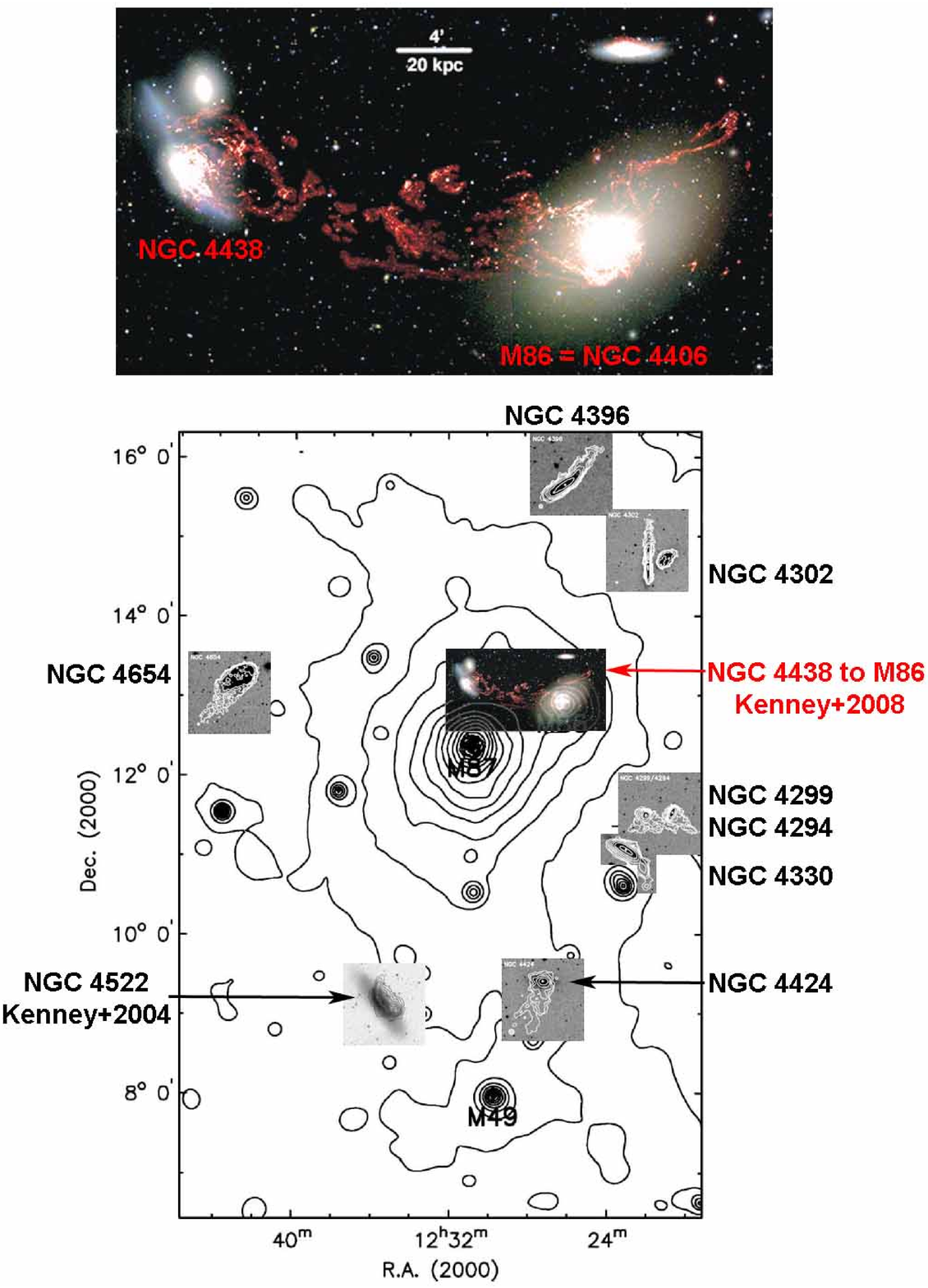}

Fig.~24~-- The large panel shows 0.5 -- 2.0 keV X-ray brightness contours in the Virgo cluster as measured with ROSAT
by B\"ohringer \etal (1994).  Superposed are grayscale images of galaxies with H{\ts}I tails indicative of
ongoing ram-pressure gas stripping ({\it white or black contours\/}).  The H{\ts}I images are from Chung \etal (2007)
and Kenney \etal (2004).  The color inset image and larger-scale image at top show the spectacular 
H$\alpha$ emission filaments that extend from NGC 4438 to NGC 4406 (Kenney \etal 2008).   Each small inset image shows 
the galaxy centered on its position in the cluster, but the panels are magnified.  This is misleading only for NGC 4438 
$+$ NGC 4406: that inset image is positioned so that the center of NGC 4406 is correct, but then the enlargement makes 
it appear as though NGC 4438 is north of M{\ts}87, whereas in reality both NGC 4438 and NGC 4406 are NWW of M{\ts}87.  
This figure is adapted from Figure 4 in Chung \etal (2007).
\pretolerance=15000  \tolerance=15000 

\eject

\doublecolumns

      Still, many authors argue that ram-pressure stripping is not the whole -- or even the main -- story, 
because it is difficult to strip dense central gas in giant galaxies 
(Farouki \& Shapiro 1980; 
Abadi~et~al.~1999;
Quilis \etal 2000;
Tonnesen \& Bryan 2009).  
Additional processes such as galaxy harassment (\S\ts8.3), starvation (\S\ts8.4), or tidal shaking (\S\ts8.5)
may be important.  Before we turn to these, we recall indirect evidence regarding stripping:
   
\vss
\cl {8.2~\it Ram-Pressure Stripping. II.}
\cl {\it The Morphology{\ts}--{\ts}Density Relation}
\vss

      We find it impossible to think about ram-pressure stripping except in the context of the
morphology-density relation.  Dressler (1980) surveyed the relative numbers of E, S0, and S$+$Im
galaxies as a function of local galaxy density in 55 clusters of galaxies; his results are
reproduced here in Figure 25.  He showed that field environments are dominated by S$+$Im galaxies,
with only $\sim$\ts10\ts\% contributions each from ellipticals and S0s; the E~and~S0 fractions then rise and
the S$+$Im fractions fall with increasing galaxy density until there are almost no S$+$Im galaxies 
in the richest cluster environments.  

      Postman \& Geller (1984) extended Dressler's results to lower-density environments, showing that the
E, S0, and~S fractions saturate at $\sim$\ts0.1, 0.2, and 0.7, respectively, over a range of low densities.  
As Dressler emphasized, there are S0s in the field.  If spirals get turned into~S0s, 
the process cannot depend completely on high galaxy densities.

      Gunn \& Gott (1972) suggested that spirals would quickly get ram-pressure stripped
when they fell into the Coma cluster.  In Fig.~25, Coma is included in the right panel.
However, based on data in Fig.\ts25, Dressler~argued: {\it ``The relationship
between population and local density appears to hold without regard to the type of cluster involved.}  This
result contradicts the interpretation that spirals have been swept of their gas to form S0's in the high
concentration clusters.  If the idea of sweeping is to be kept, it would have to be argued that the 
process is common even in regions where the space density of galaxies, and thus presumably of gas, is
$10^2${\ts}--{\ts}$10^3$ times lower than in the rich cores of the regular clusters.  This is improbable.''
\cl{\null}

\cl{\null}

      However, ram-pressure stripping is widespread in the Virgo cluster. Virgo~is~not included, but its densities 
overlap with Figure 25.~Moreover, Figure 25 
shows that the ratio of S$+$Im to S0 galaxy numbers is larger than average at all densities in low-concentration 
clusters and smaller in X-ray-emitting clusters.  Dressler noted this and suggested that 
stripping or gas evaporation could contribute a little.  Given results of \S\ts8.1, it seems more likely 
that ram-pressure stripping happens more easily than we thought and that it helps to turn spirals
into S0s even in clusters like Virgo.  We expect that Im\ts$\rightarrow${\ts}Sph conversion 
is still easier, as implied by evidence in the Local Group below.

      Observing the evolution of the morphology-density relation with cosmological lookback time should tell 
us more about how S0s evolved.  This is a big subject, mostly beyond the scope of this paper.  
We focus~on~two~results.  Dressler~et~al.~(1997) and Willmer et al.~(2009) compare the
$z$\ts=\ts0 morphology-density relation to that for clusters at $z \simeq 0.5$ and groups at $z \simeq 0.4$, 
respectively.~For clusters, Dressler~et~al.~(1997) find that the differences shown here in Figure 25 are much
larger at $z \simeq 0.5$:~then, centrally concentrated, regular clusters show a relation similar to that at $z = 0$,
but low-concentration, irregular clusters show almost~no~relation.  Dressler concludes that ``S0s are generated 
in large numbers only after cluster virialization.''  These results are consistent with ours.  In contrast, 
Wilman~et~al.~(2009) find that $z \simeq 0.4$ groups with \hbox{$\sim$\ts5\ts--\ts20} bright galaxies and velocity
dispersions \hbox{$\sim$\ts200\ts--\ts500} km s$^{-1}$ (Wilman \etal 2005) are indistinguishable from clusters in their 
galaxy populations.  I.{\ts}e., they already have larger S0 and smaller S fractions than does the field.  Then,
between $z \simeq 0.4$ and the present, the S0 fractions increase and the spiral fractions decrease in the same way
in groups and clusters.~``The S0 fraction in groups is at least as high as in $z \sim 0.4$ clusters and 
X-ray-selected groups, which have more luminous ingragroup~medium~(IGM).  Interactions with a bright X-ray-emitting 
IGM cannot be important for the formation of the majority of S0s in the universe'' (Wilman \etal 2009).  Instead, 
they conclude that ``minor mergers, galaxy
harassment, and tidal interactions are the most likely mechanisms [to make S0s].''

\singlecolumn

\cl{\null}

\vfill

\vskip 1.9truein

\includegraphics{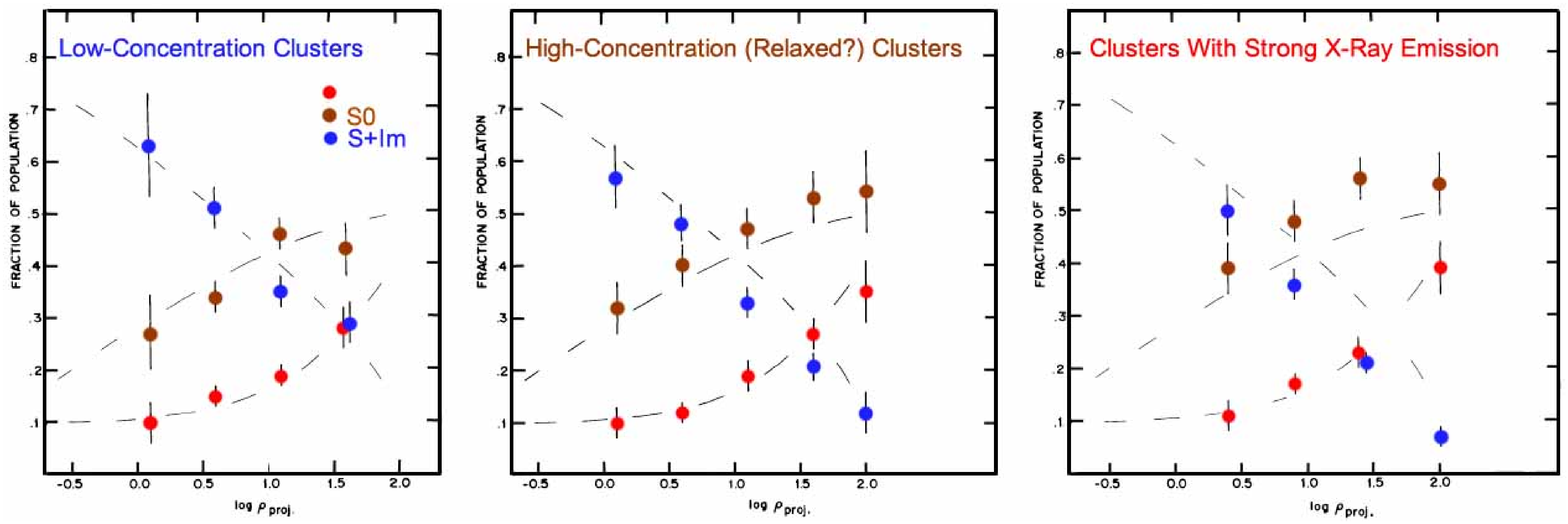}

Fig.~25~-- Morphology{\ts}--{\ts}density relation as a function of cluster richness, adapted from
Figures 8{\ts}--{\ts}10 of Dressler (1980).  Each panel shows the fraction of E, S0, and S$+$Im galaxies
as a fraction of the log of the projected density in units of galaxies Mpc$^{-2}$.  The galaxies were
classified by Dressler based mostly on  large-scale (10\sd9 mm$^{-1}$), $B$-band photographic plates
taken with the Las Campanas Observatory 2.5 m telescope.  The dashed lines show the mean fractions
for all 55 rich clusters in the sample, while the data points show the fractions for 9 low-concentration
clusters ({\it left\/}), 10 high-concentration clusters ({\it center\/}), and 8 strongly X-ray-emitting
clusters ($L_X$ \gapprox \ts$10^{44}$ erg s$^{-1}$ for $H_0 = 50$ km s$^{-1}$ Mpc$^{-1}$, {\it right\/}).

\eject

\doublecolumns

      We partly agree and we partly disagree.  The Wilman~et al.~(2009) results, like the Dressler (1980) 
results, may be explained if ram-pressure stripping happens more easily than we have thought.
At the same time, we, like Wilman, suggest (\S\ts8.5) that other transformation processes happen, too.
Also, we cannot exclude that some part of the difference~between S$+$Im and S0 galaxies is
set by proto-cluster environments in ways that do not involve galaxy transformation.  This possibility
was preferred by Dressler.  

     A compelling observation which suggests that hot~cluster gas is not necessary for
ram-pressure stripping of dwarf galaxies is shown in Figure~26.  Close dwarf companions of Local
Group giant galaxies are almost all spheroidals.  Distant companions are irregulars.  Sph/Im galaxies have
intermediate morphologies and live at intermediate distances.  There are a few exceptions -- at least
three~Sphs are free-flyers, and the Magellanic Clouds survive at Galacticentric distances at which other
companions are gas-free.  They are the largest companions illustrated.  This result has been known 
for a long time
(Einasto \etal 1974;
van den Bergh 1994b,{\ts}c;
Mateo 1998)
and is beautifully illustrated in Mateo's (2008) figure.

\vskip 2.55truein

\includegraphics{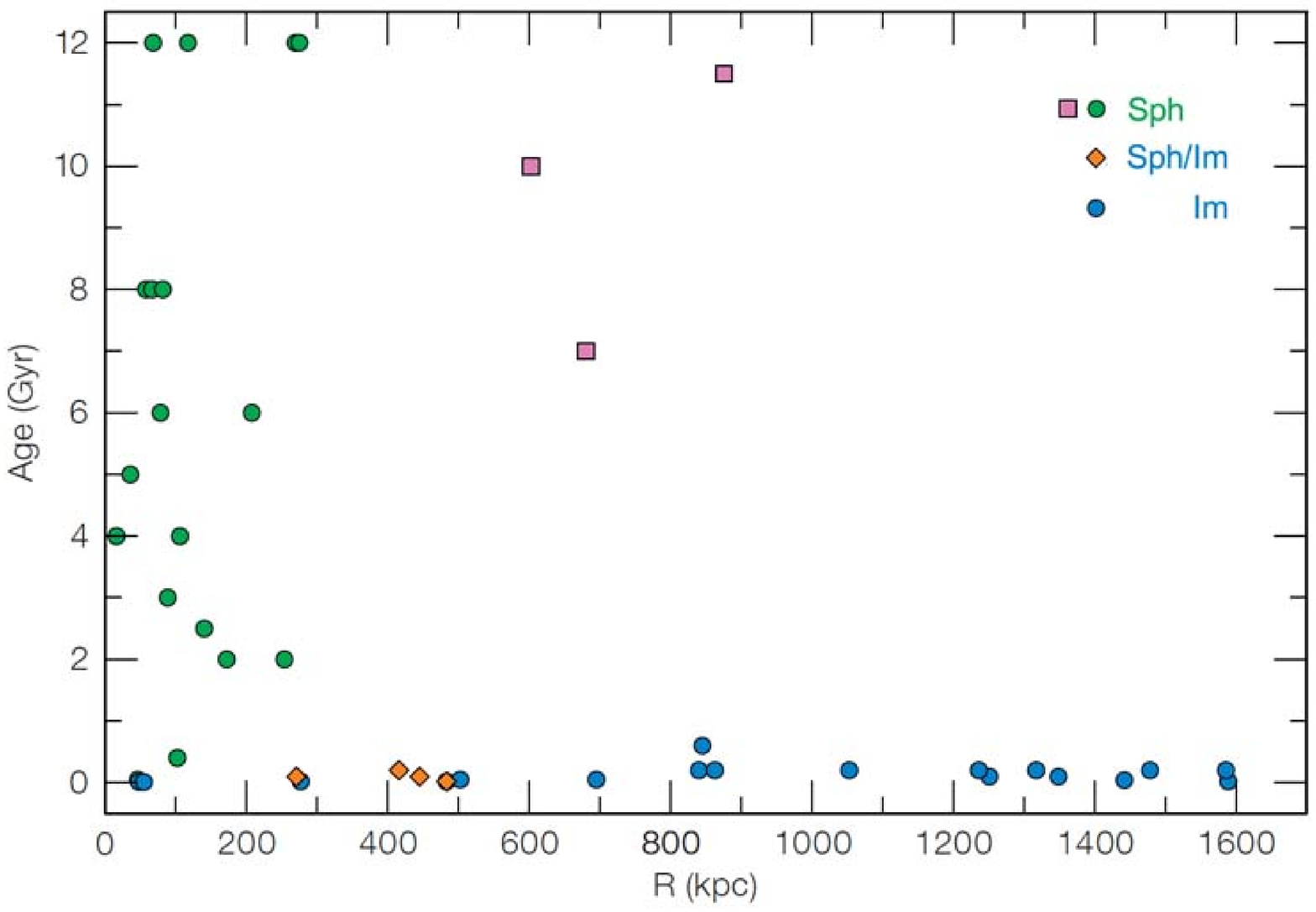}

Fig.~26~-- From Mateo (2008), the ages of the youngest stellar populations in dwarf galaxy companions
versus Galactocentric or M31centric distance $R$.  Except for the Magellanic Clouds, all close companions
of our Galaxy and of M{\ts}31 are spheroidals.  Almost all distant companions are irregulars, with the
exception of three free-flying Sphs, Cetus, Tucana, and And XVIII.  The Sph/Im transition galaxies mostly
lie at distances intermediate between those of spheroidals and irregulars.  This figure is not up-to-date
with all recent discoveries of dwarf galaxies, but the above generalizations are robust.

\vss

      Living near a giant galaxy is dangerous for \hbox{gas-rich} dwarfs, but the reason is not established
by Figure 26.  Could the Im $\rightarrow$ Sph transformation process mainly be gravitational; e.{\ts}g.,
tidal shaking that promotes star formation?  But Local Group and M{\ts}81 Group spheroidals and 
irregulars overplot almost exactly in the Figure 20 parameter correlations.  This requires fine-tuning of the 
star formation.  It is not excluded.  And it can happen concurrently with other effects.  But Figure 26 may 
be an indication that ram-pressure stripping can happen even in environments that are gentler than 
cluster centers.  It may be indirect evidence for a pervasive warm-hot intergalactic medium (WHIM: Dav\'e \etal 
2001) that is difficult to detect directly but that may be enough to convert dwarf irregulars into spheroidals.
Freeland \& Wilcots (2011) present evidence based on bent radio jets for just such gas in galaxy groups.  Still
hotter hot gas can be retained in galaxies that have total masses $M$ \gapprox \ts$M_{\rm crit} \sim 10^{12}$ $M_\odot$
(``$M_{\rm crit}$~quenching'' of star formation; 
Dekel \& Birnboim 2006; 
Cattaneo \etal 2006; 
Faber \etal 2007; 
KFCB).
This corresponds to $M_V$ \lapprox \ts$-21.5$ (KFCB).  Many of the first-ranked galaxies in these groups 
are massive enough (see Wilman \etal 2005, Table 2), and even when they are not, the potential well of a
group is determined by more than one member.  Therefore:

      We suggest that ram-pressure stripping is one of the important processes that converts 
S$+$Im galaxies into S0$+$Sph galaxies.  Hierarchical clustering simulations are becoming powerful enough 
to follow ram-pressure stripping in a cosmological context (Tonnesen \etal 2007, 2008, 2009, 2010); they 
show that other processes discussed in the following subsections happen too.  These other proceeses must 
be especially important for the small fraction of S0 galaxies that live in the field.  

      We emphasize: 
{\it It is not necessary to remove all central gas or even to prevent star formation~in~all~S0s.}
S0$_3$ galaxies with dust disks are illustrated in the {\it Hubble Atlas} (Sandage 1961). 
Optical emission lines are seen in 75\ts\% of the SAURON galaxies, especially in S0s and even in
the Virgo cluster (Sarzi \etal 2006).  H{\ts}I gas was discovered long ago 
(Balick \etal 1976; van~Woerden \etal 1983; van Driel \& van Woerden 1991).
Welch \& Sage (2003) detected CO emission from molecular gas in 78\ts\% of the field S0 galaxies that they 
surveyed, usually near the center (see also 
Sage \& Wrobel 1989; 
Thronson \etal 1989;
Devereux \& Young 1991;
Young \etal 1995).
The ATLAS3D project shows these results in particular detail and again even in the Virgo cluster 
(Davis \etal 2009; Young \etal 2011).
Also, 60\ts\% of S0s were detected by IRAS in 60 $\mu$m and 100 $\mu$m emission
(Knapp \etal 1989).  
Finally, Temi \etal (2009) observed E and S0 galaxies with the 
{\it Spitzer Space Telescope\/} and detected gas, dust, and small amounts of star formation.
S0s have more star formation than ellipticals and show different correlations between
24 $\mu$m luminosity, molecular gas mass, and other star formation indicators.  Temi \etal (2009) suggest
that ``rotationally supported cold gas in S0 galaxies may be a relic of their previous incarnation
as late-type spirals.''  Some S0s are exceedingly gas-deficient, but many contain
molecular gas, dust, and some star formation near the center, as predicted by theoretical studies of
ram-pressure stripping.  Similarly,  H{\ts}I deficient spirals in the Virgo cluster often have normal 
molecular gas content (Kenney \& Young 1986).
 
      Finally, it is important to remember that S0s occur~in the field.~E.{\ts}g., NGC 3115 is very isolated.~The
processes that clean them of gas and convert them from ``blue cloud'' to ``red sequence'' 
galaxies in the SDSS color-magnitude relation remain an ongoing puzzle (Faber \etal 2011).   NGC~3115 contains 
little X-ray gas (Li \etal 2011); ``$M_{\rm crit}$ quenching'' may help, but it is difficult to believe that it
is the whole story.  On the other hand, NGC 3115 contains an unusually massive black hole $M_\bullet \sim 10^9$ 
$M_\odot$ 
(Kormendy \& Richstone 1992;
Kormendy \etal 1996;
Emsellem \etal 1999).
It is possible that feedback from an AGN phase of accretion onto this black hole was instrumental in
cleaning the galaxy of cold gas (van den Bergh 1993, 2009b).

\vss
\cl {8.3~\it Galaxy Harassment}
\vsss\vskip 1pt

      Galaxy harassment is the cumulative effect of many, high-speed encounters with other galaxies 
in a cluster and with the overall cluster potential; numerical simulations show that it strips outer mass, 
heats disks, and promotes gas inflow toward the center 
(Moore et al.~1996; 
Moore \etal 1998, 1999; 
Lake \etal 1998).  
A variant in the Local Group is tidal stirring of dwarfs on very elliptical orbits by the Galaxy or M{\ts31
(Mayer \etal 2001a,{\ts}b,~2006)
Harassment is suggested to convert flimsy, late-type disks into spheroidals and more robust, earlier-type
spirals into much hotter systems that resemble S0s in many ways.  One success of this picture that
comes ``for free'' is that inflowing gas feeds star formation; this helps to explain why spheroidals
in which star formation stopped long ago do not have much lower surface brightnesses than current
versions of spiral galaxy progenitors (Figure 20).  The process is clean and inescapable.  And it 
provides a natural explanation  for many of the observations of Sph and S0c galaxies that we discussed 
in previous sections.  We do not repeat them all, but the following are worth emphasizing: 
\vskip 3pt

\ihi  1.{\ts\ts}Faint spheroidals are not flat, but some of the brightest contain edge-on disks
         (e.{\ts}g., VCC 2048 in \S\ts4.1.3).  NGC 4638 is similar (\S\ts4.1.4).  Also similar but less 
         extreme are the edge-on SB0s NGC 4762 (\S\ts4.1.1)
         and NGC 4452 (\S\ts4.1.2).  They have very flat inner disks but very thick outer
         disks, as expected from dynamical heating processes.  The fat outer disk of NGC 4762 is still
         warped and irregular; this is one example among many of an ongoing tidal encounter; i.{\ts}e.,
         of harassment in action.
\vss

\ihi  2.{\ts\ts}Sph and Im galaxies have similar distributions of observed axial ratios; this is
          another sign that these two types of galaxies are closely related 
          (Ferguson \& Sandage 1989; Binggeli \& Popescu 1995).  Spheroidals are also similar in
          shape to ellipticals (above papers; Ichikawa 1989), but the flattest Sphs
          are flatter than any elliptical (e.{\ts}g., Ryden \& Terndrup 1994).  
          It is difficult to turn this {\it consistency} with Im $\rightarrow$ Sph evolution into 
          an argument {\it either for or against} such evolution (Binggeli \& Popescu 1995).
          Progenitor Ims need not have the same shapes as descendent Sphs if
          harassment is one of the transformation processes.
\vss

\ihi  3.{\ts\ts}Stars and gas that are liberated by gravitational stripping form a cluster 
         intergalactic background.  Remarkably deep imaging observations by Mihos (2011) and Mihos \etal 
         (2005, 2009, 2011) reveal this background light in the Virgo cluster.  It is also detected
         via planetary nebulae (e.{\ts}g.,
         Arnaboldi \etal 1996, 2002, 2004;
         Castro-Rodrigu\'ez \etal 2009;
         see Arnaboldi \& Gerhard 2009 and Arnaboldi 2011 for reviews).  These can
         be used at surface brightnesses that are otherwise unreachable, and they provide
         velocity information.  Intracluster globular clusters are also detected in Virgo (Williams
         \etal 2007a), and metal abundance distributions have been measured in individual intracluster
         stars (Williams \etal 2007b). These observations show that the intracluster light in
         Virgo is irregular and riddled with tidal streams, indicating that it is in the early stages
         of formation.  
\vss

\iihi    Beginning with a photographic detection in Coma (Thuan \& Kormendy 1977), intracluster light
         has been detected and studied in the Coma cluster (e.{\ts}g., Adami \etal 2005; Okamura 2011)
         and in many other clusters (e.{\ts}g., Krick \& Bernstein 2007; Gonzalez~et al. 2007,
         see Arnaboldi \& Gerhard 2009 and Arnaboldi 2011 for reviews).
         In many cases, the intracluster light shows irregularities such as streams that
         are indicative of non-equilibrium, ongoing formation.
\vss

\iihi    On the smallest scales, the halos of our Galaxy (Ibata \etal 1994, 1995, 2001b)
         and M{\ts}31 (Ibada \etal 2001a, 2007; Ferguson \etal 2002)
         contain stellar streams that are dwarf galaxies being torn apart by tides.
         On the largest scales, cD halos (Morgan \& Lesh 1965; Oemler 1976; Schombert 1988)
         belong gravitationally to their clusters more than to their
         central galaxies (Dressler 1979; Kelson \etal 2002) 
         and also consist of tidally disrupted galaxies (Richstone 1976).  
\vss

\iihi    Gravity is not negotiable.  The above are all results of galaxy harassment.  If it can disrupt 
         dwarf galaxies to produce giant galaxy halos and intercluster light, 
         then it can heat individual galaxies short of disrupting them.  This is a paraphrase for the 
         production of spheroidal galaxies via harassment.
\vss\parindent=10pt

\ihi  4.{\ts\ts}Harassment should also produce outcomes that are intermediate between disk thickening
         by dynamical heating and total galaxy disruption.  
         Sandage \& Binggeli (1984) discovered ``a new class of dwarfs that are of huge size
         (10000 pc in diameter in the extreme) and of very low surface brightness of about 
         25 $B$ mag/arcsec$^2$ at the center.''  A~Sph example is shown in Figure 27.
         We interpret these galaxies as spheroidals that have been harassed almost to death.

\vfill

\includegraphics{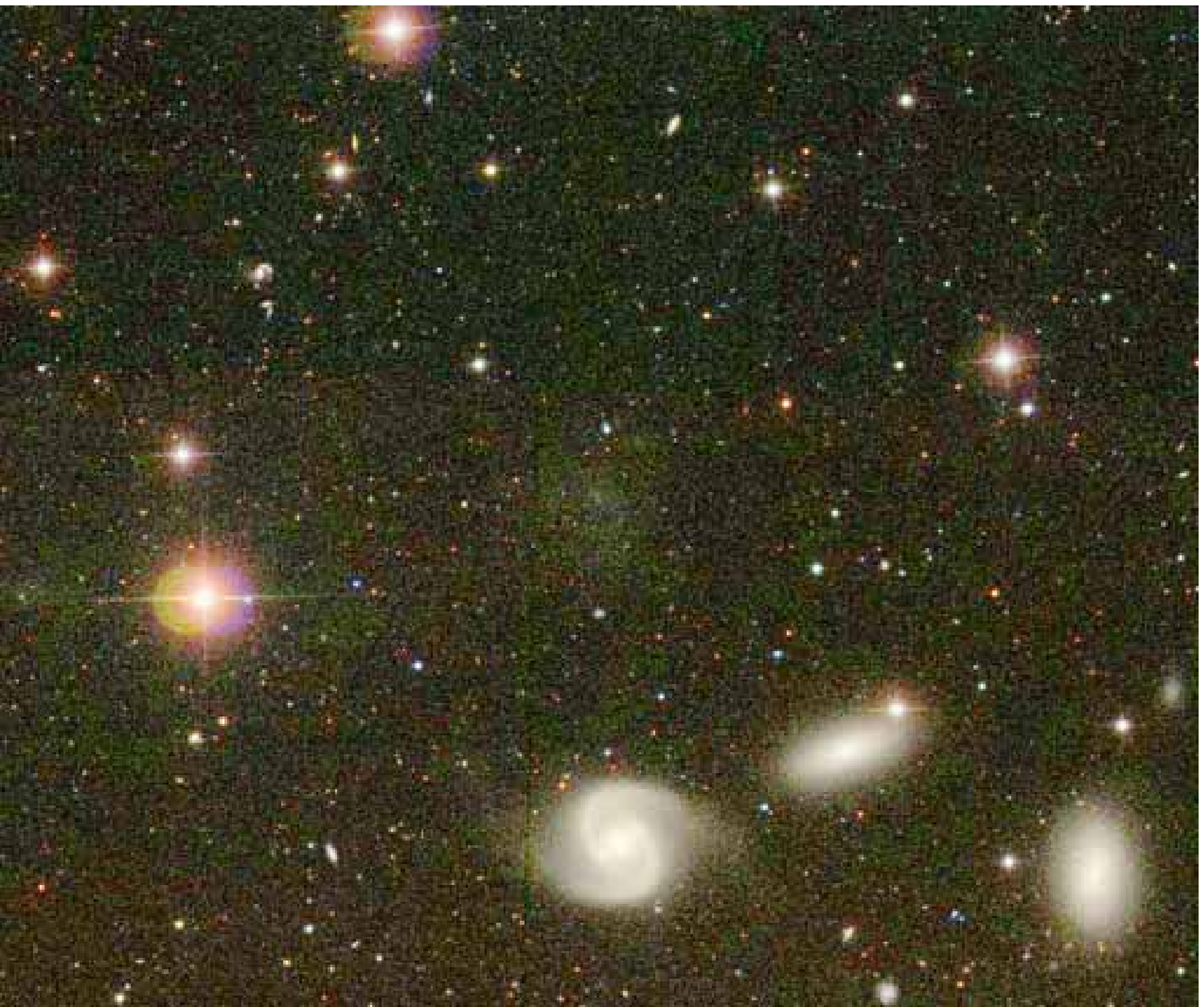}

Fig.~27~-- The ultra-low-surface brightness dwarf galaxy VF18-71 = VCC 1052 discovered by Sandage \etal (1984).
This is an SDSS color image from WIKISKY.  The contrast is set
extremely high near the sky to show the low-surface-brightness dwarf, but it is set very low at higher surface
brightnesses to show some of the internal structure of the trio of galaxies ({\it left to right:\/} NGC 4440, NGC 4436, 
and NGC 4431) to the south.  VCC 1052 is 4\md5 north of NGC 4440.  This field is in the heart of
the Virgo cluster: M{\ts}87 is almost due east ({\it to the left\/}) and NGC 4374\ts$+${\ts}NGC 4406 are to the north-west
({\it toward the upper-right\/}).
\pretolerance=15000  \tolerance=15000 

\eject

\ihi  5.{\ts\ts}Harassment produces Sphs that are triaxial and slowly rotating.  It is consistent 
         with the observation that the brightest Sphs are often the most disky ones, whereas fainter
         spheroidals often rotate much less than isotropic systems of the observed flattening.
         Furthermore, Sph flattening distributions are best explained if these galaxies are modestly triaxial
         (Binggeli \& Popescu 1995).
\vss

\ihi  6.{\ts\ts}Harassment can make Sphs that have kinematically decoupled subsystems, including 
         counter-rotation of the harassed outer parts with respect to the remnant inner galaxy
         (De Rijcke \etal 2004; Gonz\'alez-Garc\`\i a \etal 2005).~Thomas \etal (2006) see
         counter-rotation in the spheroidal galaxy VCC 510.

\vss
\cl {8.4~\it Starvation}
\vss

      Larson \etal (1980) point out that continued star formation at current rates would exhaust the
available gas in most spirals in much less than a Hubble time and suggest that their lifetimes are prolonged
by late gas infall.  A corollary is that starving star formation by cutting off the supply of cold gas 
forces the galaxy to evolve into an S0.  

      Starvation is often discussed in terms of gravitational stripping of a cold gas reservoir.
A more likely scenario~in clusters is that the only available gas is hot.~Virgo qualifies.  
Starvation naturally accompanies and assists ram-pressure stripping in turning gas-rich galaxies into S0s.      
A drawback is that starvation decreasees surface brightnesses; then Sph and S0 surface brightnesses
are surprisingly high (Boselli \etal 2009).  Nevertheless, it seems unavoidable.  We do not discuss it further
only because no observations that are within the scope of this paper constrain it very directly.

\vss
\cl {8.5~\it Everything That Is Not Forbidden Is Mandatory}
\vss

      Astronomers like clean explanations.  They debate about which of many possible processes 
make spheroidals.  We have not reviewed them all; additional examples include various tidal shaking
variants of harassment (e.{\ts}g.,
D'Onghia \etal 2009;
Kazantzidis \etal 2011)
and the failure to accrete sufficient baryons after reionization (e.{\ts}g.,
Bullock \etal 2000;
Cattaneo \etal 2011).  

  We suggest that the relevant question is not ``Which of these mechanisms is correct?'' It is:
``How can you stop any of them from happening?''  It~seems likely to us that all of the above 
processes matter.  

\vss
\cl {9.~\sc THE REVISED PARALLEL SEQUENCE CLASSIFICATION}
\cl {\sc AND GALAXY BIMODALITY}
\cl {\sc IN THE COLOR-MAGNITUDE RELATION}
\vss

\parindent=8pt

      Fig.\ts28 shows how spheroidal galaxies and more~generally the revised parallel-sequence 
classification relate to the galaxy bimodality in the SDSS color-magnitude diagram.  The bright
part of the red sequence consists of ellipticals, S0s, and early-tpe spirals.  However, the luminosity 
functions of all of these galaxy types are bounded (Binggeli \etal 1988; see also Fig.~3 here); they
drop rapidly fainter than $M_V \sim -18$ ($\log {M/M_\odot} \sim 9.7$ in the bottom panel). 
Figure 3 shows further that spheroidals with $M_V < -18$ or $\log {M/M_\odot} > 9.5$ are rare but 
that the Sph luminosity function rises rapidly at lower luminosities and masses.  At the left boundary
of Figure 28, the red sequence is already dominated by spheroidals.  The E{\ts}--{\ts}Sph dichotomy
is visible in Figure 28 as two somewhat distinct high points in the red-sequence contours.  This is
not evident in most published color-magnitude correlations because the magnitude limits of most SDSS
samples are too bright to reach the more numerous spheroidals.
 
      Figure 28(c) illustrates the broader galaxy formation context within which our results are a
small contribution.  If all baryons on the Universe were in galaxies, then the mass function
of galaxies predicted by the cold dark matter fluctuation spectrum would be the almost-straight line
(e.{\ts}g., Somerville \& Primack 1999).  The total galaxy mass function ({\it black squares\/}) never quite
reaches this line -- no galaxies have quite the universal baryon fraction $f_b \simeq 0.17$
(Komatsu \etal 2009).  This smallest shortfall is believed to be due to WHIM baryons 
(Dav\'e \etal 2001).

\parindent=10pt

      The increasing shortfalls at higher galaxy masses are believed to be related to $M_{\rm crit}$ 
quenching (see Cattaneo \etal 2009 for a review).  At these masses, galaxies and clusters of galaxies 
can hold onto hot, X-ray-emitting~gas.  Cooling is too slow to convert this gas into visible stars.  
Candidate processes to keep the hot gas hot are a combination of AGN feedback (e.{\ts}g., Cattaneo
\etal 2009) and heating from late cosmological infall (Dekel \& Birnboim 2006, 2008).  The Universe is 
not old enough for galaxy formation to be as nearly completed at these masses as it is at lower masses.  
In particular, in rich clusters, more baryons are still in hot gas than in visible galaxies (e.{\ts}g., 
Watt \etal 1992;
David \etal 1995;
Vikhlinin \etal 2006).  
These are hostile environments for gas-rich spirals, and we argue that ram-pressure stripping,
galaxy harassment, and starvation can together convert blue-cloud spirals into red and dead S0s.
In addition, processes that make S0s in the field; e.{\ts}g., AGN feedback and perhaps others that
we have not yet discovered, probably also work in clusters.

      The increasing shortfall of baryonic galaxies at $\log {M/M_\odot} < 10.5$ is relevant 
to spheroidals.  This shortfall is usually thought to be caused by supernova-driven
mass ejection (Dekel~\&~Silk~1986).  We agree, but we present evidence in this paper and
in KFCB that, in addition to baryonic blowout, other processes also help to convert blue irregulars 
into red spheroidals.  Candidate processes again are ram-pressure stripping, galaxy harassment, and 
starvation, but in these dwarfs, probably not AGN feedback.  If the color-magnitude plot were 
extended to very faint magnitudes, we expect a steep rise in the number of galaxies as spheroidals 
take over the red sequence.  

   In the smallest galaxies, the baryonic shortfall must continue to grow, as extreme dwarfs retain
only a frosting of baryons in galaxies that are dominated by dark matter (Figure 19).  Most of the 
smallest Sph galaxies may be too dark to be discovered (Kormendy \& Freeman 2004, 2011). 

\vfill\eject

\singlecolumn

\cl{\null} 

\null

\null

\vfill

\cl{\null} \vskip 4.5truein

\vfill

\includegraphics{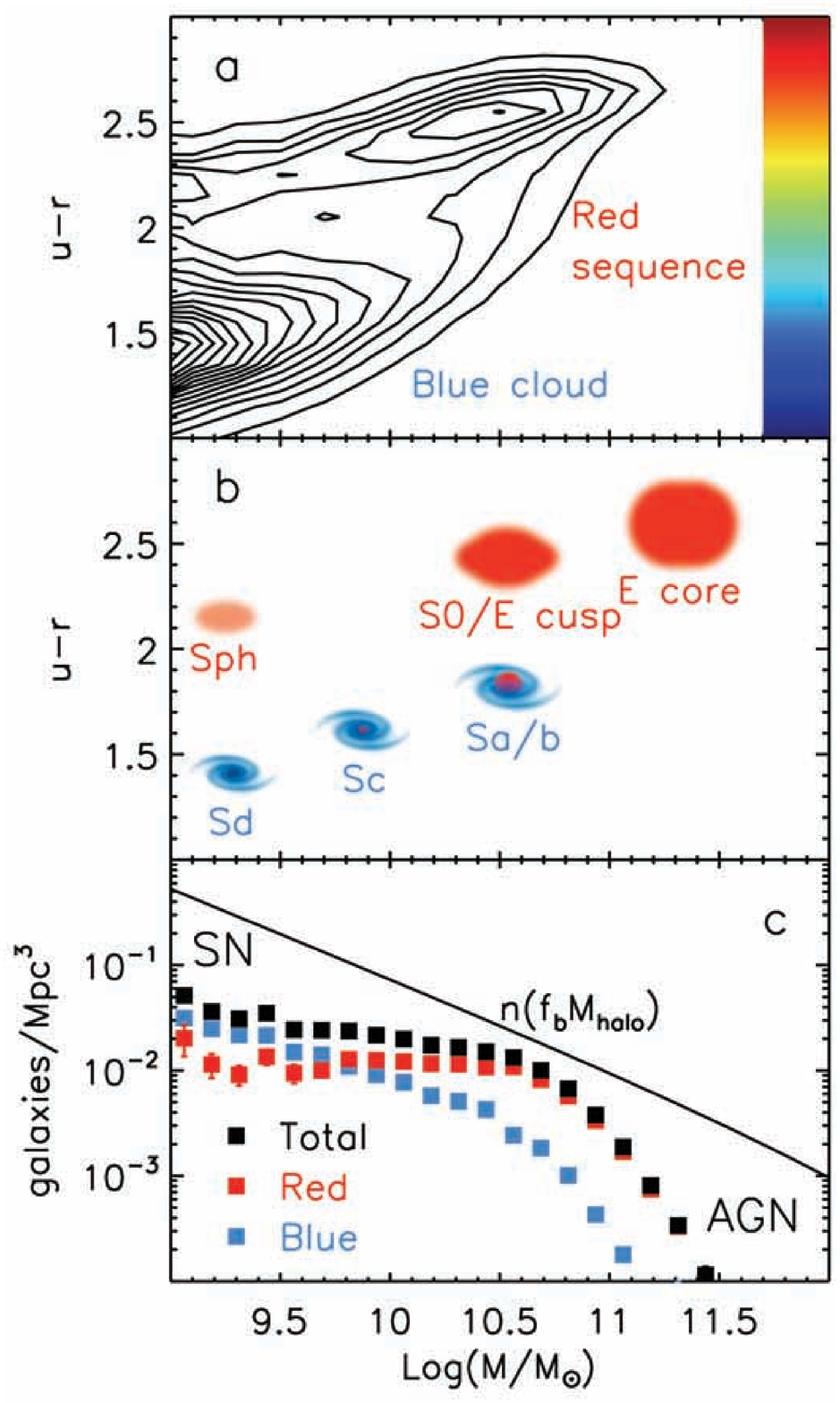}

Fig.~28~-- Correspondence between our parallel-sequence classification and color bimodality 
in the SDSS color-magnitude relation.~This figure is adapted from a draft of Fig.\ts1 in 
Cattaneo \etal (2009); we thank Andrea Cattaneo for permission to use it.  Panel (a) 
shows contours of galaxy number density in the correlation between SDSS $u - r$ 
color and galaxy baryonic mass $M/M_\odot$ (Baldry \etal 2004).  It shows the  narrow 
``red sequence''  of mostly-non-star-forming galaxies and the broader ``blue cloud'' of 
actively-star-forming galaxies.  Panel (b) shows the morphological types from our Figure~1 
that dominate in various parts of Panel (a).  The rapidly rising luminosity function of 
spheroidal galaxies at the low-mass limit of the diagram accounts for the contour around 
(9.0, 2.2) in Panel~(a).  Panel~(c) shows the baryonic mass functions of the red sequence, 
the blue cloud, and their sum (Bell \etal 2003).  The faint-end upturn of the red
sequence mass function is a statistical fluctuation; it is not a detection of spheroidals. 
The diagonal line schematically shows the prediction from the 
$\Lambda$CDM density fluctuation spectrum.  The baryonic mass is approximated by 
$f_b M_{\rm halo}$, where $f_b$ is the universal baryon fraction and $M_{\rm halo}$ is the 
halo mass.  The well known shortfall of observed galaxies with respect to this 
prediction is usually interpreted as the result of $M_{\rm crit}$ quenching aided by AGN 
feedback and continued infall at the high-mass end and supernova-driven
(``SN-driven'') baryon ejection at the low-mass end (see Cattaneo \etal 2009 for a review).
We agree, but we suggest that other processes such as ram-pressure stripping 
also transform blue-cloud galaxies into red-sequence Sphs.  The important ``take-home point''
is that the bright end of the red sequence consists of Es, S0s, and early-type spirals,
but the faint end -- beyond the magnitude limit of most SDSS studies -- is dominated 
by spheroidal galaxies.
\pretolerance=15000  \tolerance=15000 

\eject

\doublecolumns

\vss 
\cl {\sc ACKNOWLEDGMENTS}
\vss

      The inspiration for the parallel-sequence classification is due to Sidney van den Bergh (1976),
in a paper that~was well ahead of its time.  JK has been priviledged to enjoy Sidney's good friendship and
advice for more~than~40~years.  RB has benefited more indirectly through Sidney's papers, also augmented by
personal discussions whenever possible.  We~are~delighted to add a useful small twist to Sidney's classification scheme, 
and it is our pleasure to dedicate this paper to him.

      Writing this paper has been on our minds for many years, but the impetus that motivated us 
to finish it now came from the 2011 ESO Workshop on Stellar Systems in High-Density Environments.
This was an appropriate venue at which to update the observational basis for the E\ts--{\ts}Sph
dichotomy, to incorporate it into a revised parallel sequence galaxy classification, and to emphasize the
observations indicative of S $\rightarrow$ Sph transformation by ram-pressure 
stripping (Kormendy 2011a).  By the time~this paper was delivered on day 4 of the meeting~and~continuing
on the next day, so much evidence for ram-pressure stripping had been presented by other authors 
that it was clear that this was an idea whose time had come.  We thank Magda Arnaboldi and Ortwin 
Gerhard for their hard work in organizing a successful meeting and for their generous allocation 
of time for the oral version of this paper.

      It is a pleasure to thank the anonymous referee for a supportive and thorough report,
including suggestions that improved our science arguments significantly.  Helpful discussions with 
Ron Buta,
Michele Cappellari, and 
Jacqueline van Gorkom 
are also gratefully acknowledged.

      We sincerely thank Ken Freeman for permission to reproduce Figure 19 before
publication and Andrea Cattaneo for permission to reproduce Figure 28 from a draft version
of Figure 1 in Cattaneo \etal (2009).  For permission to reproduce figures, we also thank 
Bruno Binggeli (Figure 3),
Helmut Jerjen (Figure 21),
Daniel Weisz (Figures 22 and 23),
Jacqueline van Gorkom (Figure 24),
Alan Dressler (Figure 25), and
Mario Mateo (Figure 26).  Also, some galaxy images were adapted from the WIKISKY image database 
{\tt www.wikisky.org}. 

      This work makes extensive use of data products from the {\it Hubble Space Telescope\/} and
obtained from the Hubble Legacy Archive, which is a collaboration between the Space Telescope 
Science Institute (STScI/NASA), the Space Telescope European Coordinating Facility (ST-ECF/ESA) 
and the Canadian Astronomy Data Centre (CADC/NRC/CSA).  We also used the digital image database 
of the Sloan Digital Sky Survey (SDSS).  Funding for 
the SDSS and SDSS-II has been provided by the Alfred P.~Sloan Foundation, the Participating Institutions, 
the National Science Foundation, the U.~S.~Department of Energy, the National Aeronautics and Space 
Administration, the Japanese Monbukagakusho, the Max Planck Society, and the Higher
Education Funding Council for England.  The SDSS is managed by the Astrophysical Research 
Consortium for the Participating Institutions. The Participating Institutions are the 
American Museum of Natural History, Astrophysical Institute Potsdam, University of Basel, 
University of Cambridge, Case Western Reserve University, University of Chicago, Drexel 
University, Fermilab, the Institute for Advanced Study, the Japan Participation Group, 
Johns Hopkins University, the Joint Institute for Nuclear Astrophysics, the Kavli 
Institute for Particle Astrophysics and Cosmology, the Korean Scientist Group, the 
Chinese Academy of Sciences (LAMOST), Los Alamos National Laboratory, the 
Max-Planck-Institute for Astronomy (MPIA), the Max-Planck-Institute for Astrophysics 
(MPA), New Mexico State University, Ohio State University, University of Pittsburgh, 
University of Portsmouth, Princeton University, the United States Naval Observatory, 
and the University of Washington.

     This research depended critically on extensive use of NASA's  Astrophysics Data System 
bibliographic services.

     This work made extensive use of the NASA/IPAC Extragalactic~Database~(NED), 
which is operated by the Jet Propulsion Laboratory and the California Institute of Technology (Caltech)
under contract with NASA.  We also made use of Montage, funded by NASA's Earth Science Technology Office, 
Computational Technnologies Project, under Cooperative Agreement No.~NCC5-626 between NASA and Caltech.
The code is maintained by the NASA/IPAC Infrared Science Archive.
We also used the HyperLeda electronic database (Paturel et al. 2003) at
{\tt http://leda.univ-lyon1.fr} and the image display tool SAOImage DS9 developed by the Smithsonian 
Astrophysical Observatory.

      JK's work was supported by NSF grant AST-0607490 and by the Curtis T.~Vaughan, Jr.~Centennial
Chair in Astronomy.  We are most sincerely grateful to Mr.~and Mrs.~Curtis T.~Vaughan, 
Jr.~for their continuing support of Texas astronomy. 

\vss

{\it Facilities:} HST: WFPC2;
                  HST: ACS;
                  HST: NICMOS;
                  SDSS: digital image archive

\vss
\cl {APPENDIX: SURFACE PHOTOMETRY}
\cl {AND BULGE-DISK DECOMPOSITION}
\cl {OF VIRGO CLUSTER S0 GALAXIES}
\vss

      Section 4 discussed photometry results that illuminate our theme that parameter correlations 
are continuous from Sph galaxies through S0 disks.  That is, we concentrated there on
the ``missing'' S0c galaxies and on spheroidals that contain disks.  The remarkably 
large range in $B/T$ ratios in S0s is also central to our argument and is presented in \S\ts4.2,.
Table~1 there contains parameters from the decompositions illustrated in \S\ts4.1 plus those derived
in this Appendix.  Parameters from bulge-disk decompositions of 19 additional Virgo cluster S0s,
13 of which are ACS VCS galaxies, were taken from Gavazzi \etal (2000) and are plotted in Figures 17, 
18, and 20, but these are not tabulated here.  The details of our \S\ts4 photometry, together with
similar photometry and bulge-disk decompositions for the remaining 8 ACS VCS S0s are the subjects 
of this Appendix.

\vss\vsss
\cl {A.1.~\it Photometry Measurements}
\vss\vsss

      Our photometry measurement techniques were discussed in detail in KFCB.  Here, we provide a short
summary.

      Except for NGC 4638 and VCC 2048, the profiles~in~the main paper were measured using program
{\tt profile} (Lauer 1985) in the image processing system {\tt VISTA} (Stover 1988).  The interpolation algorithm
in {\tt profile} is optimized for high spatial resolution, so it is best suited to the galaxies that contain
edge-on, thin disks.  NGC 4352 and NGC 4528 in this Appendix were also measured with {\tt profile}.

      The photometry of VCC 2048, NGC 4638, and the rest of the galaxies discussed in this Appendix 
was carried out by fitting isophotes using the algorithm of 
Bender (1987), 
Bender \& M\"ollenhoff (1987), and 
Bender, D\"obereiner, \& M\"ollenhoff  (1987, 1988) 
as implemented in  the ESO image processing system {\tt MIDAS\/} (Banse \etal 1988).
The software fits ellipses to the galaxy isophotes; it calculates the ellipse
parameters and parameters describing departures of the isophotes from
ellipses.  The ellipse parameters are surface brightness, isophote center 
coordinates $X_{\rm cen}$ and $Y_{\rm cen}$, major and minor axis radii, and hence
ellipticity $\epsilon$ and position angle PA of the major axis. 
The radial deviations of the isophotes $i$ from the fitted ellipses are 
expanded in a Fourier series of the form ($\theta = $eccentric anomaly),
\vskip -5pt 
$$ 
\Delta r_i = \sum_{k=3}^{N} 
             \left [ a_k \cos (k \theta_i) + b_k \sin (k \theta_i ) 
             \right ].\eqno{ (1) } 
$$
The most important parameter is $a_4$, expressed in the figures as a percent of the major-axis 
radius $a$.  If $a_4 > 0$, the isophotes are disky-distorted; large $a_4$ at intermediate or 
large radii is the most reliable sign of an S0 disk in very bulge-dominated S0s (see KFCB 
for examples).  If $a_4 < 0$, the isophotes are boxy. The importance of boxy and disky distortions 
is discussed in Bender (1987); Bender \etal (1987, 1988, 1989); Kormendy \& Djorgovski (1989); 
Kormendy \& Bender (1996), and \S\ts2.1 here.

      We measured the ACS VCS $g$-band images of all of the galaxies discussed here and SDSS
$g$-band images for most of them.  Both profiles are illustrated in the figures.  An average
profile keeping only reliable data (ACS at small $r$, SDSS at the largest $r$, both data in
between) is used for profile decompositions.

      Photometric zeropoints are based on the ACS images and were calculated as discussed
in KFCB.  VEGAmag $g$ magnitudes were converted to $V$ using the calibration derived in KFCB
for early-type galaxies, 
\vskip -17pt
\null
$$    V = g + 0.320 - 0.399\ (g - z).   \eqno{(2)} $$
\null
\vskip -14pt
VEGAmag $(g - z)$ is taken from the Ferrarese \etal (2006) tabulation of the AB mag 
galaxy color; $(g - z) = (g - z)_{AB} + 0.663$.  SDSS $g$-band zeropoints agree very well with
HST ACS zeropoints but~were~not used.  Instead, the SDSS profiles were shifted in surface 
brightness to agree with the zeropointed ACS profiles.  Profile disagreements were pruned,
and the cleaned profiles were averaged for use in photometric decompositions.

      To tie our present photometry to that of KFCB, we remeasured the elliptical galaxy
NGC~4551 = VCC~1630.  The parameters are listed in Table 1.  Here, we find a S\'ersic index
of $n = 1.968 \pm 0.056$; KFCB got $n = 1.98 \pm 0.06$.  From our S\'ersic fit here, we
derive an effective radius in kpc of $\log {r_e} = 0.080 \pm 0.005$; KFCB got $\log {r_e} =
0.084 \pm 0.008$.  We get an effective brightness $\mu_e \equiv \mu(r_e) = 20.715 \pm 0.032$
$V$ mag arcsec$^{-2}$; KFCB measured $\mu_e = 20.75 \pm 0.04$ $V$ mag arcsec$^{-2}$.  The
agreement with KFCB is satisfactory.

      Profile decomposition into nuclei (when necessary), (pseudo)bulges, lenses (when
necessary), and disks was carried out via $\chi^2$ minimization using a combination of a 
grid-based technique (to explore wide parameter ranges) and the Levenberg-Marquardt algorithm as
implemented by Press \etal (1986). Because S0 disks often turn out to have non-exponential
surface brightness profiles, we always fitted the profiles with (at least) two S\'ersic functions.
Many S0s -- especially barred ones -- have $n < 1$; in fact, Gaussians ($n = 0.5$) or even profiles
that cut off at large radii faster than a Gaussian ($n < 0.5$) are common for lens components and 
for the outer disks of barred galaxies.  Our decompositions are based on major-axis surface brightness 
profiles, but we checked their reliability by decomposing mean axis and minor axis profiles as well as by 
visual inspection of the image.  We provide a comparison of major- and mean-axis decompositions~for~two~examples, 
NGC\ts4442 (\S\ts A.8), and
NGC\ts4483 (\S\ts A.8).

      Bulge and disk luminosities were calculated using major-axis S\'ersic parameters and an ellipticity 
for each component that is consistent with the ellipticity profile of the galaxy.  Total luminosity 
was used as a integral constraint for the calculated disk and bulge luminosities.
Thus the accuracy of the decompositions does not suffer because we used major-axis profiles rather than
a two-dimensional decomposition technique.  We benefit from the ease with which we
can avoid complications due to bars.

      The parameters for the galaxies' components are given in Table~1. Errors were derived from
the covariance matrix at the minimum and mostly are marginalized one-$\sigma$ errors. We
explored a variety of fitting ranges and weighting schemes and, when the parameters varied 
more than the one-$\sigma$ errors predicted, we increased the error estimates to be conservatively 
consistent with the parameter variation.

\vss
\cl {A.2.~\it NGC 4762 = VCC 2095}
\vss

      The photometry of the (R)SB0bc galaxy NGC 4762 is presented in \S\ts4.1.1.
We discuss the galaxy here only in order to explain the counterintuitive result that the
disk effective brightness $\mu_e$ looks almost the same as that of the whole galaxy, but
the disk effective radius $r_e$ is $\sim 7$ times larger than that of the whole galaxy.
Since the disk by itself is fainter than the whole galaxy, this looks wrong.

      This situation appears to be due to two problems with the Ferrarese \etal (2006)
photometry:

      Their $g_{AB}$ parameters for the whole galaxy are taken from their Table 4 
and plotted as the green point with the brown center in our Fig.~7.~Their parameters
are converted to VEGAmag $V$ as above and then converted to major-axis parameters by 
dividing their mean-profile $r_e$ by $\sqrt (1 - \epsilon)$.  As we do for all Ferrarese 
\etal (2006) galaxy parameters, we use the mean ellipticity for the whole galaxy from Table 4 
of their paper.  For NGC 4762 = VCC 2095, this is $\epsilon = 0.34$.  But an axial ratio of 
$b/a = 0.66$ is clearly inconsistent both with the isophotes that they illustrate in
their Figure 6 and with $b/a \simeq 0.15$ measured here (Fig.~6).  This accounts for a
factor of 2.1 difference in their $r_e$ versus any that we measure for the disk

      Second, their profile for NGC 4762 is azimuthally averaged.  Ferrarese \etal (2006)
remark that azimuthally averaged profiles of edge-on, disks should be used with caution.
We agree: we do not know how to interpret them.  What is clear is this: The profile shown
in their Figure 56 is much less sensitive to the disk.  They reach a maximum mean radius
of 100$^{\prime\prime}$, which corresponds to major-axis $r^{1/4} = 3.3$, using their
value of $\epsilon$.  Our Figure~6 shows the inner two shelves in the brightness
profile at $r^{1/4} \leq 3.3$.   There is no sign of them in Ferrarese \etal (2006) Figure 56.
The B(lens) shelves are shown in a major-axis cut profile in their Figure 107, but the cut
profile was not used to determine the galaxy parameters.
So we see a higher major-axis disk $\mu_e$ than one would derive from their photometry.
And since the bulge profile is removed before the disk $r_e$ is calculated, the effective
radius of the disk is substantially larger than that of the galaxy as a whole.  These 
effects account for the rest of the difference between the green$+$brown point in our
Figure 7 and the point that represents the disk.

\vss\vsss
\cl {A.3.~\it NGC 4452 = VCC 1125}
\vss\vsss

      Similar comments apply to the total-galaxy and disk parameters of NGC 4452 =
VCC 1125 in our Figure~9.  Ferrarese \etal (2006) measure
a mean ellipticity \hbox{$\epsilon \simeq 0.68$} or $b/a = 0.32$ (their Table 4).  We agree
at large radii (our Figure 8).  But the inner B(lens) structure that contributes much
of the disk light is flatter \hbox{($\epsilon \simeq 0.9$).}  Our measurements of the major-axis
profile show the B(lens) shelves, but their measurements of the azimuthally
averaged profile (their Figure 72) do not.  These are the reasons why we measure a
higher effective surface brightness of the pseudobulge-subtracted disk than Ferrarese
measures for the galaxy as a whole. 

\vss
\cl {A.4.~\it NGC 4352 = VCC 698}
\vss

      NGC~4352 is an unbarred and well-inclined S0 galaxy, so the photometry and decomposition are 
straightforward.  Figure 29 ({\it top\/}) shows a color image of the galaxy made
from the HST ACS images ({\it blue\/} = $g$; {\it green\/} = mean of $g$ and $z$; {\it red\/} = $z$).
The bulge-to-total luminosity ratio looks moderately small, but a contour plot of the SDSS~$g$ image
(smoothed with a 2$^{\prime\prime}$-FWHM Gaussian, {\it middle panel\/}) shows that the isophote 
ellipticity at $r \simeq 80^{\prime\prime}$ is $\epsilon \simeq 0.35$, similar to the value in the bulge.  
This suggests that the bulge dominates at both small and large radii.

      Surface photometry of the ACS $g$ image was carried out with {\tt VISTA profile}, and the 
results are shown in the bottom two panels of Figure 29.  The galaxy has a nucleus together with disky,
central ``extra light'' (both seen by Ferrarese \etal 2006) like that in the extra light ellipticals 
in KFCB.  We include this in our bulge-disk decomposition, but we take account of a fit that omits the 
nuclear disk in the parameter error estimates.  The disk is Gaussian.  There is a small pseudobulge
contribution from the nuclear disk, but the bulge is robustly classical whether the nuclear disk is
included ($n = 3.7 \pm 0.2$) or not ($n = 2.9^{+0.3}_{-0.1}$).  We adopt $n = 3.7^{+0.5}_{-0.8}$ as shown
in the key.  The bulge-to-total luminosity ratio is $B/T = 0.71^{+0.13}_{-0.05}$.  
The outer drop in $\epsilon$ is consistent with the decomposition result that the bulge
dominates at large $r$.  NGC 4352 is an S0a -- a smaller, more face-on,
less bulge-dominated version of NGC 3115.

      The parameters from the above decomposition are listed in Table 1 and plotted in Figures~17, 18, and 20. 
However, there is an important caveat with the above discussion.  It is possible that NGC 4352 resembles
NGC 4638 (Fig.~13); i.{\ts}e., that the rounder isophotes at large radii are not a sign that the bulge
becomes comparable in brightness to the disk again at large radii.  Rather, the rounder isophotes at large
radii may belong to a heavily heated outer part of the disk.  Then the disk
would be more nearly exponential and we would derive $B/T \sim 0.25$.  Our conclusions would not change,
but the evidence for dynamical thickening of disks would get stronger.
This caveat is discussed in \S{\ts}A.12.

\cl{\null} \vskip 5.7truein

\vfill

\includegraphics{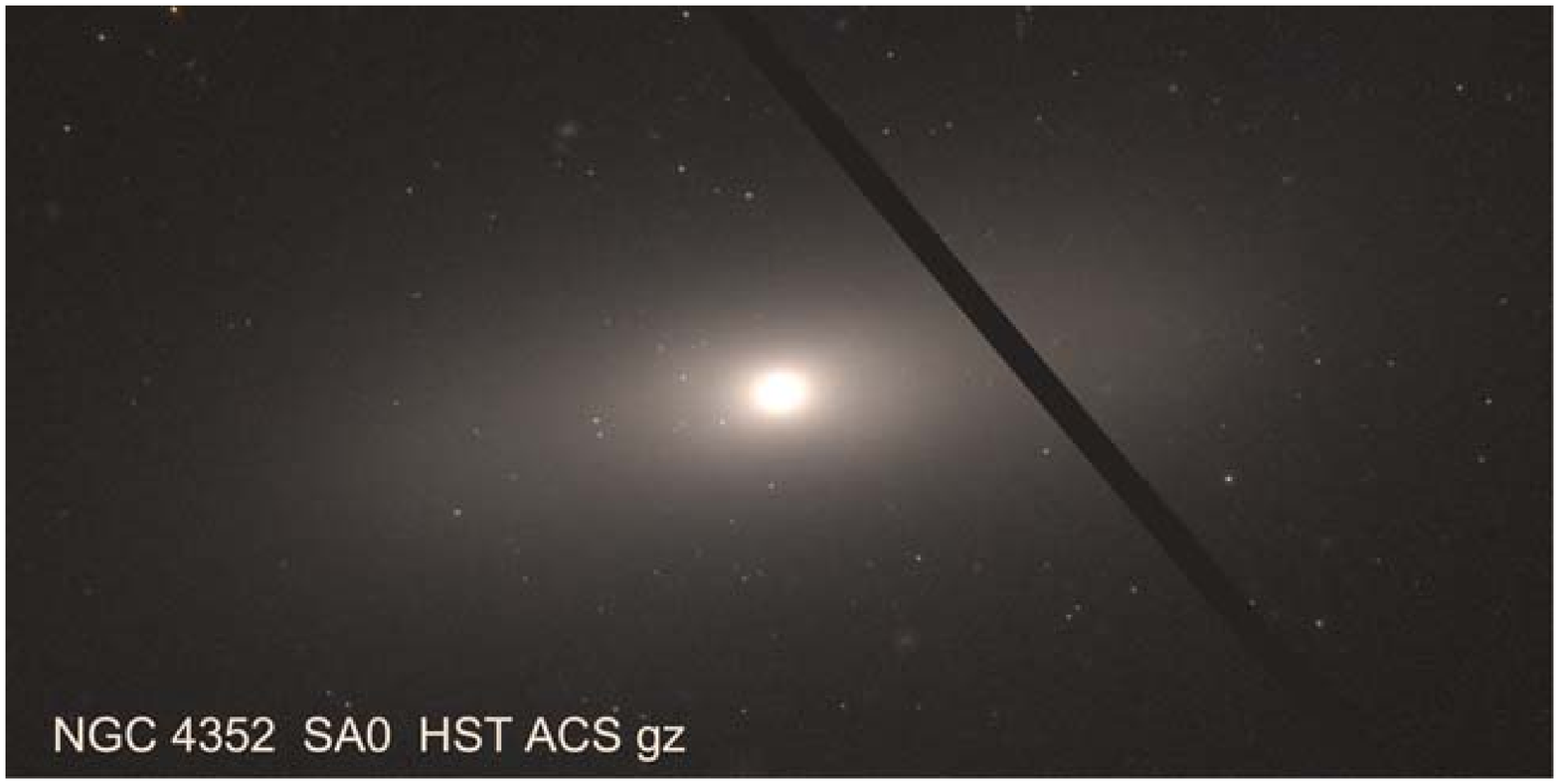}

\includegraphics{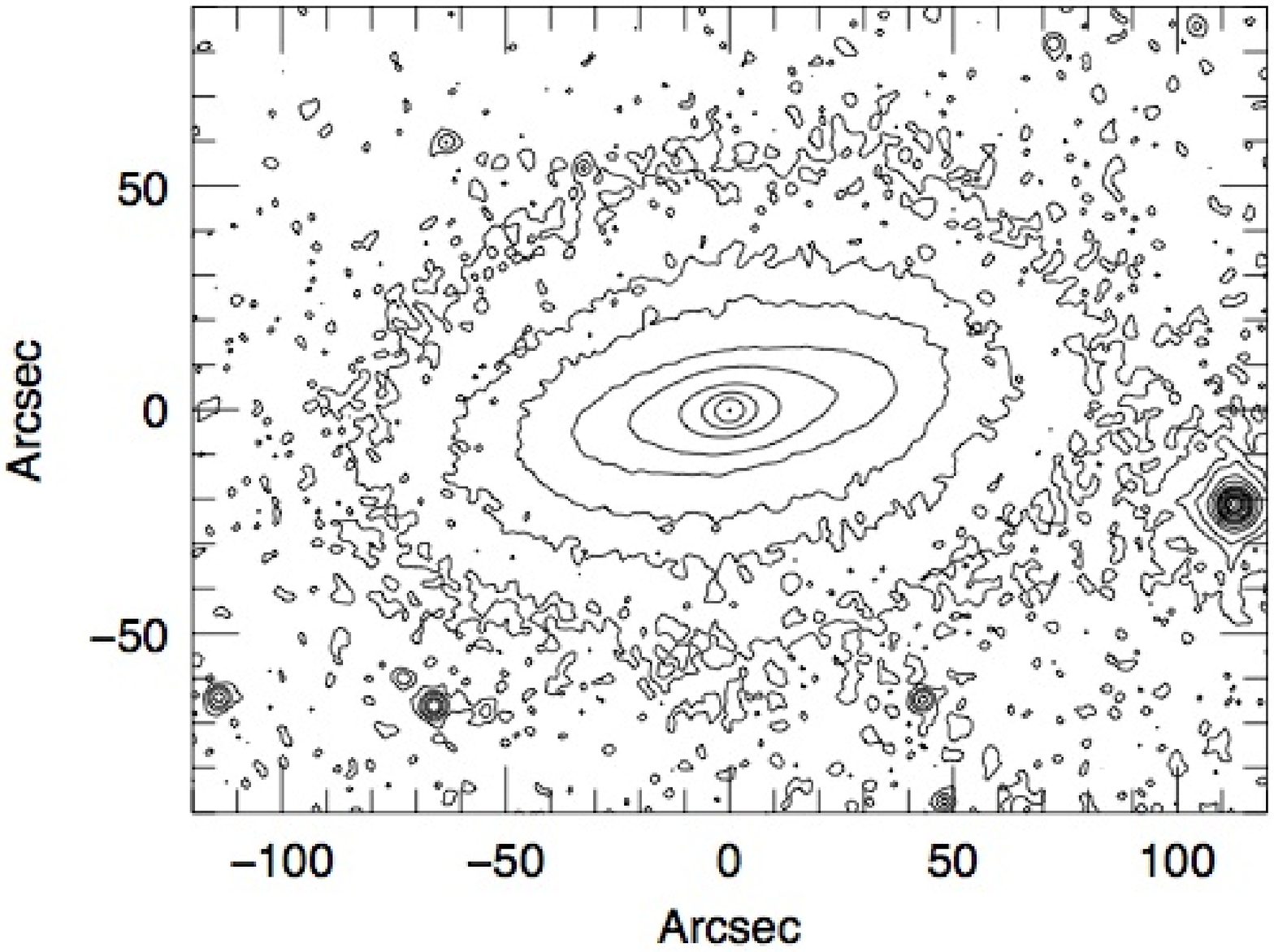}

\includegraphics{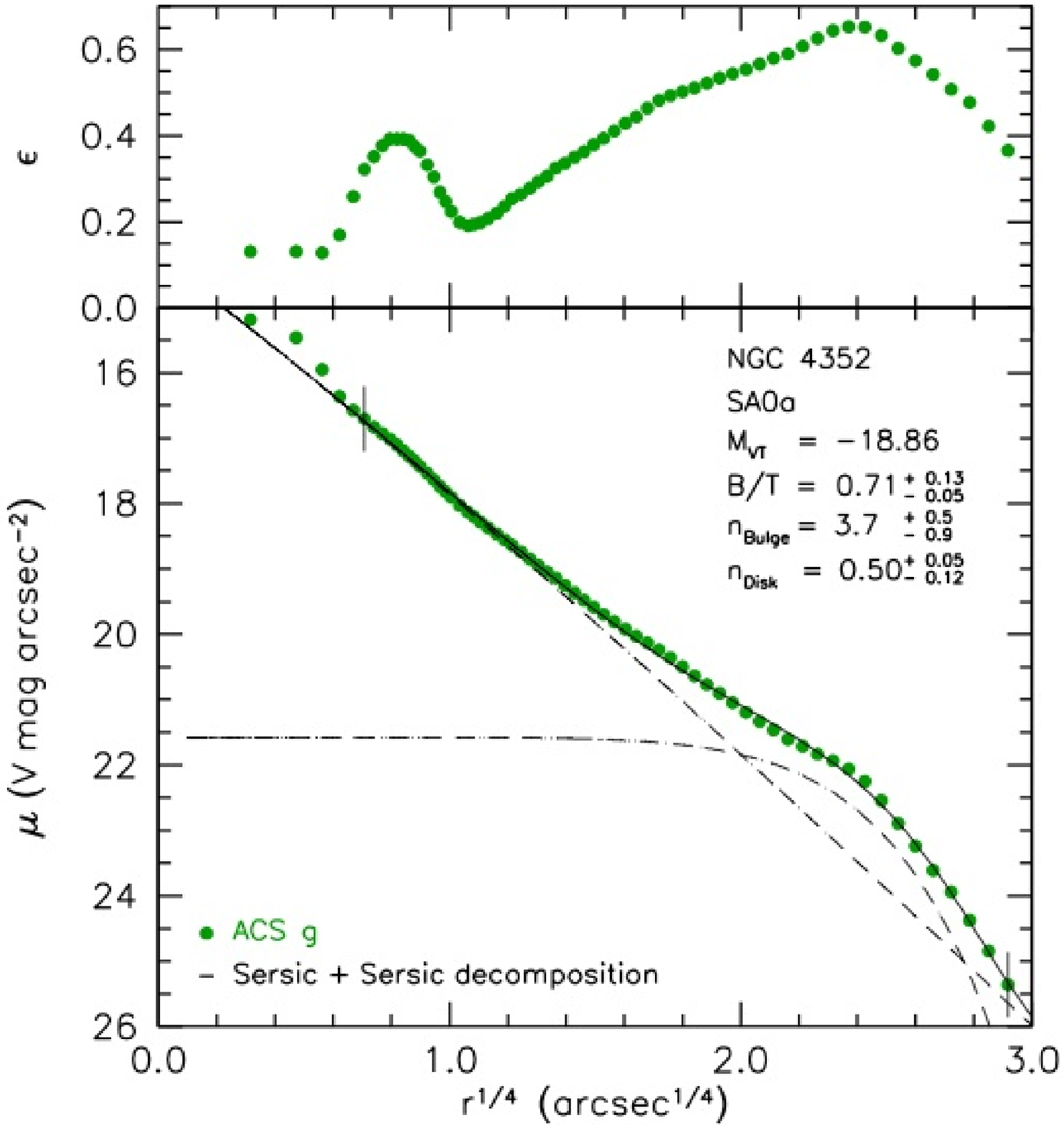}

Fig.~29~-- ({\rm top\/}) Color image of the SA0 galaxy NGC\ts4352.  North is up and east is at left 
as in all images illustrated.  The field is 130$^{\prime\prime}$ wide.  ({\it middle\/}) Brightness
contours in the SDSS $g$ image.   ({\it bottom\/}) Ellipticity and surface brightness along the 
major axis of NGC 4352.  The dashed curves show the bulge-disk decomposition inside the fit range 
shown by vertical dashes.  The sum of the components ({\it solid curve}) 
fits the data with RMS = 0.049 mag arcsec$^{-2}$. 

\eject

\vss
\cl {A.5.~\it NGC 4528 = VCC 1537}
\vss

      NGC 4528 is shown in an ACS color image in Figure~30.  HST resolution shows
what was not clear at ground-based resolution: NGC 4528 is barred, and the bar lies almost
along the minor axis (Ferrarese \etal 2006).  There is a hint of an inner ring ``(r)'' around the 
end of the bar, but mainly, the ring outlines the rim of a lens component that shows up in
our photometry (Figure 31) as a shelf in the light distribution interior to the outer disk.  To 
put it differently, the major-axis disk brightness profile consists of two parts, a lens and an
outer disk that form two distinct shelves in $\mu(r)$ in Figure 31.  This is very common behavior in 
barred galaxies (Kormendy 1979b).  In this case, the lens shelf is too abrupt to be
well fitted by a S\'ersic function; and a result; our best fits have RMS $\simeq$ 0.1
and $n \simeq 1.0 \pm 0.05$.
The outer disk is well fitted by a S\'ersic function with a very small index; our decomposition 
gives $n = 0.17$ to 0.3, but the outer disk photometry is very uncertain and the leverage is so 
small that its $n$ not well constrained.
Note that the lens and disk have the same apparent ellipticity; i,{\ts}e., the lens 
really is a part of the disk.  

      Taking $\epsilon(r)$ into account, our photometry implies that the lens-to-total
luminosity ratio is ${\rm lens}/T \simeq 0.54$ and the disk-to-total luminosity ratio is 
$D/T \sim 0.07$.  The bar fraction is not accurately determined but is a few percent.

\cl{\null} \vskip 4.7truein

\includegraphics{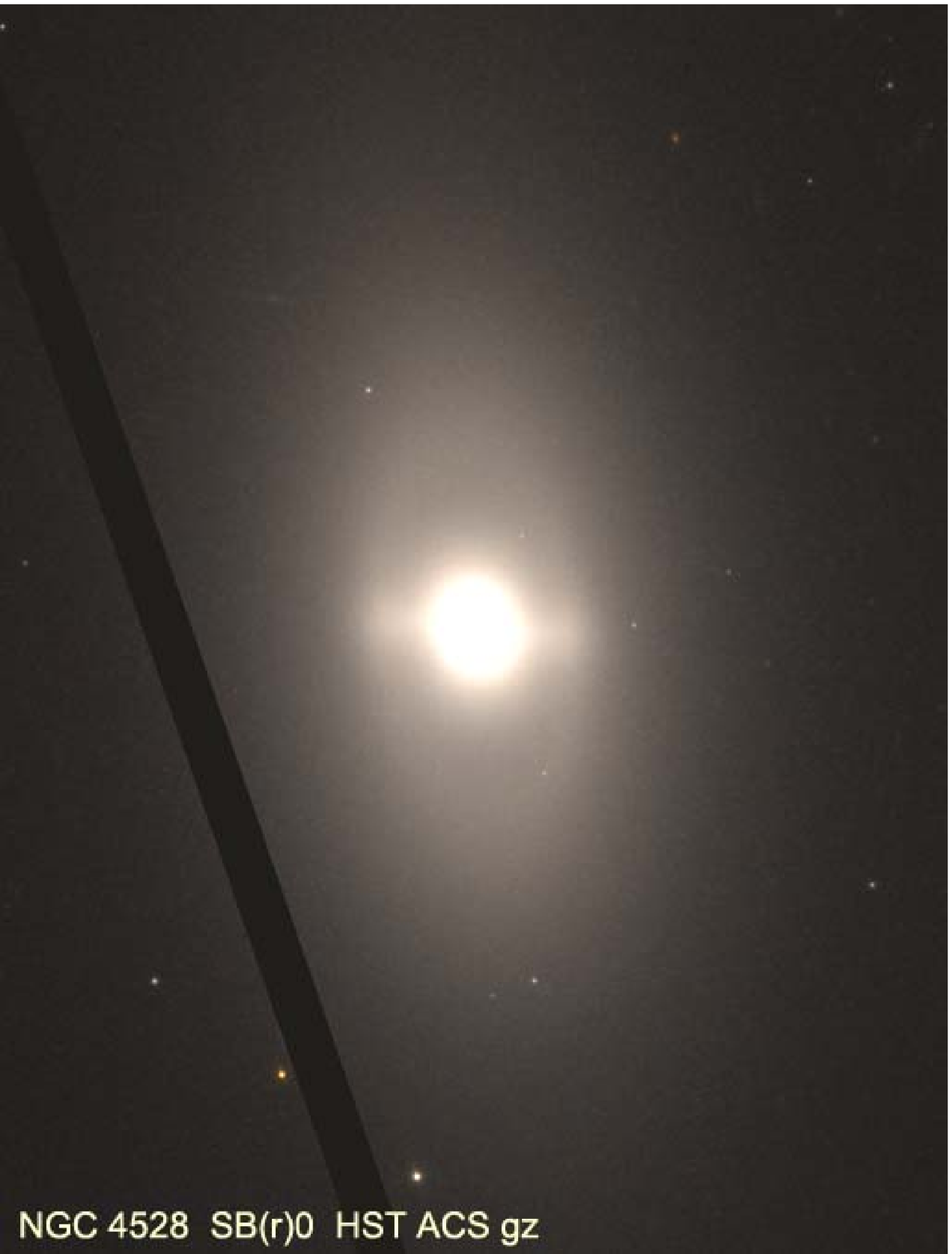}


Fig.~30~-- Color image of the SB(r)0 galaxy NGC\ts4528 constructed from the HST ACS
$g$ image ({\it blue}), the mean of the $g$ and $z$ images ({\it green\/}), and the $z$ image
({\it red\/}).  The bar is located almost along the minor axis, so its effects on our
major-axis bulge-disk decomposition are negligible.  

\vskip 12pt

      The bulge parameters are the ones that we need most.  Despite the high surface brightness of the
bulge above the lens, the bulge parameters are more uncertain than usual, in part because the photometry 
of the outer disk is very uncertain and in part because there is strong parameter coupling for three -- not
the usual two -- components.  The bulge S\'ersic index is $n = 2.55^{+0.61}_{-0.20}$.  Also, most of the 
bulge light is in an almost round distribution.  We conclude that this is at least primarily a classical bulge.
The bulge-to-total luminosity ratio is $B/T \simeq 0.38 \pm 0.13$.   In comparison, the SAb galaxy M{\ts}31 
has $B/T \simeq 0.25$, and the SAab galaxy M{\ts}81 has $B/T \simeq 0.34$ (Kormendy 2011b).
We tentatively classify NGC 4528 as SB(r,lens)0ab.  But we emphasize that the stage along the Hubble sequence
is uncertain and that no conclusions of this paper are affected if it turns out, after further work, that
a Hubble type of S0a or S0b is more appropriate.

     Figure 32 shows, for both NGC 4352 and NGC 4528, the bulge and disk parameters derived in
our photometric decomposition compared with the F2006 parameters for the whole galaxy.  Both galaxies 
are well behaved.  The small bulge of NGC 4528 helps to define the extension of the E{\ts}$+${\ts}bulge 
sequence to the left of (i.{\ts}e., toward more compact bulges than) the spheroidals.  Like other 
bulges, the explanation for the compactness of this classical bulge cannot be that it has been tidally
stripped of its outer parts.  In any case, its S\'ersic index of $n = 2.55$ is normal.

\vfill

\cl{\null} \vskip 3.80truein

\includegraphics{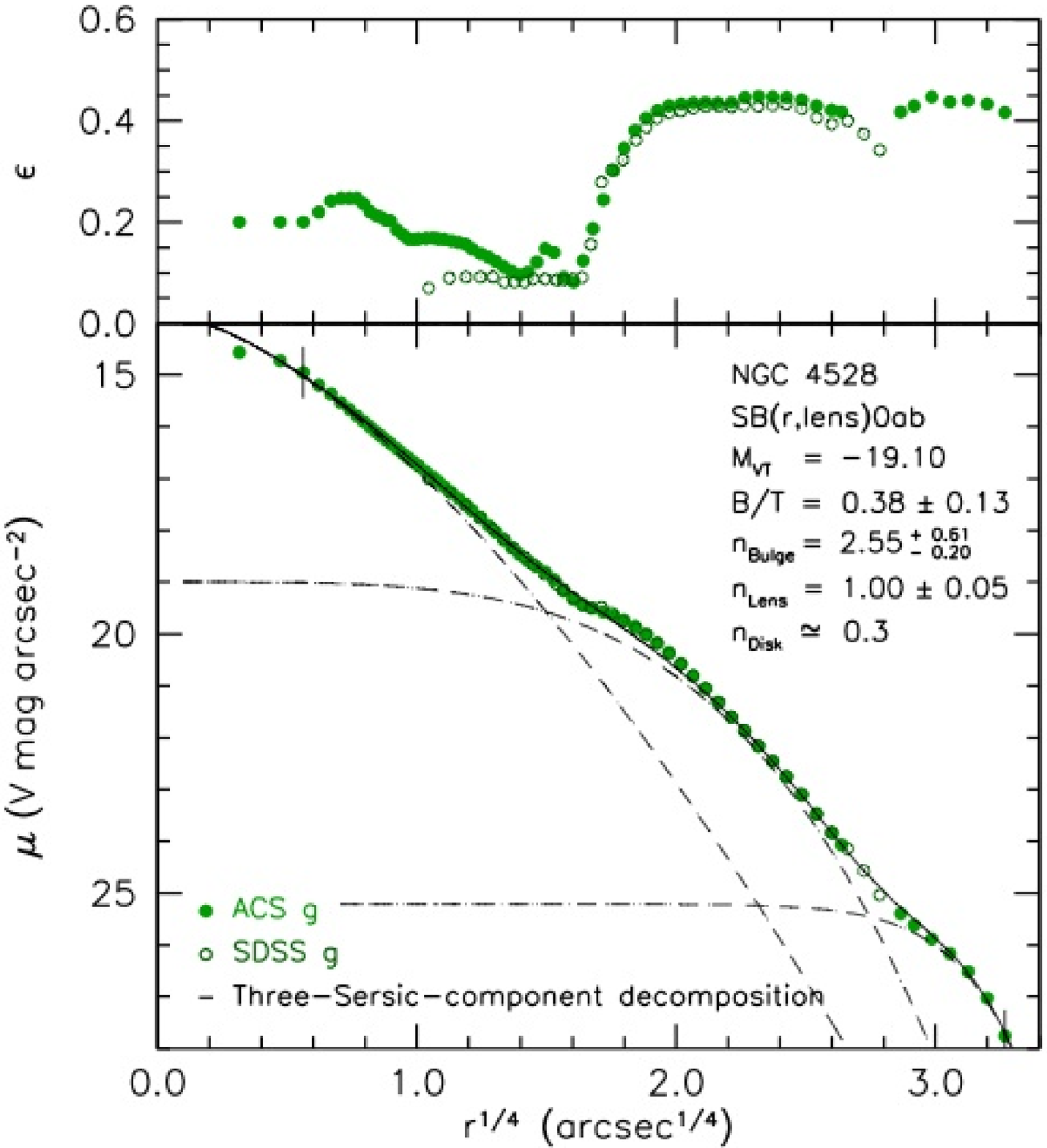}

Fig.~31~-- Ellipticity $\epsilon$
and surface brightness $\mu_V$ along the major axis of NGC 4528 measured by fitting ellipses to 
the isophotes in the ACS and SDSS $g$-band images.  The dashed curves show a three-S\'ersic-component,
bulge-lens-disk decomposition inside the fit range ({\it vertical dashes\/}).  The sum of the components 
({\it solid curve}) fits the data with RMS = 0.105 mag arcsec$^{-2}$. 

\vskip 8pt

\cl{\null} 

\eject

\cl{\null} \vskip 4.9truein

\includegraphics{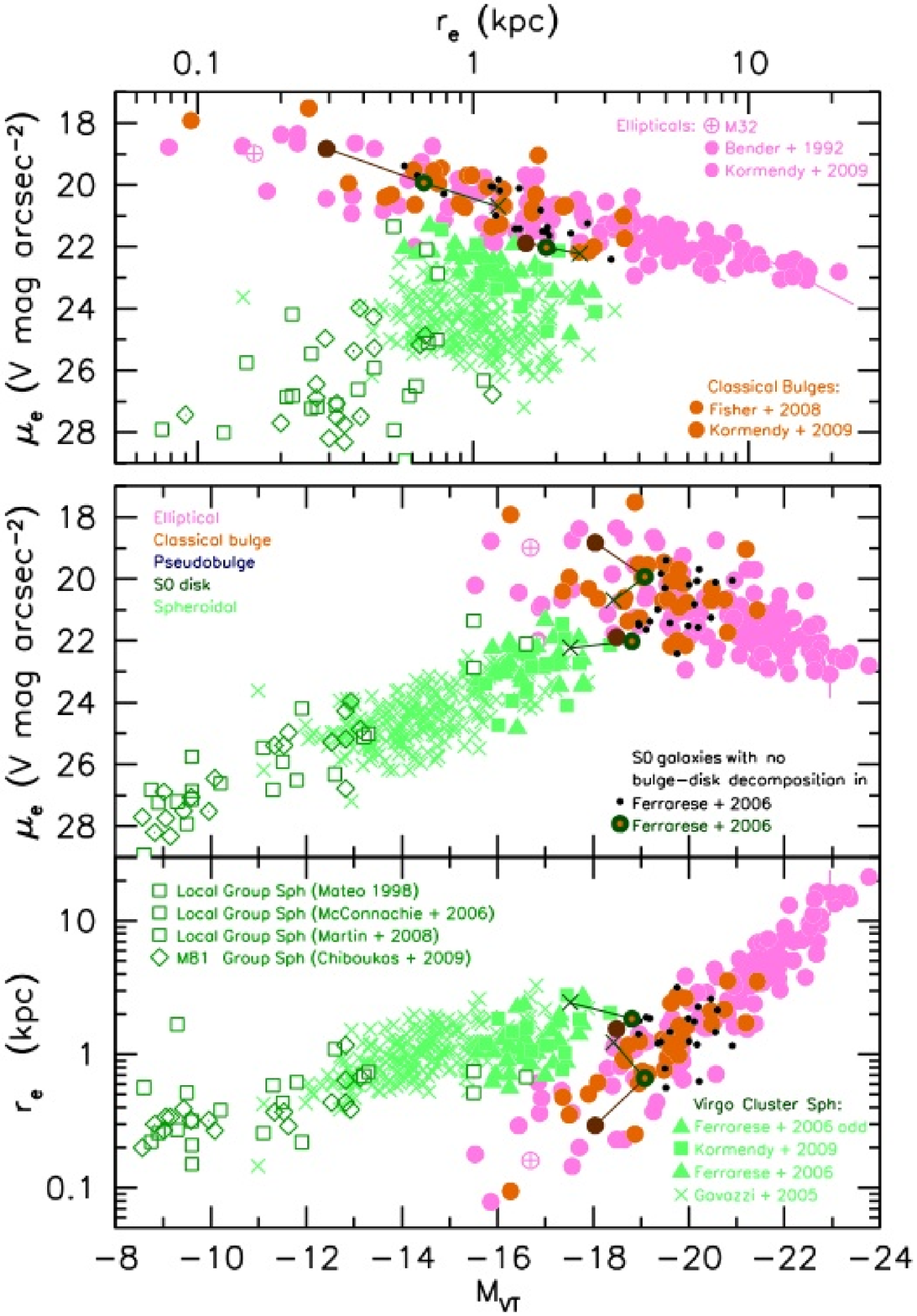}

Fig.~32~-- Parameter correlations from Figure 2 showing the results of the bulge-disk 
decompositions of the unbarred S0a galaxy NGC 4352 ({\it lower points in top panel\/}) and 
the SB(lens)0ab galaxy NGC 4528 ({\it upper points in top panel\/}).  
The dark green filled circles with the brown centers show the total parameters measured by F2006 
for the bulge and disk together.  These points are connected by straight lines to the bulge 
parameters ({\it dark brown filled circles\/}) and the disk parameters ({\it dark green crosses\/}).

\vskip 12pt

\cl {A.6.~\it NGC 4417 = VCC 944}
\vss

      NGC 4417 (Fig.~33) is very similar to NGC 4352~(\S\ts A.4): it is nearly edge-on; it has
a prominent bulge~at~small~$r$, and the flattened isophotes produced by the disk at intermediate
radii gave way to rounder isophotes again at large radii.  The usual interpretation
is that the galaxy is bulge-dominated, and the photometry and decomposition ({\it bottom panels\/} of
Figure 33) make this quantitative.  The decomposition is straightforward and robust. 
Major- and mean-axis decompositions yield consistent results.  They confirm that the bulge dominates
the light at both small and large radii.  The ellipticity maximum at 4$^{\prime\prime}$ is due to a faint central disk, and the
slightly boxy appearance of the bulge is due to a faint bar.  However, the bulge is classical
($n\approx 4$). The disk is essentially Gaussian. We measure $B/T = 0.88^{+0.06}_{-0.11}$ and classify 
the galaxy as SA0a.

      The same caveat that we discussed for NGC 4352 applies here.  It is possible that the rounder,
outer isophotes are a sign that the disk has been heated rather than that the bulge
takes over again at large radii (\S\ts A.12).

\cl{\null} 

\vskip 6.7truein

\vfill


\includegraphics{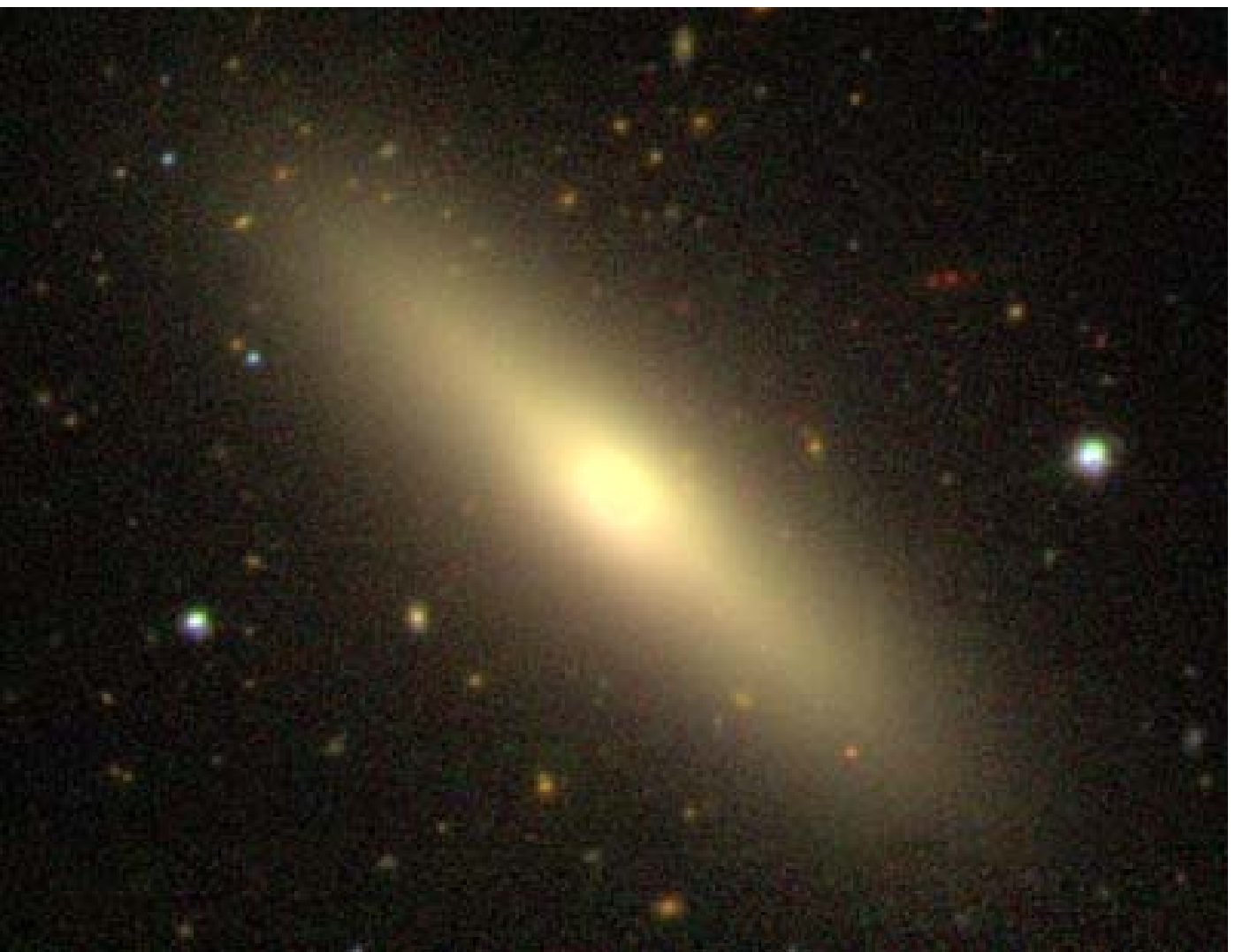} 

\includegraphics{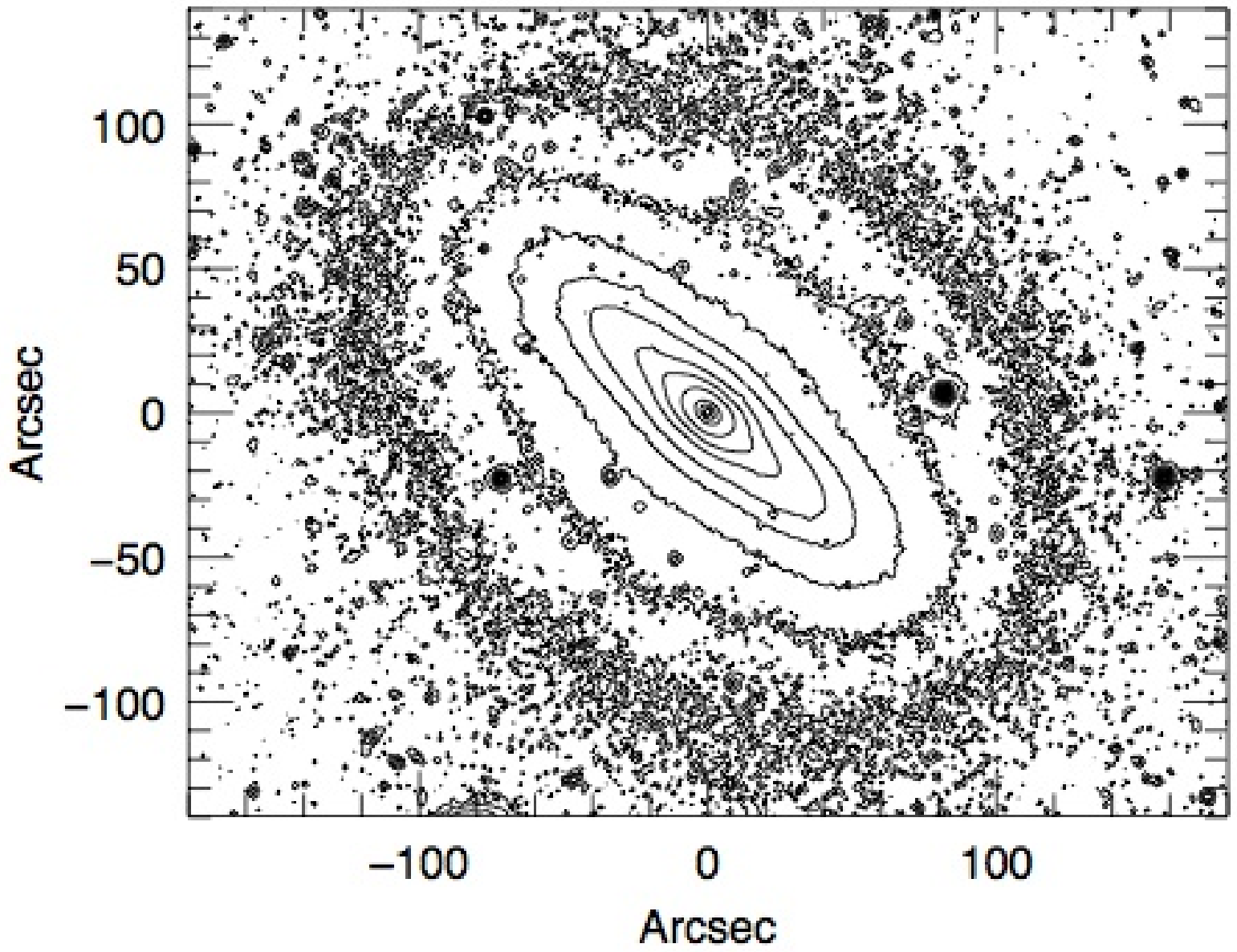}

\includegraphics{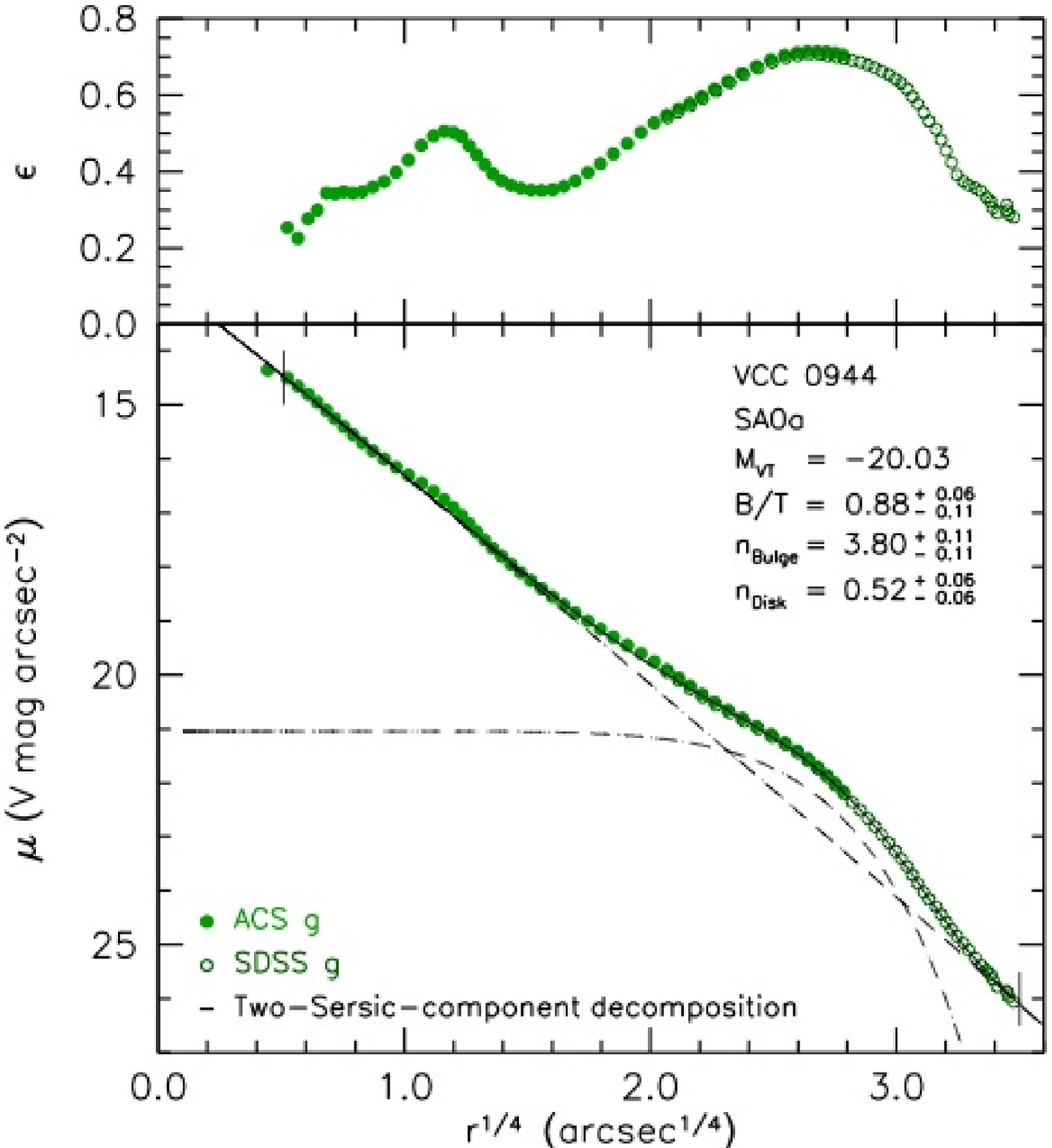}

Fig.~33~-- ({\it top\/}) {\tt WIKISKY} image of NGC\ts4417{\ts}={\ts}VCC\ts944. 
({\it middle\/}) Brightness contours of the SDSS $g+r+i$ images smoothed with a 1\sd2\ts-FWHM Gaussian.
({\it bottom\/}) Major-axis brightness and ellipticity profiles.  Dashed curves show the bulge-disk 
decomposition; the sum of the components ({\it solid curve}) fits the data with RMS = 0.043 mag arcsec$^{-2}$.

\eject

\vss
\cl {A.7.~\it NGC 4442 = VCC 1062}
\vss

NGC 4442 is a weakly barred S0. Decomposition is straightforward; our major- and mean-axis
decompositions give fully consistent parameters for the bulge. The parameters of the disk
are somewhat affected by the presence of the bar, as the comparison of major- and mean-axis
profiles show (Fig.~34 {\it vs.} Fig.~35). Nevertheless, the disk seems to be more nearly
Gaussian than exponential. The bulge is classical and contributes significantly again at the 
largest radii.  The bulge-to-total ratio is very robust and high, i.{\ts}e,. the galaxy is 
of type SB0a.

\cl{\null} \vskip 6.7truein

\includegraphics{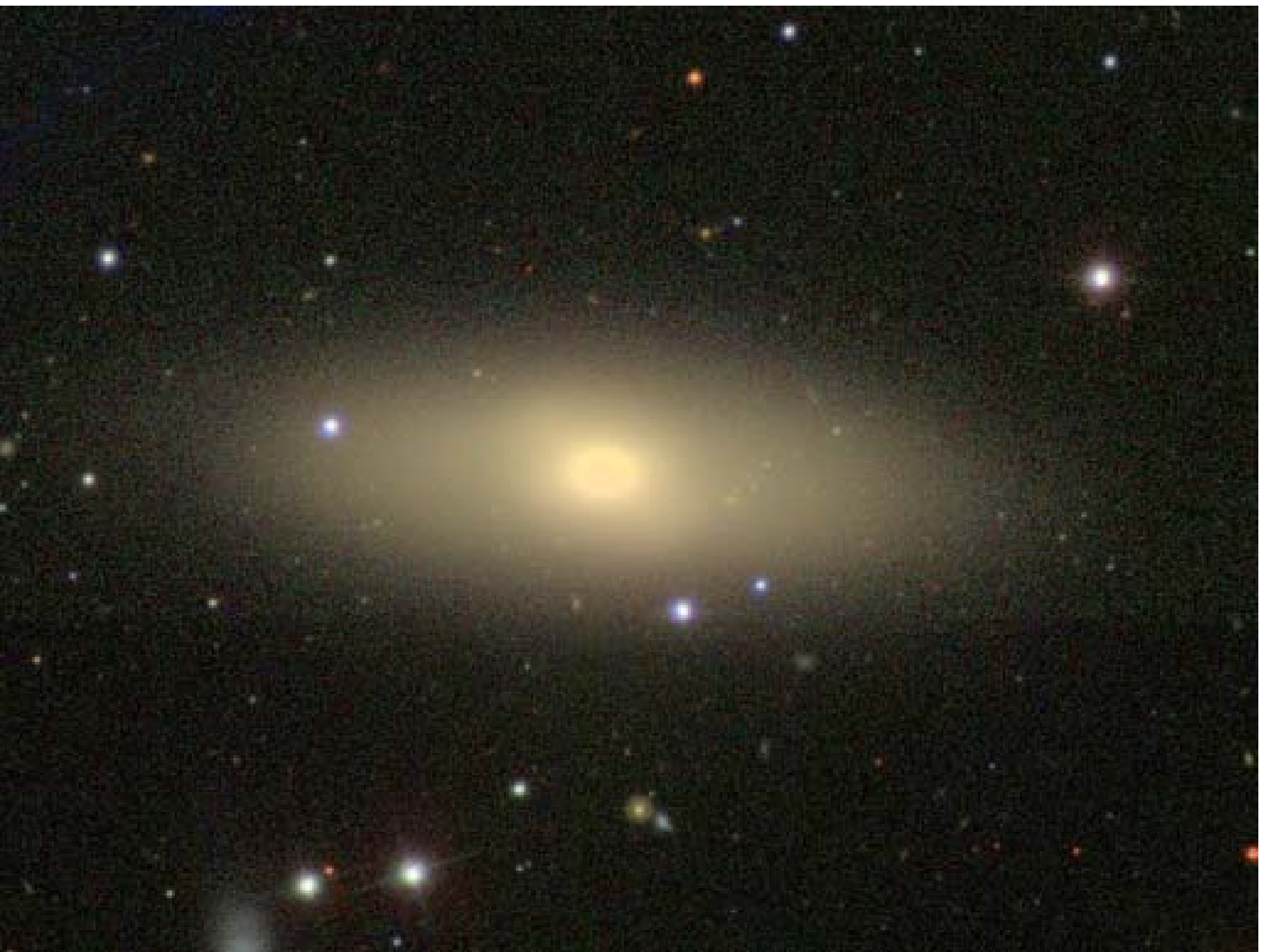}

\includegraphics{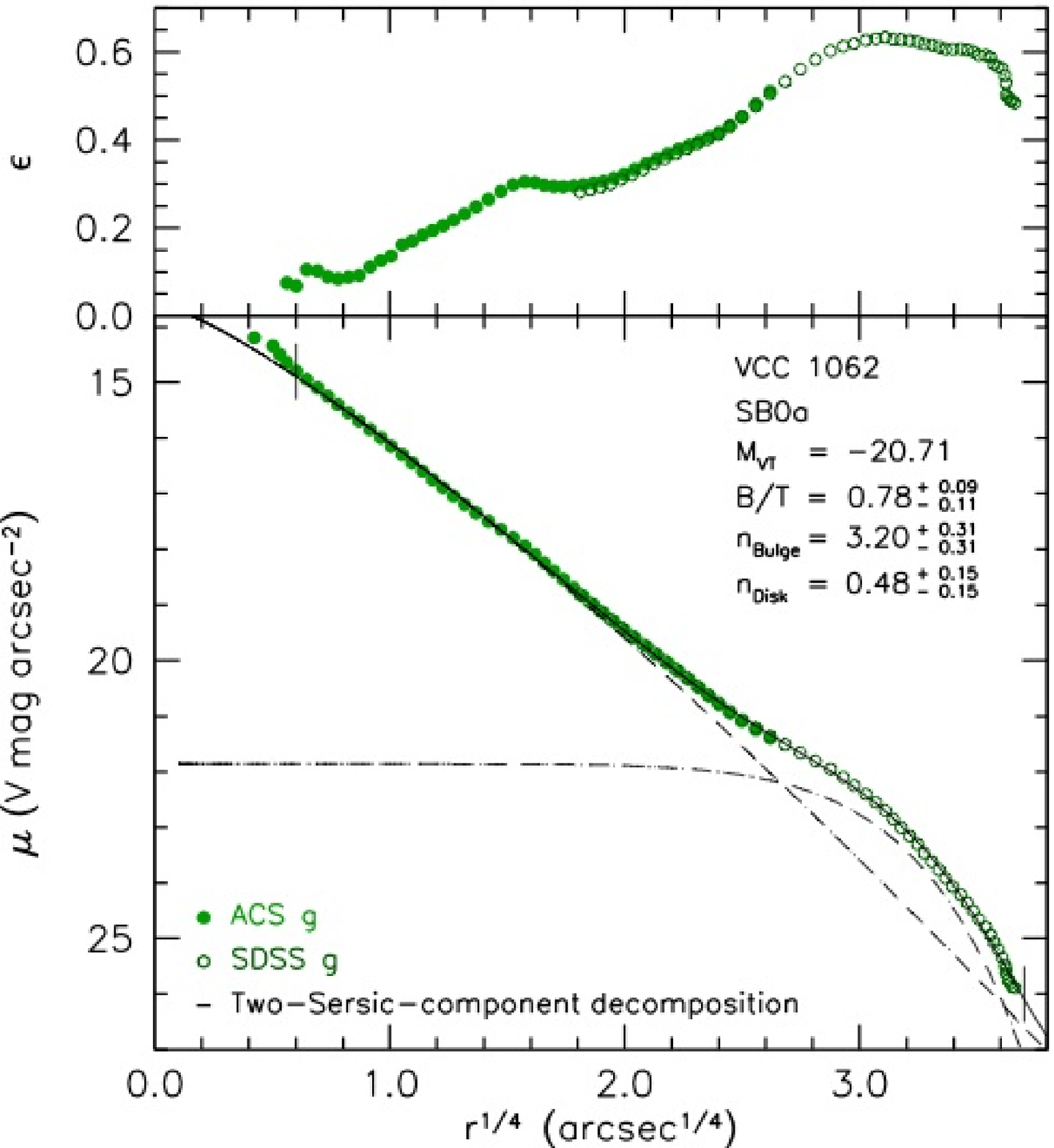}

Fig.~34~-- ({\it top\/}) Color image of the barred S0 NGC\ts4442 = VCC~1062 from {\tt
  WIKISKY}.  ({\it bottom\/}) Ellipticity~$\epsilon$ and surface brightness $\mu_V$ along the
  major axis of NGC 4442.  The dashed curves show the S\'ersic-S\'ersic, bulge-disk
  decomposition inside the fit range ({\it vertical dashes\/}).  The sum of the components
  ({\it solid curve}) fits the data with RMS = 0.060 mag arcsec$^{-2}$.

\cl{\null} 

\vfill

 \includegraphics{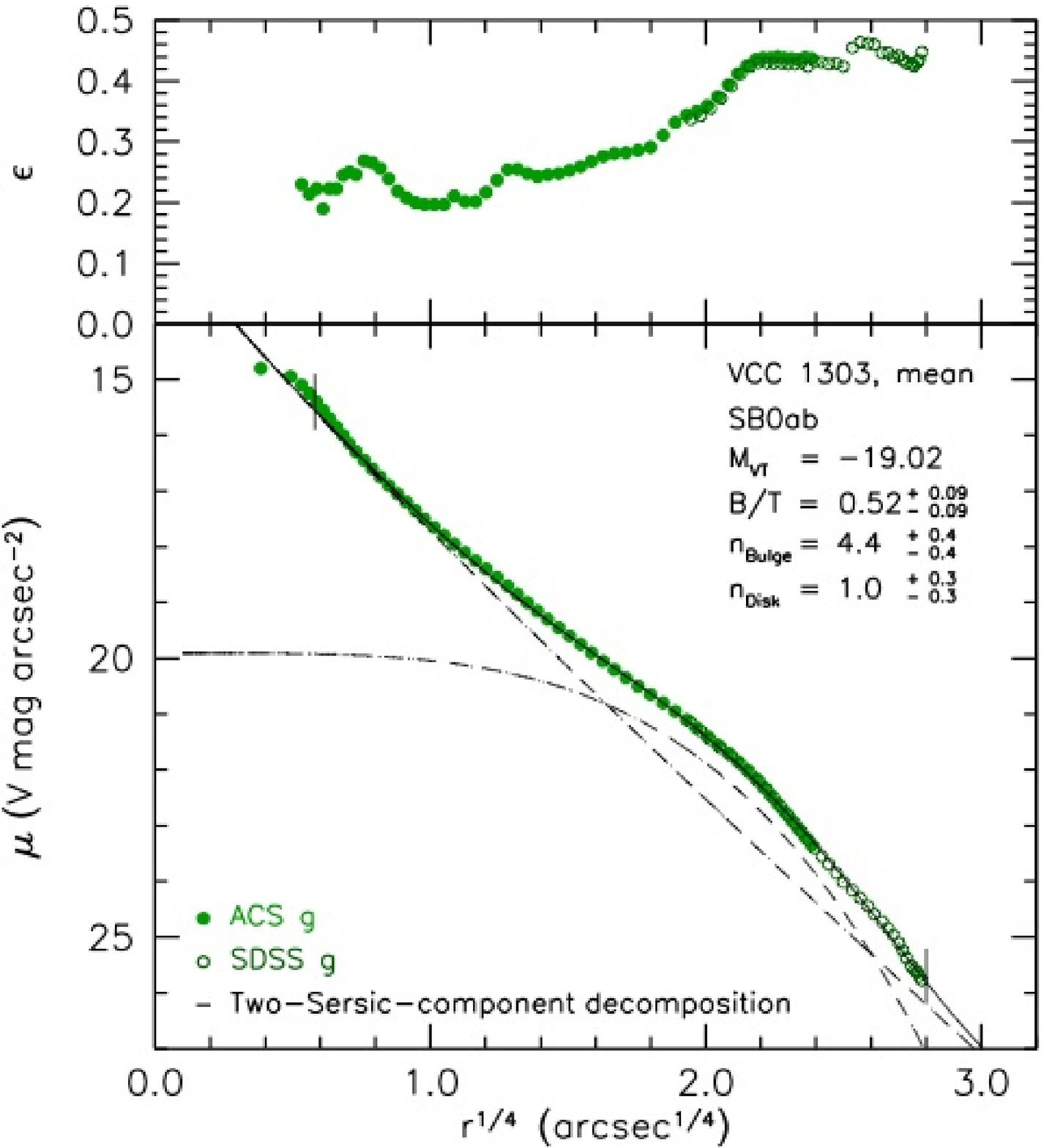}

 \includegraphics{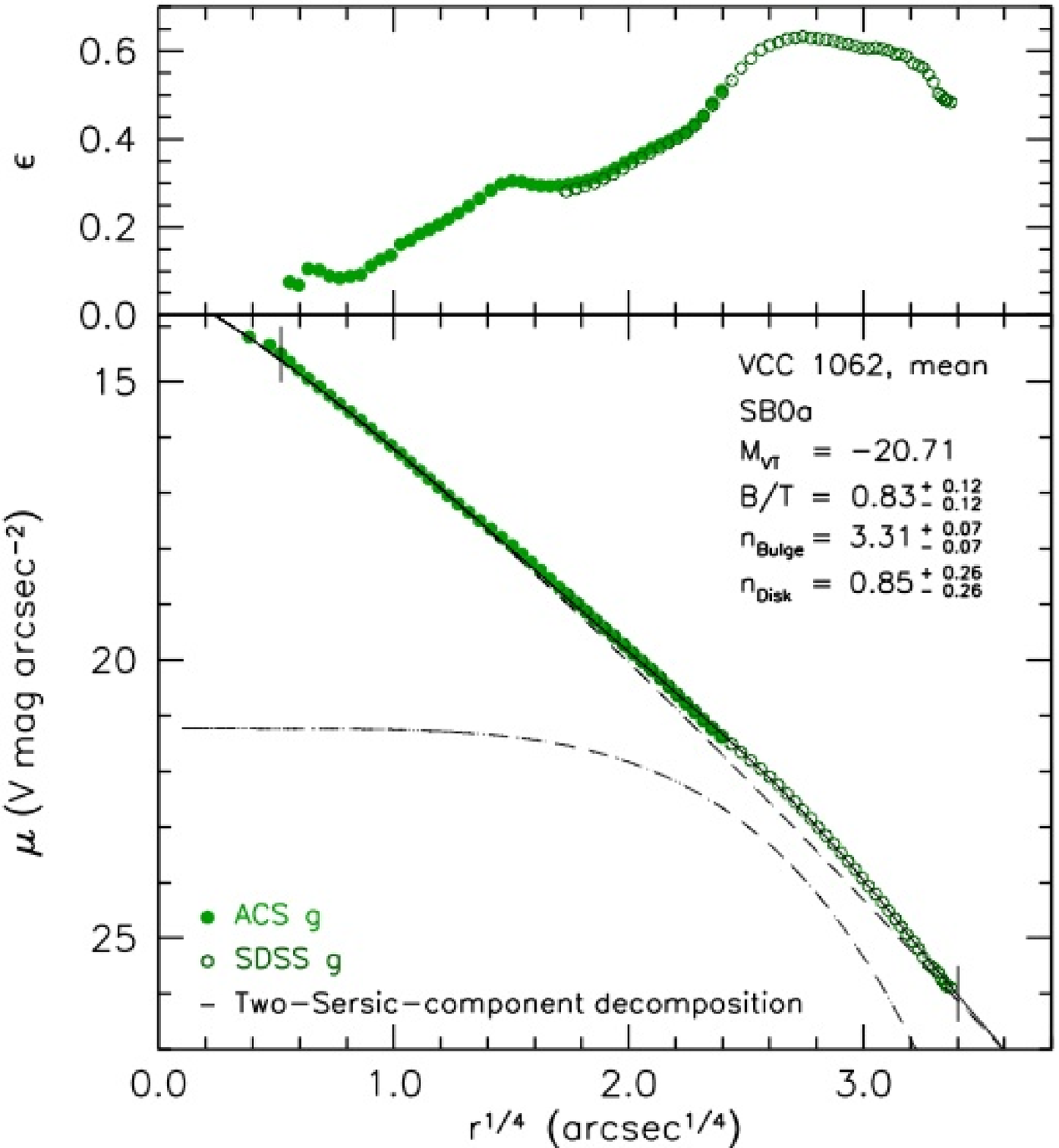}

Fig.~35~--  Mean-axis profiles of NGC\ts4442 = VCC~1062 ({\it bottom}) and 
NGC\ts4483 = VCC~1303 ({\it top}). The dashed curves show the decompositions; the component
sums ({\it solid curves}) fits the data with RMS = 0.033 mag arcsec$^{-2}$ for NGC\ts4442 and 
RMS = 0.048 mag arcsec$^{-2}$ for NGC\ts4483.  These decompositions can be compared with the 
major-axis decompositions of these galaxies in Fig.~34 and Fig.~36. The bulge-to-total ratios are robust.

\eject

\vss
\cl {A.8.~\it NGC 4483 = VCC 1303}
\vss

NGC\ts4483 is a barred S0 galaxy in which the bulge and disk contribute equal amounts of
light. There is evidence for a faint nucleus (F2006). The decompositions along the major and 
mean axes yield consistent results, both for the bulge and for the disk (cf.~Fig.~36 and
Fig.~35). The bulge is classical with an almost-de-Vaucouleurs-law profile.  The disk is
exponential.  The bulge-to-total ratio implies a type of SB0ab.

\cl{\null} \vskip 6.9truein

\includegraphics{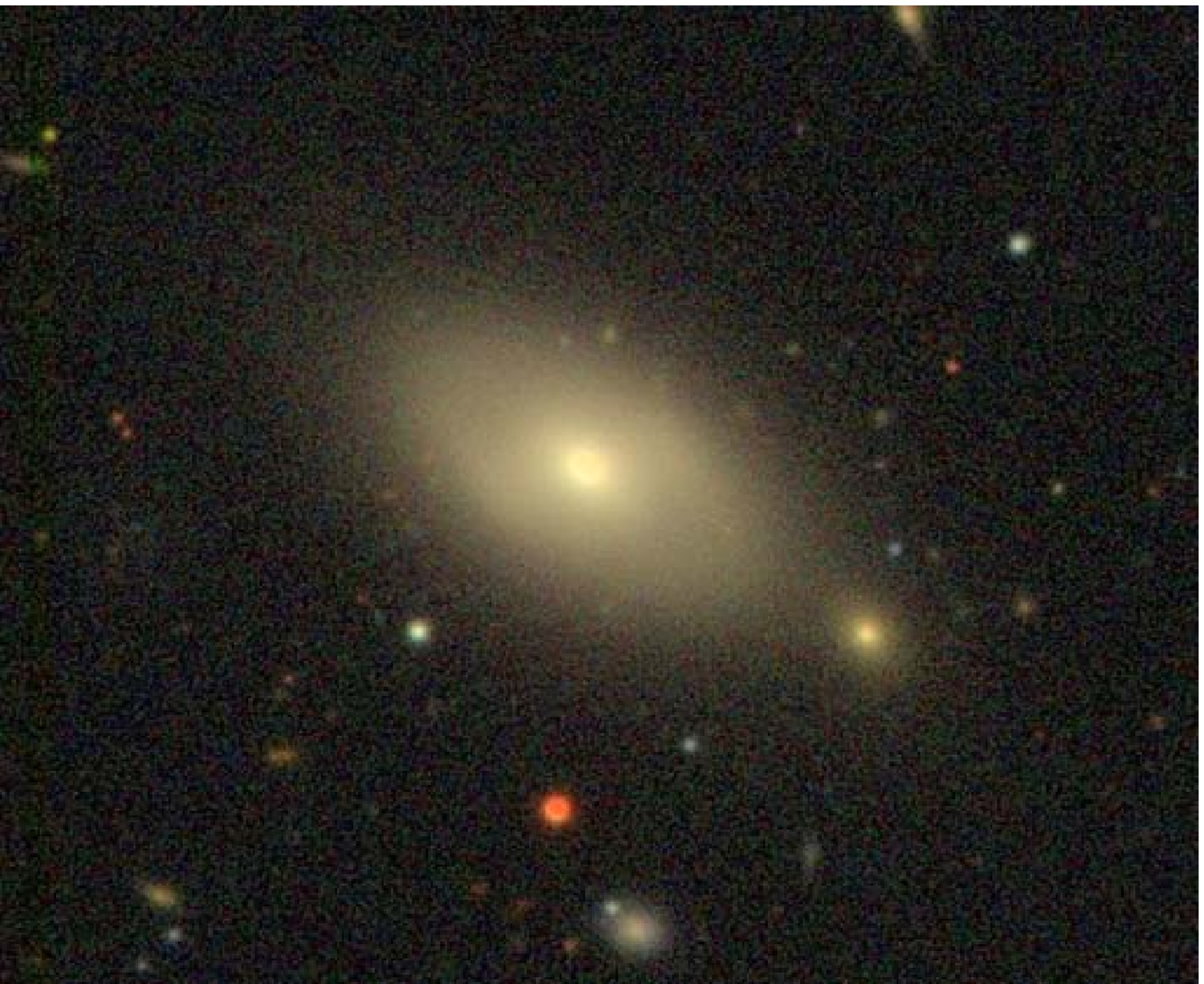}

\includegraphics{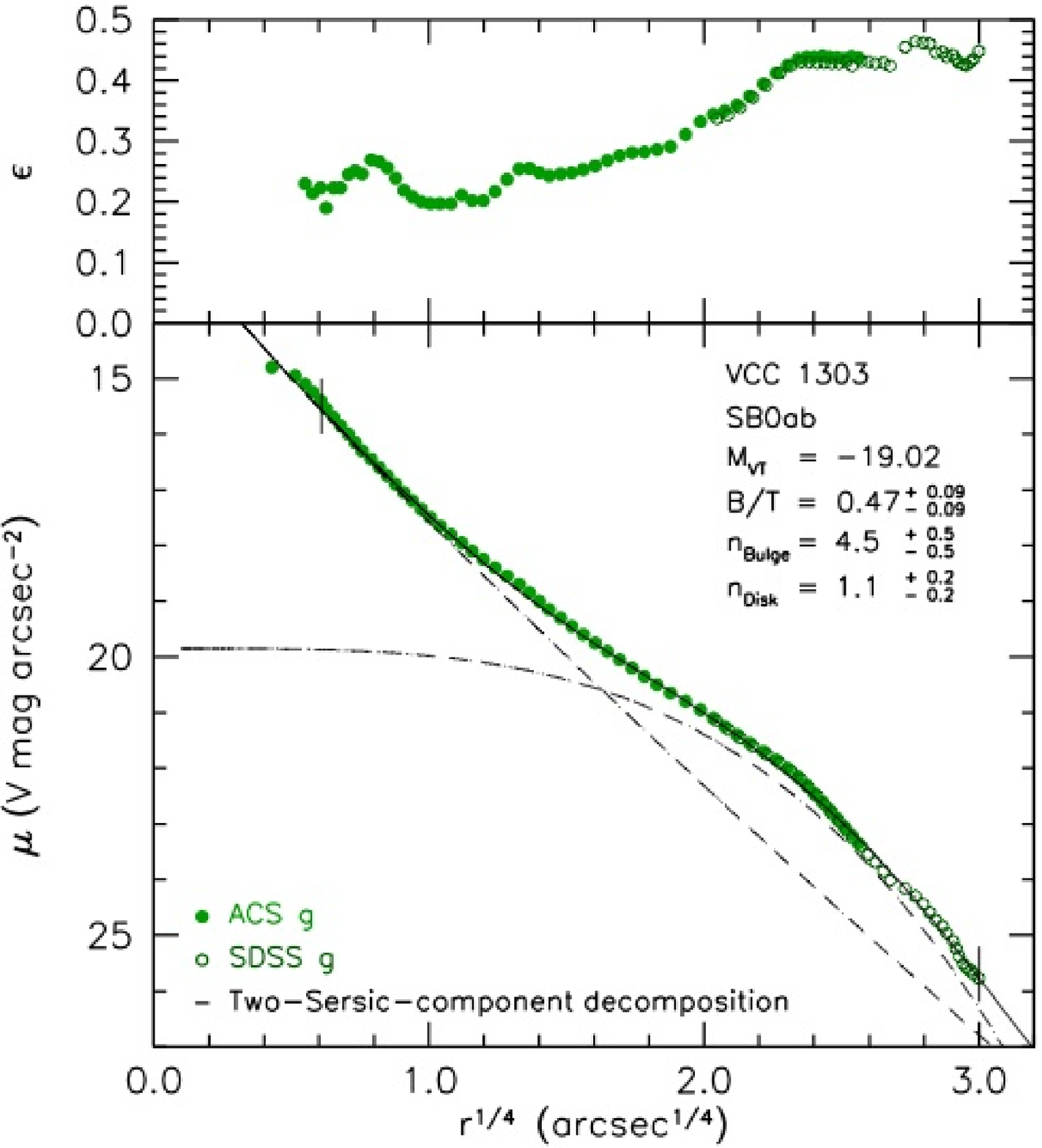}

Fig.~36~-- ({\it top\/}) Color image of NGC\ts4483 = VCC~1303 from {\tt WIKISKY}.
Again, the bar is oriented well away from the major axis and has little effect
on our decomposition.  ({\it bottom\/}) Ellipticity~$\epsilon$
and surface brightness $\mu_V$ along the major axis of NGC 4483.  The dashed curves show the
decomposition; the sum of the components ({\it solid curve}) fits the data with RMS = 0.058 mag arcsec$^{-2}$.

\vfill

\vss
\cl {A.9.~\it NGC 4550 = VCC 1619}
\vss

      NGC~4550 is unusual: it consists of two counter-rotating disks (Rubin, etal 1992; Rix \etal 1992). 
They are coplanar and largely overlap in radius. In combination, they yield a brightness profile 
with a S\'ersic index that is abnormally large for a single disk, 
i.{\ts}e., $n = 1.69^{+0.13}_{-0.08}$.  The ACS image shows dust near the center, but the absorption 
is not strong enough to explain the large $n$ . The disk surface brightness is high.
We classify the tiny central component as a pseudobulge, based on its high
flattening.  We classify NGC 4550 as an unbarred, peculiar SA0c.

\cl{\null} \vskip 6.7truein

\includegraphics{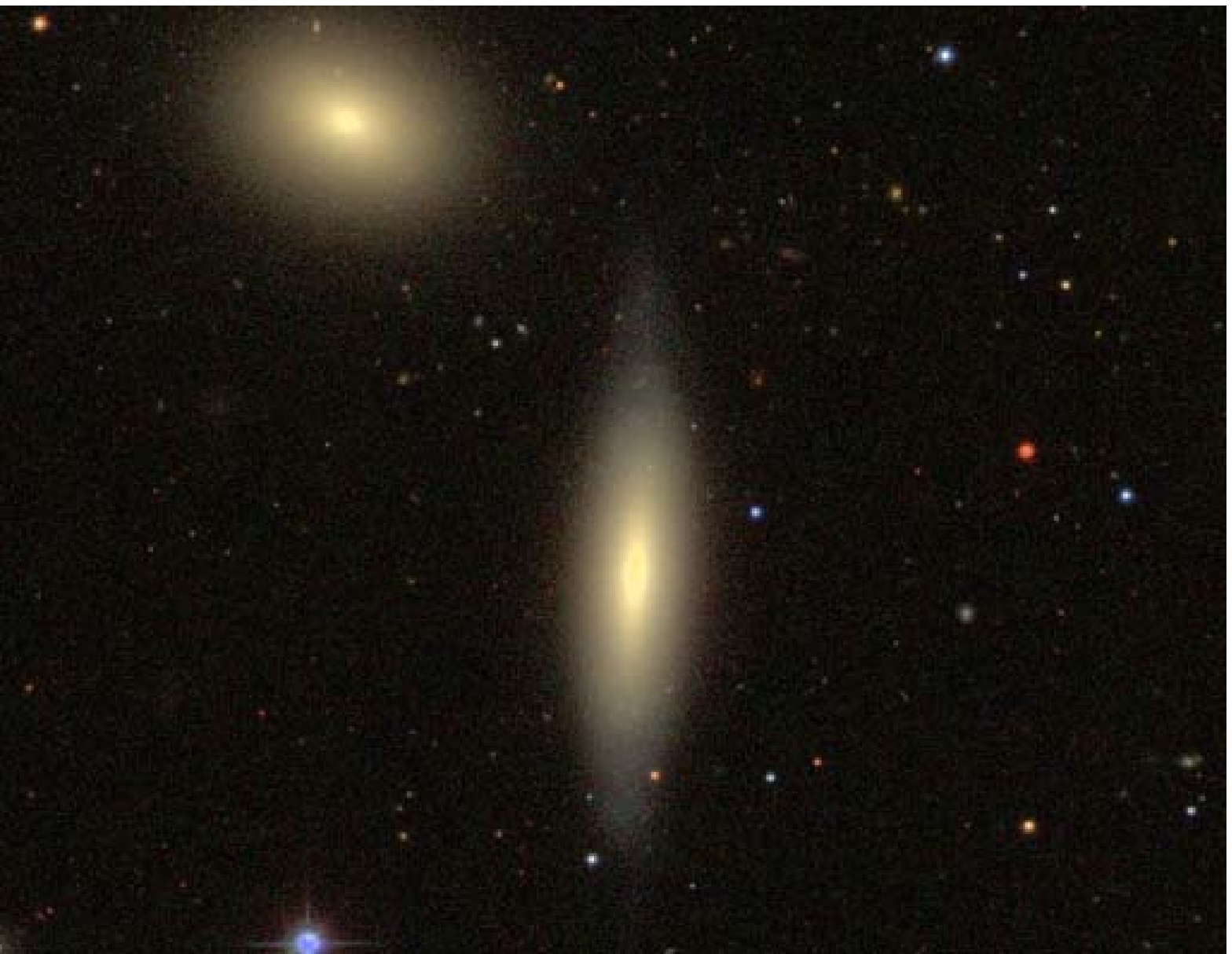}

\includegraphics{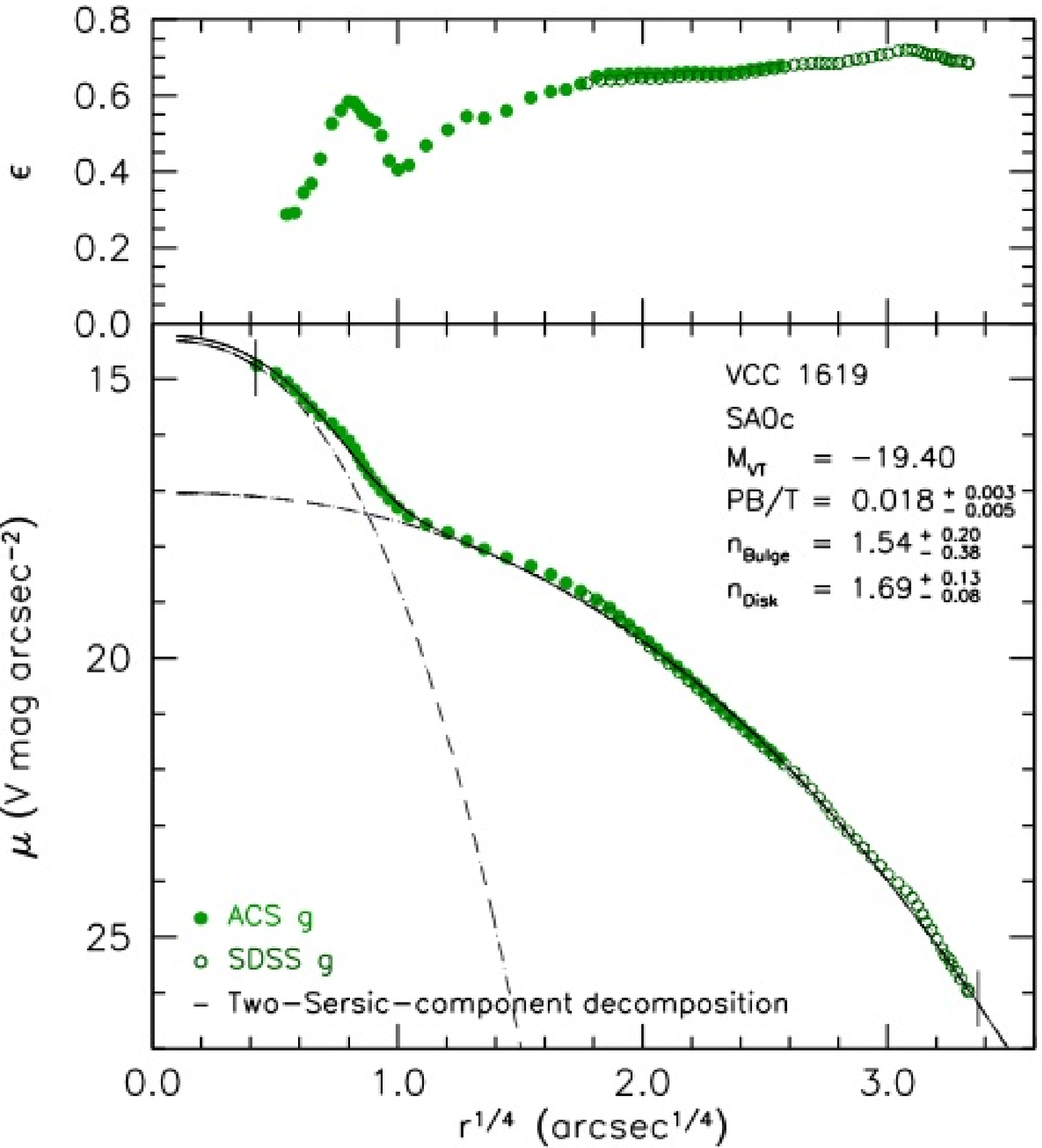}

Fig.~37~-- ({\it top\/}) Color image of NGC\ts4550 = VCC~1619 from {\tt WIKISKY}.
 ({\it bottom\/}) Ellipticity~$\epsilon$
and surface brightness $\mu_V$ along the major axis of NGC 4550.  The dashed curves show the
S\'ersic-S\'ersic, nucleus-disk decomposition inside the fit range ({\it vertical dashes\/}).  
The sum of the components ({\it solid curve}) fits the data with RMS = 0.113 mag arcsec$^{-2}$.

\vfill
\eject

\vss
\cl {A.10.~\it NGC 4578 = VCC 1720}
\vss

NGC 4578 is an almost-face-on, unbarred S0 galaxy with easily separable bulge, disk and nucleus
components.  The bulge is classical and contains $\sim$ 56\ts\% of the galaxy light.  We
classify NGC 4578 as SA0ab.  The disk is clearly inconsistent with an exponential profile; 
it requires a steeper cut-off; i.{\ts}e., $n = 0.56 \pm 0.11$, consistent with a Gaussian.
For the Virgo cluster environment, the galaxy is relatively isolated (Figure 40).

\cl{\null} \vskip 6.6truein

\includegraphics{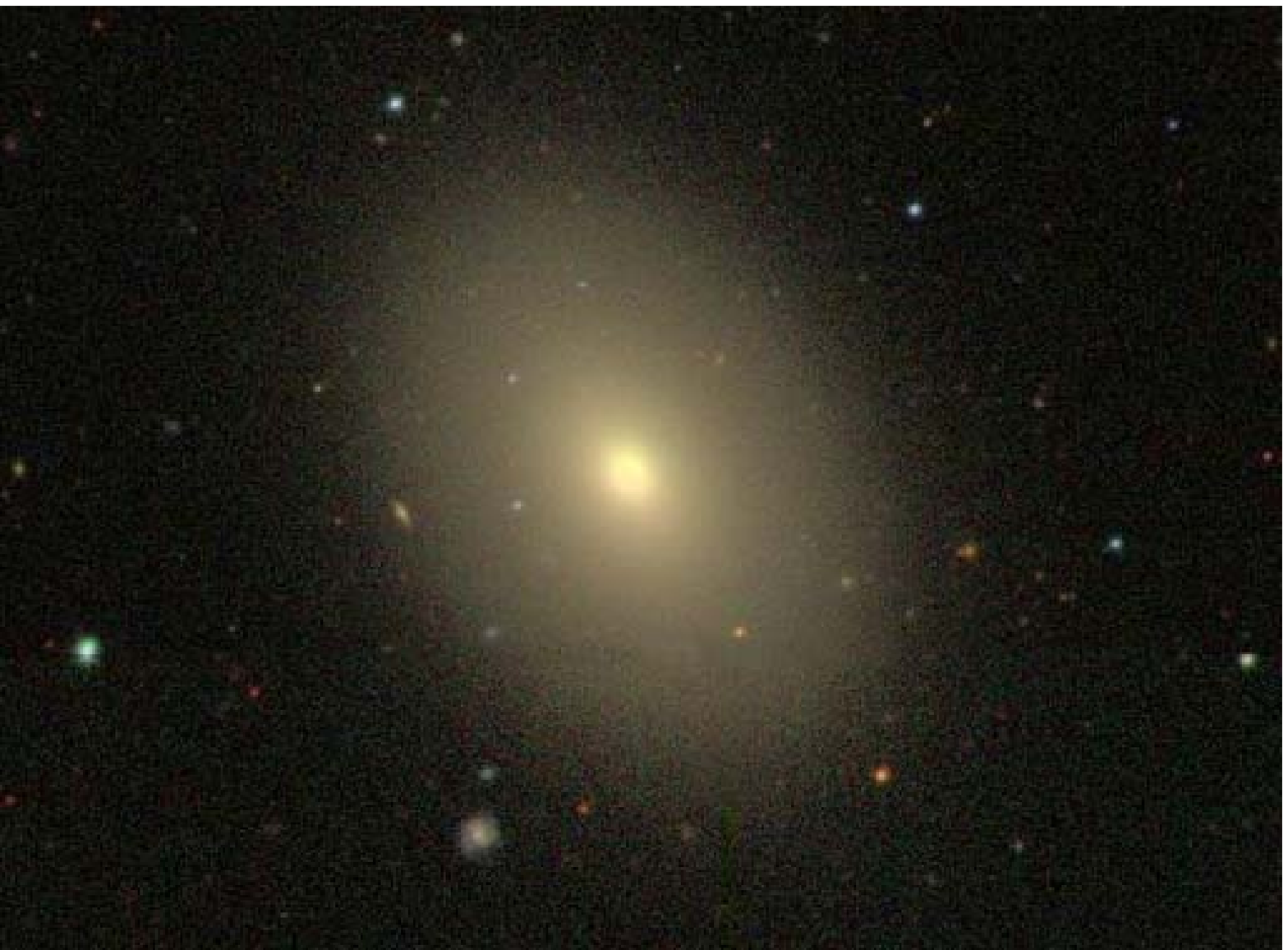}

 \includegraphics{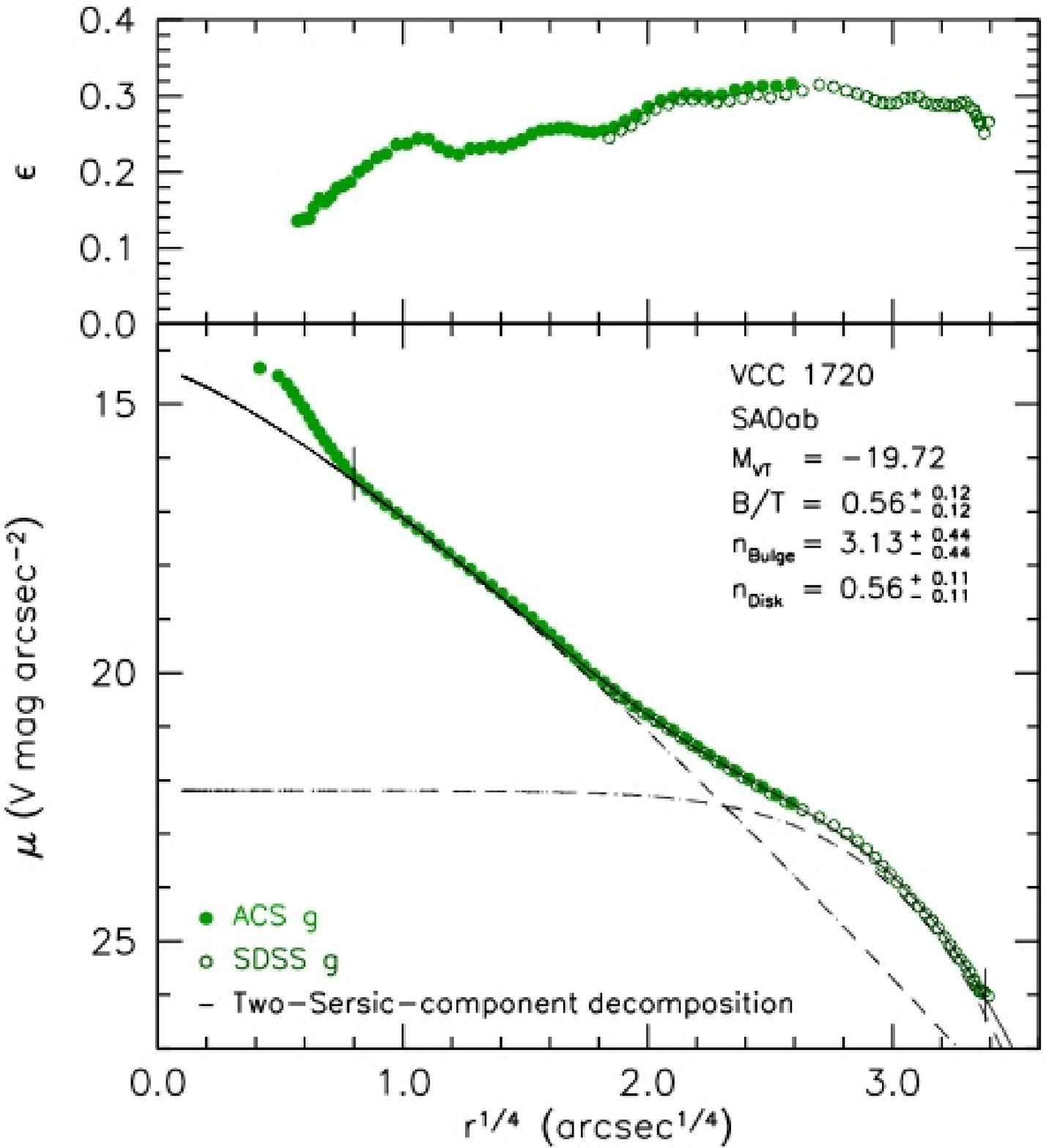}

Fig.~38~-- ({\it top\/}) Color image of NGC\ts4578 = VCC~1720 from {\tt WIKISKY}.  The galaxy 
is unbarred and an easy case for bulge-disk decomposition.  ({\it bottom\/}) Ellipticity~$\epsilon$
and surface brightness $\mu_V$ along the major axis of NGC 4578.  The dashed curves show the
bulge-disk decomposition; the sum of the components ({\it solid curve}) fits the data with 
RMS = 0.038 mag arcsec$^{-2}$.

\vskip 0.5truein

\vss
\cl {A.11.~\it NGC 4623 = VCC 1913}
\vss

NGC 4623 is a disk-dominated S0 with a small bulge.  The ellipticity profile reveals a nuclear
disk, i.{\ts}e., a pseudobulge contribution to bulge that also has a classical component (overall
$n = 3.3 \pm 0.8$).  The disky part covers almost the whole radius range of the bulge, so we
classify it as pseudo.  Its parameters are relatively uncertain, because it dominates over only a 
small brightness range.  The disk profile is consistent with exponential.  From $PB/T$, we
classify the galaxy as SA0b.

\vfill

\cl{\null} 

\includegraphics{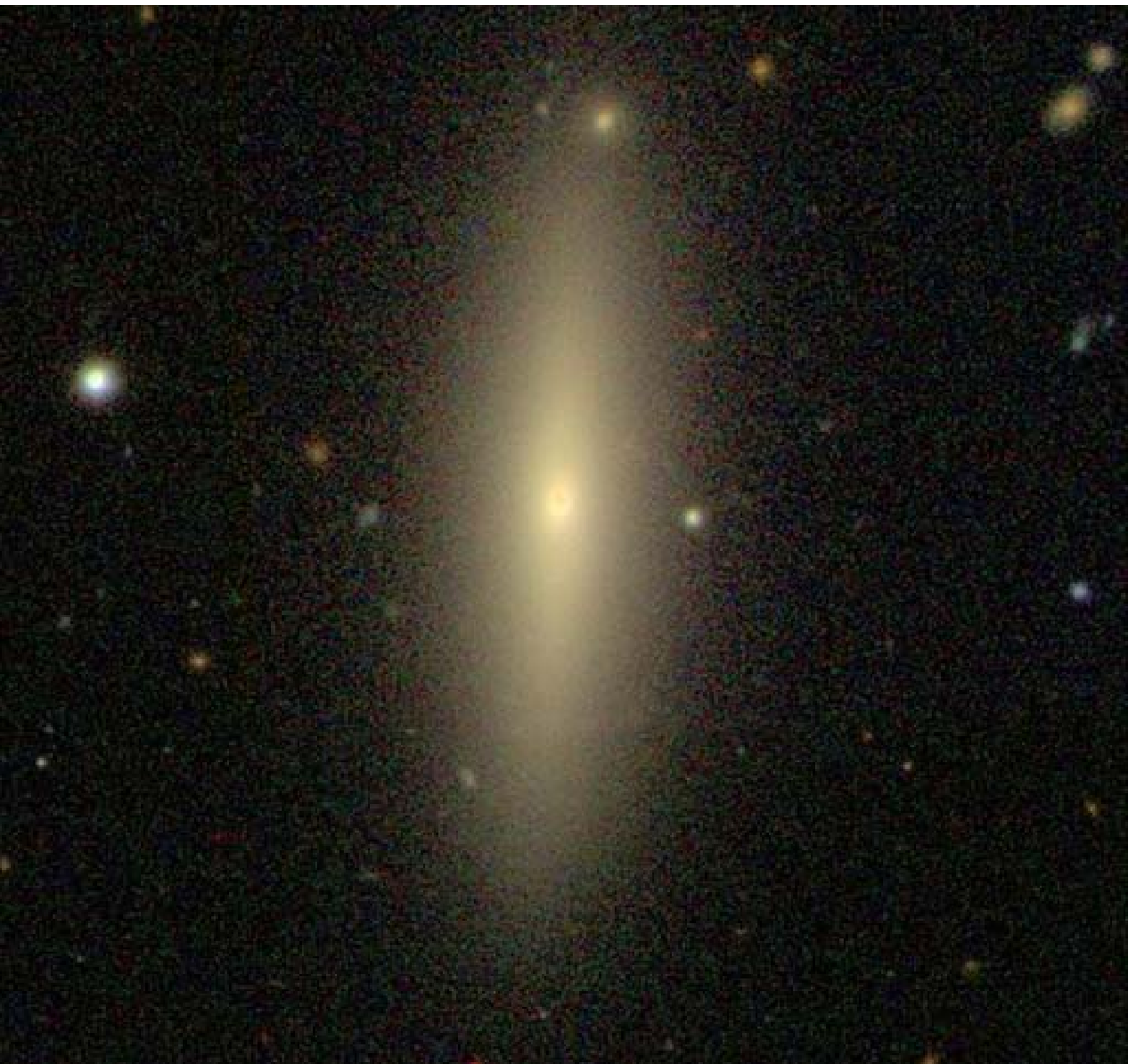}

 \includegraphics{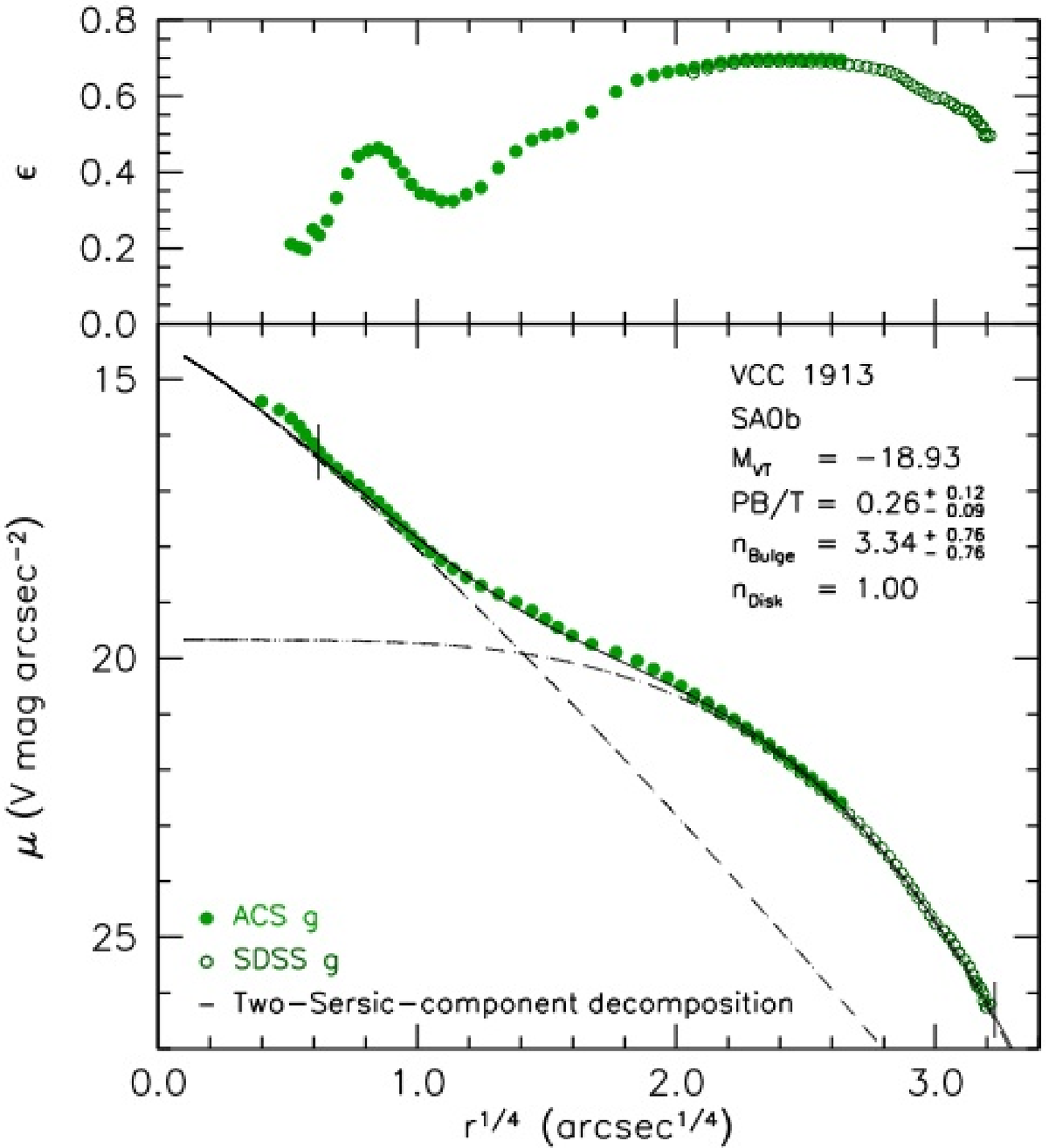}

Fig.~39~-- ({\rm top\/}) Color image of NGC\ts4623 = VCC~1913 from {\tt WIKISKY}.  ({\it
 bottom\/}) Ellipticity~$\epsilon$ and surface brightness $\mu_V$ along the major axis of NGC
 4623.  The dashed curves show the decomposition; the sum of the components ({\it solid
 curve}) fits the data with RMS = 0.056 mag arcsec$^{-2}$.

\eject

\singlecolumn

\cl{\null} \vskip -24pt

\cl {A.12.~\it Are the Disks of Many Virgo S0s Heated at Large Radii?}
\vss

      In \S\ts A.4 on NGC 4352 and \S\ts A.6 on NGC 4417, we noted a caveat with the assumption
that a bulge dominates at both small and large radii when the outer ellipticity returns from high
values back down to values like those in the inner bulge.  Such an ellipticity profile is observed
in NGC 4638 (Figure 15), and there -- because the galaxy is edge-on -- we can tell that the outer
isophotes belong to a Sph-like halo and not to the outward extrapolation of the bulge profile.
We suggested that the halo was produced by dynamical heating of the outer disk.  We used NGC 4638
as a ``proof-of-concept'' galaxy for the idea that Sph galaxies can be produced by the harassment
of disks.

      Similarly, the rounder outer isophotes of NGC 4352 and NGC 4417 may not be a sign that the 
bulge dominates there.  If we assign this outer light to the disk, then revised decompositions give 
$B/T$ values of $\sim 0.25$ to 0.35 for both galaxies.  Then we would classify them as S0ab\ts--{\ts}S0b.
Our observations cannot test this possibility; we leave this for future work.  Meanwhile, we adopt the 
conservative interpretation in \S\S{\ts}A.4 and A.6.  However, in this section, we note some hints 
that outer disk heating may indeed be more important than we suggested in the main text of this paper.

      1 -- NGC 4638 is a powerful hint, as described above.  It is located almost between two giant
ellipticals in the Virgo cluster (Figure 40), so it is not implausible that NGC 4638 has been
dynamically harassed more than most galaxies.  

      2 -- Gaussian disks are surprisingly common in our analysis.  It is no surprise to see outer 
cutoffs in the disk profiles of barred galaxies; they are characteristic of outer rings and of 
oval disks (Kormendy 1979b, 1982; Kormendy \& Kennicutt 2004;
Buta 1995, 2011; Buta \etal 2007).  Thus the Gaussian components (lenses, outer disks, or both) in 
NGC 4762 \phantom{00000000000000}

\vfill

\includegraphics{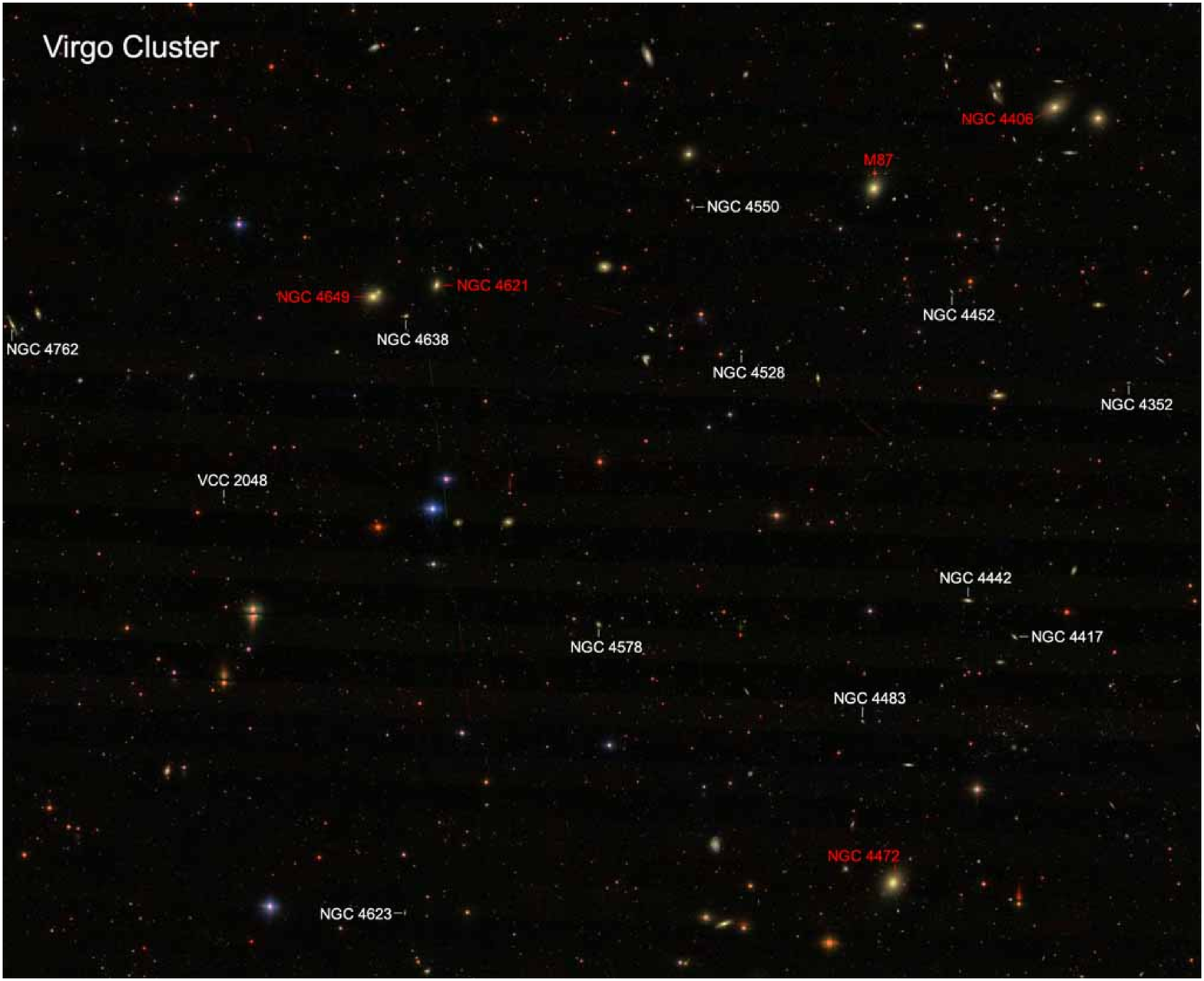}

Fig.~40~-- Central parts of the Virgo cluster showing the positions of the S0 galaxies 
measured in this paper.  Some of the brightest ellipticals in the cluster are labeled in red.  
The S0s are located near the cluster center, mostly in the region where Figure 24 shows ongoing
ram-pressure stripping.  NGC 4638, the remarkable S0 with a Sph halo (\S\ts4.1.4), is located
almost between NGC 4649 and NGC 4621.  This image is from {\tt www.wikisky.org}.

\eject

\doublecolumns

\noindent(\S\ts4.1.1), 
NGC 4452 (\S\ts4.1.2),
NGC 4528 (\S{\ts}A.5), and
NGC 4442 (\S{\ts}A.7)
are no surprise.  But it is surprising that the unbarred galaxies
NGC 4352 (\S{\ts}A.4),
NGC 4417 (\S{\ts}A.6), and
NGC 4578 (\S{\ts}A.10) have Gaussian disks, not exponential disks (Freeman 1970).  However, a disk 
with a gradual outer cutoff may be the result~if~the~disk is heated at large $r$ and 
therefore{\ts}--{\ts}\hbox{especially in edge-on} galaxies such as NGC 4352 and NGC 4417\ts--{\ts}much 
reduced in surface brightness.

      3 -- NGC 4442 and NGC 4483 have no visible outer disks.  Each galaxy has a weak bar that 
fills a strong lens in one dimension.  This is normal.  But we see no sign of an outer
disk around the B(lens) structure.  We have not seen this behavior before; one additional example
(NGC 4340, also in Virgo) is noted in the de Vaucouleurs Atlas (Buta \etal 2007).  The 
observational remark is sufficient, but 
we note that we know no way to form a bar that completely fills its disk.  This
observed situation could result if the outer disk has been heated enough so that we do not detect it.

      4 -- Exactly the sort of disk heating that we propose is in progress in NGC 4762.
The B(lens) structure is thin and flat.  But the outer disk is thicker, warped, and distorted
symmetrically into an $\int$ structure that is a signature of tidal responses.  The culprit
is presumably NGC 4754, a similar-luminosity SB0 located less than two diameters away along the 
minor axis of NGC 4762.  The warped outer disk of NGC 4762 is expected to phase-wrap in azimuth
at different rates at different radii.  Eventually, its outer disk will just look thick.
Many impulsive encounters such as the present one or the long-term result of more gradual dynamical
harassment could plausible produce an S0$+$Sph galaxy such as NGC 4638.  The same observations
and conclusions apply to NGC 4452; its minor-axis companion is the large Sph galaxy 
IC 3381 = VCC 1087. 

      These are hints that outer disk heating may be more important than the main text suggests.  
Figure 40 shows that the S0 galaxies discussed above lie in the central, ``busy'' parts of the 
Virgo cluster, where galaxy harassment is most likely.  The main conclusions of this paper would 
be strengthened by this result but do not depend on it.  In Table\ts1, we make the more conservative 
assumption that NGC 4352 and NGC 4417 are bulge-dominated.  

      We are conscious of the apparent contradiction in claiming that tidal truncation 
of small ellipticals~is~minor whereas dynamical heating of S0 disks may be important.  But:
(1) the above evidence favors disk heating, whereas observations discussed in \S\ts3.2.2 
disfavor serious tidal truncation of small Es.~Also, (2) the smallest ellipticals are a
factor of $\sim$ 10 smaller than the smallest S0 disks.  Tidal forces are more effective in
harassing larger objects.  Most important: (3) {\it There is a natural radius in S0 disks beyond
which it is relatively easy to tilt orbits vertically and hence to thicken the disk after
radius-dependent azimuthal phase wrapping.  This is the radius outside which the disk is 
non-self-gravitating within the dark halo.  At smaller radii, the structure is relatively
stiff because of disk restoring forces.  But where the dark matter dominates the potential,
there are virtually no disk restoring forces.}  The result -- we suggested -- can be seen in
NGC 4762: the bright B(lens) structure is flat and thin, whereas the much fainter outer disk is 
strongly warped.  Small ellipticals are already almost round; they have no ``handle'' at large radii, 
and scrambling their orbits further has little effect.

\vss\vsss
\cl{\vsc REFERENCES}
\vss

\frenchspacing

{\vsc\smlbaselines

\nnhi Abadi, M.~G., Moore, B., \& Bower, R.~G.~1999, MNRAS, 308, 947

\nnhi Abramson, A., Kenney, J.~D.~P., Crowl, H.~H., Chung, A., van Gorkom, J.~H., Vollmer, B., \& Schiminovich, D.~2011,
      AJ, 141, 164

\nnhi Adami, C., \etal 2005, A\&A, 429, 39  

\nnhi Arnaboldi, M.~2011, paper presented at the ESO Workshop on Fornax, Virgo, Coma et al.: 
      Stellar Systems in High Density Environments, 
      {\eightpoint\tt /sci/meetings/2011/fornax\_virgo2011/talks\_pdf/}
      {\eightpoint\tt Arnaboldi\_Magda.pdf}

\nnhi Arnaboldi, M., Gerhard, O.~2009, Highlights of Astronomy, 15, 97

\nnhi Arnaboldi, M., Gerhard, O., Aguerri, J.~A.~L., Freeman, K.~C., Napolitano, N.~R., 
      Okamura, S., \& Yasuda, N.~2004, ApJ, 614, L33

\nnhi Arnaboldi, M., \etal 1996, ApJ, 472, 145  

\nnhi Arnaboldi, M., \etal 2002, AJ, 123, 760  

\nnhi Baggett. W.~E., Baggett, S.~M., \& Anderson, K.~S.~J.~1998, AJ, 116, 1626

\nhi Bahcall, J.~N., \& Casertano, S.~1985, ApJ, 293, L7

\nnhi Baldry, I.~K., Glazebrook, K., Brinkmann, J., Ivezi\'c, \v{Z}., Lupton, R.~H.,
             Nichol, R.~C., \& Szalay, A.~S.~2004, ApJ, 600, 681

\nnhi Balick, N., Faber, S.~M., \& Gallagher, J.~S.~1976, ApJ, 209, 710

\nnhi Banse, K., Ponz, D., Ounnas, C., Grosb\o l, P., \& Warmels, R.~1988,
             in Instrumentation for Ground-Based Optical Astronomy: Present 
             and Future, ed.~L.~B.~Robinson (New York: Springer), 431 

\nnhi Barazza, F.~D., Binggeli, B., \& Jerjen, H.~2002, A\&A, 391, 823  

\nnhi Bedregal, A.~G., Arag\'on-Salamanca, A., \& Merrifield, M.~R.~2006, MNRAS, 373, 1125  

\nnhi Begelman, M.~C., Blandford, R.~D., \& Rees, M.~J,~1980, Nature, 287, 307

\nnhi Bell, E.~F., \etal 2003, ApJS, 149. 289 

\nnhi Bender, R.~1987, Mitt.~Astr.~Gesellschaft, No.~70, 226

\nnhi Bender, R., Burstein, D., \& Faber, S.~M.~1992, ApJ, 399, 462

\nnhi Bender, R., Burstein, D., \& Faber, S.~M.~1993, ApJ, 411, 153

\nnhi Bender, R., D\"obereiner, S., \& M\"ollenhoff, C.~1987, A\&A, 177, L53 

\nnhi Bender, R., D\"obereiner, S., \& M\"ollenhoff, C.~1988, A\&AS, 74, 385 

\nnhi Bender, R., \& M\"ollenhoff, C.~1987, A\&A, 177, 71 

\nnhi Bender, R., \& Nieto, J.-L.~1990, A\&A, 239, 97  

\nnhi Bender, R., Paquet, A., \& Nieto, J.-L.~1991, A\&A, 246, 349

\nnhi Bender, R., Surma, P., D\"obereiner, S., M\"ollenhoff, C., \& Madejsky,
            R. 1989, A\&A, 217, 35 

\nnhi Bernardi, M., \etal 2003, AJ, 125, 1882  

\nnhi Binggeli, B.~1994, in ESO/OHP Workshop on Dwarf Galaxies, ed.~G.~Meylan \&
            P.~Prugniel (Garching: ESO), 13

\nnhi Binggeli, B., \& Cameron, L.~M.~1991, A\&A, 252, 27

\nnhi Binggeli, B., \& Popescu, C.~C.~1995, A\&A, 298, 63

\nnhi Binggeli, B., Sandage, A., \& Tammann, G.~A.~1985, AJ, 90, 1681

\nnhi Binggeli, B., Sandage, A., \& Tammann, G.~A.~1988, ARA\&A, 26, 509

\nnhi Blanton, M.~R., \etal 2003, ApJ, 594, 186 

\nnhi Blanton, M.~R., Eisenstein, D., Hogg, D.~W., Schlegel, D.~J., \&
            Brinkmann, J.~2005, ApJ, 629, 143 

\nnhi B\"ohringer, H., Briel, U.~G., Schwarz, R.~A., Voges, W., Hartner, G., \& 
            Tr\"umper, J.~1994, Nature, 368, 828  

\nnhi Boselli, A., Boiossier, S., Cortese, L., \& Gavazzi, G.~2008, ApJ, 674, 742

\nnhi Boselli, A., Boiossier, S., Cortese, L., \& Gavazzi, G.~2009, AN, 330, 904

\nnhi Bothun, G.~D., \& Thompson, I.~B.~1988, AJ, 96, 877  

\nnhi Bullock, J.~S., Kravtsov, A.~V., \& Weinberg, D.~H.~2000, ApJ, 539, 517  

\nnhi Burstein, D.~1979, ApJ, 234, 435

\nnhi Buta, R.~1995, ApJS, 96, 39  

\nnhi Buta, R.~2011, in Planets, Stars, and Stellar Systems, Volume 6, ed.~W.~C.~Keel (New York: Springer),
      in press (arXiv:1102.0550)

\nnhi Buta, R., \& Combes, F.~1996, Fund. Cosmic Phys., 17, 95

\nnhi Buta, R., \& Crocker, D.~A.~1991, AJ, 102, 1715

\nnhi Buta, R.~J., Corwin, H.~G., \& Odewahn, S.~C.~2007, The de Vaucouleurs Atlas of Galaxies
      (Cambridge: Cambridge University Press)

\nnhi Buta, R., Laurikainen, E., Salo, H., \& Knapen, J. H. 2010, ApJ, 721, 259 

\nnhi Buta, R.~J., \etal 2010, ApJS, 190, 147  

\nnhi Byun, Y.-I., \& Freeman, K. C. 1995, ApJ, 448, 563

\nnhi Cappellari, M., \etal 2007, MNRAS, 379, 418   

\nnhi Cappellari, M., \etal 2011a, MNRAS, 413, 813   

\nnhi Cappellari, M., \etal 2011b, MNRAS, in press (arXiv:1104.3545) 

\nhi Castro-Rodrigu\'ez, N., Arnaboldi, M., Aguerri, J.~A.~L., Gerhard, O., Okamura, S., 
     Yasuda, N., \& Freeman, K.~C.~2009, A\&A, 507, 621

\nnhi Cattaneo, A., Dekel, A., Devriendt, J., Guideroni, B., \& Blaizot, J.~2006,
            MNRAS, 370, 1651  

\nnhi Cattaneo, A., Mamon, G.{\ts}A., Warnick, K., \& Knebe, A.\ts2011, A\&A, 553,~5

\nnhi Cattaneo, A., et al.~2009, Nature, 460, 213  

\nnhi Cayette, V., Kotanyi, C., Balkowski, C., \& van Gorkom, J.~H.~1994, AJ, 107, 1003

\nnhi Cayette, V., van Gorkom, J.~H., Balkowski, C., \& Kotanyi, C.~1990, AJ, 100, 604

\nnhi Chen, C.-W., C\^ot\'e, P., West, A., Peng, E.~W., \& Ferrarese, L.~2010, ApJS, 191, 1

\nnhi Chiboukas, K., Karachentsev, I.~D., \& Tully, R.~B.~2009, AJ, 137, 3009

\nnhi Chung, A., van Gorkom, J.~H., Kenney, J.~D.~P., \& Vollmer, B.~2007, ApJ, 659, L115

\nnhi Chung, A., van Gorkom, J.~H., Kenney, J.~D.~P., Crowl, H., \& Vollmer, B.~2009, AJ, 138, 1741  

\nnhi Combes, F., Dupraz, C., Casoli, F., \& Pagani, L.~1988, A\&A, 203, L9  

\nnhi C\^ot\'e, P., et al.~2007, ApJ, 671, 1456 

\nnhi Courteau, S., Dutton, A.~A., van den Bosch, F.~C., MacArthur, L.~A., Dekel, A., McIntosh, D.~H., 
      \& Dale, D.~A.~2007, ApJ, 671, 203  

\nnhi Da Costa, G.~S.~1994, in ESO/OHP Workshop on Dwarf Galaxies, ed.~G.~Meylan
            \& P.~Prugniel (Garching: ESO), 221

\nnhi Dav\'e, R., \etal 2001, ApJ, 552, 473

\nnhi David, L.~P., Jones, C., \& Forman, W.~1995, ApJ, 445, 578  

\nnhi Davis, T.~A., \etal 2009, MNRAS, 313, 968  

\nnhi Davies, R.~L., Efstathiou, G., Fall, S.~M., Illingworth, G., \& Schechter, P.~L.~1983,
             ApJ, 266, 41

\nnhi Dekel, A., \& Birnboim, Y.~2006, MNRAS, 368, 39  

\nnhi Dekel, A., \& Birnboim, Y.~2008, MNRAS, 383, 119

\nnhi Dekel, A., \& Silk, J.~1986, ApJ, 303, 39

\nnhi De Rijcke, S., Dejonghe, H., Zeiliinger, W.~W., \& Hau, G.~K.~T.~2001, ApJ, 559, L21   

\nnhi De Rijcke, S., Dejonghe, H., Zeiliinger, W.~W., \& Hau, G.~K.~T.~2003, A\&A, 400, 119  

\nnhi De Rijcke, S., Dejonghe, H., Zeiliinger, W.~W., \& Hau, G.~K.~T.~2004, A\&A, 426, 53   

\nnhi De Rijcke, S., Prugniel, P., Simien, F., \& Dejonghe, H.~2006, MNRAS, 39, 1321  

\nnhi de Souza, R. E., Gadotti, D. A., \& dos Anjos, S. 2004, ApJS, 153, 411

\nnhi de Vaucouleurs, G.~1959, Handbuch der Physik, 53, 275

\nnhi Devereux, N.~A., \& Young, J.~S.~1991, ApJ, 371, 515

\nnhi de Zeeuw, P.~T., \etal 2002, MNRAS, 329, 513  

\nnhi Djorgovski, S., \& Davis, M.~1987, ApJ, 313, 59

\nnhi Djorgovski, S., de Carvalho, R., \& Han, M.-S.~1988, in The Extragalactic 
             Distance Scale, ed.~S.~van den Bergh \& C.~J.~Pritchet (San Francisco: ASP), 329

\nnhi Dolphin, A.~E.~2002, MNRAS, 332, 91  

\nnhi D'Onghia, E., Besla, G., Cox, T.~J., \& Hernquist, L.~2009, Nature, 460, 605

\nnhi Dressler, A.~1979. ApJ, 231, 659    

\nnhi Dressler, A.~1980. ApJ, 236, 351

\nnhi Dressler, A., Oemler, A., Couch, W.~J., Smail, I., Ellis, R.~S., Barger, A.,
      Butcher, H., Poggianti, B., \& Sharples, R.~M.~1997, ApJ, 490, 577

\nnhi Ebisuzaki, T., Makino, J., Okamura, S.~K.~1991, Nature, 354, 212

\nnhi Einasto, J., Saar, E., Kaasik, A., \& Chernin, A.~D.~1974, Nature, 252, 111

\nnhi Emsellem, E., Dejonghe, H., \& Bacon, R.~1999, MNRAS, 303, 495  

\nnhi Emsellem, E., \etal 2004, MNRAS, 352, 721  

\nnhi Emsellem, E., \etal 2007, MNRAS, 379, 401  

\nnhi Emsellem, E., \etal 2011, MNRAS, 414, 888  

\nnhi Faber, S.~M.~1973, ApJ, 179, 423

\nnhi Faber, S.~M., \& Lin, D.~N.~C.~1983, ApJ, 266, L17

\nnhi Faber, S.~M., \etal 1987, in Nearly Normal Galaxies: From the Planck
                                 Time to the Present, ed. S.~M.~Faber (New York: Springer), 175 

\nnhi Faber, S.~M., \etal 1997, AJ, 114, 1771

\nnhi Faber, S.~M., \etal 2007, ApJ, 665, 265  

\nnhi Faber, S.~M., \etal 2011, in preparation

\nnhi Farouki, R., \& Shapiro, S.~L.~1980, ApJ, 241, 928

\nnhi Ferguson, A.~M.~N., Irwin, M.~J., Ibata, R.~A., Lewis, G.~F., \& Tanvir, N.~R.~2002, AJ, 124, 1452

\nnhi Ferguson, H.~C., \& Sandage, A.~1989, ApJ, 346, L53

\nnhi Ferrarese, L., \etal 2006, ApJS, 164, 334 (F2006)

\nnhi Fisher, D.~B., \& Drory, N.~2008, AJ, 136, 773

\nnhi Frebel, A.~2010, AN, 331, 474

\nnhi Freeland, E., \& Wilcots, E.~2011, ApJ, 738, 145

\nnhi Freeman, K.~C.~1970, ApJ, 160, 811

\nnhi Gavazzi, G., Donati, A., Cucciati, O., Sabatini, S., Boselli, A., Davies, J., 
             \& Zibetti, S.~2005, A\&A, 430, 411

\nnhi Gavazzi, G., Franzetti, P., Scodeggio, M., Boselli, A., \& Pierini, D.~2000, A\&A, 361, 863

\nnhi Geha, M., Guhathakurta, P., Rich, R.~M., \& Cooper, M.~C.~2006, AJ, 131, 332 

\nnhi Geha, M., Guhathakurta, P., \& van der Marel, R.~P.~2003, AJ, 126, 1794

\nnhi Geisler, D., Smith, V.~V., Wallerstein, G., Gonzalez, G., \& Charbonnel, C.~2005, AJ, 129, 1428

\nnhi Gerola, H., Carnevali, P., \& Salpeter, E.~E.~1983, ApJ, 268, L75

\nnhi Gerola, H., Seiden, P.~E., \& Schulman, L.~S.~1980, ApJ, 242, 517

\nnhi Glass, L., \etal 2011, ApJ, 726, 31  

\nnhi Gonzalez, A.~H., Zaritsky, D., \& Zabludoff, A.~I.~2007, ApJ, 666, 147 

\nnhi Gonz\'alez-Garc\'\i a, A.~C., Aguerri, J.~A.~L., \& Balcells, M.~2005, A\&A, 444, 803

\nnhi Graham, A.~W., Erwin, P., Trujillo, I., \& Asensio Ramos, A.~2003, AJ, 125, 2951  

\nnhi Graham, A.~W., \& Guzm\'an, R.~2003, AJ, 125, 2936

\nnhi Graham, A.~W., \& Guzm\'an, R.~2004, in Penetrating Bars Through Masks of Cosmic Dust: The Hubble 
              Tuning Fork Strikes a New Note, ed.~D.~L.~Block et al.~(Dordrecht: Kluwer), 723

\nnhi Grebel, E.~K., Gallagher, J.~S., \& Harbeck, D.~2003, AJ, 125, 1926

\nnhi Gunn, J.~E., \& Gott, J.~R.~1972, ApJ, 176, 1

\nnhi Gursky, H., Kellogg, E., Murray, S., Leong, C., Tananbaum, H., \& Giacconi, R.~1971, ApJ, 167, L81

\nnhi Hamabe, M.~1982, PASJ, 34, 423

\nnhi Hernandez, X., Gilmore, G., \& Valls-Gabaud, D.~2000, MNRAS, 317, 831  

\nnhi Hinz, J.~L., Rieke, G.~H., \& Caldwell, N.~2003, AJ, 126, 2622  

\nnhi Hogg, D.~W., \etal 2002, AJ, 124, 646  

\nnhi Hogg, D.~W., \etal 2004, ApJ, 601, L29 

\nnhi Hopkins, P.~F., Bundy, K., Hernquist, L., Wuyts, S., \& Cox, T.~J.~~2010, MNRAS, 401, 1099 

\nnhi Hopkins, P.~F., Cox, T.~J., Dutta, S.~N., Hernquist, L., Kormendy, J., \& Lauer, T.~R.~2009a,
             ApJS, 181, 135 (astro-ph0805.3533)  

\nnhi Hopkins, P.~F., Cox, T.~J., \& Hernquist, L.~2008, ApJ, 689, 17  

\nnhi Hopkins, P.~F., Hernquist, L., Cox, T.~J., Keres, D., \& Wuyts, S.~2009b, ApJ, 691, 1424  

\nnhi Hubble, E.~1936, The Realm of the Nebulae (New Haven: Yale University Press)

\nnhi Hurley-Keller, D., Mateo, M., \& Nemec, J.~1998, AJ, 115, 1840

\nnhi Ibata, R.~A., Gilmore, G., \& Irwin, M.~J.~1994, Nature, 370, 194  

\nnhi Ibata, R.~A., Gilmore, G., \& Irwin, M.~J.~1995, MNRAS, 277, 781   

\nnhi Ibata, R., Irwin, M., Lewis, G., Ferguson, A., \& Tanvir, N.~2001a, Nature, 412, 49

\nnhi Ibata, R., Lewis, G.~F., Irwin, M., Totten, E., \& Quinn, T.~2001b, ApJ, 551, 294  

\nnhi Ibata, R., Martin, N.~F., Irwin, M., Chapman, S., Ferguson, A.~M.~N., Lewis, G.~F.,
      \& McConnachie, A.~W.~2007, ApJ, 671, 1591  

\nnhi Ichikawa, S.-I.~1989, AJ, 97, 1600

\nnhi Irwin, M.~J., \etal 2007, ApJ, 656, L13  

\nnhi Jerjen, H., \& Binggeli, B.~1997, in The Second Stromlo Symposium:
             The Nature of Elliptical Galaxies, ed.~M. Arnaboldi \etal (San Francisco:
             ASP), 239

\nnhi Jerjen, H., Kalnajs, A., \& Binggeli, B.~2000, A\&A, 358, 845  

\nnhi Jerjen, H., Kalnajs, A., \& Binggeli, B.~2001, in ASP Conf. Ser. 230, Galaxy Disks and 
      Disk Galaxies, ed.~J. G. Funes \& E. M. Corsini (San Francisco: ASP), 239

\nnhi Jorgensen, I., Franx, M., \& Kj\ae rgaard, P.~1996, MNRAS, 280, 167

\nnhi Karachentsev, I., Aparicio, A., \& Makarova, L.~1999, A\&A, 352, 363  

\nnhi Kauffmann, G., Heckman, T.~M., \& Best, P.~N.~2008, MNRAS, 384, 953

\nnhi Kazantzides, S., \L okas, E.~L., Callegari, S., Mayer, L., \& Moustakas, L.~A.~2011, ApJ, 726, 98

\nnhi Kelson, D.~D., Zabludoff, A.~I., Williams, K.~A., Trager, S.~C., Mulchaey, J.~S., \&
             Bolte, M.~2002, ApJ, 576, 720

\nnhi Kenney, J.~D., \& Young, J.~S.~1986, ApJ, 303, L13

\nnhi Kenney, J.~D.~P., \& Koopmann, R.~A.~1999, AJ, 117, 181  

\nnhi Kenney, J.~D.~P., Tal, T., Crowl, H.~H., Feldmeier, J., \& Jacoby, G.~H.~2008, ApJ, 687, L69

\nnhi Kenney, J.~D.~P., van Gorkom, J.~H., \& Vollmer, B.~2004, AJ, 127, 3361  

\nnhi Kent, S.~M.~1985, ApJS, 59, 115

\nnhi King, I.~R.~1966, AJ, 71, 64

\nnhi Kirby, E.~M., Jerjen, H., Ryder, S.~D., \& Driver, S.~P.~2008, AJ, 136, 1866

\nnhi Knapp, G.~R., Guhathakurta, P., Kim, D.-W., \& Jura, M.~1989, ApJS, 70, 329 

\nnhi Komatsu, E., \etal 2009, ApJS, 180, 330  

\nnhi Kormendy, J.~1977a, ApJ, 217, 406

\nnhi Kormendy, J.~1977b, ApJ, 218, 333

\nnhi Kormendy, J.~1979a, in Photometry, Kinematics and Dynamics of Galaxies, ed.~D.~S.~Evans
     (Austin: Department of Astronomy, University of Texas at Austin) 341

\nnhi Kormendy, J.~1979b, ApJ, 227, 714

\nnhi Kormendy, J.~1981, in The Structure and Evolution of Normal Galaxies, ed.~S.~M.~Fall \&
      D.~Lynden-Ball (Cambridge: Cambridge University Press), 85

\nnhi Kormendy, J.~1982, in Twelfth Advanced Course of the Swiss Society of Astronomy and Astrophysics, 
      Morphology and Dynamics of Galaxies, ed.~L.~Martinet \& M.~Mayor (Sauverny: Geneva Ob.), 113 

\nnhi Kormendy, J.~1985, ApJ, 295, 73

\nnhi Kormendy, J.~1987, in Nearly Normal Galaxies: From the Planck Time
             to the Present, ed.~S.~M.~Faber (New York: Springer), 163

\nnhi Kormendy, J.~1989, ApJ, 342, L63

\nnhi Kormendy, J.~1999, in Galaxy Dynamics: A Rutgers Symposium, ed.~D. Merritt,
            J.~A.~Sellwood, \& M.~Valluri (San Francisco: ASP), 124

\nnhi Kormendy, J.~2004, in Penetrating Bars Through Masks of Cosmic Dust: The Hubble 
              Tuning Fork Strikes a New Note, ed.~D.~L.~Block et al.~(Dordrecht: Kluwer), 816

\nnhi Kormendy, J.~2009, in ASP Conference Series, Vol.~419, Galaxy Evolution:
              Emerging Insights and Future Challenges, ed.~S.~Jogee, I.~Marinova,
              L.~Hao, \& G.~A.~Blanc (San Francisco: ASP), 87

\nnhi Kormendy, J.~2011a, paper presented at the ESO Workshop on Fornax, Virgo, Coma et al.: 
      Stellar Systems in High Density Environments, 
      {\eightpoint\tt /sci/meetings/2011/fornax\_virgo2011/talks\_pdf/}
      {\eightpoint\tt Kormendy\_John.pdf}

\nnhi Kormendy, J.~2011b, ApJS, in preparation  


\nnhi Kormendy, J., \& Bender, R.~1994, in ESO/OHP Workshop on Dwarf Galaxies, 
                 ed.~G.~Meylan \& P.~Prugniel (Garching: ESO), 161

\nnhi Kormendy, J., \& Bender, R.~1996, ApJ, 464, L119

\nnhi Kormendy, J., \& Bender, R.~2009, ApJ, 691, L142

\nnhi Kormendy, J., \& Bender, R.~2011, Nature, 469, 374

\nnhi Kormendy, J., Drory, N., Bender, R., \& Cornell, M.~E.~2010, ApJ, 723, 54

\nnhi Kormendy, J., \& Djorgovski, S.~1989, ARA\&A, 27, 235

\cl{\null}

\nnhi Kormendy, J., \& Fisher, D. B.~2008, in Formation and Evolution of Galaxy Disks,
      ed.~J.~G.~Funes \& E.~M.~Corsini (San Francisco: ASP), 297

\nnhi Kormendy, J., Fisher, D.~B., Cornell, M.~E., \& Bender, R.~2009, ApJS, 182, 216 (KFCB)

\nnhi Kormendy, J., \& Freeman, K.~C.~2004, in IAU Symposium 220, Dark Matter
      in Galaxies, ed.~S.~D.~Ryder, D.~J.~Pisano, M.~A.~Walker, \& K.~C.~Freeman
      (San Francisco: ASP), 377

\nnhi Kormendy, J., \& Freeman, K.~C.~2011, ApJ, in preparation

\nnhi Kormendy, J., \& Kennicutt, R.~C.~2004, ARA\&A, 42, 603

\nnhi Kormendy, J., \& Richstone, D.~1992, ApJ, 393; 559

\nnhi Kormendy, J., \etal 1996, ApJ, 459, L57  

\nnhi Kotanyi, C., van Gorkom, J.~H., \& Ekers, R.~D.~1983, ApJ, 273, L7  

\nnhi Krajnovi\'c, D.~2011, paper presented at the ESO Workshop on Fornax, Virgo, Coma et al.: 
      Stellar Systems in High Density Environments, 
      {\eightpoint\tt /sci/meetings/2011/fornax\_virgo2011/talks\_pdf/}
      {\eightpoint\tt Krajnovic\_Davor.pdf}

\nnhi Krajnovi\'c, D., \etal 2008, MNRAS, 390, 93    

\nnhi Krajnovi\'c, D., \etal 2011, MNRAS, 414, 2923  

\nnhi Krick, J.~E., \& Bernstein, R.~A.~2007, AJ, 134, 466  

\nnhi Lake, G., Katz, N., \& Moore, B.~1998, ApJ, 495, 152

\nnhi Larson, R.~B.~1974, MNRAS, 169, 229

\nnhi Larson, R.~B., Tinsley, B.~M., \& Caldwell, C.~N.~1980, ApJ, 237, 692  

\nnhi Lauer, T.~R.~1985, ApJS, 57, 473

\nnhi Laurikainen, E., Salo, H., \& Buta, R. 2005, MNRAS, 362, 1319

\nnhi Laurikainen, E., Salo, H., Buta, R., \& Knapen, J., Speltincx, T., \& Block, D.~2006, AJ, 132, 2634

\nnhi Laurikainen, E., Salo, H., Buta, R., \& Knapen, J. H. 2007, MNRAS, 381, 401

\nnhi Laurikainen, E., Salo, H., Buta, R., \& Knapen, J. H. 2011, MNRAS, in press (arXiv:1110.1996)

\nnhi Laurikainen, E., Salo, H., Buta, R., \& Knapen, J. H., \& Comer\'on, S.~2010, MNRAS, 405, 1089

\nnhi Li, J.-T., Wang, Q.~D., Li, Z., \& Chen, Y.~2011, ApJ, 737, 41

\nnhi Lin, D.~N.~C., \& Faber, S.~M.~1983, ApJ, 266, L21

\nnhi Lisker, T., Brunngr\"aber, R., \& Grebel, E.~K.~2009, AN, 330, 966          

\nnhi Lisker, T., Grebel, E.~K., \& Binggeli, B.~2006, AJ, 132, 497               

\nnhi Lisker, T., Grebel, E.~K., Binggeli, B., \& Glatt, K.~2007, ApJ, 660, 1186  

\nnhi Makarova, L.~1999, A\&AS, 139, 491

\nnhi Makino, J., \& Ebisuzaki, T.~1996, ApJ, 465, 527

\nnhi Martin, N.~F., de Jong, J.~T.~A., \& Rix, H.-W.~2008, ApJ, 684, 1075  

\nnhi Mateo, M.~1998, ARA\&A, 36, 435

\nnhi Mateo, M.~2008, The Messenger, 134, 3

\nnhi Mayer, L., Governato, F., Colpi, M., Moore, B., Quinn, T., Wadsley, J., Stadel, J., 
      \& Lake, G.~2001a, ApJ, 547, L123  

\nnhi Mayer, L., Governato, F., Colpi, M., Moore, B., Quinn, T., Wadsley, J., Stadel, J., 
      \& Lake, G.~2001b, ApJ, 559, 754  

\nnhi Mayer, L., Mastropietro, C., Wadsley, J., Stadel, J., \& Moore, B.~2006, MNRAS, 369, 1021  

\nnhi McConnachie, A.~W., \& Irwin, M.~J.~2006, MNRAS, 365, 1263  

\nnhi McDermid, R.~M., \etal 2006, MNRAS, 373, 906  

\nnhi Meekins, J.~F., Fritz, G., Chubb, T.~A., Friedman, H., \& Henry, R.~C.~1971, Nature, 231, 107

\nnhi Mei, S., \etal 2007, ApJ, 655, 144  

\nnhi M\'endez-Abreu, J., Aguerri, J.~A.~L., Corsini, E.~M., \& Simonneau, E.~2008, A\&A, 478, 353  

\nnhi Merritt, D.~2006, ApJ,, 648, 976

\nnhi Mihos, J.~C.~2011, paper presented at the ESO Workshop on Fornax, Virgo, Coma et al.: Stellar
      Systems in High Density Environments, 
      {\eightpoint\tt /sci/meetings/2011/fornax\_virgo2011/talks\_pdf/}
      {\eightpoint\tt Mihos\_Chris.pdf}

\nnhi Mihos, J.~C., Harding, P., Feldmeier, J., \& Morrison, H.~2005, ApJ, 631, L41

\nnhi Mihos, J.~C., \& Hernquist, L.~1994, ApJ, 437, L47

\nnhi Mihos, J.~C., \etal 2011, in preparation

\nnhi Mihos, J.~C., Janowiecki, S., Feldmeier, J.~J., Harding, P., \& Morrison, H.~2009, ApJ, 698, 1879

\nnhi Milosavljevi\'c, M., \& Merritt, D.~2001, ApJ, ApJ, 563, 34

\nnhi Milosavljevi\'c, M., Merritt, D., Rest, A., \& van den Bosch, F.~C.~2002,
            MNRAS, 331, L51

\nnhi Moore, B., Ghigna, S., Governato, F., Lake, G., Quinn, T., Stadel, J.,
           \& Tozzi, P.~1999, ApJ, 524, L19

\nnhi Moore, B., Katz, N., Lake, G., Dressler, A., \& Oemler, A.~1996, Nature,
             379, 613  

\nnhi Moore, B., Lake, G., \& Katz, N.~1998, ApJ, 495, 139  

\nhi Morgan, W.~W.~1951, Publ.~Obs.~Univ.~of Michigan, 10, 33

\nnhi Morgan, W.~W., \& Lesh, J.~R.~1965, ApJ, 142, 1364

\nnhi Mori, M., \& Burkert, A.~2000, ApJ, 538, 559  

\nnhi Naab, T., Johansson, P.~H., \& Ostriker, J.~P.~2009, ApJ, 699, L178 

\nnhi Neistein, E., Maoz, D., Rix, H.-W., \& Tonry, J.~L.~1999, AJ, 117, 2666  

\nnhi Oemler, A.~1976, ApJ, 209, 693

\nnhi Okamura, S.~2011, paper presented at the ESO Workshop on Fornax, Virgo, Coma et al.: 
      Stellar Systems in High Density Environments,
      {\eightpoint\tt http://www.eso.org/sci/meetings/2011/fornax\_virgo2011/}
      {\eightpoint\tt posters.html}

\nnhi Oser, L., Ostriker, J.~P., Naab, T., Johansson, P.~H., \& Burkert, A.~2010, ApJ, 725, 2312 

\nnhi Oser, L., Naab, T., Ostriker, J.~P., \& Johansson, P.~H.~2011, arXiv:1106.5490

\nnhi Oosterloo, T., \& van Gorkom, J.~H.~2005, A\&A, 437, L19  

\nnhi Paturel, G., Petit, C., Prugniel, P., Theureau, G., Rousseau, J., Brouty, M.,
            Dubois, P., \& Cambr\'esy, L.~2003, A\&A, 412, 45

\nnhi Pedraz, S., Gorgas, J., Cardiel, N., S\'anchez-Bl\'azquez, P., \& Guzm\'an, R.~2002, MNRAS, 332, L59  

\nnhi Peng, C. Y., Ho, L. C., Impey, C. D., \& Rix, H.-W. 2002, AJ, 124, 266

\nnhi Pildis, R.~A., Schombert, J.~M., \& Eder, J.~A.~1997, ApJ, 481, 157

\nnhi Postman, M., \& Geller, M.~J.~1984, ApJ, 281, 95

\nnhi Press, W.~H., Flannery, B.~P., Teukolsky, S.~A., \& Vetterling, W.~T.~1986,
            Numerical Recipes: The Art of Scientific Computing (Cambridge: Cambridge
            University Press)

\nnhi Quilis, V., Moore, B., \& Bower, R.~2000, Science, 288, 1617  

\nnhi Quinlan, G.~D., \& Hernquist, L.~1997, NewA, 2, 533

\nnhi Richstone, D.~O.~1976, ApJ, 204, 642

\nnhi Rix, H.-W., Franx, M., Fisher, D., \& Illingworth, G.~1992, ApJ, 400, L5

\nnhi Robertson, B., Cox, T.~J., Hernquist, L., Franx, M., Hopkins, P.~F., 
            Martini, P., \& Springel, V.~2006, ApJ, 641, 21

\nnhi Roediger, E., \& Hensler, G.~2005, A\&A, 433, 875

\nnhi Rubin, V.~C., Graham, J.~A., \& Kenney, J.~D.~P.~1992, ApJ, 394, L9

\nnhi Ryden, B.~S., \& Terndrup. D.~M.~1994, ApJ, 425, 43

\nnhi Sage, L.~J., \& Wrobel, J.~M.~1989, ApJ, 344, 204

\nnhi Saglia, R.~P, Bender, R., \& Dressler, A.~1993, A\&A, 279, 75

\nnhi Saito, M.~1979, PASJ, 31, 193  

\nhi Sancisi, R. \& van Albada, T.~S.~1987, in IAU Symposium 117,
              Dark Matter in the Universe, ed.~J.~Kormendy \&
              G.~R.~Knapp (Dordrecht: Reidel), 67

\nnhi Sandage, A.~1961, The Hubble Atlas of Galaxies (Washington: Carnegie 
             Institution of Washington)

\nhi Sandage, A.~1975, in Galaxies and the Universe, ed.~A.~Sandage, M.~Sandage,
     \& J.~Kristian \hbox{(Chicago:~University of Chicago Press), 1}

\nhi Sandage, A.~2004, in Penetrating Bars Through Masks of Cosmic Dust: The Hubble 
              Tuning Fork Strikes a New Note, ed.~D.~L.~Block et al.~(Dordrecht: Kluwer), 39

\nnhi Sandage, A., \& Bedke, J.~1994, The Carnegie Atlas of Galaxies
              (Washington, DC: Carnegie Institution of Washington)

\nnhi Sandage, A., \& Binggeli, B.~1984, AJ, 89, 919

\nnhi Sandage, A., Binggeli, B., \& Tammann, G.~A.~1985a, in ESO Workshop on the
             Virgo Cluster, ed. O.-G.~Richter \& B.~Binggeli (Garching: ESO), 239

\nnhi Sandage, A., Binggeli, B., \& Tammann, G.~A.~1985b, AJ, 90, 1759

\nnhi Sarzi, M., \etal 2006, MNRAS, 366, 1151  

\nnhi Schombert, J.~M.~1988, ApJ, 328, 475  

\nnhi Scorza, C., \& Bender, R.~1995, A\&A, 293, 20

\nnhi Scorza, C., Bender, R., Winkelmann, C., Capaccioli, M., \& 
             Macchetto, D.~F.~1998, A\&AS, 131, 265

\nnhi Sellwood, J.~A., \& Wilkinson, A.~1993, Rep. Prog. Phys., 56, 173

\nnhi S\'ersic, J.~L.~1968, Atlas de Galaxias Australes (Cordoba: 
             Observatorio Astronomico, Universidad de Cordoba)

\nnhi Shen, S., Mo, H.~J., White, S.~D.~M., Blanton, M.~R., Kauffmann, G., Voges, W., 
      Brinkmann, J., \& Csabai, I.~2003, MNRAS, 343, 978

\nnhi Shetrone, M.~D., Bolte, M., \& Stetson, P.~B.~1998, AJ, 115, 1888

\nnhi Shetrone, M.~D., C\^ot\'e, P., \& Sargent, W.~L.~W.~2001, ApJ, 548, 592

\nnhi Shetrone, M.~D., Siegel, M.~H., Cook, D.~O., \& Bosler, T.~2009, AJ, 137, 62

\nnhi Shetrone, M.~D., Venn, K.~A., Tolstoy, E., Primas, F., Hill, V., \& Kaufer, A.~2003,
      AJ, 125, 684

\nnhi Simard, L., et al. 2002, ApJS, 142, 1

\nnhi Simien, F., \& Prugniel, Ph.~2002, A\&A, 384, 371

\nnhi Simien, F., \& de Vaucouleurs, G.~1986, ApJ, 302, 564

\nnhi Somerville, R.~S., \& Primack, J.~R.~1999, MNRAS, 310, 1087

\nnhi Stover, R.~J.~1988, in Instrumentation for Ground-Based Optical Astronomy:
            Present and Future, ed.~L.~B.~Robinson (New York: Springer-Verlag), 443

\nnhi Strateva, I., \etal 2001, AJ, 122, 1861 

\nnhi Temi, P., Brighenti, F., \& Matthews, W.~G.~2009, ApJ, 695, 1

\nnhi Thomas, D., Brimioulle, F., Bender, R., Hopp, U., Greggio, L., Maraston, C., \& 
            Saglia, R.~P.~2006, A\&A, 445, L19

\nnhi Thronson, H.~A., Tacconi, L., Kenney, J., Greenhouse, M.~A., Margulis, M., 
      Tacconi-Garman, L., \& Young, J.~S.~1989, ApJ, 344, 747

\nnhi Thuan, T.~X., \& Kormendy, J.~1977, PASP, 89, 466

\nnhi Tolstoy, E., Hill, V., \& Tosi, M.~2009, ARA\&A, 47, 371

\nnhi Tolstoy, E., Venn, K.~A., Shetrone, M., Primas, F., Hill, V., Kaufer, A., \&
      Szeifert, T.~2003, AJ, 125, 707

\nnhi Tonnesen S., \& Bryan G. L., 2008, ApJ, 684, L9

\nnhi Tonnesen S., \& Bryan G. L., 2009, ApJ, 694, 789

\nnhi Tonnesen S., \& Bryan G. L., 2010, ApJ, 709, 1203

\nnhi Tonnesen S., Bryan G. L., \& van Gorkom, J.~H.~2007, ApJ, 671, 1434

\nnhi Tremaine, S.~1981, in The Structure and Evolution of Normal Galaxies, 
            ed.~S.~M.~Fall \& D.~Lynden-Bell (Cambridge: Cambridge University Press), 67

\nnhi Tully, R.~B., \& Fisher, J.~R.~1977, A\&A, 54, 661

\nnhi van Albada, T.~S.~1982, MNRAS, 201, 939

\nnhi van Albada, T.~S. \& Sancisi, R.~1986, Phil.~Trans.~R.~Soc.~London A, 320, 447

\nnhi van den Bergh, S. 1976, ApJ, 206, 883  

\nnhi van den Bergh, S. 1977. in Nearly Nearly Normal Galaxies: From the Planck Time
                              to the Present, ed.~S.~M.~Faber (New York: Springer), 19

\nnhi van den Bergh, S.~1993, ApJ, 411, 178  

\nnhi van den Bergh, S.~1994a, AJ, 107, 153

\nnhi van den Bergh, S.~1994b, AJ, 107, 1328

\nnhi van den Bergh, S.~1994c, ApJ, 428, 617

\nnhi van den Bergh, S. 2007, The Galaxies of the Local Group (Cambridge: Cambridge University Press)

\nnhi van den Bergh, S. 2009a, ApJ, 694, L120  

\nnhi van den Bergh, S. 2009b, ApJ, 702, 1502 

\nnhi van den Bosch, R.~C.~E.~, van de Ven, G., Verolme, E.~K., Cappellari, M.,
            and de Zeeuw, P.~T.~2008, MNRAS, 385, 647

\nnhi van der Kruit, P.~C., \& Freeman, K.~C.~2011, ARA\&A, 49, 301

\nnhi van Dokkum, P.~G., \etal 2010, ApJ, 709, 1018 

\nnhi van Driel, W., \& van Woerden, H.~1991, A\&A, 243, 71

\nnhi van Gorkom, J.~H., \& Kenney, J.~D.~P.~2011, ARA\&A, in preparation

\nnhi van Woerden, H., van Driel, W., \& Schwarz, U.~J.~1983, in IAU Symposium 100,
      Internal Kinematics and Dynamics of Galaxies, ed.~E.~Athanassoula (Dordrecht: Reidel), 99

\nnhi van Zee, L., Skillman, E.~D., \& Haynes, M.~P.~2004, AJ, 128, 121  

\nnhi Veilleux, S., Bland-Hawthorn, J., Cecil, G., Tully, R.~B., \& Miller, S.~T.~1999, ApJ, 520, 111

\nnhi Venn, K.~A., \& Hill, V.~M.~2008, The Messenger, 134, 23

\nnhi Venn, K.~A., Irwin, M., Shetrone, M.~D., Tout, C.~A., \& Hill, V.~M., 
      \& Tolstoy, E.~2004, AJ, 128, 1177

\nnhi Vikhlinin, A., Kravtsov, A., Forman, W., Jones, C., Markevich, M., Murray, S.~S.,
      \& van Speybroeck, L.~2006, ApJ 640, 691  

\nnhi Vollmer, B., Braine, J., Combes, F., \& Sofue, Y.~2005, A\&A, 441, 473  

\nnhi Vollmer, B., Soida, M., Chung, A., van Gorkom, J.~H., Otmianowska-Mazur, K., Beck, R., 
      Urbanik, M., \& Kenney, J.~D.~P.~2008, A\&A, 483, 89  

\nnhi Wakamatsu, K.-I., \& Hamabe, M. 1984, ApJS, 56, 283

\nnhi Watt, M.~P., Ponman, T.~J., Bertram, D., Eyles, C.~J., Skinner, G.~K., \& 
      Willmore, A.~P.~1992, MNRAS, 258, 738  

\nnhi Weinzirl, T., Jogee, S., Khochfar, S., Burkert, A., \& Kormendy, J.~2009, ApJ, 696, 411

\nnhi Weisz, D.~R., \etal 2011a, ApJ, 739, 5

\nnhi Weisz, D.~R., \etal 2011b, ApJ (arXiv:1101.1301)

\nnhi Welch, G.~A., \& Sage, L.~J.~2003, ApJ, 584, 260

\nnhi Wilman, D.~J., Balogh, M.~L., Bower, R.~G., Mulchaey, J.~S., Oemler, A., Carlberg, R.~G.,
      Morris, S.~L., \& Whitaker, R.~J.~2005, MNRAS, 358, 71

\nnhi Wilman, D.~J., Oemler, A., Mulchaey, J.~S., McGee, S.~L., Balogh, M.~L., \& 
      Bower, R.~G.~2009, ApJ, 692, 298

\nnhi Williams, B.~F., \etal 2007a, ApJ, 654, 835  

\nnhi Williams, B.~F., \etal 2007b, ApJ, 656, 756  

\nnhi Williams, M.~J., Bureau, M., \& Cappellari, M.~2010, MNAS, 409, 1330  

\nnhi Wirth, A., \& Gallagher, J.~S.~1984, ApJ, 282, 85

\nnhi Yoshida, M., \etal 2004, AJ, 127, 90  

\nnhi Young, J.~S., \etal 1995, ApJS, 98, 219

\nnhi Young, L.~M., \etal 2011, MNRAS, 414, 940  

}

\vfill \eject\end